\documentclass[a4paper,11pt,oneside]{article}

\usepackage[margin=1in]{geometry}
\usepackage{physics}
\usepackage{enumerate}
\usepackage{empheq}
\usepackage{booktabs}
\usepackage{amsmath,amssymb,amsbsy,amstext, amsthm, simplewick}
\usepackage{hyperref}
\usepackage{graphicx}
\usepackage{amsfonts}
\usepackage{color}
\usepackage{wasysym}
\usepackage{tikz}
\usepackage{upgreek}
\usepackage{floatrow}
\usepackage{subfig}
\usepackage{xcolor}
\usepackage{dsfont}
\usepackage{tikz}
\usetikzlibrary{decorations.markings}

\usepackage{colortbl}
\definecolor{summersky}{cmyk}{0.71,0.33,0,0.5}
\definecolor{flamingo}{cmyk}{0,0.51,0.71,0.5}
\definecolor{rp}{cmyk}{0.2, 1, 0.6, 0}
\definecolor{pacificblue}{cmyk}{0.95,0.3,0, 0.5}
\definecolor{gray60}{cmyk}{0.4,0.4,0,0.8}

\hypersetup{
    pdftoolbar=true,        
    pdfmenubar=true,        
    pdffitwindow=true,     
    pdfstartview={FitH},    
    pdfnewwindow=true,      
    colorlinks=true,       
    linkcolor=pacificblue,          
    citecolor=flamingo,        
    filecolor=magenta,      
    urlcolor=pacificblue           
}

\theoremstyle{definition}

\numberwithin{equation}{section}

\newcommand{\ex}[1]{\langle #1 \rangle}
\newcommand{\expi}[1]{\langle \! \langle #1 \rangle \! \rangle}
\newcommand{\be}{\begin{eqnarray} }
\newcommand{\ee}{\end{eqnarray} }
\newcommand{\bs}{\begin{split} }
\newcommand{\es}{\end{split} }

\newcommand{\R}{\mathcal{R}}
\renewcommand{\L}{\mathcal{L}}
\renewcommand{\H}{\mathcal{H}}

\newcommand{\then}{\quad \Rightarrow\quad}
\renewcommand{\O}{\mathcal{O}}

\newcommand{\bfx}{\mathbf{x}}

\newcommand{\fpi}{f_{\pi}}
\newcommand{\pir}{\pi_{r}}
\newcommand{\pia}{\pi_{a}}
\newcommand{\dpir}{\dot{\pi}_{r}}
\newcommand{\dpia}{\dot{\pi}_{a}}

\begin{document}

\begin{titlepage}
\thispagestyle{empty}

\vspace*{2.0cm}

\begin{flushright}
{\scshape Cambridge, 2026}
\end{flushright}

\vspace{3.0cm}



{\sffamily\LARGE\bfseries Lectures on Open Systems and Cosmology\par}

\vspace{3mm}

{\large Master Equation, Schwinger--Keldysh and Open Effective Field Theories\par}

\vspace{8mm}

\noindent\rule{\textwidth}{0.8pt}

\vspace{8mm}

\noindent
{\sffamily\large Enrico Pajer}

\vspace{3mm}

\noindent
{\itshape
Department of Applied Mathematics and Theoretical Physics,\\
University of Cambridge, Wilberforce Road, Cambridge, CB3 0WA, UK
}

\vspace{5mm}

\noindent
{\itshape E-mail:} \href{mailto:enrico.pajer@gmail.com}{enrico.pajer@gmail.com}

\vspace{1cm}

\noindent
{\scshape Abstract:}
Open systems are ubiquitous in physics. Many realistic systems interact, at least weakly, with environmental degrees of freedom that may be too numerous, too complicated, inaccessible, or unknown. When only a subset of degrees of freedom is observed, its reduced dynamics can differ qualitatively from that of a closed system, displaying dissipation, noise, decoherence, memory effects, or loss of information into unobserved sectors. When the microscopic description is also unknown, one is led to an open effective description, in which the relevant degrees of freedom are treated systematically while the environment and the microscopics are parametrized rather than solved for explicitly.

This perspective is especially important in gravity and cosmology. The main open problems of cosmology, including inflation, dark matter, and dark energy, involve spacetime-filling sectors whose microscopic nature is unknown and whose observed effects are primarily gravitational. At the same time, gravitational systems often lack a preferred notion of conserved energy because they are time dependent, and naturally display out-of-equilibrium dynamics. These lecture notes introduce the operator formalism and the Schwinger--Keldysh path-integral as tools to study open systems, with emphasis on open effective field theories and inflation. They are aimed at master students, PhD students, and researchers approaching these topics for the first time.

\vfill

\end{titlepage}

\tableofcontents

 
 
\section*{Introduction}

The theory of open systems should be part of the standard toolkit of every physicist. No realistic system is perfectly isolated: every system interacts, at least weakly, with an environment whose degrees of freedom are either too numerous, too complicated, inaccessible, or simply unknown. This is such a general situation that open-system techniques have applications across essentially all branches of physics, from quantum mechanics and quantum information to condensed matter, statistical physics, particle physics, cosmology, and general relativity. The goal of these notes is to introduce some of the basic tools that allow one to describe the dynamics of a subsystem without having to keep track of the full microscopic state of everything it interacts with.

\paragraph{An open-system approach to gravity and cosmology.} Open systems are especially natural in \textit{cosmology}. The main open problems of cosmology can be roughly summarized by the questions of inflation, dark matter, and dark energy. These problems, which have been studied intensely for decades, share some striking features. 
First, they all involve spacetime-filling components, sectors, or effective media. Inflation is the name we give to the physics that filled the universe before the hot big bang and drove an early period of accelerated expansion, while dark energy denotes the component responsible for the present accelerated expansion. Dark matter, although clustered rather than completely homogeneous at late times, is also a spacetime-filling component whose presence is inferred almost entirely through its gravitational effects. Because of this, these problems are not naturally approached as ordinary particle-physics problems, where one imagines particles propagating on a vacuum that can be prepared and studied in isolation. They are closer in spirit to condensed matter problems, where the relevant (in this case gravitational) dynamics takes place inside a medium.

Second, in all these cases the microscopic physics is unknown. This is not merely the familiar situation of QCD or of complex materials, where the microscopic laws may be known but the resulting dynamics is too hard to solve in practice. In cosmology we often do not know the microscopic degrees of freedom at all. Despite the many models of inflation, dark matter, and dark energy, we do not know what these substances really are. This almost forces us to adopt an effective-field-theory point of view: we identify the relevant degrees of freedom and the symmetries that constrain them, while remaining as agnostic as possible about the detailed microphysics. 

Third, as far as present observations tell us, these sectors interact with the visible world mostly, and perhaps only, through gravity. We suspect, for example, that the inflaton must eventually interact with the Standard Model in order for reheating to occur, but we do not have direct observational access to this interaction. Similarly, despite enormous experimental effort, we do not yet have positive evidence for non-gravitational interactions between dark matter and Standard Model particles. In this sense cosmology is naturally an open-system problem: the gravitational degrees of freedom, namely the metric itself, are the accessible system, while the unknown ingredients that fill the universe act as an environment. More generally, the precise split between system and environment may depend on the question: the system may be the metric, the curvature perturbation, a set of long-wavelength modes, or any other accessible set of degrees of freedom. The resulting description is therefore an \textit{open effective field theory}: it is effective because the microscopic structure of the environment is unknown, and it is open because the observed gravitational degrees of freedom need not evolve unitarily once the environmental degrees of freedom are ignored.

Open-system ideas are also deeply relevant to general relativity more broadly. Gravity interacts with everything, and a truly closed-system description would in principle require us to describe all degrees of freedom in the universe, which is neither feasible nor useful. Moreover, once matter is introduced, gravity typically reacts by producing time-dependent spacetimes. In such backgrounds there need not be a timelike Killing vector, and hence no conserved energy associated with time translations. This is a serious obstruction to the most familiar Wilsonian intuition, where heavy degrees of freedom can be integrated out because energy conservation prevents them from being efficiently populated at low energies. In a time-dependent gravitational background this argument can fail: heavy fields may be excited by the background evolution and then act as environmental degrees of freedom for the light sector. General relativity also contains horizons, which introduce further obstructions, sometimes even in principle, to accessing parts of the system. Although these notes will not focus on horizons, they provide an important additional motivation for developing open quantum system techniques in gravitational physics.\\

A closely related theme is that open systems are generically \textit{out of equilibrium}. This is essential in cosmology, where the time dependence of the background is not a small perturbation but the phenomenon one is trying to describe. Equilibrium thermodynamics has received an enormous amount of attention, and in many contexts it provides the correct language. However, the cosmological situations that motivate these notes are usually not equilibrium situations. For this reason we will avoid specializing to thermal equilibrium unless explicitly stated, and will instead keep the formalism flexible enough to describe general out-of-equilibrium processes. This point is also important in gravitational physics. When a spacetime has a timelike Killing vector, or enough symmetry to define a suitable analytic continuation, one can often Wick rotate to a Euclidean geometry and extract a thermal interpretation, as in the cases of black holes and of suitable de Sitter patches. But many realistic gravitational and cosmological questions do not have this degree of symmetry. They should be treated as Lorentzian, out-of-equilibrium dynamical problems. After all, our universe is very far from equilibrium, as the formation of structure throughout cosmic history makes manifest.

\paragraph{Aims and goals.} The main purpose of these lecture notes is to bring together three subjects that are often discussed separately: open quantum systems in the operator formalism, the Schwinger--Keldysh path integral, and cosmology. The operator formalism and the quantum master equation provide a natural language for many discussions of open quantum systems, especially in quantum information and quantum optics. The path integral formalism is, in principle, equivalent, but it is often used in different communities and is indispensable in quantum field theory, gauge theory, and gravity. Cosmology adds a further layer of structure, because it combines quantum fields, time-dependent backgrounds, gravitational constraints, and observational questions about primordial fluctuations. These three subjects have much to teach each other. Progress can be made by importing results and intuitions from one area into another, but doing so is difficult because the necessary material is rarely presented in a unified notation and at a common pedagogical level.\\

The \textit{cross-sectional perspective} taken in these notes is especially valuable in the present era. The rapid development of artificial intelligence is changing the way researchers learn and work. It is becoming increasingly realistic for physicists to cross disciplinary boundaries, acquire unfamiliar tools, and test ideas outside their original area of expertise. With recently available AI tools, researchers have increasingly powerful assistance in navigating unfamiliar literatures, checking calculations, and acquiring the language of neighbouring fields. This should encourage us to be less afraid of interdisciplinary research and bolder in tackling ambitious questions. Rather than proceeding along increasingly specialized tracks that rarely communicate with each other, we can try to identify which tools are genuinely needed for a problem and then learn them efficiently. These notes are written in that spirit. They do not aim to be exhaustive, but to provide a bridge between communities and to make it easier for students and researchers to move between operator methods, path integrals, and cosmological applications.\\

The notes are aimed at master students, junior PhD students, and researchers approaching open quantum systems, the Schwinger--Keldysh formalism, or out-of-equilibrium cosmology for the first time. For this reason they include some introductory material, such as the basic theory of density matrices and elements of standard cosmology, which many readers may already have encountered elsewhere. The purpose of including this material is not only to make the notes self-contained, but also to establish notation and conventions before moving to more advanced topics. The emphasis is on building a practical toolkit: the main definitions, equations, examples, and conceptual structures that are needed to enter the literature. More specialized and research-level developments are mostly left to the references.

These notes were prepared for lectures given at three schools in 2026: the \href{https://scuolagalileiana.unipd.it/wp-content/uploads/2026/03/Scuola-tematica-fisica-DEF.pdf}{Scuola Tematica} on ``Theoretical Physics of the Fundamental Interactions'' at the Scuola Galileiana di Studi Superiori in Padova, Italy; the \href{https://indico.math.cnrs.fr/event/15597/}{2026 IHES summer school} on ``Cosmological Correlators''; and \href{https://indico.global/event/16455/overview}{La Ricotta Summer School} on ``The Disordered Universe'' in As Barreiras, Castro Caldelas, Spain. Their intended use is therefore pedagogical. The hope is that they provide a robust foundation from which interested readers can approach the modern literature on open quantum systems, Schwinger--Keldysh effective actions, and cosmological applications, and eventually use these tools in their own research.


\paragraph{Structure of the notes.} Section \ref{sec:1} reviews the quantum mechanics of open systems, including density matrices, composite systems, partial traces, entanglement and decoherence. Section \ref{sec:2} introduces reduced dynamics, quantum channels, dynamical maps and the GKSL master equation, including its microscopic origin and its extension to time-dependent maps and CP divisibility. Section \ref{sec:3} discusses explicit solutions of the Lindblad equation, with examples involving qubits and the dissipative harmonic oscillator.

Section \ref{sec:4} develops the Schwinger--Keldysh formalism, from the doubled time contour and the Keldysh basis to the Feynman--Vernon influence functional and its general consistency conditions. Section \ref{sec:SKmap} relates local Schwinger--Keldysh actions to dynamical maps and complete positivity, and discusses jump operators, Langevin dynamics and dissipative scalar fields. Section \ref{sec:5} reviews the cosmological background, inflationary correlators and primordial non-Gaussianity. Section \ref{sec:6} constructs the open effective field theory of inflation and applies it to the power spectrum and bispectrum. Section \ref{sec:stochastic} introduces stochastic inflation and uses the Langevin and Fokker--Planck descriptions to study secular growth and late-time behaviour in de Sitter. Appendix A collects useful results on the inversion of quadratic Schwinger--Keldysh kernels.


\subsection*{References and resources}

Here I summarise references that I used to compile these notes and highlight a selected list of useful resources among the many available in the literature.

Introductory material on the density matrix can be found in almost any book on quantum mechanics. Section \ref{sec:1} follows the course Principles of Quantum Mechanics of the Cambridge Maths Tripos and the nice lecture notes by David Skinner \cite{Skinner}.

A classic and comprehensive textbook on open quantum systems is Breuer and Petruccione \cite{breuerTheoryOpenQuantum2002}. The presentation in Section \ref{sec:2} and Section \ref{sec:3} partly follows this reference. The focus there is more on the operator formalism and stochastic methods, but it contains a vast array of topics and applications. 

A great reference on the Schwinger-Keldysh path integral for open systems is Alex Kamenev's book \cite{kamenev_2011}. This textbook builds from the ground up and presents a large number of applications to quantum many-body systems. Since the language is that of quantum many-body physics, it may take a little while getting used to for people with a high-energy background. Parts of Section \ref{sec:4} were based on this reference.

The review of cosmology and inflation are based on my own lecture notes for the Cosmology \cite{PajerCosmologyNotes} and Field Theory in Cosmology \cite{PajerFieldTheoryCosmologyNotes} course in Part III of the Cambridge Maths Tripos. A nice reference for the perturbative expansion of the Schwinger-Keldysh path integral in closed systems is \cite{Chen:2017ryl}. Section \ref{sec:6} is mostly based on the original papers \cite{Creminelli:2006xe,Cheung:2007st,LopezNacir:2011kk,Salcedo:2024smn,Salcedo:2026sdn}.

For the discussion of light fields in de Sitter and stochastic inflation, there are many very good reviews \cite{Woodard:2025cez,Cruces:2022imf,Vennin:2020kng}, which moreover discuss much more advanced aspects compared to the elementary discussions presented in these lectures. There are hundreds of papers on stochastic inflation and on improving the leading order description to capture systematically all quantum effects. We don't discuss any of these issues in detail and we refer the reader to the above reviews and references therein. \\

There are many good reviews of open quantum systems, which discuss different aspects of the problem from various perspectives. A complementary field-theoretic treatment, connecting influence functionals, noise, decoherence and cosmology, is given by Calzetta and Hu
\cite{Calzetta:2008iqa}. Modern Schwinger--Keldysh effective theories and non-Markovian quantum dynamics are reviewed respectively in
\cite{Liu:2018kfw,RivasHuelgaPlenio2014}. The pioneering work of Hong Liu and Paolo Glorioso has been particularly influential for my understand of this subject. An excellent very recent review by T.~Colas \cite{Colas:2025app} discusses the use of open quantum systems and the Schwinger-Keldysh path integral for inflation and gravity in the late universe. It emphasises the construction of open effective field theories and covers some of the latest developments in the field. Another excellent reference is the review by L.~M.~Sieberer, M.~Buchhold and S.~Diehl \cite{2016RPPh...79i6001S}, which focuses on applications to driven open quantum systems and, in particular, cold atoms and optical cavities.


\section{The quantum mechanics of open systems}\label{sec:1}

The standard formulation of quantum mechanics often begins with a closed system described by a state vector evolving unitarily in an isolated Hilbert space. In many physical situations, however, this description is either impractical or incomplete. We may only have probabilistic information about the microscopic state, we may be interested in a subsystem of a larger composite system, or the degrees of freedom we observe may interact with an environment that is inaccessible or too complicated to model in detail. These situations motivate the density matrix formalism and, more generally, the framework of open quantum systems. In this framework, the basic object is no longer necessarily a pure state, but a density matrix, which allows one to describe pure states, statistical mixtures, and later on reduced states obtained by tracing over unobserved degrees of freedom.

The goal of this section is to introduce the basic quantum-mechanical tools needed for the rest of these notes. We begin with the density matrix and its interpretation, then discuss simple examples such as the qubit and the Bloch vector, before turning to composite systems, tensor products, partial traces, entanglement, and decoherence. These notions are not only conceptually important; they also provide the natural language in which open-system dynamics is formulated. In particular, they clarify how a subsystem can evolve from a pure state into a mixed state even when the full system remains in a pure state and evolves unitarily. Many readers may already be familiar with this material and may want to skip directly to the next section. 

 
 \subsection{The density matrix}

In many physical situations, the description of a quantum system in terms of a single state vector is either inconvenient or insufficient. This may happen because the system is prepared only probabilistically in different states, because we do not know its exact microscopic state, or because later on we will want to describe subsystems of a larger quantum system. These considerations motivate the introduction of the density operator, also called the density matrix.

Suppose we have only classical probabilistic information about the quantum state of the system. This additional uncertainty comes on top of the standard probabilistic nature of quantum mechanics. Hence, let's imagine the system is in the normalized state $|\psi_\alpha\rangle \in \mathcal H$ with \textit{classical} probability $p_\alpha$, where
\begin{equation}
0 \le p_\alpha \le 1,
\qquad
\sum_\alpha p_\alpha = 1,
\qquad
\langle \psi_\alpha | \psi_\alpha \rangle = 1 .
\end{equation}
Note that the states $|\psi_\alpha\rangle$ need not be orthogonal. It is important to stress that we are \textit{not} saying that the system is in a linear superposition of states $ |\psi_\alpha\rangle $, since that would be simply another state in the Hilbert space. 

How do we extract observables in this case? If $Q$ is a linear Hermitian operator on the Hilbert space, corresponding to a physical observable, its expectation value is then
\begin{equation}
\langle Q \rangle
=
\sum_\alpha p_\alpha \, \langle \psi_\alpha | Q | \psi_\alpha \rangle .
\end{equation}
It is convenient to package the information we have about the state of the system into a single quantity. It turns out that the following operator does the job
\begin{equation}
\rho \equiv \sum_\alpha p_\alpha \, |\psi_\alpha\rangle \langle \psi_\alpha | ,
\end{equation}
We call $ \rho$ the density matrix\footnote{An equivalent name is density operator. Unfortunately, both terminologies have shortcomings. The term ``matrix'' suggests a finite-dimensional Hilbert space; however, we will use it interchangeably with ``linear operator'', also in the case of infinite-dimensional Hilbert spaces. On the other hand, the term ``operators'' is reminiscent of physical observables, while here $ \rho$ represents the \textit{state} of the system.}. 

The advantage of this definition is that expectation values can be written in the compact form
\begin{equation}\label{traceQ}
\langle Q \rangle = \Tr_{\mathcal H}(\rho Q) \,,
\end{equation}
where $ \Tr$ is the trace operation. For finite-dimensional Hilbert spaces, the trace is very familiar; however, it may become subtle in infinite-dimensional spaces. More formally, a trace is a map $\Tr$ defined on a suitable linear space of operators $\mathcal A$ such that, for all $A,B\in\mathcal A$ and all complex numbers $a,b$, it satisfies linearity,
\begin{equation}
\Tr(aA+bB)=a\,\Tr(A)+b\,\Tr(B) ,
\end{equation}
cyclicity,
\begin{equation}
\Tr(AB)=\Tr(BA) ,
\end{equation}
whenever both products lie in the domain of $\Tr$, and positivity\footnote{Often one also requires faithfulness, namely that $\Tr(A^\dagger A)=0$ implies $A=0$.},
\begin{equation}
\Tr(A^\dagger A)\ge 0 .
\end{equation}
In infinite dimensions the trace is in general not defined on all bounded operators, but only on a suitable subclass, such as the ``trace-class'' operators. We will not concern ourselves with these issues in this note, and we will always assume the trace is well defined. 

To verify that \eqref{traceQ} computes the expectation value of $ Q$, let $\{|q_n\rangle\}$ be an orthonormal basis of eigenstates of $Q$, so that
\begin{equation}
Q = \sum_n q_n |q_n\rangle\langle q_n| .
\end{equation}
Then
\begin{align}
\Tr_{\mathcal H}(\rho Q)
&= \sum_n \langle q_n | \rho Q | q_n \rangle \\
&= \sum_{n,\alpha} p_\alpha \, \langle q_n | \psi_\alpha \rangle \langle \psi_\alpha | Q | q_n \rangle \\
&= \sum_{n,\alpha} p_\alpha \, q_n \, |\langle q_n | \psi_\alpha \rangle|^2 \\
&= \sum_\alpha p_\alpha \, \langle \psi_\alpha | Q | \psi_\alpha \rangle = \langle Q \rangle .
\end{align}
Thus the density matrix contains all the information needed to compute the expectation value of any observable.

A particularly important special case is when the system is described by a single normalized state $|\psi\rangle$. In that case,
\begin{equation}
\rho = |\psi\rangle\langle\psi| ,
\end{equation}
and the trace formula reduces to the usual expression
\begin{equation}
\Tr(\rho Q) = \langle \psi | Q | \psi \rangle .
\end{equation}

We now summarize the general properties of a density matrix. A density operator $\rho : \mathcal H \to \mathcal H$ must satisfy\footnote{Sometimes positive semi-definiteness is indicated with the shorthand notation $ \rho \geq 0$.}
\begin{align}
\rho^\dagger &= \rho , \\
\langle \phi | \rho | \phi \rangle &\ge 0
\qquad \text{for all } |\phi\rangle \in \mathcal H , \\
\Tr \rho &= 1 .
\end{align}
These conditions mean, respectively, that $\rho$ is Hermitian, positive semidefinite, and normalized to unit trace. 

Conversely, any operator satisfying these three properties is a legitimate density matrix. Indeed, assuming finite dimension for simplicity, since $\rho$ is Hermitian, it admits a spectral decomposition
\begin{equation}
\rho = \sum_n p_n |n\rangle\langle n| ,
\end{equation}
where the eigenvalues $p_n$ are real. Positivity implies
\begin{equation}
p_n \ge 0 ,
\end{equation}
while the trace condition gives
\begin{equation}
\sum_n p_n = 1 .
\end{equation}
Therefore $\rho$ can always be interpreted as a probabilistic mixture of orthonormal states $|n\rangle$.

It is also useful to note that the matrix elements of $\rho$, namely
\begin{align}
\rho_{nm}:=\langle n | \rho | m\rangle\,,
\end{align} 
are bounded in any orthonormal basis. The diagonal entries satisfy
\begin{equation}
\rho_{nn} \ge 0,
\qquad
\sum_n \rho_{nn} = 1,
\qquad
\Rightarrow
\qquad
\rho_{nn} \le 1 .
\end{equation}
Moreover, positivity implies the bound
\begin{align}
|\rho_{nm}|
&:= |\langle n|\rho|m\rangle|
\nonumber\\
&= |\langle n|A^\dagger A|m\rangle|
\qquad
\text{with } \rho \ge 0 \;\Rightarrow\; \rho = A^\dagger A
\nonumber\\
&= |\langle An|Am\rangle|
\nonumber\\
&\le \sqrt{\langle An|An\rangle}\,\sqrt{\langle Am|Am\rangle}
\qquad \text{(Cauchy--Schwarz)}
\nonumber\\
&= \sqrt{\langle n|A^\dagger A|n\rangle}\,\sqrt{\langle m|A^\dagger A|m\rangle}
\nonumber\\
&= \sqrt{\langle n|\rho|n\rangle}\,\sqrt{\langle m|\rho|m\rangle}
\nonumber\\
&= \sqrt{\rho_{nn}\rho_{mm}} .
\end{align}
and therefore $  |\rho_{nm}| \le \frac12  $ for $ n\neq m$.

\paragraph{Pure vs mixed states} We now turn to the distinction between pure and mixed states. A state is called \textit{pure} if there exists a normalized vector $|\chi\rangle \in \mathcal H$ such that
\begin{equation}
\rho = |\chi\rangle\langle\chi| .
\end{equation}
Otherwise the state is called \textit{mixed}.

This definition admits several equivalent characterizations. First, if $\rho = |\psi\rangle\langle\psi|$, then
\begin{equation}
\rho^2
=
|\psi\rangle\langle\psi|\psi\rangle\langle\psi|
=
|\psi\rangle\langle\psi|
=
\rho .
\end{equation}
So every pure state satisfies $ \rho^2 = \rho   $. Conversely, suppose that
\begin{equation}
\rho^2 = \rho .
\end{equation}
In the eigenbasis of $\rho$,
\begin{equation}
\rho |n\rangle = p_n |n\rangle ,
\end{equation}
so the condition $\rho^2=\rho$ implies
\begin{equation}
p_n^2 = p_n .
\end{equation}
Hence each eigenvalue is either $0$ or $1$. Since $\Tr \rho = 1$, exactly one eigenvalue is equal to $1$, while all the others vanish. Therefore
\begin{equation}
\rho = |\chi\rangle\langle\chi|
\end{equation}
for some normalized state $|\chi\rangle$, and the state is pure. We have therefore shown that
\begin{equation}
\rho \text{ pure}
\qquad \Longleftrightarrow \qquad
\rho^2 = \rho .
\end{equation}
This has the following trivial corollary: for integer $ n\geq 2$, one has
\begin{equation}
\rho \text{ pure}
\qquad \Longleftrightarrow \qquad
\rho^n = \rho \,.
\end{equation}
A further equivalent condition is
\begin{equation}
\Tr \rho^2 = 1 .
\end{equation}
Indeed, if $\rho$ is pure, then $\rho^2=\rho$, and so
\begin{equation}
\Tr \rho^2 = \Tr \rho = 1 .
\end{equation}
Conversely, in the diagonal basis,
\begin{equation}
\Tr \rho^2 = \sum_n p_n^2 .
\end{equation}
Since $p_n \ge 0$ and $\sum_n p_n = 1$, one has
\begin{equation}
\sum_n p_n^2 \le 1 ,
\end{equation}
with equality if and only if one eigenvalue equals $1$ and all the others vanish. Thus
\begin{equation}
\Tr \rho^2 = 1
\qquad \Longleftrightarrow \qquad
\rho \text{ is pure}.
\end{equation}

More generally, for a mixed state one has
\begin{equation}
0 < \Tr \rho^2 < 1 .
\end{equation}

We may summarize the equivalent criteria for purity as
\begin{align}
\rho \text{ is pure}
&\Longleftrightarrow
\rho = |\chi\rangle\langle\chi| \text{ for some normalized } |\chi\rangle \nonumber\\
&\Longleftrightarrow
\rho^2 = \rho \nonumber\\
&\Longleftrightarrow
\Tr \rho^2 = 1 .
\end{align}
If these conditions fail, the state is mixed. Density matrices will be the natural starting point for our discussion of open quantum systems.


\paragraph{Von Neumann entropy} Given a density matrix $\rho$, we define its von Neumann entropy by
\begin{equation}
S(\rho) := - \Tr(\rho \ln \rho) \,.
\end{equation}
This apparently arbitrary operation is rooted in the classical \textit{information theory} and closely related to Shannon entropy. The von Neumann entropy of a given density matrix captures essential properties of the corresponding quantum state. While it is by no means the only notion of entropy (in fact, we will encounter a few more in the next subsection), it plays somewhat of a prominent role in quantum information theory and the study of open systems.

Since $ \rho$ is a Hermitian operator, it can be diagonalized. If $\rho$ has spectral decomposition
\begin{equation}
\rho = \sum_n p_n |n\rangle\langle n| \,,
\qquad
p_n \ge 0 \,,
\qquad
\sum_n p_n = 1 \,,
\end{equation}
then, since $\ln \rho = \sum_n (\ln p_n) |n\rangle\langle n|$ on the support of $\rho$, the entropy becomes
\begin{equation}\label{Sdiag}
S(\rho) = - \sum_n p_n \ln p_n \,,
\end{equation}
with the standard convention that $x \ln x \to 0$ as $x \to 0^+$. Thus the von Neumann entropy depends only on the eigenvalues of $\rho$ and is independent of the basis in which it is calculated. The form \eqref{Sdiag} makes it explicit that the von Neumann entropy may be regarded as the quantum analogue of the Shannon entropy.

Since the eigenvalues of a density matrix satisfy $0 \le p_n \le 1$, one has $\ln p_n \le 0$, and therefore
\begin{equation}
S(\rho) \ge 0 \,.
\end{equation}
Moreover,
\begin{equation}
S(\rho)=0
\qquad \Longleftrightarrow \qquad
\rho \text{ is pure}.
\end{equation}
Indeed, if $\rho$ is pure, then its eigenvalues are $1,0,0,\dots$, so
\begin{equation}
S(\rho) = - 1 \cdot \ln 1 = 0 \,.
\end{equation}
Conversely, if $S(\rho)=0$, then
\begin{equation}
0 = - \sum_n p_n \ln p_n \,.
\end{equation}
Each term in the sum is non-negative, since $-x \ln x \ge 0$ for $0\le x \le 1$. Hence every term must vanish. But $-p_n \ln p_n =0$ only for $p_n=0$ or $p_n=1$. Since the eigenvalues sum to $1$, exactly one eigenvalue must be equal to $1$ and all the others must vanish. Therefore $\rho$ is pure.

If the Hilbert space has finite dimension
\begin{equation}
d := \dim \mathcal H < \infty \,,
\end{equation}
then the entropy is bounded above by
\begin{equation}
S(\rho) \le \ln d \,.
\end{equation}
This upper bound may be obtained by extremizing the entropy subject to the normalization constraint $\Tr \rho =1$. Introducing a Lagrange multiplier $\lambda$, we extremize
\begin{equation}
\Phi[\rho,\lambda] := S(\rho) - \lambda \bigl(1-\Tr \rho \bigr) \,.
\end{equation}
Varying with respect to $\rho$ gives
\begin{align}
0 = \delta \Phi
&= - \Tr\!\left[\delta \rho \bigl(\ln \rho + 1 - \lambda \bigr)\right] .
\end{align}
Since this must hold for arbitrary $\delta \rho$, we find
\begin{equation}
\ln \rho + 1 - \lambda = 0 \,,
\end{equation}
and hence
\begin{equation}
\rho = e^{\lambda-1}\,\mathds{1} _{\mathcal H} \,.
\end{equation}
Imposing $\Tr \rho =1$ gives
\begin{equation}
1 = \Tr \rho = e^{\lambda-1}\Tr \mathds{1} _{\mathcal H} = e^{\lambda-1} d \,,
\end{equation}
so
\begin{equation}
\rho_{\max} = \frac{1}{d}\,\mathds{1} _{\mathcal H} \,.
\end{equation}
Evaluating the entropy on this state, we obtain
\begin{equation}
S_{\max}
=
-\Tr\!\left(\frac{1}{d}\mathds{1} _{\mathcal H}\ln \frac{1}{d}\mathds{1} _{\mathcal H}\right)
=
-\Tr\!\left(\frac{1}{d}\mathds{1} _{\mathcal H}\right)\ln \frac{1}{d}
=
\ln d \,.
\end{equation}
Thus the von Neumann entropy is maximal for the maximally-mixed state
\begin{equation}
\rho = \frac{1}{d}\,\mathds{1} _{\mathcal H} \,,
\end{equation}
for which all states are equally likely. This result builds the intuition that a larger entropy corresponds to having more uncertainty, or equivalently less information, about the exact quantum state of the system. 


\paragraph{Thermal density matrix and the Gibbs distribution} There is one density matrix playing a prominent role in physics, the thermal density matrix. This particular choice plays such an important role that sometimes one feels the need to specify that a given density matrix is non-thermal, even though that is true for all but a measure-zero set of density matrices. The density matrix formalism is essential in quantum statistical mechanics. Here we provide only a lightning discussion. Suppose that, besides the normalization condition
\begin{equation}
\Tr \rho = 1 \,,
\end{equation}
we also fix the average energy
\begin{equation}
\Tr(\rho H)=U \,.
\end{equation}
We now extremize the entropy subject to both constraints. Introducing Lagrange multipliers $\lambda$ and $\beta$, we consider
\begin{equation}
\Phi[\rho,\lambda,\beta]
:=
S(\rho)-\lambda\bigl(1-\Tr \rho \bigr)-\beta\bigl(\Tr(\rho H)-U\bigr) \,.
\end{equation}
Its variation is
\begin{align}
0=\delta \Phi
&=
\delta S(\rho)+\lambda\,\delta(\Tr \rho)-\beta\,\delta\bigl(\Tr(\rho H)\bigr)
\nonumber\\
&=
-\Tr\!\left[\delta \rho\,(\ln \rho +1)\right]
+\lambda\,\Tr(\delta \rho)
-\beta\,\Tr(\delta \rho\, H)
\nonumber\\
&=
-\Tr\!\left[\delta \rho\,\bigl(\ln \rho +1-\lambda+\beta H\bigr)\right] .
\end{align}
Since this must hold for arbitrary $\delta\rho$, we obtain
\begin{equation}
\ln\rho +1-\lambda+\beta H = 0 \,,
\end{equation}
and therefore
\begin{equation}
\rho = e^{\lambda-1}\,e^{-\beta H} \,.
\end{equation}
The constant $e^{\lambda-1}$ is fixed by the normalization condition:
\begin{equation}
1 = \Tr \rho = e^{\lambda-1}\Tr(e^{-\beta H}) \,.
\end{equation}
Therefore
\begin{equation}
\rho
=
\frac{e^{-\beta H}}{\Tr(e^{-\beta H})}
=: \frac{e^{-\beta H}}{Z(\beta)} \,,
\end{equation}
where
\begin{equation}
Z(\beta) := \Tr(e^{-\beta H})
\end{equation}
is the \textit{partition function}, which here emerges simply as a normalisation factor for the trace. In a basis of energy eigenstates with $H|n\rangle = E_n |n\rangle$, 
\begin{equation}
\rho
=
\frac{1}{Z(\beta)}
\sum_n e^{-\beta E_n} |n\rangle\langle n| \,.
\end{equation}
This is the \textit{thermal density matrix}, or Gibbs state. The parameter $\beta$ is determined implicitly by the condition
\begin{equation}
\Tr(\rho H)=U \,,
\end{equation}
and in thermodynamics is identified with the inverse temperature,
\begin{equation}
\beta = \frac{1}{T}
\end{equation}
in units where Boltzmann's constant is set to one. As a closing remark, it should be noted that the thermal density matrix represents an equilibrium state only when the Hamiltonian is time-independent or, under the adiabatic approximation, when it varies very slowly.  


\subsection{Example: the qubit and the Bloch vector} 

It is important to cement the above concepts with the help of a toy model. Here we will discuss possibly the simplest quantum mechanical system: the qubit. The qubit has a two-dimensional Hilbert space, $ \mathcal{H}= \mathbb{C}^{2}$, which is the simplest non-trivial possibility in quantum mechanics. We will use up and down arrows to denote the two basis elements of this Hilbert space
\begin{align}
\mathbb{C}^{2}=\text{Span}(|\uparrow\rangle,|\downarrow \rangle)\,,
\end{align} 
Thus an arbitrary pure state is given by the complex linear combination 
\begin{align}
\ket{\psi}=\alpha \ket{\uparrow}+\beta\ket{\downarrow}\,.
\end{align}
Normalization requires
\begin{equation}
|\alpha|^{2}+|\beta|^{2}=1 \,.
\end{equation}
The density matrix associated with this pure state is therefore
\begin{align}
\rho
&:= |\psi\rangle\langle\psi|
\nonumber\\
&= (\alpha |\uparrow\rangle+\beta |\downarrow\rangle)
(\alpha^{*}\langle\uparrow|+\beta^{*}\langle\downarrow|)
\nonumber\\
&= |\alpha|^{2} |\uparrow\rangle\langle\uparrow|
+\alpha\beta^{*} |\uparrow\rangle\langle\downarrow|
+\alpha^{*}\beta |\downarrow\rangle\langle\uparrow|
+|\beta|^{2} |\downarrow\rangle\langle\downarrow| .
\end{align}
In the basis $\{|\uparrow\rangle,|\downarrow\rangle\}$ this becomes
\begin{equation}
\rho =
\begin{pmatrix}
|\alpha|^{2} & \alpha\beta^{*} \\
\alpha^{*}\beta & |\beta|^{2}
\end{pmatrix} .
\end{equation}
One may immediately verify that this matrix is Hermitian, has unit trace, and satisfies
\begin{equation}
\rho^{2}=\rho ,
\end{equation}
as expected for a pure state.

Let us next consider a mixed state. Suppose, for example, that the system is in the state $|\uparrow\rangle$ with probability $1/2$ and in the state $|\downarrow\rangle$ with probability $1/2$. Then
\begin{equation}
\rho
:= \frac12 |\uparrow\rangle\langle\uparrow|
+\frac12 |\downarrow\rangle\langle\downarrow|
= \frac12 \mathbf{1}_{\mathcal H} .
\end{equation}
In matrix form,
\begin{equation}
\rho =
\frac12
\begin{pmatrix}
1 & 0 \\
0 & 1
\end{pmatrix} .
\end{equation}
This state is not pure, since
\begin{equation}
\rho^{2}=\frac14 \mathbf{1}_{\mathcal H}\neq \rho ,
\end{equation}
or equivalently
\begin{equation}
\Tr \rho^{2}=\frac12 <1 .
\end{equation}
It is in fact the maximally mixed state, in the sense that it assigns equal weight to every direction in the two-dimensional Hilbert space.

More generally, any Hermitian $2\times 2$ matrix can be expanded in the basis formed by the identity and the Pauli matrices,
\begin{equation}
\sigma_{x}:=
\begin{pmatrix}
0 & 1 \\
1 & 0
\end{pmatrix},
\qquad
\sigma_{y}:=
\begin{pmatrix}
0 & -i \\
i & 0
\end{pmatrix},
\qquad
\sigma_{z}:=
\begin{pmatrix}
1 & 0 \\
0 & -1
\end{pmatrix}.
\end{equation}
Since a density matrix is Hermitian and has unit trace, it can always be written in the form
\begin{equation}
\rho
:= \frac12 \bigl(\mathbf{1}_{\mathcal H}+\vec b\cdot \vec \sigma\bigr) ,
\end{equation}
where
\begin{equation}
\vec b := (b_x,b_y,b_z)\in \mathbb{R}^{3},
\qquad
\vec \sigma := (\sigma_x,\sigma_y,\sigma_z) .
\end{equation}
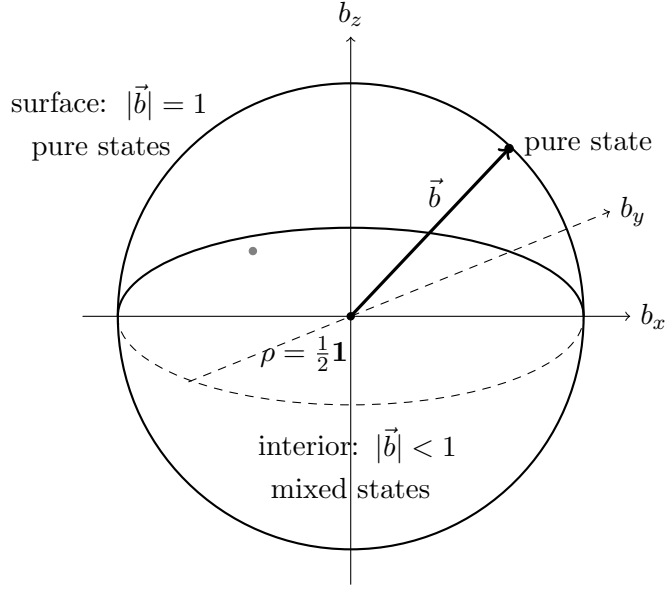
\begin{figure}[t]
\centering
\begin{tikzpicture}[scale=2.2, line cap=round, line join=round]

\def\R{1.4}
\def\ang{22}

\draw[thick] (0,0) circle (\R);

\draw[dashed] (-\R,0) arc[start angle=180,end angle=360,x radius=\R,y radius={0.38*\R}];
\draw[thick] (\R,0) arc[start angle=0,end angle=180,x radius=\R,y radius={0.38*\R}];

\draw[->] (-1.15*\R,0) -- (1.2*\R,0) node[right] {$b_x$};
\draw[->] (0,-1.15*\R) -- (0,1.2*\R) node[above] {$b_z$};
\draw[->, dashed] ({-0.75*\R*cos(\ang)},{-0.75*\R*sin(\ang)})
-- ({1.2*\R*cos(\ang)},{1.2*\R*sin(\ang)}) node[right] {$b_y$};

\coordinate (P) at ({0.68*\R},{0.72*\R});
\draw[->, very thick] (0,0) -- (P);
\node at (0.36*\R,0.53*\R) {$\vec b$};
\filldraw[black] (P) circle (0.025);
\node at (1.02*\R,0.74*\R) {pure state};

\coordinate (M) at ({-0.42*\R},{0.28*\R});
\filldraw[gray] (M) circle (0.022);

\filldraw[black] (0,0) circle (0.022);
\node at (-0.20*\R,-0.16*\R) {$\rho=\frac12 \mathbf{1}$};

\node at (-0.18*\R-1.5,0.92*\R) {surface:};
\node at (0.28*\R-1.5,0.92*\R) {$|\vec b|=1$};
\node at (0-1.5,0.72*\R) {pure states};

\node at (-0.18*\R,-0.84*\R+0.4) {interior:};
\node at (0.28*\R,-0.84*\R+0.4) {$|\vec b|<1$};
\node at (0,-1.02*\R+0.4) {mixed states};

\end{tikzpicture}
\caption{The Bloch ball for a two-level quantum system. Any qubit density matrix can be written as
\(
\rho := \frac12 \bigl(\mathbf{1}+\vec b\cdot \vec \sigma\bigr)
\)
with $|\vec b|\leq 1$. Points on the surface, for which $|\vec b|=1$, correspond to pure states, while points in the interior, for which $|\vec b|<1$, correspond to mixed states. The center $\vec b=0$ is the maximally mixed state.}
\label{fig:bloch-ball}
\end{figure}
In matrix form this is
\begin{equation}
\rho
=
\frac12
\begin{pmatrix}
1+b_z & b_x-i b_y \\
b_x+i b_y & 1-b_z
\end{pmatrix} .
\end{equation}
The vector $\vec b$ is called the \textit{Bloch vector}. Positivity of $\rho$ constrains its norm. Indeed,
\begin{align}
\det \rho
&=
\frac14 \Bigl[(1+b_z)(1-b_z)-(b_x-i b_y)(b_x+i b_y)\Bigr]
\nonumber\\
&=
\frac14 \bigl(1-b_x^{2}-b_y^{2}-b_z^{2}\bigr)
\nonumber\\
&=
\frac14 \bigl(1-|\vec b|^{2}\bigr) \geq 0 .
\end{align}
Hence
\begin{equation}
|\vec b|\leq 1 .
\end{equation}
The space of qubit density matrices is therefore the unit ball in $\mathbb{R}^{3}$, known as the Bloch ball.

The distinction between pure and mixed states has a simple geometric interpretation in this language. Using the identity
\begin{equation}
(\vec b\cdot \vec \sigma)^{2}=|\vec b|^{2}\mathbf{1}_{\mathcal H} ,
\end{equation}
one finds
\begin{align}
\rho^{2}
&=
\frac14
\bigl(\mathbf{1}_{\mathcal H}+2\vec b\cdot\vec \sigma+(\vec b\cdot\vec \sigma)^{2}\bigr)
\nonumber\\
&=
\frac14
\bigl((1+|\vec b|^{2})\mathbf{1}_{\mathcal H}+2\vec b\cdot\vec \sigma\bigr) .
\end{align}
It follows that $\rho^{2}=\rho$ if and only if
\begin{equation}
|\vec b|=1 .
\end{equation}
Therefore pure states lie on the surface of the Bloch ball, namely the Bloch sphere, while mixed states lie in its interior. The maximally mixed state corresponds to the center,
\begin{equation}
\vec b = 0,
\qquad
\rho = \frac12 \mathbf{1}_{\mathcal H} .
\end{equation}

The expectation value of the spin operators is directly related to the Bloch vector. Indeed,
\begin{equation}
\langle \sigma_i\rangle
:= \Tr(\rho \sigma_i)=b_i ,
\qquad i=x,y,z .
\end{equation}
Thus the Bloch vector provides a convenient geometric parametrization of the state of a qubit.

It is also instructive to express the von Neumann entropy of a qubit directly in terms of its Bloch vector. Since
\begin{equation}
\rho := \frac12 \bigl(\mathbf{1}_{\mathcal H}+\vec b\cdot \vec \sigma\bigr) ,
\end{equation}
and $(\vec b\cdot \vec \sigma)^2 = |\vec b|^2 \mathbf{1}_{\mathcal H}$, the eigenvalues of $\rho$ are
\begin{equation}
p_{\pm} = \frac{1 \pm |\vec b|}{2} .
\end{equation}
Indeed, after a suitable unitary change of basis one may always align $\vec b$ with the $z$-axis, so that
\begin{equation}
\rho \sim \frac12
\begin{pmatrix}
1+|\vec b| & 0 \\
0 & 1-|\vec b|
\end{pmatrix} ,
\end{equation}
from which the eigenvalues are immediate. Since the von Neumann entropy depends only on the eigenvalues of the density matrix, we obtain
\begin{equation}\label{Sqbit}
S(\rho)
=
-\frac{1+|\vec b|}{2}\ln\frac{1+|\vec b|}{2}
-\frac{1-|\vec b|}{2}\ln\frac{1-|\vec b|}{2} .
\end{equation}
Thus, for a qubit, the entropy is not sensitive to the direction of $\vec b$, but only to its magnitude. This formula makes the geometry of the Bloch ball even more transparent. On the surface of the Bloch ball, where $|\vec b|=1$, the eigenvalues are $(1,0)$ and therefore
\begin{equation}
S(\rho)=0 ,
\end{equation}
as expected for a pure state. At the center of the Bloch ball, where $\vec b=0$, the eigenvalues are both equal to $1/2$, and the entropy reaches its maximum value
\begin{equation}
S(\rho)= -2\cdot \frac12 \ln\frac12 = \ln 2 .
\end{equation}
More generally, the entropy decreases monotonically as $|\vec b|$ increases from $0$ to $1$. Therefore states closer to the center of the Bloch ball are more mixed and carry larger entropy, while states closer to the surface are less mixed and carry smaller entropy. For a qubit, the distance from the center of the Bloch ball thus provides a direct geometric measure of purity and, equivalently, of the lack of entropy.
 
 
 \subsection{Composite systems, tensor products, and partial trace}

So far we have discussed the state of a single quantum system. In many situations of physical interest, however, the system under consideration is made of several parts. This is especially important for open quantum systems, where one is typically interested in a subsystem, denoted for example by $A$, interacting with an environment $B$. The appropriate mathematical framework for such composite systems is the tensor product of Hilbert spaces. 

Let $\mathcal H_A$ and $\mathcal H_B$ be the Hilbert spaces associated with two quantum systems $A$ and $B$. The Hilbert space of the combined system is
\begin{equation}
\mathcal H_{AB} = \mathcal H_A \otimes \mathcal H_B .
\end{equation}
If $\{|e_a\rangle\}$ is an orthonormal basis of $\mathcal H_A$ and $\{|f_\alpha\rangle\}$ is an orthonormal basis of $\mathcal H_B$, then a basis of $\mathcal H_A \otimes \mathcal H_B$ is given by the tensor-product vectors
\begin{equation}
|e_a\rangle \otimes |f_\alpha\rangle ,
\end{equation}
and any state $|\Psi\rangle \in \mathcal H_A \otimes \mathcal H_B$ can be expanded as
\begin{equation}
|\Psi\rangle
=
\sum_{a,\alpha} c_{a\alpha} \, |e_a\rangle \otimes |f_\alpha\rangle .
\end{equation}
It is important to emphasize that a \textit{generic} state of the composite system cannot be written as a single tensor product
\begin{equation}\label{factorised}
|\Psi\rangle \neq |\phi\rangle \otimes |\chi\rangle
\qquad \text{in general}.
\end{equation}
Indeed, the dimension of the tensor product Hilbert space is 
\begin{align}
\text{dim}(\mathcal H_A \otimes \mathcal H_B)=\text{dim}(\mathcal H_A) \times\text{dim}(\mathcal H_B)\,.
\end{align}
States that can be written in the factorized form in \eqref{factorised} are called \textit{product states}, or separable states. States that cannot be written this way are called \textit{entangled states}. Note that both entangled and product states are pure states, not mixed. 

The inner product on the tensor-product space is defined on basis elements by
\begin{equation}
\bigl(
\langle e_a| \otimes \langle f_\alpha|
\bigr)
\bigl(
|e_b\rangle \otimes |f_\beta\rangle
\bigr)
=
\langle e_a|e_b\rangle \, \langle f_\alpha|f_\beta\rangle ,
\end{equation}
and is then extended to arbitrary states by linearity. The Dirac notation visually helps us here by guiding us to glue together bras and kets of the same subsystem.

Operators on composite systems are constructed similarly. If
\begin{equation}
\O_{A} : \mathcal H_A \to \mathcal H_A,
\qquad
\O_{B} : \mathcal H_B \to \mathcal H_B,
\end{equation}
then the tensor-product operator
\begin{equation}
\O_{A} \otimes \O_{B} : \mathcal H_A \otimes \mathcal H_B \to \mathcal H_A \otimes \mathcal H_B
\end{equation}
is defined by
\begin{equation}
(\O_{A} \otimes \O_{B})(|\phi\rangle \otimes |\chi\rangle)
=
(\O_{A}|\phi\rangle)\otimes(\O_{B}|\chi\rangle).
\end{equation}
Then, a generic operator on the composite system is a linear combination of such product operators, not of the simple product form $\O_A\otimes \O_B$. Indeed, if $\{|e_a\rangle\}$ is a basis of $\mathcal H_A$ and $\{|f_\alpha\rangle\}$ is a basis of $\mathcal H_B$, then the rank-one operators
\begin{equation}
|e_a\rangle\langle e_b| \otimes |f_\alpha\rangle\langle f_\beta|
\end{equation}
form a basis of the space of linear operators on $\mathcal H_A\otimes \mathcal H_B$. Therefore any operator $\O$ on the composite Hilbert space can be written as
\begin{equation}
\O
=
\sum_{a,b,\alpha,\beta}
\O_{a\alpha,b\beta}\,
\Bigl(|e_a\rangle\langle e_b| \otimes |f_\alpha\rangle\langle f_\beta|\Bigr) .
\end{equation}
In this sense, product operators are the elementary building blocks of general operators on tensor-product Hilbert spaces.

A particular case, which will be important for us later, consists of an operator acting only on subsystem $A$. This is represented as
\begin{equation}
\O_{A} \otimes \mathds{1} _B ,
\end{equation}
while an operator acting only on subsystem $B$ is represented as
\begin{equation}
\mathds{1} _A \otimes \O_{B} .
\end{equation}
Such operators commute:
\begin{equation}
[\O_{A} \otimes \mathds{1} _B,\mathds{1} _A \otimes \O_{B}]=0 .
\end{equation}
This simply expresses the fact that operations acting on different subsystems are independent.

\paragraph{Partial trace and the reduced density matrix} We are now ready to state one of the central questions of open systems. Suppose we are interested only in operators acting on subsystem $A$. This may be the case, for a variety of reasons. For example, system $ B$ may be too big so that we cannot effectively measure it. Or it may be so complicated or difficult to describe so that we are not technically able to make precise predictions about it. How do we then formulate our mathematical description of the quantum mechanics of system $A $, which in general is allowed to talk to and interact with system $ B$? The answer is the \textit{reduced density matrix} obtained by taking the \textit{partial trace} over the degrees of freedom of subsystem $B$.

Let $\{|n\rangle_B\}$ be an orthonormal basis of $\mathcal H_B$. The reduced density matrix of subsystem $A$ is defined by
\begin{equation}
\rho_A \equiv \Tr_B(\rho_{AB})
=
\sum_n {}_B\langle n| \rho_{AB} |n\rangle_B .
\end{equation}
Similarly, the reduced density matrix of subsystem $B$ is
\begin{equation}
\rho_B \equiv \Tr_A(\rho_{AB}) .
\end{equation}
Note that the partial trace is basis independent, even though its definition is often written using a chosen orthonormal basis.

The reduced density matrix is the correct object because it reproduces the expectation values of all observables acting only on subsystem $A$. If an observable has the form
\begin{equation}
Q = Q_A \otimes \mathds{1} _B ,
\end{equation}
then
\begin{equation}
\langle Q\rangle
=
\Tr_{A\otimes B}\bigl(\rho_{AB}(Q_A\otimes \mathds{1} _B)\bigr)
=
\Tr_A(\rho_A Q_A) .
\end{equation}
Thus, as far as measurements on subsystem $A$ are concerned, all the information contained in the full state $\rho_{AB}$ is encoded in the reduced density matrix $\rho_A$.

Let us verify this formula explicitly. Writing the partial trace in a basis $\{|n\rangle_B\}$ of $\mathcal H_B$,
\begin{align}
\Tr_{A\otimes B}\bigl(\rho_{AB}(Q_A\otimes \mathds{1} _B)\bigr)
&=
\sum_n
\Tr_A\Bigl(
{}_B\langle n|
\rho_{AB}(Q_A\otimes \mathds{1} _B)
|n\rangle_B
\Bigr)
\nonumber\\
&=
\sum_n
\Tr_A\Bigl(
{}_B\langle n|\rho_{AB}|n\rangle_B \, Q_A
\Bigr)
\nonumber\\
&=
\Tr_A\Bigl(
\bigl(\sum_n {}_B\langle n|\rho_{AB}|n\rangle_B\bigr) Q_A
\Bigr)
\nonumber\\
&=
\Tr_A(\rho_A Q_A) .
\end{align}
The reduced density matrix is the central object of study of open quantum systems. In these notes we will discuss some of its many properties and different formalisms to describe its dynamics.

We will see in the next section that the partial trace is of great importance for open quantum systems. In that context one typically decomposes the total Hilbert space as
\begin{equation}
\mathcal H = \mathcal H_S \otimes \mathcal H_E ,
\end{equation}
where $S$ denotes the system of interest and $E$ its environment. The full system may evolve unitarily, as in standard closed-system quantum mechanics, but the observer only measures the subsystem $S$. The relevant object is then not the full density matrix $\rho_{SE}$ but the reduced density matrix
\begin{equation}
\rho_S = \Tr_E(\rho_{SE}) .
\end{equation}
Even if $\rho_{SE}$ describes a pure state, $\rho_S$ will in general be mixed once the system becomes entangled with the environment. As we will see shortly, entanglement may also be generated by time evolution, even when it is not present at the beginning.

\paragraph{Example: tracing over a qubit.} Let's discuss an example using qubits. Consider a full system consisting of the tensor product of two qubits $ A$ and $ B$, and the separable two-qubit state
\begin{equation}
|\Psi\rangle
=
\frac{1}{\sqrt{2}}\bigl(|\uparrow\rangle + |\downarrow\rangle\bigr)_A
\otimes
|\downarrow\rangle_B .
\end{equation}
The full density matrix describing this pure, product state is
\begin{equation}
\rho_{AB}=|\Psi\rangle\langle\Psi| .
\end{equation}
Tracing over subsystem $B$ gives
\begin{align}
\rho_A
&=
\Tr_B(\rho_{AB})
\nonumber\\
&=
{}_B\langle\uparrow|\rho_{AB}|\uparrow\rangle_B
+
{}_B\langle\downarrow|\rho_{AB}|\downarrow\rangle_B .
\end{align}
Since the state has support only on $|\downarrow\rangle_B$, only the second term contributes, and one obtains
\begin{equation}
\rho_A
=
\frac{1}{2}
\bigl(|\uparrow\rangle+|\downarrow\rangle\bigr)_A
\bigl(\langle\uparrow|+\langle\downarrow|\bigr)_A .
\end{equation}
In matrix form, in the basis $\{|\uparrow\rangle,|\downarrow\rangle\}$,
\begin{equation}
\rho_A
=
\frac12
\begin{pmatrix}
1 & 1 \\
1 & 1
\end{pmatrix} .
\end{equation}
This is still a pure state. As we will prove shortly in full generality, if the total state is a product state, then each subsystem is also in a pure state. Intuitively, since there was no entanglement between the two subsystems, not knowing one did not prevent us from knowing the other one exactly.

Now consider instead the entangled (pure) state
\begin{equation}
|\mathrm{EPR}\rangle
=
\frac{1}{\sqrt{2}}
\Bigl(
|\uparrow\rangle_A |\downarrow\rangle_B
-
|\downarrow\rangle_A |\uparrow\rangle_B
\Bigr) ,
\end{equation}
known as Bell state\footnote{``EPR'' stands for Einstein, Podolsky and Rosen. }.  The full state is pure:
\begin{equation}
\rho_{AB}=|\mathrm{EPR}\rangle\langle \mathrm{EPR}| .
\end{equation}
However, tracing out subsystem $B$ yields
\begin{equation}
\rho_A
=
\Tr_B(\rho_{AB})
=
\frac12
\Bigl(
|\uparrow\rangle\langle\uparrow|
+
|\downarrow\rangle\langle\downarrow|
\Bigr)_A ,
\end{equation}
or in matrix form
\begin{equation}
\rho_A
=
\frac12
\begin{pmatrix}
1 & 0 \\
0 & 1
\end{pmatrix}_A .
\end{equation}
This reduced density matrix is mixed:
\begin{equation}
\rho_A^2 \neq \rho_A .
\end{equation}
Thus a subsystem of a larger system can be in a mixed state even if the total state is pure. This is one of the basic signatures of entanglement. 


\subsection{Entanglement, Schmidt decomposition, and entropy}

Entanglement is one of the most characteristic features of quantum mechanics, and in the context of open systems it is precisely what makes a subsystem behave as if it were in a mixed state even when the total state is pure.

We will almost always assume that the full system is in a pure state $|\Psi\rangle$, so that $ \rho_{AB} := |\Psi\rangle\langle\Psi|   $. The reduced density matrices obtained by tracing over one of the two subsystems are
\begin{equation}
\rho_A := \Tr_B \rho_{AB},
\qquad
\rho_B := \Tr_A \rho_{AB} .
\end{equation}
The von Neumann entropy associated with $\rho_A$ or $\rho_B$ is known as \textit{entanglement entropy} and measures the entanglement of the pure state $|\Psi\rangle$. As we'll see shortly, one finds
\begin{align}
S_{A}=-\Tr \rho_{A} \log \rho_{A}=-\Tr \rho_{B} \log \rho_{B}=S_{B}\,.
\end{align}
Before discussing it in detail, a couple of extremely useful technical results need to be covered. 

\paragraph{Schmidt decomposition and Schmidt number}

A basic theorem for bipartite pure states is the Schmidt decomposition. Let
\begin{equation}
|\Psi\rangle \in \mathcal H_A \otimes \mathcal H_B .
\end{equation}
Then there exist orthonormal sets $\{|i_A\rangle\}$ in $\mathcal H_A$ and $\{|i_B\rangle\}$ in $\mathcal H_B$, together with non-negative real numbers $\lambda_i \ge 0$, such that
\begin{equation}\label{schmidtdec}
|\Psi\rangle = \sum_{i=1}^r \lambda_i \, |i_A\rangle \otimes |i_B\rangle ,
\qquad
\sum_{i=1}^r \lambda_i^2 = 1 .
\end{equation}
We will sketch the proof of this result shortly. The integer
\begin{equation}
r := \#\{i \, : \, \lambda_i \neq 0\}
\end{equation}
is called the \textit{Schmidt number}, or Schmidt rank, of the state. Since the Schmidt number $ r$ fixes the range of the sum both over the index $ i_{A}$ and $ i_{B}$, this means that when one of the two subsystems, let's say B, is enormously bigger than A, $\text{dim}\mathcal{H}_{A} \ll \text{dim}\mathcal{H}_{B} $ , we never need to use more than $\text{dim}\mathcal{H}_{A} $ orthonormal vectors from it. 

The Schmidt decomposition may be obtained as follows. Choose orthonormal bases $\{|e_a\rangle\}$ of $\mathcal H_A$ and $\{|f_\alpha\rangle\}$ of $\mathcal H_B$, and expand
\begin{equation}
|\Psi\rangle = \sum_{a,\alpha} c_{a\alpha} \, |e_a\rangle \otimes |f_\alpha\rangle .
\end{equation}
The coefficients $c_{a\alpha}$ may be regarded as a matrix $C$. From linear algebra, recall the singular value decomposition, which states that there exist unitary square matrices $U$, of size $ \text{dim}\mathcal{H}_{A}\times\text{dim}\mathcal{H}_{A}$ and $V$, of size $ \text{dim}\mathcal{H}_{B}\times\text{dim}\mathcal{H}_{B}$, such that
\begin{equation}
C = U D V^\dagger ,
\end{equation}
where $D$, of size $ \text{dim}\mathcal{H}_{A}\times \text{dim}\mathcal{H}_{B}$, is rectangular diagonal with non-negative diagonal entries $D_{ii}=\lambda_i$. Defining
\begin{equation}
|i_A\rangle := \sum_a U_{ai} |e_a\rangle ,
\qquad
|i_B\rangle := \sum_\alpha V^*_{\alpha i} |f_\alpha\rangle ,
\end{equation}
one obtains \eqref{schmidtdec}. The normalization of $|\Psi\rangle$ implies
\begin{equation}
1 = \langle \Psi|\Psi\rangle = \sum_i \lambda_i^2 .
\end{equation}

The Schmidt decomposition makes the structure of bipartite pure states completely transparent. A state is a product state if and only if it has Schmidt number $r=1$. If $r>1$, the state is entangled.

\paragraph{Reduced density matrices in Schmidt form}

Using the Schmidt decomposition, the density matrix of the full system is
\begin{equation}
\rho_{AB}
=
|\Psi\rangle\langle\Psi|
=
\sum_{i,j} \lambda_i \lambda_j
\bigl(|i_A\rangle\langle j_A|\bigr)
\otimes
\bigl(|i_B\rangle\langle j_B|\bigr) .
\end{equation}
Tracing over subsystem $B$ gives
\begin{align}
\rho_A
&:= \Tr_B \rho_{AB}=
\sum_{i,j} \lambda_i \lambda_j
|i_A\rangle\langle j_A| \,
\Tr\bigl(|i_B\rangle\langle j_B|\bigr)
\nonumber\\
&=
\sum_{i,j} \lambda_i \lambda_j
|i_A\rangle\langle j_A| \,
\langle j_B|i_B\rangle=
\sum_i \lambda_i^2 |i_A\rangle\langle i_A| .
\end{align}
Similarly,
\begin{equation}
\rho_B = \sum_i \lambda_i^2 |i_B\rangle\langle i_B| .
\end{equation}
This shows why the Schmidt decomposition is so important when discussing partial traces: the reduced density matrices are automatically diagonal in this basis. Therefore the non-zero eigenvalues of $\rho_A$ and $\rho_B$ are the same, namely
\begin{equation}
p_i := \lambda_i^2 ,
\qquad
p_i \ge 0,
\qquad
\sum_i p_i = 1 .
\end{equation}

This has several immediate consequences.

First, the rank of $\rho_A$ and of $\rho_B$ is the Schmidt number:
\begin{equation}
\mathrm{rank}(\rho_A)=\mathrm{rank}(\rho_B)=r .
\end{equation}

Second, all spectral quantities of $\rho_A$ and $\rho_B$ coincide. In particular, the von Neumann entropies are equal:
\begin{equation}\label{sameent}
S(\rho_A) = S(\rho_B) .
\end{equation}
Indeed,
\begin{equation}
S(\rho_A)
:=
-\Tr(\rho_A \ln \rho_A)
=
-\sum_i p_i \ln p_i ,
\end{equation}
and similarly
\begin{equation}
S(\rho_B)
=
-\sum_i p_i \ln p_i .
\end{equation}
For a bipartite pure state, this common entropy is called the entanglement entropy:
\begin{equation}
S_E(|\Psi\rangle)
:=
S(\rho_A)
=
S(\rho_B) .
\end{equation}

\paragraph{Pure reduced state if and only if product state}

We now prove the fundamental relation between entanglement and mixedness of the reduced state.

Suppose first that $|\Psi\rangle$ is a product state:
\begin{equation}
|\Psi\rangle = |\phi\rangle_A \otimes |\chi\rangle_B .
\end{equation}
Then
\begin{equation}
\rho_{AB}
=
\bigl(|\phi\rangle\langle\phi|\bigr)_A
\otimes
\bigl(|\chi\rangle\langle\chi|\bigr)_B ,
\end{equation}
and tracing over $B$ gives
\begin{equation}
\rho_A
=
|\phi\rangle\langle\phi| \, \Tr\bigl(|\chi\rangle\langle\chi|\bigr)
=
|\phi\rangle\langle\phi| .
\end{equation}
Hence $\rho_A$ is pure. Likewise, $\rho_B$ is pure.

Conversely, suppose that $\rho_A$ is pure. Since the non-zero eigenvalues of $\rho_A$ are $\lambda_i^2$, purity implies that exactly one eigenvalue is equal to $1$ and all others vanish. Therefore there is only one non-zero Schmidt coefficient, so the Schmidt number is $r=1$, and
\begin{equation}
|\Psi\rangle = |1_A\rangle \otimes |1_B\rangle .
\end{equation}
Thus the original pure state is a product state.

We have therefore shown:
\begin{equation}
\rho_A \text{ pure}
\qquad \Longleftrightarrow \qquad
|\Psi\rangle \text{ is a product state}.
\end{equation}
Equivalently,
\begin{equation}
|\Psi\rangle \text{ entangled}
\qquad \Longleftrightarrow \qquad
\rho_A \text{ is mixed},
\end{equation}
and similarly for $\rho_B$. In terms of the Schmidt number, this is simply the statement that
\begin{equation}
r=1
\qquad \Longleftrightarrow \qquad
\text{product state},
\end{equation}
while
\begin{equation}
r>1
\qquad \Longleftrightarrow \qquad
\text{entangled state}.
\end{equation}
For example, for the Bell state
\begin{equation}
|\mathrm{EPR}\rangle
=
\frac{1}{\sqrt{2}}
\Bigl(
|\uparrow\rangle_A |\downarrow\rangle_B
-
|\downarrow\rangle_A |\uparrow\rangle_B
\Bigr) ,
\end{equation}
the Schmidt coefficients are
\begin{equation}
\lambda_1=\lambda_2=\frac{1}{\sqrt{2}} ,
\end{equation}
so
\begin{equation}
S_E = -2 \cdot \frac12 \ln \frac12 = \ln 2 .
\end{equation}


\paragraph{Purity, linear entropy, and R\'enyi entropies}

Besides the von Neumann entropy, it is sometimes useful to consider other quantities that probe the mixedness of a state.

The \textit{purity} of a density matrix is defined by
\begin{equation}
\mathcal P(\rho) := \Tr(\rho^2) .
\end{equation}
It satisfies
\begin{equation}
0<\mathcal P(\rho)\le 1 ,
\end{equation}
with
\begin{equation}
\mathcal P(\rho)=1
\qquad \Longleftrightarrow \qquad
\rho \text{ is pure}.
\end{equation}
For a state with eigenvalues $p_i$,
\begin{equation}
\mathcal P(\rho)=\sum_i p_i^2 .
\end{equation}

A closely related quantity is the \textit{linear entropy}, defined by
\begin{equation}
S_L(\rho) := 1-\Tr(\rho^2) = 1-\mathcal P(\rho) .
\end{equation}
It vanishes for pure states and is positive for mixed states. It is often technically simpler than the von Neumann entropy\footnote{Strictly speaking, the von Neumann entropy cannot be Taylor expanded around a pure state, since $S(\rho)$ is non-analytic when $\rho$ has zero eigenvalues, so the word ``linear'' is a bit misleading. Rather, for a nearly pure state with eigenvalues $\{1-\epsilon,\epsilon p_1,\ldots,\epsilon p_{d-1}\}$, where $\sum_i p_i=1$ and $\epsilon\ll1$, one finds
\[
S(\rho)=-\epsilon\ln\epsilon+\mathcal{O}(\epsilon),\qquad
S_L(\rho)=1-\Tr(\rho^2)=2\epsilon+\mathcal{O}(\epsilon^2).
\]
Thus, both entropies vanish as the state approaches purity, although $  S $ exhibits the characteristic non-analytic $-\epsilon\ln\epsilon$ behaviour. Actually $  S_{L} $ approximates $  S $ near the maximally mixed state. Writing $ \rho = \frac{\mathds{1} }{d}+\delta\rho  $, with $ \Tr(\delta\rho)=0   $ and expanding the logarithm, one finds that
\begin{align}
S(\rho)&=\ln d - \frac{d}{2}\Tr(\delta\rho^2)+\cdots , & 1-\Tr(\rho^2)&=1-\frac{1}{d}-\Tr(\delta\rho^2) .
\end{align}
Thus, up to additive and multiplicative constants, $  S_{L} $ measures the leading deviation of $  S $ away from its maximum.}.


More generally, the \textit{R\'enyi entropies} are defined for $n>0$, $n\neq 1$, by
\begin{equation}
S_n(\rho)
:=
\frac{1}{1-n}\ln \Tr(\rho^n) .
\end{equation}
In terms of the eigenvalues $p_i$,
\begin{equation}
S_n(\rho)
=
\frac{1}{1-n}\ln\Bigl(\sum_i p_i^n\Bigr) .
\end{equation}
The overall normalization is chosen such that the limit $n\to 1$ reproduces the von Neumann entropy:
\begin{equation}
\lim_{n\to 1} S_n(\rho)= -\Tr(\rho\ln\rho)=S(\rho) .
\end{equation}
The case $n=2$ is directly related to purity:
\begin{equation}
S_2(\rho) = -\ln \Tr(\rho^2) = -\ln \mathcal P(\rho) .
\end{equation}
Thus purity, linear entropy, R\'enyi entropies, and von Neumann entropy all provide different ways of quantifying mixedness.

\paragraph{Subadditivity and mutual information} We will only make two brief forays in the beautiful field of quantum information theory. This one is about subadditivity and mutual information. For a bipartite system with density matrix $\rho_{AB}$ and reduced states $\rho_A$, $\rho_B$, the von Neumann entropy satisfies the subadditivity inequality
\begin{equation}\label{subadditivity}
S(\rho_{AB}) \le S(\rho_A)+S(\rho_B) .
\end{equation}
Equivalently,
\begin{equation}
S(\rho_A)+S(\rho_B)-S(\rho_{AB}) \ge 0 .
\end{equation}
This non-negative quantity is called the \textit{mutual information}:
\begin{equation}
I(A:B)
:=
S(\rho_A)+S(\rho_B)-S(\rho_{AB}) .
\end{equation}
It measures the total amount of correlations, both classical and quantum, between the two subsystems. For a product state $\rho_{AB}=\rho_A\otimes\rho_B$, one has
\begin{equation}
I(A:B)=0 .
\end{equation}
For a bipartite pure state, since $S(\rho_{AB})=0$, the mutual information becomes
\begin{equation}
I(A:B)=2S_E .
\end{equation}

We now sketch a proof of subadditivity based on the positivity of relative entropy and Jensen's inequality. Define the \textit{relative entropy}
\begin{equation}
S(\rho\|\sigma)
:=
\Tr(\rho \ln \rho)-\Tr(\rho \ln \sigma) .
\end{equation}
A standard result, sometimes called Klein's inequality, states that
\begin{equation}
S(\rho\|\sigma)\ge 0 .
\end{equation}
One way to prove this is to use the convexity of the function
\begin{equation}
f(x):=x\ln x ,
\end{equation}
together with Jensen's inequality\footnote{Jensen's inequality states that, for a convex function $f$ and non-negative weights $p_i$ satisfying $\sum_i p_i=1$,
\begin{equation}
f\!\left(\sum_i p_i x_i\right) \le \sum_i p_i f(x_i) \,.
\end{equation}}, namely applied in the eigenbasis of $\sigma$.

To prove subadditivity, choose
\begin{equation}
\sigma := \rho_A \otimes \rho_B .
\end{equation}
Then positivity of relative entropy gives
\begin{equation}
S(\rho_{AB}\|\rho_A\otimes\rho_B)\ge 0 .
\end{equation}
Expanding this,
\begin{align}
0
&\le
\Tr(\rho_{AB}\ln\rho_{AB})
-
\Tr\bigl(\rho_{AB}\ln(\rho_A\otimes\rho_B)\bigr)
\nonumber\\
&=
\Tr(\rho_{AB}\ln\rho_{AB})
-
\Tr\bigl(\rho_{AB}(\ln\rho_A\otimes \mathds{1} _B)\bigr)
-
\Tr\bigl(\rho_{AB}(\mathds{1} _A\otimes \ln\rho_B)\bigr) .
\end{align}
Using the defining property of the partial trace,
\begin{equation}
\Tr\bigl(\rho_{AB}(\ln\rho_A\otimes \mathds{1} _B)\bigr)
=
\Tr_A(\rho_A\ln\rho_A) ,
\end{equation}
and similarly for $B$, one finds
\begin{equation}
0
\le
\Tr(\rho_{AB}\ln\rho_{AB})
-
\Tr_A(\rho_A\ln\rho_A)
-
\Tr_B(\rho_B\ln\rho_B) .
\end{equation}
Multiplying by $-1$ gives
\begin{equation}
S(\rho_{AB}) \le S(\rho_A)+S(\rho_B) ,
\end{equation}
which is the desired result.

\paragraph{Concavity of entropy from Jensen's inequality}

Another important property is that the von Neumann entropy is concave:
\begin{equation}
S\Bigl(\sum_\alpha p_\alpha \rho_\alpha\Bigr)
\ge
\sum_\alpha p_\alpha S(\rho_\alpha) .
\end{equation}
A useful special case is obtained by diagonalizing a density matrix,
\begin{equation}
\rho = \sum_i p_i |i\rangle\langle i| ,
\end{equation}
for which
\begin{equation}
S(\rho)= -\sum_i p_i \ln p_i .
\end{equation}
Since the function
\begin{equation}
g(x):= -\ln x
\end{equation}
is convex on $x>0$, Jensen's inequality implies
\begin{equation}
-\ln\Bigl(\sum_i p_i x_i\Bigr)\le \sum_i p_i (-\ln x_i) .
\end{equation}
This elementary inequality underlies many entropy inequalities, including positivity of relative entropy and the classical version of subadditivity.

\paragraph{Purification}

So far, we have seen how to start from the state of the full system and obtain the reduced density matrix by performing partial traces. It is natural to ask if there is a way (of course not unique) to take the trip in the opposite direction. This is indeed possible and it turns out that any mixed state on a Hilbert space $\mathcal H_A$ can always be regarded as the reduced state of a pure state in a larger Hilbert space. This construction is called \textit{purification}.

Let
\begin{equation}
\rho_A = \sum_i p_i |i_A\rangle\langle i_A|
\end{equation}
be the spectral decomposition of a density matrix on $\mathcal H_A$, with $p_i\ge 0$ and $\sum_i p_i=1$. Introduce an auxiliary Hilbert space $\mathcal H_R$ with orthonormal basis $\{|i_R\rangle\}$ of dimension at least equal to the rank of $\rho_A$. Then the state
\begin{equation}\label{purification}
|\Psi\rangle
:=
\sum_i \sqrt{p_i}\, |i_A\rangle \otimes |i_R\rangle \in \H_{A}\otimes \H_{R}
\end{equation}
is pure and satisfies
\begin{equation}
\Tr_R\bigl(|\Psi\rangle \langle\Psi|\bigr)=\rho_A .
\end{equation}
Indeed,
\begin{align}
|\Psi\rangle\langle\Psi|
&=
\sum_{i,j}\sqrt{p_i p_j}\,
|i_A\rangle\langle j_A|
\otimes
|i_R\rangle\langle j_R| ,
\end{align}
so tracing over $R$ gives
\begin{align}
\Tr_R\bigl(|\Psi\rangle\langle\Psi|\bigr)
&=
\sum_{i,j}\sqrt{p_i p_j}\,
|i_A\rangle\langle j_A| \,
\langle j_R|i_R\rangle
\nonumber\\
&=
\sum_i p_i |i_A\rangle\langle i_A| =\rho_A .
\end{align}
Thus every mixed state may be thought of as arising from entanglement with an enlarged system.

The purification \eqref{purification} is not unique. Different purifications are related by unitary transformations acting on the auxiliary Hilbert space $\mathcal H_R$. This freedom reflects the fact that many different larger pure states may give rise to the same reduced density matrix.

To summarize, the Schmidt decomposition shows that every bipartite pure state is characterized by a set of non-negative Schmidt coefficients. These coefficients determine the reduced density matrices, the entanglement entropy, the Schmidt number, the purity, and the whole family of R\'enyi entropies. In particular, the reduced state is pure if and only if the original state is a product state, while it is mixed if and only if the original state is entangled. Finally, purification shows that every mixed state may be embedded into a larger pure system, a point of view that will be central in the study of open quantum systems.


\subsection{Decoherence}

We now come to one of the central physical consequences of entanglement in composite systems: decoherence. We only discuss the basic idea, which is very simple but far reaching. Even if the combined system evolves unitarily and remains in a pure state, a subsystem may evolve from a pure state into a mixed state because it becomes entangled with degrees of freedom that are not observed. In this sense, decoherence is not a failure of quantum mechanics, but rather a direct consequence of applying quantum mechanics to a larger system while only observing part of it.

Consider therefore a bipartite system with Hilbert space
\begin{equation}
\mathcal H = \mathcal H_A \otimes \mathcal H_B ,
\end{equation}
where $A$ is the system of interest and $B$ is its environment, measuring device, or bath. Suppose that at time $t=0$ the total state is pure and separable,
\begin{equation}
|\Psi_0\rangle = |\phi\rangle_A \otimes |\chi\rangle_B ,
\end{equation}
so that
\begin{equation}
\rho_{AB}(0) = |\Psi_0\rangle\langle\Psi_0| ,
\qquad
\rho_A(0) = |\phi\rangle\langle\phi| .
\end{equation}
In particular, the initial reduced state of subsystem $A$ is pure:
\begin{equation}
\rho_A(0)^2 = \rho_A(0) .
\end{equation}

As we will see in the next section, and is probably familiar from standard quantum mechanics, the full system evolves unitarily under some operator $U_{AB}(t)$,
\begin{equation}
\rho_{AB}(t)
=
U_{AB}(t)\,\rho_{AB}(0)\,U_{AB}^\dagger(t) .
\end{equation}
If there is no interaction between $A$ and $B$, so that the evolution factorizes as
\begin{equation}
U_{AB}(t) = U_A(t)\otimes U_B(t) ,
\end{equation}
then the reduced density matrix of $A$ remains pure for all times, since
\begin{equation}
\rho_A(t)
=
U_A(t)\,\rho_A(0)\,U_A^\dagger(t) .
\end{equation}
Thus pure states remain pure if the subsystem does not become entangled with its environment.

The situation changes qualitatively when $A$ interacts with $B$. In that case the full unitary evolution need not factorize, and the reduced density matrix becomes
\begin{equation}
\rho_A(t)
:=
\Tr_B \rho_{AB}(t)
=
\Tr_B\!\left(U_{AB}(t)\,\rho_{AB}(0)\,U_{AB}^\dagger(t)\right) .
\end{equation}
Assuming the initial product state above, one may rewrite this in a useful form. Choose an orthonormal basis $\{|\beta\rangle\}$ of $\mathcal H_B$, and define
\begin{equation}\label{def:kraus}
M_\beta(t) := \langle \beta|U_{AB}(t)|\chi\rangle ,
\end{equation}
where $M_\beta(t)$ are operators on $\mathcal H_A$ known as Kraus operators. Then
\begin{align}\label{Krausoperatorrepresentation}
\rho_A(t)
&=
\sum_\beta
M_\beta(t)\,\rho_A(0)\,M_\beta^\dagger(t) .
\end{align}
This is the operator-sum, or Kraus, representation of the reduced evolution. The operators $M_\beta(t)$ satisfy
\begin{equation}
\sum_\beta M_\beta^\dagger(t) M_\beta(t) = \mathds{1} _A ,
\end{equation}
as a consequence of the unitarity of $U_{AB}(t)$. Indeed,
\begin{align}
\sum_\beta M_\beta^\dagger M_\beta
&=
\sum_\beta
\langle \chi|U_{AB}^\dagger|\beta\rangle
\langle \beta|U_{AB}|\chi\rangle
\nonumber\\
&=
\langle \chi|U_{AB}^\dagger
\left(\sum_\beta |\beta\rangle\langle\beta|\right)
U_{AB}|\chi\rangle
\nonumber\\
&=
\langle \chi|U_{AB}^\dagger U_{AB}|\chi\rangle =\mathds{1} _A .
\end{align}
Therefore the reduced evolution preserves the trace of $\rho_A$.

The key physical point is that, for interacting systems, the reduced state $\rho_A(t)$ is in general no longer pure:
\begin{equation}
\rho_A(t)^2 \neq \rho_A(t) .
\end{equation}
This loss of purity reflects the entanglement generated between $A$ and $B$. One says that subsystem $A$ has decohered. In practice, decoherence often manifests itself as the suppression of off-diagonal entries of the density matrix in a preferred basis.

\paragraph{A simple measurement model}

A particularly instructive example is obtained by taking $A$ to be a qubit, with basis $\{|\uparrow\rangle,|\downarrow\rangle\}$, and $B$ to be a measuring apparatus with orthonormal pointer states $\{|0\rangle,|1\rangle,|2\rangle\}$. Let the initial state of the apparatus be $|0\rangle$, and consider an idealized ``one-step'' unitary evolution such that
\begin{align}
U\bigl(|\uparrow\rangle\otimes|0\rangle\bigr)
&=
|\uparrow\rangle\otimes
\left(\sqrt{1-p}\,|0\rangle+\sqrt{p}\,|1\rangle\right) ,
\\
U\bigl(|\downarrow\rangle\otimes|0\rangle\bigr)
&=
|\downarrow\rangle\otimes
\left(\sqrt{1-p}\,|0\rangle+\sqrt{p}\,|2\rangle\right) ,
\end{align}
where $0\le p\le 1$. In words, the apparatus sometimes remains in its ready state $|0\rangle$, and sometimes registers the outcome $|\uparrow\rangle$ by moving to $|1\rangle$ or the outcome $|\downarrow\rangle$ by moving to $|2\rangle$.

Since the environment starts in $|0\rangle$, the Kraus operators are
\begin{equation}
M_\beta = \langle \beta|U|0\rangle ,
\qquad
\beta=0,1,2 .
\end{equation}
From the definition of $U$ above one immediately finds
\begin{align}
M_0 &= \sqrt{1-p}\,\mathds{1} _A , &
M_1 &= \sqrt{p}\,|\uparrow\rangle\langle\uparrow| ,  &
M_2 &= \sqrt{p}\,|\downarrow\rangle\langle\downarrow| .
\end{align}
It is easy to verify that
\begin{equation}
M_0^\dagger M_0 + M_1^\dagger M_1 + M_2^\dagger M_2
=
(1-p)\mathds{1} _A + p|\uparrow\rangle\langle\uparrow| + p|\downarrow\rangle\langle\downarrow|
=
\mathds{1} _A .
\end{equation}

Let the initial density matrix of the qubit be
\begin{equation}
\rho_A =
\begin{pmatrix}
\rho_{\uparrow\uparrow} & \rho_{\uparrow\downarrow} \\
\rho_{\downarrow\uparrow} & \rho_{\downarrow\downarrow}
\end{pmatrix} .
\end{equation}
After the interaction, the reduced density matrix becomes
\begin{equation}
\rho_A \mapsto \sum_{\beta=0}^2 M_\beta \rho_A M_\beta^\dagger .
\end{equation}
Using the explicit form of the Kraus operators, one finds
\begin{equation}
\rho_A \mapsto
\begin{pmatrix}
\rho_{\uparrow\uparrow} & (1-p)\rho_{\uparrow\downarrow} \\
(1-p)\rho_{\downarrow\uparrow} & \rho_{\downarrow\downarrow}
\end{pmatrix} .
\end{equation}
Thus the diagonal entries remain unchanged, while the off-diagonal entries are suppressed by a factor of $(1-p)$. This is the hallmark of decoherence: coherence between $|\uparrow\rangle$ and $|\downarrow\rangle$ is gradually lost, while the probabilities for the two outcomes are preserved.

\paragraph{Phase damping}

Suppose the above one-step process acts repeatedly over short time intervals $\delta t$, and define a rate
\begin{equation}
\Gamma := \frac{p}{\delta t} .
\end{equation}
After $N$ steps, with total time $t=N\delta t$, the off-diagonal terms are multiplied by
\begin{equation}
(1-p)^N
=
\left(1-\frac{\Gamma t}{N}\right)^N
\approx e^{-\Gamma t}
\qquad
\text{for } N\gg 1 .
\end{equation}
Therefore the reduced density matrix evolves approximately as
\begin{equation}
\rho_A(t)
=
\begin{pmatrix}
\rho_{\uparrow\uparrow}(0) & e^{-\Gamma t}\rho_{\uparrow\downarrow}(0) \\
e^{-\Gamma t}\rho_{\downarrow\uparrow}(0) & \rho_{\downarrow\downarrow}(0)
\end{pmatrix} .
\end{equation}
This process is often called phase damping or dephasing. We will encounter it again as a solution of the quantum master equation. 

For example, if the initial state of the qubit is pure,
\begin{equation}
|\psi\rangle = a|\uparrow\rangle + b|\downarrow\rangle ,
\end{equation}
then initially
\begin{equation}
\rho_A(0)
=
\begin{pmatrix}
|a|^2 & ab^* \\
a^*b & |b|^2
\end{pmatrix} ,
\end{equation}
while at late times
\begin{equation}
\lim_{t\to\infty}\rho_A(t)
=
\begin{pmatrix}
|a|^2 & 0 \\
0 & |b|^2
\end{pmatrix} .
\end{equation}
The final state is diagonal in the basis singled out by the apparatus, namely $\{|\uparrow\rangle,|\downarrow\rangle\}$. This basis is often called the preferred basis, or pointer basis. The probability of finding the system up or down has not changed. What has changed is that the system never behaves as being both up and down at the same time, as typical of quantum interference. Rather, there is only a classical probability $|a|^2 $ to be up and a classical probability $ |b|^{2}$ to be down. The loss of coherence between these two states reduces quantum mechanics to a classical statistical ensemble. 

The cause of this behaviour is clear. The subsystem $A$ becomes entangled with the apparatus $B$, and once the degrees of freedom of $B$ are ignored, the reduced state of $A$ loses its phase coherence. Decoherence therefore explains why coherent superpositions are hard to observe in macroscopic settings: interactions with the environment rapidly suppress off-diagonal terms in the basis selected by the dominant coupling to that environment.


\section{The quantum master equation}\label{sec:2}

Having introduced the density matrix and the basic kinematics of composite quantum systems, we now turn to dynamics. Even when the full system evolves unitarily, the reduced density matrix of a subsystem generally does not. Once one traces over environmental degrees of freedom, the subsystem is described by an effective evolution law that can include dissipation, decoherence, and noise. The central goal of this section is to understand how such reduced dynamics can be characterized in a mathematically consistent and physically transparent way.

We begin by contrasting the unitary evolution of the full system with the effective evolution of the open subsystem, and then introduce the reduced dynamical map, its Kraus representation, and the notions of positivity, complete positivity, and trace preservation. We next study the conditions under which a continuous time evolution can be generated by a Liouvillian and, in the Markovian limit, takes the universal Gorini--Kossakowski--Sudarshan--Lindblad form. Finally, we discuss the microphysical assumptions that lead to this description and explain how the usual semigroup picture must be generalized when the generator is explicitly time dependent, as is often the relevant situation in cosmology.


 
\subsection{Full system vs open system evolution} So far we have discussed the density matrix as a kinematical object encoding the state of a quantum system. We move on to discuss its time evolution. To begin with, let's assume we have control of the full system and can use standard unitary evolution from quantum mechanics. If the system evolves unitarily, then in the Schr\"odinger picture a pure state evolves as
\begin{equation}
|\psi(t)\rangle = U(t)|\psi(0)\rangle ,
\end{equation}
where $U(t)$ is a unitary operator, satisfying
\begin{equation}
U^\dagger(t)U(t)=U(t)U^\dagger(t)=\mathbf{1}_{\mathcal H} .
\end{equation}
It follows immediately that the density matrix evolves according to
\begin{equation}\label{rhotunitary}
\rho(t) = U(t)\,\rho(0)\,U^\dagger(t) .
\end{equation}
Indeed, for a pure state $\rho(0)=|\psi(0)\rangle\langle\psi(0)|$, one has
\begin{align}
\rho(t)
&:= |\psi(t)\rangle\langle\psi(t)|
\nonumber\\
&= U(t)|\psi(0)\rangle\langle\psi(0)|U^\dagger(t)
\nonumber\\
&= U(t)\rho(0)U^\dagger(t) .
\end{align}
By linearity, the same formula also holds for mixed states.

Equation \eqref{rhotunitary} preserves all the defining properties of a density matrix. Hermiticity is preserved because
\begin{equation}
\rho^\dagger(t)
=
\bigl(U(t)\rho(0)U^\dagger(t)\bigr)^\dagger
=
U(t)\rho^\dagger(0)U^\dagger(t)
=
\rho(t) .
\end{equation}
Positivity is preserved since, for any $|\phi\rangle\in\mathcal H$,
\begin{equation}
\langle \phi|\rho(t)|\phi\rangle
=
\langle \phi|U(t)\rho(0)U^\dagger(t)|\phi\rangle
=
\langle \phi'(t)|\rho(0)|\phi'(t)\rangle
\geq 0 ,
\end{equation}
where we defined $|\phi'(t)\rangle := U^\dagger(t)|\phi\rangle$. Finally, the trace is preserved by cyclicity:
\begin{equation}
\Tr\rho(t)
=
\Tr\bigl(U(t)\rho(0)U^\dagger(t)\bigr)
=
\Tr\bigl(\rho(0)U^\dagger(t)U(t)\bigr)
=
\Tr\rho(0)
=
1 .
\end{equation}
Therefore unitary evolution maps density matrices to density matrices.

If the Hamiltonian is time independent, the unitary evolution operator is
\begin{equation}
U(t)=e^{-iHt} ,
\end{equation}
where we have set $\hbar=1$. Differentiating \eqref{rhotunitary} with respect to time then gives the \textit{von Neumann equation}
\begin{equation}\label{vneq}
i\,\frac{d\rho}{dt} = [H,\rho] .
\end{equation}
Indeed,
\begin{align}
\frac{d\rho}{dt}
&=
\frac{dU}{dt}\rho(0)U^\dagger
+
U\rho(0)\frac{dU^\dagger}{dt}
\nonumber\\
&=
(-iHU)\rho(0)U^\dagger
+
U\rho(0)(iU^\dagger H)
\nonumber\\
&=
-iH\rho(t)+i\rho(t)H ,
\end{align}
which is equivalent to \eqref{vneq}. This equation is the density-matrix analogue of the Schr\"odinger equation, and may be viewed as the quantum counterpart of Liouville's equation in classical mechanics\footnote{In classical statistical mechanics, the Liouville equation governs the time evolution of a phase-space probability density $f(q,p,t)$ under Hamiltonian flow:
\begin{equation}
\frac{\partial f}{\partial t} = \{H,f\} ,
\end{equation}
where $\{\cdot,\cdot\}$ denotes the Poisson bracket. This is equivalent to saying that $ f(q(t),p(t),t)$ is constant along the solutions $ (q(t),p(t))$ of Hamilton's equations. The analogy with the von Neumann equation is that both describe the deterministic evolution of a statistical state under Hamiltonian dynamics: in the classical case a probability distribution on phase space, and in the quantum case a density matrix on Hilbert space.}.

A useful consequence of unitary evolution is that the expectation value of an observable $Q$ may be computed equivalently in different pictures. In the Schr\"odinger picture the state evolves and the operator is fixed, so
\begin{equation}
\langle Q\rangle_t = \Tr\bigl(\rho_S(t)\,Q_S\bigr) .
\end{equation}
In the Heisenberg picture the state is fixed and the operator evolves as
\begin{equation}
Q_H(t) := U^\dagger(t)\,Q_S\,U(t) ,
\qquad
\rho_H := \rho(0) .
\end{equation}
Using cyclicity of the trace, one finds
\begin{equation}
\Tr\bigl(\rho_S(t)Q_S\bigr)
=
\Tr\bigl(U(t)\rho(0)U^\dagger(t)Q_S\bigr)
=
\Tr\bigl(\rho(0)U^\dagger(t)Q_SU(t)\bigr)
=
\Tr\bigl(\rho_H Q_H(t)\bigr) .
\end{equation}
Thus, for closed systems, the density matrix formalism is completely equivalent to the usual formulations of quantum mechanics, but has the advantage of treating pure states and mixed states in a unified way.

Finally, unitary evolution preserves the spectrum of $\rho$, since \eqref{rhotunitary} is a similarity transformation. Therefore quantities that depend only on the eigenvalues of $\rho$, such as $\Tr(\rho^n)$ and the von Neumann entropy $S(\rho)$, remain constant under closed-system evolution. In particular, a pure state remains pure under unitary evolution, while a mixed state remains mixed. Only when we later restrict attention to a subsystem, and trace out unobserved degrees of freedom, will purity and entropy be able to change non-trivially.


\paragraph{Open systems and reduced dynamics}

So far we have discussed the unitary time evolution of a closed quantum system. In many physical situations, however, the system of interest is not isolated. Rather, it interacts with additional degrees of freedom that are unknown, uncontrolled, unobserved, or simply too complicated to describe in detail. One then speaks of an open quantum system. The basic strategy is to identify a subsystem $S$, called the system, and to separate the remaining degrees of freedom into another subsystem $E$, generically called the environment. The total Hilbert space is then written as
\begin{equation}
\mathcal H = \mathcal H_S \otimes \mathcal H_E .
\end{equation}
Even if the combined system $S+E$ evolves unitarily, the subsystem $S$ alone will in general not do so.

The term \textit{environment} is the most general one, and we will mostly use it in what follows. In the literature one also often encounters the words \textit{bath} and \textit{reservoir}. Roughly speaking, a reservoir usually suggests an environment so large that its macroscopic properties are essentially unaffected by the interaction with $S$, while a bath often refers to an environment with additional physical properties, for example thermal equilibrium in the case of a thermal bath. These distinctions are often useful physically, but for the formal development below the important point is simply that there are degrees of freedom external to the subsystem of interest, and that we do not keep track of them explicitly.

Let the density matrix of the total system be $\rho_{SE}(t)$. Its unitary time evolution is
\begin{equation}
\rho_{SE}(t) = U(t)\,\rho_{SE}(0)\,U^\dagger(t) ,
\end{equation}
where $U(t)$ acts on $\mathcal H_S\otimes \mathcal H_E$. The physically relevant object for the subsystem $S$ is the reduced density matrix
\begin{equation}
\rho_S(t) := \Tr_E \rho_{SE}(t) .
\end{equation}
This expression already makes clear why open-system dynamics is generically more complicated than closed-system dynamics. Even though the evolution of $\rho_{SE}$ is unitary and therefore reversible, the evolution of $\rho_S$ obtained after tracing over the environment need not be either unitary or reversible.

If the initial density matrix factorizes\footnote{This assumption is more general than taking an initially factorized \textit{pure} state, $\rho_{SE}(0)=|\psi_S\rangle\langle\psi_S|\otimes|\chi_E\rangle\langle\chi_E|$. Here $\rho_S(0)$ and $\rho_E(0)$ may themselves be mixed; what matters for the following derivation is only the absence of initial system--environment correlations, namely the factorization of the density matrix.},
\begin{equation}\label{factorizedinitial}
\rho_{SE}(0) = \rho_S(0)\otimes \rho_E(0) ,
\end{equation}
then the reduced evolution may be written as a map from initial system states to final system states:
\begin{equation}
\rho_S(t) = \Phi_t\bigl[\rho_S(0)\bigr] ,
\end{equation}
where
\begin{equation}\label{dynamicalmap}
\Phi_t[\rho] := \Tr_E\!\Bigl(U(t)\,(\rho\otimes \rho_E(0))\,U^\dagger(t)\Bigr) .
\end{equation}
The family of maps $\{\Phi_t\}$ is called the \textit{quantum dynamical map}, or reduced dynamical map, of the subsystem. It is the basic object of open-system theory.

The assumption \eqref{factorizedinitial} is stronger than it may first appear. If instead the initial state $\rho_{SE}(0)$ already contains correlations between $S$ and $E$, then the reduced state at later times need not be determined by $\rho_S(0)$ alone. In that case the evolution of the subsystem generally cannot be described by a state-independent map of the form \eqref{dynamicalmap}. Thus the very possibility of describing open-system dynamics by a well-defined map $\Phi_t$ already rests on assumptions about the initial system--environment correlations.

Several remarks are in order. First, the map $\Phi_t$ is linear in $\rho_S(0)$, because both the unitary evolution and the partial trace are linear operations. Second, unlike the unitary evolution of the total system, the reduced evolution of the subsystem may change the purity and entropy of $\rho_S$. This is precisely because information may flow between the system and the environment, and because by construction we do not keep track of the environmental degrees of freedom. This is the formal origin of decoherence, dissipation, and entropy production in reduced descriptions.

Finally, the map \eqref{dynamicalmap} sends density matrices to density matrices: it preserves Hermiticity, positivity, and unit trace. Rather than proving these properties abstractly at this stage, it is useful first to rewrite the map in a more explicit form. This leads naturally to the Kraus representation, which will make these structural properties manifest.


\subsection{Kraus representation of the reduced dynamical map}

We now rewrite the reduced dynamical map \eqref{dynamicalmap} in a more explicit operator form. This will make its structure much more transparent and will also prepare the ground for the more abstract discussion of admissible quantum maps. The logical content of this subsection is that unitary evolution of the full system plus tracing over an environment implies that the dynamical map can be written in Kraus representation 
\begin{align}
\text{Partial trace over environment} \then \text{Kraus representation}\,.
\end{align}
We start from the map
\begin{equation}\label{facto}
\Phi_t[\rho]
:=
\Tr_E\!\Bigl(U(t)\,(\rho\otimes \rho_E)\,U^\dagger(t)\Bigr) ,
\end{equation}
where, to lighten notation, we have written $\rho_E$ instead of $\rho_E(0)$. At this stage we assume only that the initial density matrix factorizes, \eqref{factorizedinitial}, with $\rho_E$ an arbitrary density matrix on $\mathcal H_E$, possibly mixed.

Let us diagonalize the environment density matrix,
\begin{equation}
\rho_E = \sum_\beta p_\beta |\beta\rangle\langle \beta| ,
\qquad
p_\beta \geq 0,
\qquad
\sum_\beta p_\beta = 1 ,
\end{equation}
where $\{|\beta\rangle\}$ is an orthonormal basis of $\mathcal H_E$ consisting of eigenvectors of $\rho_E$. We also choose an arbitrary orthonormal basis $\{|\alpha\rangle\}$ of $\mathcal H_E$ in which to perform the partial trace. Then
\begin{align}
\Phi_t[\rho]
&=
\Tr_E\!\left(U(t)\,(\rho\otimes \rho_E)\,U^\dagger(t)\right)
\nonumber\\
&=
\sum_\alpha
\langle \alpha|
U(t)\,(\rho\otimes \rho_E)\,U^\dagger(t)
|\alpha\rangle
\nonumber\\
&=
\sum_{\alpha,\beta}
p_\beta\,
\langle \alpha|
U(t)\,(\rho\otimes |\beta\rangle\langle\beta|)\,U^\dagger(t)
|\alpha\rangle .
\end{align}
Since $\rho$ acts only on the system Hilbert space, this can be rewritten as
\begin{align}
\Phi_t[\rho]
&=
\sum_{\alpha,\beta}
\Bigl(\sqrt{p_\beta}\,\langle \alpha|U(t)|\beta\rangle\Bigr)
\,\rho\,
\Bigl(\sqrt{p_\beta}\,\langle \beta|U^\dagger(t)|\alpha\rangle\Bigr)
\nonumber\\
&=
\sum_{\alpha,\beta}\label{Krausrepresenation}
M_{\alpha\beta}(t)\,\rho\,M_{\alpha\beta}^\dagger(t) ,
\end{align}
where we have defined the \textit{Kraus operators}
\begin{equation}\label{twoindexkraus}
M_{\alpha\beta}(t)
:=
\sqrt{p_\beta}\,\langle \alpha|U(t)|\beta\rangle .
\end{equation}
Here $M_{\alpha\beta}(t)$ acts only on the system Hilbert space $\mathcal H_S$. The expression \eqref{Krausrepresenation} for the dynamical map $ \Phi_{t}$ is the \textit{Kraus representation} of the map.

The normalization condition satisfied by the Kraus operators follows directly from the unitarity of $U(t)$ and the fact that $\Tr \rho_E = 1$. Indeed,
\begin{align}
\sum_{\alpha,\beta} M_{\alpha\beta}^\dagger(t) M_{\alpha\beta}(t)
&=
\sum_{\alpha,\beta}
p_\beta\,
\langle \beta|U^\dagger(t)|\alpha\rangle
\langle \alpha|U(t)|\beta\rangle
\nonumber\\
&=
\sum_\beta
p_\beta\,
\langle \beta|U^\dagger(t)\left(\sum_\alpha |\alpha\rangle\langle\alpha|\right)U(t)|\beta\rangle
\nonumber\\
&=
\sum_\beta
p_\beta\,
\langle \beta|U^\dagger(t)U(t)|\beta\rangle=
\sum_\beta p_\beta\,\mathds{1} _S=
\mathds{1} _S .
\end{align}
The pair $(\alpha,\beta)$ is simply a label for the Kraus operators, and can always be combined into a single composite index, $\mu := (\alpha,\beta)  $. Then the dynamical map may be rewritten in the more common form
\begin{equation}\label{krausoneindex}
\Phi_t[\rho] = \sum_\mu M_\mu(t)\,\rho\,M_\mu^\dagger(t) ,
\end{equation}
with normalization
\begin{equation}\label{krausnorm}
\sum_\mu M_\mu^\dagger(t) M_\mu(t)=\mathds{1} _S .
\end{equation}
Note that this normalisation condition is an operator statement, not a trace. Thus the two-index and one-index versions are not structurally different: they are simply two ways of labelling the same operator-sum representation. The two-index form is useful because it keeps track of how the Kraus operators arise from the spectral decomposition of the initial environment state and from the partial trace over the final environment basis states.

It is worth emphasizing that the Kraus representation is not unique. Different sets of Kraus operators may describe the same dynamical map. In particular, even if a straightforward microscopic derivation produces certain Kraus operators, it may be possible to rewrite the same map using fewer of them. The number of indices is therefore not itself a fundamental property of the map, but rather part of a particular representation.

As a special case, suppose now that the environment starts in a pure state,
\begin{equation}
\rho_E = |\chi\rangle\langle\chi| .
\end{equation}
Then the spectral decomposition contains only one non-vanishing eigenvalue, equal to $1$, and the sum over $\beta$ collapses. The Kraus operators reduce to
\begin{equation}
M_\alpha(t) := \langle \alpha|U(t)|\chi\rangle ,
\end{equation}
so that the reduced map takes the simpler form
\begin{equation}
\Phi_t[\rho] = \sum_\alpha M_\alpha(t)\,\rho\,M_\alpha^\dagger(t) .
\end{equation}
This is the form that is often written in elementary discussions. It is valid in the special case of an initial factorized density matrix \eqref{facto}.

The conclusion of this subsection is that any reduced dynamical map obtained from an initially factorized density matrix and unitary evolution of the full system always admits a Kraus representation. 


\subsection{Positive, completely positive, and trace-preserving maps}

Having derived the Kraus representation of the reduced dynamical map, we now step back and ask a more abstract question. Suppose we are simply given a linear map
\begin{equation}
\Phi : \mathcal L(\mathcal H_S) \to \mathcal L(\mathcal H_S)
\end{equation}
acting on operators on the Hilbert space of the system. What properties must such a map satisfy in order to describe a physically admissible evolution of density matrices? The answer will lead us to the notions of positivity, complete positivity, and trace preservation.

\paragraph{Positivity}

A first obvious requirement is that $\Phi$ should send positive operators to positive operators. Recall that a density matrix $\rho$ is positive if
\begin{equation}
\langle \psi | \rho | \psi \rangle \geq 0
\qquad
\text{for all } |\psi\rangle \in \mathcal H_S .
\end{equation}
A linear map $\Phi$ is therefore called \textit{positive} if
\begin{equation}
\rho \geq 0
\qquad \Longrightarrow \qquad
\Phi[\rho] \geq 0 .
\end{equation}
This is clearly necessary if $\Phi$ is to map density matrices to density matrices.

However, positivity alone turns out not to be sufficient. The reason is that the system $S$ may itself be part of a larger system, and a map that acts physically on $S$ should remain physically sensible even when applied only to one factor of a composite Hilbert space.

\paragraph{Complete positivity}

Let $\mathcal H_A$ be an auxiliary Hilbert space, often called an ancilla, and let $\mathds{1} _A$ denote the identity map on operators acting on $\mathcal H_A$. Given a map $\Phi$ on $\mathcal L(\mathcal H_S)$, one may define the extended map
\begin{equation}
\mathds{1} _A \otimes \Phi
\end{equation}
acting on operators on $\mathcal H_A \otimes \mathcal H_S$ by leaving the ancilla untouched and applying $\Phi$ only to the system.

A map $\Phi$ is called \textit{completely positive} if, for every auxiliary Hilbert space $\mathcal H_A$,
\begin{equation}
X \geq 0
\qquad \Longrightarrow \qquad
(\mathds{1} _A \otimes \Phi)[X] \geq 0
\end{equation}
for all positive operators $X$ on $\mathcal H_A \otimes \mathcal H_S$. Thus complete positivity requires positivity not only for the system by itself, but also when the system is viewed as part of an arbitrarily larger quantum system.

This requirement is physically natural. If $\Phi$ is to describe an admissible evolution of subsystem $S$, then it should not produce an unphysical operator when $S$ happens to be entangled with some spectator degrees of freedom that are not themselves evolved. Every completely positive map is positive, but the converse is false. 

A famous counterexample is the transpose map. In a fixed basis, it is defined by
\begin{equation}
T[\rho] := \rho^T .
\end{equation}
This map is positive, since the transpose of a positive matrix has the same eigenvalues and is therefore still positive. However, it is not completely positive. To see this, consider the maximally entangled Bell state
\begin{equation}
|\Phi^+\rangle := \frac{1}{\sqrt{2}}\bigl(|00\rangle+|11\rangle\bigr) ,
\qquad
\rho_{\Phi^+}:=|\Phi^+\rangle\langle\Phi^+| .
\end{equation}
If one applies the transpose only to the second subsystem, the resulting operator
\begin{equation}
(\mathbf{1}\otimes T)[\rho_{\Phi^+}]
\end{equation}
is the partial transpose of $\rho_{\Phi^+}$, and it has a negative eigenvalue. Therefore it is not a positive operator. This shows that $T$ is positive on a single system but fails to remain positive when extended trivially to a larger system, and hence it is not completely positive.
\paragraph{Trace preservation}

A second basic requirement is that total probability remain normalized. The map $\Phi$ should preserve the trace:
\begin{equation}
\Tr \Phi[\rho] = \Tr \rho .
\end{equation}
A map with this property is called \textit{trace preserving}. If $\Phi$ is trace preserving and sends density matrices to positive operators, then it maps density matrices to density matrices.

A linear map that is both completely positive and trace preserving is called a \textit{CPTP map}. In finite-dimensional quantum information theory such maps are also often called \textit{quantum channels}. This is the appropriate abstract notion of a physically admissible evolution map for a quantum system.

Note that we didn't require that the dynamical map should preserve the Hermitian property of the density matrix. In fact, this is already implied by the stronger requirement of positivity, and therefore also by complete positivity. Indeed, any Hermitian operator $H$ can be written as the difference of two positive operators,
\begin{equation}
H = H_{+} - H_{-} ,
\end{equation}
for example by separating the positive and negative parts of its spectral decomposition. If $\Phi$ is a linear positive map, then both $\Phi[H_{+}]$ and $\Phi[H_{-}]$ are positive, and hence Hermitian. By linearity,
\begin{equation}
\Phi[H] = \Phi[H_{+}] - \Phi[H_{-}] ,
\end{equation}
which is therefore again Hermitian. Thus positivity already guarantees that Hermitian operators are mapped to Hermitian operators.


\paragraph{Kraus representation implies CPTP}

The importance of the Kraus representation derived in the previous subsection is that these properties become immediate. Indeed, suppose the map is written in Kraus representation \eqref{krausoneindex}. Let us first show positivity. If $\rho \geq 0$, then for any $|\psi\rangle \in \mathcal H_S$,
\begin{align}
\langle \psi | \Phi[\rho] | \psi \rangle
&=
\sum_\mu
\langle \psi | M_\mu \rho M_\mu^\dagger | \psi \rangle
\nonumber\\
&=
\sum_\mu
\langle \phi_\mu | \rho | \phi_\mu \rangle
\geq 0 ,
\end{align}
where we defined
\begin{equation}
|\phi_\mu\rangle := M_\mu^\dagger |\psi\rangle .
\end{equation}
Therefore
\begin{equation}
\rho \geq 0
\qquad \Longrightarrow \qquad
\Phi[\rho] \geq 0 .
\end{equation}

In fact, the same argument immediately shows complete positivity. For any auxiliary Hilbert space $\mathcal H_A$, the extended map takes the form
\begin{equation}
(\mathds{1} _A \otimes \Phi)[X]
=
\sum_\mu
(\mathds{1} _A \otimes M_\mu)\,X\,(\mathds{1} _A \otimes M_\mu^\dagger) .
\end{equation}
If $X \geq 0$, then for any vector $|\Psi\rangle \in \mathcal H_A \otimes \mathcal H_S$,
\begin{align}
\langle \Psi | (\mathds{1} _A \otimes \Phi)[X] | \Psi \rangle
&=
\sum_\mu
\langle \Psi_\mu | X | \Psi_\mu \rangle
\geq 0 ,
\end{align}
where
\begin{equation}
|\Psi_\mu\rangle := (\mathds{1} _A \otimes M_\mu^\dagger)|\Psi\rangle .
\end{equation}
Hence
\begin{equation}
X \geq 0
\qquad \Longrightarrow \qquad
(\mathds{1} _A \otimes \Phi)[X] \geq 0 ,
\end{equation}
which proves complete positivity.

Trace preservation follows directly from cyclicity of the trace:
\begin{align}
\Tr \Phi[\rho]
=
\Tr\!\left(\sum_\mu M_\mu \rho M_\mu^\dagger\right)
=
\Tr\!\left(\rho \sum_\mu M_\mu^\dagger M_\mu\right)
=
\Tr \rho ,
\end{align}
where in the last step we used \eqref{krausnorm}. Thus any map written in Kraus form with the normalization condition \eqref{krausnorm} is automatically CPTP.

As a corollary, we conclude that tracing over an environment with full system unitary evolution gives a dynamical map that is completely positive and trace-preserving.

The result above shows that Kraus form implies complete positivity and trace preservation. The converse is also true: every CPTP map on a finite-dimensional Hilbert space admits a normalized Kraus representation. This is a basic structure theorem of quantum theory. We will not prove it in full detail here, but we will use it repeatedly in what follows. Thus, in finite dimensions, the notions of CPTP map and Kraus map are equivalent. \\

Summarising the logical connections 
\begin{align}
\text{Partial trace over env.} \then \text{Kraus representation} \quad \Leftrightarrow \quad \text{CPTP}\,.
\end{align}


\subsection{One-parameter dynamical maps, semigroups, and the Liouvillian}

So far we thought of $\Phi_t$ as a single map from time $t = 0 $ to some time $t $. In practice, however, we want to be able to evolve from all times to all times. This simple consideration leads us to uncover an important mathematical property with deep physical implications. Let's start with the maths. We want to consider a one-parameter family of maps
\begin{equation}
\{\Phi_t\}_{t\geq 0},
\end{equation}
where each $\Phi_t$ acts on density matrices of the system and describes the evolution from the initial time $0$ to the later time $t$:
\begin{equation}
\rho(t) = \Phi_t[\rho(0)] .
\end{equation}
In the previous subsections we motivated such maps microscopically by coupling the system to an environment, evolving the total system unitarily, and tracing over the environment. We now abstract away from that construction and study the properties of the family $\{\Phi_t\}$ directly.

A basic consistency condition is that doing nothing should leave the state unchanged. Therefore one requires
\begin{equation}
\Phi_0 = \mathbf{1} ,
\end{equation}
where $\mathbf{1}$ denotes the identity map on operators. In addition, we will consider the following \textit{semigroup} property
\begin{equation}\label{semigroupprop}
\Phi_{t+s} = \Phi_t \circ \Phi_s ,
\qquad
t,s\geq 0 .
\end{equation}
This means that evolving first for a time $s$ and then for a further time $t$ is equivalent to evolving once for the total time $t+s$. We will derive this property from a series of approximations and assumptions about the property of the environment. For the moment, we simply think about what this property tells us: the law of evolution depends only on the duration of the time interval, not on the absolute initial time. 

Equation \eqref{semigroupprop} is the natural analogue, for open systems, of the familiar group property of unitary evolution in closed systems. For a closed system one has
\begin{equation}
U(t+s)=U(t)U(s) ,
\end{equation}
where the inverse $U(-t)=U^\dagger(t)$ also exists. By contrast, for an open system the maps $\Phi_t$ need not be invertible, since information may be lost to the environment in the reduced description. One therefore obtains in general only a semigroup rather than a full group.

The semigroup property has an important physical interpretation. It encodes the idea that the future evolution depends only on the present state and not on the detailed past history of the system. This property is the quantum analogue of Markovianity for classical stochastic processes, where the probability of the next step depends exclusively on the current configuration rather than on the history of previous steps. Not every open-system evolution satisfies this property exactly, but it is a central idealization and the starting point of the Lindblad theorem.

Assuming now that the family $\{\Phi_t\}$ depends continuously on time, one may define its infinitesimal generator. This is a linear map $\mathcal L$ acting on operators, defined by
\begin{equation}\label{Liouvilliandef}
\mathcal L[\rho]
:=
\lim_{\delta t\to 0^+}
\frac{\Phi_{\delta t}[\rho]-\rho}{\delta t} .
\end{equation}
The map $\mathcal L$ is called the \textit{Liouvillian}, Liouville superoperator\footnote{Given the familiar mathematical structure of states as vectors on a Hilbert space and observables as operators, this nomenclature reminds us that $ \Phi_{t}$ is a map from operators to operators, rather than from vectors to vectors. In equations, if $ \mathcal{B}(\H)$ is the space of operators on a Hilbert space $ \H$, then a superoperator maps $ \L:\mathcal{B}(\H) \to \mathcal{B}(\H)$. Note that $ \L$ acts on generic operators, not just density matrices. In fact the space of density matrices is not a vector space but a compact convex subset of $\mathcal B(\mathcal H) $.}, or simply the generator of the dynamical semigroup. It plays for open systems the role that the Hamiltonian plays for closed systems.

Using the semigroup property, one may derive a local differential equation for the evolution. Indeed,
\begin{align}
\frac{d}{dt}\rho(t)
&=
\lim_{\delta t\to 0^+}
\frac{\rho(t+\delta t)-\rho(t)}{\delta t}
\nonumber\\
&=
\lim_{\delta t\to 0^+}
\frac{\Phi_{t+\delta t}[\rho(0)]-\Phi_t[\rho(0)]}{\delta t}
\nonumber\\
&=
\lim_{\delta t\to 0^+}
\frac{\Phi_{\delta t}\bigl[\Phi_t[\rho(0)]\bigr]-\Phi_t[\rho(0)]}{\delta t}
\nonumber\\
&=
\mathcal L[\rho(t)] .
\end{align}
Thus the reduced dynamics satisfies the \textit{master equation}
\begin{equation}\label{mastereqL}
\frac{d\rho}{dt} = \mathcal L[\rho] .
\end{equation}
This is the general form of a time-homogeneous quantum master equation. 

Formally, the solution of \eqref{mastereqL} may be written as
\begin{equation}
\rho(t)=e^{t\mathcal L}[\rho(0)] ,
\end{equation}
so that
\begin{equation}
\Phi_t = e^{t\mathcal L} .
\end{equation}
Here the exponential is understood as the exponential of a linear superoperator acting on the space of operators. In complete analogy with ordinary differential equations, the generator $\mathcal L$ determines the full semigroup.

For a closed system, the Liouvillian reduces to the commutator with the Hamiltonian:
\begin{equation}
\mathcal L[\rho] = -i[H,\rho] ,
\end{equation}
so that the master equation becomes the von Neumann equation \eqref{vneq}.
For an open system, however, the Liouvillian can contain additional terms describing decoherence, dissipation, and noise. The crucial question is then the following: what is the most general form of $\mathcal L$ such that the maps $\Phi_t=e^{t\mathcal L}$ remain completely positive and trace preserving for all $t\geq 0$? The answer is provided by the Gorini--Kossakowski--Sudarshan--Lindblad theorem, to which we now turn.


\subsection{The GKSL theorem and the Lindblad equation}

We are now ready to state the central structure theorem for Markovian open-system dynamics, which is due to Gorini, Kossakowski, Sudarshan and Lindblad (GKSL)\footnote{The theorem goes back to two independent 1976 papers: Gorini, Kossakowski and Sudarshan \cite{10.1063/1.522979} gave the finite-dimensional $N$-level-system result, while Lindblad \cite{Lindblad:1975ef} gave a closely related operator-algebraic characterization of generators of completely positive quantum dynamical semigroups.  At the time, Sudarshan was the senior figure at Texas; Gorini and Kossakowski were younger/mid-career mathematical physicists visiting Texas; and Lindblad had only recently completed his PhD in Stockholm.}. It characterizes the most general generator of a one-parameter quantum dynamical semigroup.

\paragraph{GKSL theorem} Let $\mathcal H_S$ be a finite-dimensional Hilbert space, and let $\{\Phi_t\}_{t\geq 0}$ be a one-parameter family of linear maps on $\mathcal L(\mathcal H_S)$ such that

\begin{enumerate}
\item $\Phi_0=\mathds{1} $,
\item $\Phi_{t+s}=\Phi_t\circ \Phi_s$ for all $t,s\geq 0$,
\item each $\Phi_t$ is completely positive and trace preserving,
\item the family is continuous in $t$.
\end{enumerate}

Then there exists a linear superoperator $\mathcal L$ such that
\begin{equation}
\Phi_t = e^{t\mathcal L} ,
\qquad
\frac{d\rho}{dt} = \mathcal L[\rho] ,
\end{equation}
and the generator $\mathcal L$ necessarily takes the form
\begin{equation}\label{GKSLgeneral}
\mathcal L[\rho]
=
-i[H,\rho]
+
\sum_{a,b} c_{ab}
\left(
F_a \rho F_b^\dagger
-\frac12 \{F_b^\dagger F_a,\rho\}
\right) .
\end{equation}
Here $H=H^\dagger$ is a Hermitian operator on $\mathcal H_S$, the $ F_a$ are elements of a basis of traceless operators on $\mathcal H_S$, $ \{ .,.\}$ indicates the anti-commutator and the so-called \textit{Kossakovski matrix} $c_{ab}$ is positive semidefinite. Conversely, any generator of the form \eqref{GKSLgeneral} defines a completely positive trace-preserving semigroup.

It is important to distinguish the mathematical assumptions of the theorem from the physical approximations often used in microscopic derivations. The theorem itself assumes semigroup evolution and CPTP from the start. By contrast, when deriving a master equation from a system coupled to an environment, one typically invokes additional physical assumptions such as weak coupling, large environment, short environmental correlation time, and often the Born, Markov, and secular approximations. These physical assumptions are meant to justify the mathematical framework in which the theorem applies. We will stick to the maths for this section and come back to the physical underpinning of these assumptions in later sections. 

\paragraph{Diagonal Lindblad form} Since the matrix $c_{ab}$ is positive semidefinite, it may be diagonalized by a unitary transformation. One may therefore rewrite \eqref{GKSLgeneral} in the equivalent and more familiar form
\begin{equation}\label{Lindbladform}
\frac{d\rho}{dt}
=
-i[H,\rho]
+
\sum_\mu
\left(
L_\mu \rho L_\mu^\dagger
-\frac12 \{L_\mu^\dagger L_\mu,\rho\}
\right) .
\end{equation}
This is the \textit{Lindblad equation}, or Lindblad master equation. The operators $L_\mu$ are called \textit{jump operators}, Lindblad operators, or sometimes noise operators.

The interpretation of the various terms is straightforward. The commutator term
\begin{equation}
-i[H,\rho]
\end{equation}
is the Hamiltonian part of the evolution. By itself it would generate ordinary unitary time evolution. The remaining terms describe the genuinely open-system effects. The piece
\begin{equation}
L_\mu \rho L_\mu^\dagger
\end{equation}
accounts for transitions, jumps, or noise induced by the environment. The anticommutator term
\begin{equation}
-\frac12 \{L_\mu^\dagger L_\mu,\rho\}
\end{equation}
is required to preserve the trace and to ensure that the total generator is compatible with complete positivity. Indeed, it is easy to verify that \eqref{Lindbladform} preserves the trace:
\begin{align}
\frac{d}{dt}\Tr\rho
&=
-i\,\Tr[H,\rho]
+
\sum_\mu
\left(
\Tr(L_\mu \rho L_\mu^\dagger)
-\frac12 \Tr(L_\mu^\dagger L_\mu \rho)
-\frac12 \Tr(\rho L_\mu^\dagger L_\mu)
\right)=
0 ,
\end{align}
where we used cyclicity of the trace. Hermiticity is also preserved manifestly. The highly nontrivial part of the theorem is that the form \eqref{Lindbladform} is precisely what is needed to guarantee complete positivity for the full semigroup.

\paragraph{Sketch of the derivation in finite dimensions} We now give a brief physics-style sketch of why the generator must take the form \eqref{GKSLgeneral}. Since $\Phi_t$ is a semigroup, its generator is determined by the short-time behavior of the map. For a very small time step $\delta t$, one may write
\begin{equation}
\Phi_{\delta t} = \mathds{1}  + \delta t\,\mathcal L + O(\delta t^2) .
\end{equation}
Because each $\Phi_{\delta t}$ is CPTP, it admits a Kraus representation,
\begin{equation}
\Phi_{\delta t}[\rho]
=
\sum_\mu M_\mu(\delta t)\,\rho\,M_\mu^\dagger(\delta t) ,
\qquad
\sum_\mu M_\mu^\dagger(\delta t) M_\mu(\delta t)=\mathds{1}  .
\end{equation}
The idea is now to expand the Kraus operators for small $\delta t$.

Choose one Kraus operator, say $M_0$, to contain the leading identity contribution, and write
\begin{equation}
M_0 = \mathds{1}  + \delta t\,K + O(\delta t^2) .
\end{equation}
All the remaining Kraus operators must start at order $\sqrt{\delta t}$,
\begin{equation}
M_\mu = \sqrt{\delta t}\,L_\mu + O(\delta t)
\qquad (\mu\ge 1) .
\end{equation}
This scaling is forced by the normalization condition: contributions from the nontrivial Kraus operators must enter at order $\delta t$, not order $1$.

Substituting these expansions into the operator-sum formula gives
\begin{align}
\Phi_{\delta t}[\rho]
&=
M_0 \rho M_0^\dagger
+
\sum_{\mu\ge 1} M_\mu \rho M_\mu^\dagger
\nonumber\\
&=
\rho
+
\delta t\,(K\rho+\rho K^\dagger)
+
\delta t\sum_{\mu\ge 1} L_\mu \rho L_\mu^\dagger
+
O(\delta t^2) .
\end{align}
Hence the generator has the provisional form
\begin{equation}
\mathcal L[\rho]
=
K\rho+\rho K^\dagger
+
\sum_{\mu\ge 1} L_\mu \rho L_\mu^\dagger .
\end{equation}

We must now impose trace preservation. Expanding
\begin{equation}
\sum_\mu M_\mu^\dagger M_\mu = \mathds{1} 
\end{equation}
to first order in $\delta t$, one finds
\begin{equation}
K+K^\dagger+\sum_{\mu\ge 1} L_\mu^\dagger L_\mu = 0 .
\end{equation}
Writing $K $ as a Hermitian and an anti-Hermitian part, we see that the general solution may be written as
\begin{equation}
K = -\frac12 \sum_{\mu\ge 1} L_\mu^\dagger L_\mu - iH ,
\end{equation}
where $H=H^\dagger$ is Hermitian. Substituting this back into the generator yields
\begin{align}\label{LindbladS}
\mathcal L[\rho]
&=
-i[H,\rho]
+
\sum_{\mu\ge 1}
\left(
L_\mu \rho L_\mu^\dagger
-\frac12 L_\mu^\dagger L_\mu \rho
-\frac12 \rho L_\mu^\dagger L_\mu
\right)
\nonumber\\
&=
-i[H,\rho]
+
\sum_{\mu\ge 1}
\left(
L_\mu \rho L_\mu^\dagger
-\frac12 \{L_\mu^\dagger L_\mu,\rho\}
\right) .
\end{align}
This is precisely the Lindblad form. This derivation explains the basic logic of the theorem. The Hamiltonian commutator arises from the anti-Hermitian part of the short-time correction to the leading Kraus operator, while the dissipative terms arise from the Kraus operators of order $\sqrt{\delta t}$. The full theorem refines this argument and shows that the coefficient matrix in a fixed operator basis must be positive semidefinite. The theorem also proves that the conjecture is true, namely that the map generated by the Lindblad equation is a CPTP semigroup. For a time-independent Lindblad generator, the semigroup property is immediate from $\Phi_t=e^{t\mathcal L}$, and trace preservation follows directly from $\Tr \mathcal L[\rho]=0$. The genuinely nontrivial part of the theorem is that this form also guarantees complete positivity of the full evolution; ordinary positivity then follows as a consequence. 

A few comments are worth making. First, the Lindblad representation is not unique: different choices of Lindblad operators may generate the same dynamical semigroup. Second, the theorem applies to time-homogeneous Markovian evolution. More general time-dependent master equations also occur frequently in physics, and they need not define semigroups. We will return to such distinctions later if needed. For now, \eqref{Lindbladform} gives the most general master equation describing continuous-time Markovian CPTP evolution.\\

If desired, one may rescale the jump operators and write $L_\mu=\sqrt{\gamma_\mu}\,A_\mu$ with $\gamma_\mu\ge 0$ and the normalization
\begin{equation}
\Tr(A^{\dagger}_{\mu}A_{\mu})=1 \,.
\end{equation}
The Lindblad equation then takes the form
\begin{equation}
\frac{d\rho}{dt}
=
-i[H,\rho]
+
\sum_\mu \gamma_\mu
\left(
A_\mu \rho A_\mu^\dagger
-\frac12 \{A_\mu^\dagger A_\mu,\rho\}
\right) .
\end{equation}
In this way, the operator $A_\mu$ characterizes the \textit{nature} of the dissipative process, namely which transition, decoherence channel, or noise mechanism is acting on the system, while the coefficient $\gamma_\mu$ measures the corresponding \textit{rate}, that is, the timescale or strength with which that process acts. This separation is often convenient physically, since it isolates the structure of the jump from the positive coefficient that controls how rapidly it affects the evolution.\\


It is worth noting a useful special case. If a jump operator is Hermitian, $  L ^\dagger = L  $, then its dissipative contribution may be rewritten as a double commutator
\begin{equation}\label{doublecommutator}
\frac{d\rho}{dt}
=
-i[H,\rho]
-\frac12 
\left[L ,[L ,\rho]\right]\,.
\end{equation}
This form is often useful because it makes the structure of dephasing and diffusion more transparent, and it also makes unitality manifest, since double commutators annihilate the identity operator.\footnote{More generally, if one has a family of Hermitian operators $L_\mu^\dagger=L_\mu$ with a real symmetric positive semidefinite coefficient matrix $c_{\mu\nu}$, then
\begin{equation}
\sum_{\mu,\nu} c_{\mu\nu}
\left(
L_\mu \rho L_\nu - \frac12\{L_\nu L_\mu,\rho\}
\right)
=
-\frac12
\sum_{\mu,\nu} c_{\mu\nu}[L_\mu,[L_\nu,\rho]] .
\end{equation}}


\paragraph{Redundancies in the Lindblad representation}

The Lindblad form of the generator is not unique. The first redundancy is a unitary rotation among the jump operators. If
\begin{equation}
L_\mu \to L'_\mu=\sum_\nu U_{\mu\nu}L_\nu ,
\qquad
U^\dagger U=1 ,
\end{equation}
then the Lindblad equation is unchanged. Indeed,
\begin{equation}
\sum_\mu L'_\mu \rho L_\mu^{\prime\dagger}
=
\sum_\mu L_\mu \rho L_\mu^\dagger ,
\qquad
\sum_\mu L_\mu^{\prime\dagger}L'_\mu
=
\sum_\mu L_\mu^\dagger L_\mu .
\end{equation}
The second redundancy is a shift of the jump operators by multiples of the identity. If
\begin{equation}
L_\mu \to L'_\mu=L_\mu+\alpha_\mu  ,
\qquad
\alpha_\mu\in\mathbb C ,
\end{equation}
then the same evolution is obtained provided one simultaneously shifts the Hamiltonian as
\begin{equation}\label{Hshift}
H \to H'
=
H-\frac{i}{2}\sum_\mu\left(\alpha_\mu^*L_\mu-\alpha_\mu L_\mu^\dagger\right) .
\end{equation}
A short calculation shows that
\begin{equation}
-i[H',\rho]
+
\sum_\mu
\left(
L'_\mu \rho L_\mu^{\prime\dagger}
-\frac12\{L_\mu^{\prime\dagger}L'_\mu,\rho\}
\right)
=
-i[H,\rho]
+
\sum_\mu
\left(
L_\mu \rho L_\mu^\dagger
-\frac12\{L_\mu^\dagger L_\mu,\rho\}
\right) .
\end{equation}
This makes precise the statement that part of the coherent Hamiltonian evolution can be shuffled into the jump operators and vice versa. A third and trivial redundancy is that the Hamiltonian may be shifted by a multiple of the identity.

These ambiguities are often summarized by the combined transformations
\begin{align}\label{combinedredundancy}
L_\mu &\to L'_\mu=\sum_\nu U_{\mu\nu}L_\nu+\alpha_\mu \mathds{1} ,\\
H &\to H'
=
H-\frac{i}{2}\sum_\mu\left(\alpha_\mu^*L_\mu-\alpha_\mu L_\mu^\dagger\right)+c \mathds{1}\,,
\end{align}
under which the Liouvillian is invariant. To remove this redundancy, in finite dimension one may choose them to be traceless, or more generally orthogonal to the identity.


\paragraph{Non-diagonal form and Kossakowski matrix}

The diagonal Lindblad form \eqref{Lindbladform} is not the only useful way of writing the dissipative part of the generator. Let ${F_a}$ be a fixed basis of traceless operators on the system Hilbert space, for example an orthonormal basis with respect to the Hilbert-Schmidt inner product\footnote{The discussion below assumes that the identity component has been separated from the dissipative operator basis, as in \eqref{GKSLgeneral}. If identity pieces are allowed in the jump operators, part of the dissipative-looking contribution can be moved into the Hamiltonian by the redundancy discussed below. For this reason it is often cleanest to work with traceless basis operators $F_a$ when defining the Kossakowski matrix.}. Then the most general dissipative term may be written as
\begin{equation}\label{KossakowskiForm}
\mathcal D[\rho]=\sum_{a,b} c_{ab}\left(F_a \rho F_b^\dagger-\frac12 \{F_b^\dagger F_a,\rho \}\right) ,
\end{equation}
where the matrix $C=(c_{ab})$ is called the Kossakowski matrix. The condition that the evolution be completely positive is encoded in the positivity of this matrix,
\begin{equation}
C\geq 0 .
\end{equation}
In this form, the labels $a,b$ do not yet correspond to independent physical jump processes. Rather, they label the basis operators used to expand the dissipative part of the generator. The off-diagonal entries of $C$ describe correlations, or interference, between these different operator directions.

Since $C$ is positive semidefinite, it can be diagonalized. Writing
\begin{equation}
c_{ab}=\sum_\mu U_{\mu a}\,\gamma_\mu\,U_{\mu b}^{*} ,\qquad \gamma_\mu\geq 0 .
\end{equation}
one may define
\begin{equation}\label{diagonaljumps}
L_\mu=\sqrt{\gamma_\mu}\sum_a U_{\mu a}F_a .
\end{equation}
Substituting this expression into \eqref{KossakowskiForm}, one recovers the diagonal Lindblad form \eqref{Lindbladform}. Thus the usual jump operators are the eigenvectors of the positive matrix $C$, multiplied by the square roots of the corresponding eigenvalues. The eigenvalues $\gamma_\mu$ are the associated positive rates.

This also clarifies the meaning of the number of jump operators. If some eigenvalues of $C$ vanish, the corresponding $L_\mu$ vanish and do not contribute to the master equation. Therefore the minimal number of nonzero jump operators in diagonal Lindblad form is
\begin{equation}
\#_{\rm jumps}={\rm rank}(C).
\end{equation}
Equivalently, the rank of the Kossakowski matrix counts the number of independent dissipative directions in operator space. A master equation may look as if it contains many coupled dissipative terms in a chosen basis ${F_a}$, but after diagonalizing $C$ only ${\rm rank}(C)$ independent jump operators remain.


\paragraph{The Heisenberg-picture master equation}

So far we have discussed the Lindblad equation in the Schr\"odinger picture, where the density matrix evolves in time while observables are kept fixed. It is often equally useful to describe the same dynamics in the Heisenberg picture, where instead the state is held fixed and the observables evolve.  The expectation value of an observable $\O$ is
\begin{equation}
\langle \O \rangle_t := \Tr\bigl(\rho(t)\,\O\bigr) .
\end{equation}
We now ask whether this time dependence may instead be attributed to the observable rather than to the state.

To this end, we define the adjoint superoperator $\mathcal L^\dagger$ by the relation
\begin{equation}\label{adjointdef}
\Tr\bigl(\mathcal L[\rho]\,\O\bigr)
=
\Tr\bigl(\rho\,\mathcal L^\dagger[\O]\bigr)
\end{equation}
for all density matrices $\rho$ and observables $\O$. This is simply the analogue, for superoperators, of the usual adjoint of an operator with respect to the Hilbert--Schmidt inner product.

Using \eqref{LindbladS}, one finds
\begin{align}
\Tr\bigl(\mathcal L[\rho]\,\O\bigr)
&=
-i\,\Tr\bigl([H,\rho]\O\bigr)
+
\sum_\mu
\Tr\!\left(L_\mu \rho L_\mu^\dagger \O\right)
-
\frac12
\sum_\mu
\Tr\!\left(\{L_\mu^\dagger L_\mu,\rho\}\O\right)
\nonumber\\
&=
i\,\Tr\bigl(\rho[H,\O]\bigr)
+
\sum_\mu
\Tr\!\left(\rho\,L_\mu^\dagger \O L_\mu\right)
-
\frac12
\sum_\mu
\Tr\!\left(\rho\,\{L_\mu^\dagger L_\mu,\O\}\right) ,
\end{align}
where we used cyclicity of the trace. Therefore the adjoint Liouvillian acts on observables as
\begin{equation}\label{LindbladH}
\mathcal L^\dagger[\O]
=
i[H,\O]
+
\sum_\mu
\left(
L_\mu^\dagger \O L_\mu
-\frac12 \{L_\mu^\dagger L_\mu,\O\}
\right) .
\end{equation}
The Heisenberg-picture master equation is then
\begin{equation}\label{Heisenbergmaster}
\frac{d\O}{dt} = \mathcal L^\dagger[\O] .
\end{equation}

This equation is the direct open-system analogue of the ordinary Heisenberg equation
\begin{equation}
\frac{d\O}{dt}= i[H,\O]
\end{equation}
for closed systems. The difference is the presence of the extra terms involving the Lindblad operators, which encode the effect of dissipation, noise, and decoherence on observables. The equivalence between the Schr\"odinger and Heisenberg pictures follows immediately from \eqref{adjointdef}. 

It is useful to note one immediate consequence. Since the Schr\"odinger-picture evolution preserves the trace of the density matrix, the identity operator must be a fixed point of the Heisenberg-picture evolution $ \mathcal L^\dagger[\mathds 1]=0$, as is easy to check directly from \eqref{LindbladH}. This is the Heisenberg-picture reflection of trace preservation in the Schr\"odinger picture.

For later use, it is also convenient to write the Heisenberg equation in the equivalent form
\begin{equation}
\frac{d\O}{dt}
=
i[H,\O]
+
\frac12
\sum_\mu
\left(
L_\mu^\dagger[\O,L_\mu]
+
[L_\mu^\dagger,\O]L_\mu
\right) .
\end{equation}
This form is obtained by a simple rearrangement of the dissipative terms, and sometimes makes the interpretation more transparent.

In applications, the Heisenberg-picture master equation is sometimes the most efficient way to derive evolution equations for expectation values of observables, such as occupation numbers, spin components, or correlation functions. Rather than solving for the full density matrix and then computing traces, one may evolve the relevant observables directly.


\subsection{Microphysical assumptions leading to the Lindblad equation}

So far we have discussed the Lindblad equation from a structural point of view: it is the most general generator of a continuous Markovian CPTP semigroup. We now explain, at a more microscopic level, under which physical assumptions such an equation arises from an underlying unitary evolution of a system coupled to an environment.

\paragraph{Microscopic setup}

We consider a total Hilbert space
\begin{equation}
\mathcal H = \mathcal H_S \otimes \mathcal H_E
\end{equation}
and a total Hamiltonian of the form
\begin{equation}\label{Htotmicro}
H_{\rm tot} = H_S + H_E + H_{\rm int} ,
\end{equation}
where $H_S$ acts on the system, $H_E$ on the environment, and the interaction Hamiltonian is taken to be
\begin{equation}\label{Hintmicro}
H_{\rm int} = \sum_\alpha S_\alpha \otimes X_\alpha .
\end{equation}
Here the operators $S_\alpha$ act on $\mathcal H_S$ and the operators $X_\alpha$ act on $\mathcal H_E$. The density matrix of the total system evolves unitarily according to
\begin{equation}
\frac{d\rho_{SE}}{dt} = -i[H_{\rm tot},\rho_{SE}] .
\end{equation}
Our goal is to derive from this a closed equation for the reduced density matrix
\begin{equation}
\rho_S(t) := \Tr_E \rho_{SE}(t) .
\end{equation}

\paragraph{Interaction picture and exact reduced dynamics}

It is convenient to work in the interaction picture with respect to the free Hamiltonian
\begin{equation}
H_0 := H_S + H_E .
\end{equation}
Defining
\begin{equation}
\tilde \rho_{SE}(t) := e^{iH_0 t}\rho_{SE}(t)e^{-iH_0 t} ,
\qquad
\tilde H_{\rm int}(t) := e^{iH_0 t}H_{\rm int}e^{-iH_0 t} ,
\end{equation}
the evolution equation becomes
\begin{equation}\label{intpicexact}
\frac{d\tilde \rho_{SE}}{dt} = -i[\tilde H_{\rm int}(t),\tilde \rho_{SE}(t)] .
\end{equation}
Integrating once and substituting back gives the exact integro-differential equation
\begin{equation}\label{born-exact-pre}
\frac{d\tilde \rho_{SE}}{dt}
=
-i[\tilde H_{\rm int}(t),\tilde \rho_{SE}(0)]
-
\int_0^t ds\,
[\tilde H_{\rm int}(t),[\tilde H_{\rm int}(s),\tilde \rho_{SE}(s)]] .
\end{equation}
Tracing over the environment then yields an exact but not closed equation for $\tilde \rho_S(t)$. In this exact form, the system dynamics is in general non-Markovian and retains memory of its past through the integral kernel and through the full state $\tilde \rho_{SE}(s)$ inside the integral.

\paragraph{Factorized initial state}

The first standard assumption is that the initial density matrix factorizes,
\begin{equation}\label{microfactorized}
\rho_{SE}(0)=\rho_S(0)\otimes \rho_E .
\end{equation}
As discussed earlier, this excludes initial system--environment correlations. It is already enough to guarantee that the reduced evolution may be described by a state-independent map on $\rho_S(0)$. It is more general than assuming a factorized pure state: the environment state $\rho_E$ may be mixed, and in many physical applications it is thermal.

We will also assumes that the environment state is stationary with respect to the free environment Hamiltonian,
\begin{equation}\label{rhostationary}
[H_E,\rho_E]=0 .
\end{equation}
This ensures that environment correlation functions depend only on time differences and not on absolute time. It is also convenient to shift the interaction so that
\begin{equation}
\Tr_E(\rho_E X_\alpha)=0 ,
\end{equation}
which removes the first-order term in the reduced evolution.

\paragraph{Born approximation}

The next assumption is weak coupling between system and environment. Physically, one assumes that the environment is so large that its state is only weakly perturbed by the interaction, and that system--environment correlations remain small. One then replaces the exact total state inside the memory kernel by the factorized approximation
\begin{equation}\label{Bornapprox}
\tilde \rho_{SE}(s) \approx \tilde \rho_S(s)\otimes \rho_E .
\end{equation}
This is the \textit{Born approximation}. Substituting \eqref{Bornapprox} into the exact reduced equation produces an equation that is already closed in $\tilde \rho_S$, but still nonlocal in time:
\begin{equation}\label{BornME}
\frac{d\tilde \rho_S}{dt}
=
-\int_0^t ds\,
\Tr_E\Bigl(
[\tilde H_{\rm int}(t),[\tilde H_{\rm int}(s),\tilde \rho_S(s)\otimes \rho_E]]
\Bigr) .
\end{equation}
At this stage one has a weak-coupling master equation with memory.

\paragraph{Markov approximation}

The next step is the Markov approximation. The physical idea is that the environment has a very short correlation time $\tau_E$, while the system evolves on a much longer timescale $\tau_S$. Equivalently, the bath quickly forgets the information it acquires from the system. Then, inside the memory kernel, one may approximate
\begin{equation}
\tilde \rho_S(s) \approx \tilde \rho_S(t)
\end{equation}
whenever the bath correlation functions are appreciable only for $t-s\lesssim \tau_E$. As a second step, we also extend the upper limit of the integral to infinity:
\begin{equation}
\int_0^t ds \;\longrightarrow\; \int_0^\infty ds .
\end{equation}
With these approximations, the evolution becomes local in time:
\begin{equation}\label{BornMarkovME}
\frac{d\tilde \rho_S}{dt}
=
-\int_0^\infty ds\,
\Tr_E\Bigl(
[\tilde H_{\rm int}(t),[\tilde H_{\rm int}(t-s),\tilde \rho_S(t)\otimes \rho_E]]
\Bigr) .
\end{equation}
This is the \textit{Born--Markov master equation} . It is already first order in time and time local, but it is not yet guaranteed to be in Lindblad form.

\paragraph{Decomposition into energy differences}

To go further, one resolves the system operators into components of definite energy difference. Writing the spectral decomposition of the system Hamiltonian as
\begin{equation}
H_S = \sum_n E_n |n\rangle\langle n| ,
\end{equation}
one decomposes each system operator $S_\alpha$ as
\begin{equation}\label{Somegadef}
S_\alpha = \sum_\omega A_\alpha(\omega) ,
\end{equation}
where
\begin{equation}\label{Aomegadef}
A_\alpha(\omega)
:=
\sum_{E_n-E_m=\omega}
|m\rangle\langle m|S_\alpha|n\rangle\langle n| .
\end{equation}
By construction, these operators satisfy
\begin{equation}\label{definitebohr}
[H_S,A_\alpha(\omega)] = -\omega\,A_\alpha(\omega) .
\end{equation}
Equivalently, in the interaction picture,
\begin{equation}
e^{iH_S t}A_\alpha(\omega)e^{-iH_S t}
=
e^{-i\omega t}A_\alpha(\omega) .
\end{equation}
Thus the interaction-picture system operators split into harmonics labeled by the energy differences of the system.

\paragraph{Secular approximation}

After inserting the decomposition \eqref{Somegadef} into the Born--Markov equation, one encounters terms oscillating as
\begin{equation}\label{secularapprox}
e^{i(\omega-\omega')t} .
\end{equation}
If distinct energy differences are well separated compared with the dissipative rates, these rapidly oscillating terms average to nearly zero over the coarse-grained timescale relevant to the reduced dynamics. One therefore discards all contributions with $\omega\neq\omega'$, keeping only the resonant terms with equal energy differences. This is the \textit{secular approximation}.

Physically, the secular approximation means that only processes with the same energy transfer interfere coherently over long times. Mathematically, it is this step that typically ensures that the generator takes genuine GKSL form and that populations and coherences decouple in the energy basis.

\paragraph{Emergence of the Lindblad equation}

After the Born, Markov, and secular approximations, one arrives at a master equation of the form
\begin{equation}\label{secular-gksl}
\frac{d\tilde \rho_S}{dt}
=
\sum_{\omega}\sum_{\alpha,\beta}
\gamma_{\alpha\beta}(\omega)
\left(
A_\beta(\omega)\tilde \rho_S A_\alpha^\dagger(\omega)
-\frac12
\left\{
A_\alpha^\dagger(\omega)A_\beta(\omega),\tilde \rho_S
\right\}
\right) ,
\end{equation}
The coefficients $\gamma_{\alpha\beta}(\omega)$ are determined by the environment two-point correlation matrix
\begin{equation}
C_{\alpha\beta}(t)
:=
\Tr_E\!\left(\rho_E\,X_\alpha^\dagger(t) X_\beta(0)\right),
\qquad
X_\alpha(t):=e^{iH_E t}X_\alpha e^{-iH_E t} .
\end{equation}
Its Fourier transform at the transition frequency $\omega$ gives
\begin{equation}
\gamma_{\alpha\beta}(\omega)
:=
\int_{-\infty}^{+\infty} dt\,e^{i\omega t}\,C_{\alpha\beta}(t) .
\end{equation}
The matrix $\gamma_{\alpha\beta}(\omega)$ is positive semidefinite for every fixed $\omega$. Indeed, for any complex numbers $v_\alpha$,
\begin{align}
\sum_{\alpha,\beta} v_\alpha^*\,\gamma_{\alpha\beta}(\omega)\,v_\beta
&=
\int_{-\infty}^{+\infty} dt\,e^{i\omega t}\,
\Tr_E\!\left(\rho_E\,X^\dagger(t)X(0)\right),
\qquad
X:=\sum_\beta v_\beta X_\beta ,
\end{align}
and by Bochner's theorem this Fourier transform is non-negative because the kernel
\begin{equation}
f(t):=\Tr_E\!\left(\rho_E\,X^\dagger(t)X(0)\right)
\end{equation}
is of positive type. This positivity guarantees that the dissipative part can be written in Lindblad form. These dissipative terms come together with an additional Hamiltonian correction, the so-called \textit{Lamb-shift term},
\begin{equation}
-i[H_{\rm LS},\tilde \rho_S] .
\end{equation}
The Lamb-shift Hamiltonian may be written explicitly in terms of the transition-frequency operators as
\begin{equation}
H_{\rm LS}
=
\sum_{\omega}\sum_{\alpha,\beta}
S_{\alpha\beta}(\omega)\,
A_\alpha^\dagger(\omega)A_\beta(\omega) ,
\end{equation}
where the coefficients $S_{\alpha\beta}(\omega)$ are determined by the imaginary, or principal-value, part of the bath correlation functions. Thus the same operators $A_\alpha(\omega)$ that appear in the dissipator also determine the Hamiltonian renormalization induced by the environment.

Transforming back to the Schr\"odinger picture, one obtains
\begin{equation}\label{microscopic-Lindblad}
\frac{d\rho_S}{dt}
=
-i[H_S+H_{\rm LS},\rho_S]
+
\sum_{\omega}\sum_{\alpha,\beta}
\gamma_{\alpha\beta}(\omega)
\left(
A_\beta(\omega)\rho_S A_\alpha^\dagger(\omega)
-\frac12
\left\{
A_\alpha^\dagger(\omega)A_\beta(\omega),\rho_S
\right\}
\right) .
\end{equation}
This is of GKSL form, provided that the matrices $\gamma_{\alpha\beta}(\omega)$ are positive semidefinite. In microscopic derivations, these coefficients are determined by Fourier transforms of bath correlation functions, and their positivity follows from positivity properties of those correlators.

\paragraph{Summary of assumptions}

Let us summarize the microphysical assumptions that typically lead from unitary system--environment dynamics to the Lindblad equation:

\begin{enumerate}
\item \textit{Factorized initial state:}
\begin{equation}
\rho_{SE}(0)=\rho_S(0)\otimes \rho_E .
\end{equation}

\item \textit{Stationary environment state:}
\begin{equation}
[H_E,\rho_E]=0 .
\end{equation}

\item \textit{Weak coupling / Born approximation:} the environment is only weakly perturbed and one neglects the build-up of system--environment correlations,
\begin{equation}
\tilde \rho_{SE}(t)\approx \tilde \rho_S(t)\otimes \rho_E .
\end{equation}

\item \textit{Short bath memory / Markov approximation:} the bath correlation time is much shorter than the system evolution time, so the memory kernel becomes local in time.

\item \textit{Secular approximation:} rapidly oscillating terms with different energy differences are dropped, keeping only resonant contributions.
\end{enumerate}

The first four assumptions already lead to a time-local Born--Markov master equation. The last one is what typically turns that equation into manifest Lindblad form. Thus the Lindblad equation should be understood not as the exact reduced dynamics of a generic open system, but as the effective description obtained when the environment is weakly coupled, rapidly forgetting, and the system dynamics is viewed on coarse-grained timescales for which non-resonant oscillatory terms average out.


\subsection{Time-dependent dynamical maps, composition law and CP divisibility}\label{sec:CPdiv}

In cosmology and gravity, explicit time dependence is the rule rather than the exception. The background geometry evolves in time, the Hamiltonian of the subsystem is generally time dependent, and the same is true, in general, for the effective couplings induced by the environment. For this reason, the notion of a one-parameter dynamical map is too restrictive: $ \Phi$ does not only depend on the length of the time interval $ t_{1}-t_{0}$ by which we evolve, but also on the starting time $ t_{0}$. Therefore, we need to discuss the more general framework of two-parameter dynamical maps.

The basic point is already visible in the simplest unitary setting. Consider a closed system with time-dependent Hamiltonian \(H(t)\). Its density matrix obeys
\begin{equation}
\rho(t)=U(t,t_0)\,\rho(t_0)\,U^\dagger(t,t_0) ,
\end{equation}
with
\begin{equation}
U(t,t_0)
=
T\exp\!\left(
-i\int_{t_0}^{t}ds\,H(s)
\right) .
\end{equation}
This defines a \textit{two}-parameter family of maps
\begin{equation}
\Phi(t,t_0)[\rho]
:=
U(t,t_0)\,\rho\,U^\dagger(t,t_0) .
\end{equation}
These maps always satisfy the \textit{composition law} 
\begin{equation}\label{compositionlaw}
\Phi(t_2,t_0)=\Phi(t_2,t_1)\circ \Phi(t_1,t_0) ,
\qquad
t_2\ge t_1\ge t_0 .
\end{equation}
However, in general they do not define a semigroup. Indeed, the semigroup property would require that the map depend only on the elapsed time,
\begin{equation}
\Phi(t,t_0)=\Phi(t-t_0) ,
\end{equation}
and hence satisfy
\begin{equation}
\Phi(t+s)=\Phi(t)\circ \Phi(s) .
\end{equation}
This is lost as soon as the generator depends explicitly on time. 


This motivates the following general framework. Let \(\Phi(t,t_0)\) be a differentiable family of linear maps acting on density matrices, satisfying
\begin{equation}
\rho(t)=\Phi(t,t_0)\bigl[\rho(t_0)\bigr] .
\end{equation}
Assuming differentiability with respect to the upper time argument, one defines the time-local generator \(\mathcal L_t\) by
\begin{equation}\label{timelocalgenerator}
\partial_t \Phi(t,t_0)=\mathcal L_t\circ \Phi(t,t_0) .
\end{equation}
Formally, the solution is the time-ordered exponential
\begin{equation}\label{timeorderedexp}
\Phi(t,t_0)
=
T\exp\!\left(
\int_{t_0}^{t} ds\,\mathcal L_s
\right) .
\end{equation}
Equation \eqref{timeorderedexp} is the natural generalization of \(e^{(t-t_0)\mathcal L}\) when the generator is time dependent. In the special case in which \(\mathcal L_t\) is actually independent of time, one recovers the semigroup case studied previously.

\paragraph{CP divisibility} The natural time-dependent replacement of the semigroup notion is \textit{CP divisibility}. One says that a family of maps \(\Phi(t,t_0)\) is CP divisible if, for every triple of times \(t_2\ge t_1\ge t_0\), there exists an intermediate map \(\Phi(t_2,t_1)\) such that
\begin{equation}
\Phi(t_2,t_0)=\Phi(t_2,t_1)\circ \Phi(t_1,t_0) ,
\end{equation}
and such that the intermediate map \(\Phi(t_2,t_1)\) is completely positive and trace preserving. In other words, not only the full evolution from \(t_0\) to \(t_2\), but every time step in between, in particular infinitesimal time steps, must itself define a physical quantum channel.

This notion is especially useful because it admits a precise characterization in terms of the time-local generator. Under the assumptions of differentiability and invertibility of the map, one has the following theorem: the family \(\Phi(t,t_0)\) is CP divisible if and only if the generator \(\mathcal L_t\) is of GKSL form at every time \(t\). Explicitly, this means that for each \(t\),
\begin{equation}\label{timedepGKSL}
\mathcal L_t[\rho]
=
-i[H(t),\rho]
+
\sum_{\alpha,\beta} c_{\alpha\beta}(t)
\left(
L_\alpha(t)\rho L_\beta^\dagger(t)
-\frac12\{L_\beta^\dagger(t)L_\alpha(t),\rho\}
\right) ,
\end{equation}
with the Kossakowski matrix \(c_{\alpha\beta}(t)\) a positive semidefinite matrix for every \(t\). Equivalently, after diagonalizing  \(c_{\alpha\beta}(t)\), one may write
\begin{equation}
\mathcal L_t[\rho]
=
-i[H(t),\rho]
+
\sum_\mu \gamma_\mu(t)
\left(
A_\mu(t)\rho A_\mu^\dagger(t)
-\frac12\{A_\mu^\dagger(t)A_\mu(t),\rho\}
\right) ,
\end{equation}
with $  \gamma_\mu(t)\ge 0  $. Thus the time-dependent generalization of Lindblad evolution consists simply in allowing both the Hamiltonian and the jump operators, or equivalently the canonical rates, to depend explicitly on time.

From the point of view of cosmology, this is the appropriate framework. One should generically expect a two-parameter propagator \(\Phi(t,t_0)\), not a one-parameter semigroup. If the reduced dynamics is sufficiently local in time and the instantaneous generator is of GKSL form with non-negative rates, then the evolution is CP divisible. These considerations will also be important later from the Schwinger--Keldysh point of view. A local-in-time effective action naturally leads to a time-local generator, but in a cosmological background that generator will in general be explicitly time dependent.


\section{Solutions of the Lindblad equation}\label{sec:3}

Having derived the general form of Markovian open-system evolution, we now turn to solving the Lindblad equation in concrete situations. This is the stage at which the abstract structure of the previous section becomes physically transparent: one can identify stationary states, track the evolution of populations and coherences, and understand explicitly how dissipation, decoherence, and noise manifest themselves in time. Even simple examples already display a rich interplay between the Hamiltonian part of the evolution and the genuinely open-system effects encoded in the jump operators.

The aim of this section is twofold. First, we develop some general intuition for the qualitative behaviour of Lindblad evolution, including the role of stationary states and unitality. Second, we illustrate these ideas in explicit examples, focusing on the qubit and the dissipative harmonic oscillator. These examples will serve as a bridge between the general formalism of the master equation and the field-theoretic language introduced later through the Schwinger--Keldysh formalism.


\subsection{Perturbative generators and secular resummation}\label{ssec:resummation}

In this subsection, we critically distinguish between knowing the Liouvillian generator to a given order in a perturbative expansion and knowing the corresponding solution to the same order. The two notions need not coincide at late times, because perturbative solutions may contain secular terms that grow with time and eventually invalidate the fixed-order expansion. We will explain how solving a truncated generator non-perturbatively can resum a distinguished class of such secular contributions.

In many applications, the Liouvillian governing the reduced dynamics is known only as an expansion in some small parameter. We therefore write
\begin{equation}\label{generator expansion}
\mathcal{L}
=
\epsilon \mathcal{L}_{1}
+
\epsilon^{2}\mathcal{L}_{2}
+
\epsilon^{3}\mathcal{L}_{3}
+\cdots ,
\end{equation}
where $\epsilon\ll 1$ may denote a weak system-environment coupling, a ratio of energy/time scales, or some other effective expansion parameter. For simplicity, we begin by assuming that the generators $\mathcal{L}_{n}$ are time independent. The master equation is
\begin{equation}
\frac{d\rho}{dt}
=
\mathcal{L}\rho .
\end{equation}

One possible approach is to solve this equation perturbatively by expanding
\begin{equation}
\rho(t)
=
\rho^{(0)}(t)
+
\epsilon \rho^{(1)}(t)
+
\epsilon^{2}\rho^{(2)}(t)
+\cdots .
\end{equation}
For a time-independent generator and initial condition $\rho(0)=\rho_{0}$, the exact formal solution is
\begin{equation}
\rho(t)
=
e^{t\mathcal{L}}\rho_{0}.
\end{equation}
Expanding both the generator and the exponential gives
\begin{align}
\rho(t)
={}&
\rho_{0}
+
\epsilon t\,\mathcal{L}_{1}\rho_{0}+
\epsilon^{2}
\left(
\frac{t^{2}}{2}\mathcal{L}_{1}^{2}
+
t\mathcal{L}_{2}
\right)\rho_{0}
+\cdots .
\end{align}
The terms proportional to increasing powers of time are called \emph{secular terms}. At any fixed time they are suppressed by powers of $\epsilon$, but the perturbative expansion breaks down when
\begin{equation}
\epsilon t
\sim
1.
\end{equation}
In this regime, all terms of the form
\begin{equation}
\frac{(\epsilon t)^{n}}{n!}\mathcal{L}_{1}^{n}\rho_{0}
\end{equation}
are of comparable size and should be resummed.

A natural and often useful approximation is therefore to truncate the generator at leading order,
\begin{equation}
\mathcal{L}
\simeq
\epsilon \mathcal{L}_{1},
\end{equation}
but solve the resulting evolution equation exactly:
\begin{equation}
\boxed{
\rho_{\mathrm{LO}}(t)
=
e^{\epsilon t\mathcal{L}_{1}}\rho_{0}.
}
\end{equation}
This expression resums all repeated insertions of the leading generator $\mathcal{L}_{1}$. It does not, however, determine the contributions involving $\mathcal{L}_{2},\mathcal{L}_{3},\ldots$. In particular, at second order it contains
\begin{equation}
\frac{\epsilon^{2}t^{2}}{2}\mathcal{L}_{1}^{2}\rho_{0},
\end{equation}
but not
\begin{equation}
\epsilon^{2}t\,\mathcal{L}_{2}\rho_{0}.
\end{equation}
The leading resummation is therefore useful in a regime in which
\begin{equation}
\epsilon t
=
O(1),
\qquad
\epsilon^{2}t
\ll
1.
\end{equation}
In this window, the repeated action of $\mathcal{L}_{1}$ has an order-one effect, while the corrections generated by the omitted higher-order Liouvillians remain perturbatively small.


\paragraph{Elementary decay.}
The simplest and canonical example of this phenomenon is the (classical, deterministic) equation for decay
\begin{equation}
\dot p(t)
=
-\epsilon\gamma p(t),
\qquad
\gamma>0.
\end{equation}
Comparing this very simple example to \eqref{generator expansion} we see that only the leading order generator $ -\gamma$ is non-vanishing. Keeping only the first correction in a strict perturbative expansion gives
\begin{equation}
p(t)
\simeq
p_{0}
\left(
1-\epsilon\gamma t
\right),
\end{equation}
which breaks down for $t\sim(\epsilon\gamma)^{-1}$ and eventually becomes negative. Solving exactly the equation with the leading (and only) generator instead gives
\begin{equation}
p(t)
=
p_{0}e^{-\epsilon\gamma t}.
\end{equation}
In this example, the leading equation is the full equation, so the resummation happens to be exact.


\paragraph{A nonlinear higher-order correction.}
To see more clearly what the leading resummation does and does not capture, consider
\begin{equation}
\dot p(t)
=
-\epsilon\gamma p(t)
-
\epsilon^{2}\delta \, p(t)^{2},
\qquad
p(0)=p_{0},
\end{equation}
with $ \gamma$ and $ \delta$ some parameters and $ \epsilon$ an expansion parameter. If the generator is truncated at leading order in $ \epsilon$, one finds
\begin{equation}
p_{\mathrm{LO}}(t)
=
p_{0}e^{-\epsilon\gamma t}.
\end{equation}
This resums all repeated effects of the linear term proportional to $\gamma$, but it does not include the nonlinear evolution generated by the term proportional to $\delta$.

The full equation is still simple enough to solve exactly\footnote{Defining \(q(t)
:= 1/p(t)\), one obtains
\begin{equation}
\dot q(t)
=
\epsilon\gamma q(t)
+
\epsilon^{2}\delta .
\end{equation}
It follows that
\begin{equation}
q(t)
=
e^{\epsilon\gamma t}
\left[
\frac{1}{p_{0}}
+
\frac{\epsilon\delta}{\gamma}
\left(
1-e^{-\epsilon\gamma t}
\right)
\right],
\end{equation}}
\begin{equation}
p(t)
=
\frac{
p_{0}e^{-\epsilon\gamma t}
}{
1+
\dfrac{\epsilon\delta p_{0}}{\gamma}
\left(
1-e^{-\epsilon\gamma t}
\right)
}.
\end{equation}
Expanding around the leading resummed result gives
\begin{equation}
p(t)
=
p_{\mathrm{LO}}(t)
\left[
1
-
\frac{\epsilon\delta p_{0}}{\gamma}
\left(
1-e^{-\epsilon\gamma t}
\right)
+
O(\epsilon^{2})
\right].
\end{equation}
Thus the exponential resummation captures the secular terms generated by the leading linear decay, but it does not reconstruct the genuinely new nonlinear dynamics entering at higher order. In this example the higher-order correction remains uniformly perturbative and does not generate a new secular combination proportional to $\epsilon^{2}t$.


\paragraph{An exactly soluble nonlinear quantum example.}
The same idea applies directly to the master equation of a density matrix. Consider a harmonic oscillator with number operator \( N := a^{\dagger}a \), and take both the Hamiltonian and a Hermitian jump operator to be arbitrary functions of $N$:
\begin{equation}
H
=
h(N),
\qquad
L
=
f(N).
\end{equation}
The Lindblad equation is
\begin{equation}
\dot\rho
=
-i[h(N),\rho]
+
\gamma
\left[
f(N)\rho f(N)
-
\frac{1}{2}
\left\{
f(N)^{2},\rho
\right\}
\right].
\end{equation}
Although $h$ and $f$ may be nonlinear functions, the equation is exactly soluble because $H$ and $L$ are diagonal in the number basis. Defining
\begin{equation}
h_{n}
:=
h(n),
\qquad
f_{n}
:=
f(n),
\end{equation}
one finds that each matrix element evolves independently 
\begin{equation}
\dot\rho_{nm}
=
\left[
-i\left(h_{n}-h_{m}\right)
-
\frac{\gamma}{2}
\left(f_{n}-f_{m}\right)^{2}
\right]
\rho_{nm}.
\end{equation}
Therefore
\begin{equation}
\rho_{nm}(t)
=
\exp\left[
-i\left(h_{n}-h_{m}\right)t
-
\frac{\gamma t}{2}
\left(f_{n}-f_{m}\right)^{2}
\right]
\rho_{nm}(0).
\end{equation}
The matrix units $|n\rangle\langle m|$ are eigenoperators of the Liouvillian, and the exponential resums the repeated action of the full nonlinear generator on each of them.

For example, one may choose
\begin{equation}
h(N)
=
\omega N
+
\frac{\chi}{2}N(N-1),
\qquad
f(N)
=
N^{2}.
\end{equation}
The Hamiltonian then contains a so-called Kerr interaction, while the jump operator produces nonlinear dephasing. The exact solution is nevertheless obtained immediately from the formula above. This example illustrates that an interacting or nonlinear Liouvillian can be explicitly resummed whenever it admits a sufficiently simple algebraic structure, such as a basis of Liouvillian eigenoperators.

\paragraph{Time-dependent generators.}
The discussion above assumed a time-independent Liouvillian. The generalization to an explicitly time-dependent generator,
\begin{equation}
\frac{d\rho}{dt}
=
\mathcal{L}(t)\rho,
\end{equation}
is conceptually straightforward. If the generator commutes with itself at different times,
\begin{equation}
\left[
\mathcal{L}(t),
\mathcal{L}(t')
\right]
=
0,
\end{equation}
the evolution map is
\begin{equation}
\rho(t)
=
\exp\left[
\int_{t_{0}}^{t}dt'\,
\mathcal{L}(t')
\right]
\rho(t_{0}).
\end{equation}
This is the case in the decay equation above. In this case, the relevant secular quantity is then the integrated rate
\begin{equation}
\epsilon
\int_{t_{0}}^{t}dt'\,
\gamma(t'),
\end{equation}
rather than simply $\epsilon(t-t_{0})$.

For a general time-dependent Liouvillian, the generators at different times need not commute. The solution is then written as the time-ordered exponential
\begin{equation}
\rho(t)
=
\mathcal{T}
\exp\left[
\int_{t_{0}}^{t}dt'\,
\mathcal{L}(t')
\right]
\rho(t_{0}).
\end{equation}
If
\begin{equation}
\mathcal{L}(t)
=
\epsilon\mathcal{L}_{1}(t)
+
\epsilon^{2}\mathcal{L}_{2}(t)
+\cdots,
\end{equation}
solving the equation with $\mathcal{L}_{1}(t)$ retained exactly resums all time-ordered repeated insertions of the leading generator, while it does not include contributions involving $\mathcal{L}_{2}(t),\mathcal{L}_{3}(t),\ldots$. When the resummed generator, for example $\mathcal{L}_{1}(t)$, has instantaneous GKSL form with non-negative rates at every time, the resulting evolution map is completely positive and divisible, as discussed in Section \ref{sec:CPdiv}. This will be an important consideration in justifying the positivity-improved form of Schwinger-Keldysh effective actions.  


\subsection{Stationary states and unitality}

A central notion in the study of open-system dynamics is that of a stationary state, or fixed point, of the evolution. Given a Lindblad generator $\mathcal L$, a density matrix $\rho_\ast$ is called stationary if it satisfies
\begin{equation}
\mathcal L[\rho_\ast] = 0 .
\end{equation}
As a consequence $\dot \rho_{\ast}=0 $. In terms of the dynamical semigroup $\Phi_t=e^{t\mathcal L}$, this is the same as
\begin{equation}
\Phi_t[\rho_\ast]=\rho_\ast
\qquad
\text{for all } t\geq 0 .
\end{equation}
Thus stationary states are the fixed points of the open-system evolution.

For a closed system, stationarity reduces to the familiar condition
\begin{equation}
-i[H,\rho_\ast]=0 ,
\end{equation}
so that $\rho_\ast$ must commute with the Hamiltonian. In particular, any energy eigenstate, and more generally any density matrix diagonal in the energy basis, is stationary under purely unitary evolution. For an open system, by contrast, the dissipative part of the Liouvillian plays an equally important role, and the stationary state is determined by the full equation
\begin{equation}
-i[H,\rho_\ast]
+
\sum_\mu
\left(
L_\mu \rho_\ast L_\mu^\dagger
-\frac12\{L_\mu^\dagger L_\mu,\rho_\ast\}
\right)
=0 .
\end{equation}
The resulting stationary state need not commute with $H$, and in general reflects a balance between coherent dynamics driven by $ H$ and environmental effects driven by $ L_{\mu}$.

The existence and uniqueness of stationary states is an important qualitative property of the Lindblad evolution. If there exists a unique stationary state $\rho_\ast$ and if the dynamics is sufficiently relaxing, then one expects to find the system in that state after a long time for a large class of initial states. In that case the open system loses memory of its initial condition and approaches a universal late-time state. If instead there are several stationary states, or conserved quantities that obstruct relaxation, then the late-time behavior may retain partial memory of the initial state. 


\paragraph{Unital maps and the non-decreae of entropy} The special case in which the stationary state is the identity operator deserves particular attention. A linear map $\Phi$ acting on operators on a finite-dimensional Hilbert space is called \textit{unital} if it preserves the identity operator,
\begin{equation}
\Phi[\mathds 1] = \mathds 1 .
\end{equation}
Recall that, if the Hilbert space has dimension $d$, then the maximally mixed state is $ \mathbf{1}/d $. Thus unitality means that a state carrying no information, namely the maximally mixed state, remains unchanged under the action of the map.

This property is not required for a physical map to be admissible. Many physically important channels are not unital. A basic example, to be discussed at the end of this section, is relaxation of a qubit toward its ground state: such a process is trace preserving and completely positive, but it does not preserve the maximally mixed state and is therefore not unital. By contrast, pure dephasing is unital, since it suppresses coherences without selecting a preferred pure state as late-time attractor. In this sense, unitality is closely related to the distinction between decoherence and dissipation. Unital channels typically describe mixing or dephasing dynamics, while non-unital channels are needed to describe genuine relaxation toward a non-maximally mixed stationary state.

It is useful to compare unitality with trace preservation. Trace preservation is the statement
\begin{equation}
\Tr\Phi[\rho] = \Tr\rho ,
\end{equation}
and expresses conservation of total probability in the Schr\"odinger picture. Unitality is, in a precise sense, the dual notion. Let $\Phi^\dagger$ denote the adjoint map with respect to the Hilbert--Schmidt pairing,
\begin{equation}
\Tr\bigl(\Phi[\rho]\O\bigr) = \Tr\bigl(\rho\,\Phi^\dagger[\O]\bigr) .
\end{equation}
Then, choosing $ \O = \mathds{1} $ we find that 
\begin{equation}
\Phi \text{ trace preserving}
\qquad \Longleftrightarrow \qquad
\Phi^\dagger[\mathds{1} ]=\mathds{1}  ,
\end{equation}
while
\begin{equation}
\Phi \text{ unital}
\qquad \Longleftrightarrow \qquad
\Phi^\dagger \text{ trace preserving}.
\end{equation}
Thus, in the Heisenberg picture, unitality of the Schr\"odinger-picture map corresponds to conservation of the trace for the adjoint map acting on observables.

In the special case of a one-parameter semigroup generated by a Liouvillian $\mathcal L$, unitality is equivalent to $ \mathcal L[\mathds{1} ]=0 $. For a Lindblad generator
\begin{equation}
\mathcal L[\rho]
=
-i[H,\rho]
+
\sum_\mu
\left(
L_\mu \rho L_\mu^\dagger
-\frac12\{L_\mu^\dagger L_\mu,\rho\}
\right) ,
\end{equation}
this condition becomes
\begin{equation}
\sum_\mu [L_\mu,L_\mu^\dagger]=0 .
\end{equation}
Therefore a Lindblad evolution is unital precisely when the dissipative terms preserve the identity operator.

A particularly important consequence of unitality in finite dimensions is that it prevents the von Neumann entropy from decreasing. The detailed proof of the statement is involved, so we will present an abridged version. Let the Hilbert space dimension be $d$, and define the maximally mixed state
\begin{equation}
\tau := \frac{\mathds{1} }{d} .
\end{equation}
For any density matrix $\rho$, the relative entropy with respect to $\tau$ is
\begin{align}\label{omitted}
D(\rho\|\tau)
&:=
\Tr\!\left(\rho\ln\rho - \rho\ln\tau\right)
=
\Tr(\rho\ln\rho) - \Tr\!\left(\rho\ln\frac{\mathds{1} }{d}\right)
\nonumber\\
&=
\Tr(\rho\ln\rho) + (\ln d)\Tr\rho
=
-\;S(\rho)+\ln d .
\end{align}
Now let $\Phi$ be a unital CPTP map. Since $\Phi$ is CPTP, quantum relative entropy is monotone under $\Phi$:
\begin{equation}
D(\rho\|\sigma)\geq D(\Phi[\rho]\|\Phi[\sigma]) .
\end{equation}
This is a powerful result, which we invoke without proof. Applying this with $\sigma=\tau$, and using unitality, we obtain
\begin{equation}
D(\rho\|\tau)\geq D(\Phi[\rho]\|\tau) .
\end{equation}
Using \eqref{omitted} on both sides gives
\begin{equation}
\ln d - S(\rho) \geq \ln d - S(\Phi[\rho]) ,
\end{equation}
and therefore
\begin{equation}
S(\Phi[\rho]) \geq S(\rho) .
\end{equation}
Thus a unital CPTP map cannot decrease the von Neumann entropy.

This result gives a sharp information-theoretic characterization of unital evolution. In finite dimensions, unital maps are precisely those CPTP maps that leave the maximally mixed state unchanged. They cannot decrease the von Neumann entropy and, when relaxing, may drive states toward more mixed configurations rather than toward a preferred pure state. This is why unitality naturally appears in the discussion of decoherence and entropy growth, whereas non-unitality is the hallmark of dissipative relaxation toward distinguished stationary states.


\subsection{Populations, the Pauli master equation, and dissipation}

We now return to the Lindblad equation and ask how it acts on the diagonal and off-diagonal entries of the density matrix in a physically distinguished basis. A natural choice is the energy eigenbasis of the system Hamiltonian,
\begin{equation}
H|n\rangle = E_n |n\rangle .
\end{equation}
Other choices of basis may be justified by physical considerations, but we will stick to energy because of its universal appeal. In this basis, the diagonal matrix elements
\begin{equation}
p_n(t) := \rho_{nn}(t) = \langle n|\rho(t)|n\rangle
\end{equation}
are interpreted as the \textit{populations} of the corresponding energy levels, while the off-diagonal elements
\begin{equation}
\rho_{mn}(t) := \langle m|\rho(t)|n\rangle ,
\qquad m\neq n ,
\end{equation}
represent the \textit{coherences}.

Starting from the Lindblad equation, we first derive an equation for the populations. Taking the matrix element $\langle n| \cdots |n\rangle$, one finds
\begin{align}\label{RHSparenthesis}
\dot p_n
&=
-i\langle n|[H,\rho]|n\rangle
+
\sum_\mu
\left(
\langle n|L_\mu \rho L_\mu^\dagger|n\rangle
-\frac12 \langle n|\{L_\mu^\dagger L_\mu,\rho\}|n\rangle
\right) .
\end{align}
Since the Hamiltonian is diagonal in this basis,
\begin{equation}
\langle n|[H,\rho]|n\rangle = E_n \rho_{nn}-\rho_{nn}E_n = 0 ,
\end{equation}
so the Hamiltonian part does not contribute directly to the evolution of the populations, as expected. In general, the term in parentheses on the right-hand side of \eqref{RHSparenthesis} still depends on the off-diagonal components of $\rho$.

A major simplification occurs when the diagonal sector is closed under the evolution, so that the populations evolve independently of the coherences. This is precisely what happens when the Lindblad operators may be chosen to induce transitions between definite energy levels. Let us assume that the jump operators are proportional to transitions between individual energy eigenstates%
\footnote{More generally, this closure holds whenever the jump operators have a definite charge (i.e.~energy) with respect to the Hamiltonian, in the absence of degeneracies, namely when they satisfy
\begin{equation}
[H,L_\mu]=-\Omega_\mu L_\mu ,
\end{equation}
so that they connect only energy eigenspaces whose energies differ by the fixed amount $\Omega_\mu$.
}%
,
\begin{equation}\label{chargedjumps}
L_{nm} := \sqrt{W_{nm}}\,|n\rangle\langle m| ,
\qquad n\neq m ,
\end{equation}
where $W_{nm}\geq 0$ is the rate for the transition $|m\rangle\to |n\rangle$. In microphysical derivations of the Lindblad equation, this assumption follows from the so-called secular approximation, as discussed around \eqref{secularapprox}. 
Taking the diagonal matrix element $\langle n|\cdots|n\rangle$ of the dissipator, the first term gives
\begin{align}
\sum_{k\neq \ell}\langle n|L_{k\ell}\rho L_{k\ell}^\dagger|n\rangle=
\sum_{k\neq \ell}
W_{k\ell}\,
\langle n|k\rangle\langle \ell|\rho|\ell\rangle\langle k|n\rangle
=
\sum_{\ell\neq n} W_{n\ell}\,\rho_{\ell\ell}
=
\sum_{m\neq n} W_{nm}\,p_m .
\end{align}
Similarly, since
\begin{equation}
L_{k\ell}^\dagger L_{k\ell}
=
W_{k\ell}\,|\ell\rangle\langle \ell| ,
\end{equation}
the anticommutator term contributes
\begin{align}
-\frac12
\sum_{k\neq \ell}
\langle n|\{L_{k\ell}^\dagger L_{k\ell},\rho\}|n\rangle
&=
-\frac12
\sum_{k\neq \ell}
W_{k\ell}\,
\langle n|\{|\ell\rangle\langle\ell|,\rho\}|n\rangle
\nonumber\\
&=
-\frac12
\sum_{k\neq \ell}
W_{k\ell}\,
\bigl(\delta_{n\ell}\rho_{\ell n}+\rho_{n\ell}\delta_{\ell n}\bigr)
\nonumber\\
&=
-\sum_{k\neq n} W_{kn}\,\rho_{nn}
=
-\sum_{m\neq n} W_{mn}\,p_n .
\end{align}
Thus the diagonal sector is closed: the right-hand side depends only on the populations and not on the coherences. Combining gain and loss terms gives 
\begin{equation}\label{PauliME}
\dot p_n
=
\sum_m
\left(
W_{nm} p_m - W_{mn} p_n
\right) .
\end{equation}
This is the \textit{Pauli master equation}. It has the same structure as an ordinary classical rate equation: the population of level $n$ increases by inflow from other levels $m$ and decreases by outflow from level $n$ to those same levels.

The normalization of the density matrix is preserved automatically. Indeed,
\begin{align}
\sum_n \dot p_n
&=
\sum_{n,m}
\left(
W_{nm}p_m - W_{mn}p_n
\right)
=0 \,.
\end{align}
Equation \eqref{PauliME} makes the notion of \textit{dissipation} very concrete. Dissipation refers to the irreversible redistribution of populations among the energy levels, typically accompanied by exchange of energy with the environment. In the presence of dissipation, an initially excited state need not remain excited. Instead, the population flows among levels according to the transition rates $W_{nm}$. In many physically relevant situations, especially for a cold environment, the net flow is from more excited states toward less excited ones. In that sense dissipation is directly associated with the decay of excited-state populations and the relaxation of the system toward a stationary state.


\paragraph{Decoherence} The energy-basis discussion also clarifies the role of coherences. When the jump operators are as in \eqref{chargedjumps}, the coherences decouple from the populations and satisfy separate linear equations. In many common cases one finds
\begin{equation}
\dot \rho_{mn}
=
\left[
-i(E_m-E_n)-\Gamma_{mn}
\right]\rho_{mn},
\qquad
m\neq n ,
\end{equation}
so that
\begin{equation}
\rho_{mn}(t)
=
e^{-i(E_m-E_n)t}e^{-\Gamma_{mn}t}\rho_{mn}(0) .
\end{equation}
Thus the off-diagonal elements oscillate with the energy difference and decay with decoherence rates $\Gamma_{mn}$, while the populations obey the classical-looking rate equation \eqref{PauliME}. This is the sense in which, under suitable conditions, the Lindblad equation separates into a dissipative evolution for populations and a decohering evolution for coherences. Note also that when decoherence has suppressed all off-diagonal elements, the dynamics is effectively classical, only affecting the populations. 


\paragraph{Thermal bath, detailed balance, and KMS-type relations}

We now specialize further and ask what happens when the environment is thermal. Suppose the environment is a thermal bath at inverse temperature $\beta$. One then expects the open-system dynamics to admit the Gibbs state
\begin{equation}\label{Gibbs-state}
\rho_\beta
:=
\frac{e^{-\beta H}}{Z} ,
\qquad
Z := \Tr(e^{-\beta H}) ,
\end{equation}
as a stationary state, at least in the absence of external driving and under suitable ergodicity assumptions. In the energy eigenbasis, the corresponding stationary populations are
\begin{equation}
p_n^{(\beta)}
=
\frac{e^{-\beta E_n}}{Z} .
\end{equation}
Demanding that these populations solve the Pauli master equation imposes constraints on the transition rates. A sufficient local condition is detailed balance,
\begin{equation}\label{detailed-balance}
W_{nm}\,p_m^{(\beta)} = W_{mn}\,p_n^{(\beta)} .
\end{equation}
Substituting the Gibbs weights gives\footnote{Relation \eqref{detailed-balance-ratio} is the population-level manifestation of the more general Kubo--Martin--Schwinger, or KMS, condition satisfied by equilibrium correlation functions of a thermal bath, to be discussed later. }
\begin{equation}\label{detailed-balance-ratio}
\frac{W_{nm}}{W_{mn}}
=
e^{-\beta(E_n-E_m)} .
\end{equation}
Thus upward transitions are Boltzmann suppressed relative to downward ones. This is the precise sense in which a thermal bath favors relaxation toward lower energies while still allowing thermally activated upward jumps. The physical content is transparent. A thermal bath does not merely cause arbitrary dissipation: it constrains the relative strength of excitation and relaxation processes in such a way that the system can settle into thermal equilibrium at the temperature of the bath.


\subsection{Example: the qubit as an open system}

We now return to the qubit and study how the general Lindblad equation acts in this simplest non-trivial example. Without loss of generality\footnote{Any Hermitian $2\times 2$ Hamiltonian can be written as $H=a\,\mathds{1} +\vec h\cdot\vec \sigma$. The term proportional to the identity only shifts all energies by the same constant and has no effect on the dynamics, while a suitable unitary change of basis can always rotate the vector $\vec h$ to the $z$-axis.}
, we take the Hamiltonian to be
\begin{equation}
H = \frac{\omega_0}{2}\sigma_z ,
\end{equation}
so that
\begin{equation}
H|\uparrow\rangle = \frac{\omega_0}{2}|\uparrow\rangle ,
\qquad
H|\downarrow\rangle = -\frac{\omega_0}{2}|\downarrow\rangle .
\end{equation}
Thus $|\uparrow\rangle$ is the excited state and $|\downarrow\rangle$ the ground state. In this basis the density matrix is
\begin{equation}
\rho =
\begin{pmatrix}
\rho_{\uparrow\uparrow} & \rho_{\uparrow\downarrow} \\
\rho_{\downarrow\uparrow} & \rho_{\downarrow\downarrow}
\end{pmatrix},
\qquad
\rho_{\uparrow\uparrow}+\rho_{\downarrow\downarrow}=1 .
\end{equation}
The diagonal entries are the populations, while the off-diagonal entries are the coherences.

For a qubit, a generic jump operator is simply an arbitrary complex $2\times 2$ matrix. Equivalently, it may be expanded in the basis
\begin{equation}
\mathds{1} ,\qquad
\sigma_+ := |\uparrow\rangle\langle\downarrow| = \frac{\sigma_x+i\sigma_y}{2},
\qquad
\sigma_- := |\downarrow\rangle\langle\uparrow| = \frac{\sigma_x-i\sigma_y}{2},
\qquad
\sigma_z ,
\end{equation}
as
\begin{equation}
L = a\,\mathds{1}  + b\,\sigma_+ + c\,\sigma_- + d\,\sigma_z .
\end{equation}
The identity piece is dynamically inessential, since it may be absorbed into a redefinition of the Hamiltonian and of the traceless part of the jump operator. It is therefore natural to single out the physically most relevant processes by choosing jump operators adapted to the energy basis. The three elementary cases are excitation, relaxation, and dephasing:
\begin{equation}
L_\uparrow := \sqrt{\gamma_\uparrow}\,\sigma_+,
\qquad
L_\downarrow := \sqrt{\gamma_\downarrow}\,\sigma_-,
\qquad
L_\phi := \sqrt{\gamma_\phi}\,\sigma_z .
\end{equation}
Here $\gamma_\uparrow$, $\gamma_\downarrow$, and $\gamma_\phi$ are non-negative coefficients. The first raises the energy of the qubit, the second lowers it, and the third produces dephasing in the energy basis without changing the populations.

It is convenient to include all three at once. The Lindblad equation then takes the form
\begin{align}\label{qubitLindblad}
\frac{d\rho}{dt}
&=
-i[H,\rho]
+\gamma_\downarrow
\left(
\sigma_- \rho \sigma_+
-\frac12\{\sigma_+\sigma_-,\rho\}
\right)
+\gamma_\uparrow
\left(
\sigma_+ \rho \sigma_-
-\frac12\{\sigma_-\sigma_+,\rho\}
\right)
\nonumber\\
&\hspace{1cm}
+\gamma_\phi
\left(
\sigma_z \rho \sigma_z
-\rho
\right) .
\end{align}
We now discuss the physical role of each term, first individually and then in combination.

\paragraph{Relaxation}

Suppose first that only
\begin{equation}
L=\sqrt{\gamma_\downarrow}\,\sigma_-
\end{equation}
is present. Then the master equation becomes
\begin{equation}
\frac{d\rho}{dt}
=
-i[H,\rho]
+
\gamma_\downarrow
\left(
\sigma_- \rho \sigma_+
-\frac12\{\sigma_+\sigma_-,\rho\}
\right) .
\end{equation}
A straightforward matrix computation gives
\begin{align}
\dot\rho_{\uparrow\uparrow} &= -\gamma_\downarrow\,\rho_{\uparrow\uparrow} ,
\\
\dot\rho_{\downarrow\downarrow} &= \gamma_\downarrow\,\rho_{\uparrow\uparrow} ,
\\
\dot\rho_{\uparrow\downarrow}
&=
-\left(i\omega_0+\frac{\gamma_\downarrow}{2}\right)\rho_{\uparrow\downarrow} ,
\\
\dot\rho_{\downarrow\uparrow}
&=
\left(i\omega_0-\frac{\gamma_\downarrow}{2}\right)\rho_{\downarrow\uparrow} .
\end{align}
Thus
\begin{equation}
\rho_{\uparrow\uparrow}(t)=e^{-\gamma_\downarrow t}\rho_{\uparrow\uparrow}(0),
\end{equation}
while trace preservation implies
\begin{equation}
\rho_{\downarrow\downarrow}(t)=1-\rho_{\uparrow\uparrow}(t)
=1-e^{-\gamma_\downarrow t}\rho_{\uparrow\uparrow}(0).
\end{equation}
The coherences oscillate and decay as
\begin{equation}
\rho_{\uparrow\downarrow}(t)
=
e^{-i\omega_0 t}e^{-\gamma_\downarrow t/2}\rho_{\uparrow\downarrow}(0) .
\end{equation}
This is the simplest example of dissipation: population is irreversibly transferred from the excited state to the ground state, and the stationary state is
\begin{equation}
\rho_\ast = |\downarrow\rangle\langle\downarrow| .
\end{equation}
If the initial state is pure and excited, the entropy initially increases as the state becomes mixed, then decreases again and asymptotes zero at late times, when the system settles into the pure ground state. Thus the entropy does not change monotonically under relaxation. The case of just $  L_{\uparrow} $ is similar with the excited and ground states switched.


\paragraph{Pure dephasing}

Now suppose that only $ L=\sqrt{\gamma_\phi}\,\sigma_z  $ is present. Then the dissipative contribution is
\begin{equation}
\gamma_\phi(\sigma_z\rho\sigma_z-\rho) \,,
\end{equation}
and one finds
\begin{align}
\dot\rho_{\uparrow\uparrow} &= 0 ,
\\
\dot\rho_{\downarrow\downarrow} &= 0 ,
\\
\dot\rho_{\uparrow\downarrow}
&=
-\left(i\omega_0+2\gamma_\phi\right)\rho_{\uparrow\downarrow} ,
\\
\dot\rho_{\downarrow\uparrow}
&=
\left(i\omega_0-2\gamma_\phi\right)\rho_{\downarrow\uparrow} .
\end{align}
Note that this induces a unital map since $ \rho_{\downarrow\downarrow}=\rho_{\uparrow\uparrow} =1/2$ and $ \rho_{\uparrow\downarrow}=\rho_{\downarrow\uparrow} =0$ is a solution. We see that the populations remain constant, while the coherences oscillate and decay as
\begin{equation}
\rho_{\uparrow\downarrow}(t)
=
e^{-i\omega_0 t}e^{-2\gamma_\phi t}\rho_{\uparrow\downarrow}(0) .
\end{equation}
This is the cleanest example of decoherence without dissipation. There is no energy exchange with the environment and hence no redistribution of populations, but phase information is lost. The preferred basis, or pointer basis, is precisely the energy basis $\{|\uparrow\rangle,|\downarrow\rangle\}$ singled out by $\sigma_z$.

In this case the late-time state is
\begin{equation}
\rho(\infty)
=
\begin{pmatrix}
\rho_{\uparrow\uparrow}(0) & 0 \\
0 & \rho_{\downarrow\downarrow}(0)
\end{pmatrix},
\end{equation}
which is generally mixed unless the initial state already happened to be an energy eigenstate. Therefore pure dephasing typically increases the entropy from zero to a positive asymptotic value and does so monotonically for a generic initial pure superposition.

\paragraph{All three processes together}

Let us now combine the three elementary processes. From \eqref{qubitLindblad}, one finds
\begin{align}
\dot\rho_{\uparrow\uparrow}
&=
-\gamma_\downarrow \rho_{\uparrow\uparrow}
+\gamma_\uparrow \rho_{\downarrow\downarrow} ,
\\
\dot\rho_{\downarrow\downarrow}
&=
\gamma_\downarrow \rho_{\uparrow\uparrow}
-\gamma_\uparrow \rho_{\downarrow\downarrow} ,
\\
\dot\rho_{\uparrow\downarrow}
&=
-\left[
i\omega_0+\frac{\gamma_\downarrow+\gamma_\uparrow}{2}+2\gamma_\phi
\right]\rho_{\uparrow\downarrow} ,
\\
\dot\rho_{\downarrow\uparrow}
&=
\left[
i\omega_0-\frac{\gamma_\downarrow+\gamma_\uparrow}{2}-2\gamma_\phi
\right]\rho_{\downarrow\uparrow} .
\end{align}
The populations therefore obey the Pauli master equation
\begin{equation}\label{qubitPauli}
\dot p_\uparrow
=
-\gamma_\downarrow p_\uparrow + \gamma_\uparrow p_\downarrow ,
\qquad
\dot p_\downarrow
=
\gamma_\downarrow p_\uparrow - \gamma_\uparrow p_\downarrow ,
\end{equation}
with
\begin{equation}
p_\uparrow := \rho_{\uparrow\uparrow},
\qquad
p_\downarrow := \rho_{\downarrow\downarrow}.
\end{equation}
This is precisely the classical rate equation for a two-state Markov process. The diagonal sector has become autonomous. At the same time, the coherences decay according to
\begin{equation}
\rho_{\uparrow\downarrow}(t)
=
e^{-i\omega_0 t}
\exp\!\left[
-\left(\frac{\gamma_\downarrow+\gamma_\uparrow}{2}+2\gamma_\phi\right)t
\right]
\rho_{\uparrow\downarrow}(0) .
\end{equation}
Thus the coherence time is controlled by both the dephasing channel and the population-changing channels.

The stationary populations follow from setting the left-hand side of \eqref{qubitPauli} to zero:
\begin{equation}
\gamma_\downarrow p_\uparrow^{(\ast)}
=
\gamma_\uparrow p_\downarrow^{(\ast)} ,
\qquad
p_\uparrow^{(\ast)}+p_\downarrow^{(\ast)}=1 .
\end{equation}
Hence
\begin{equation}\label{qubitfixedpoint}
p_\uparrow^{(\ast)}
=
\frac{\gamma_\uparrow}{\gamma_\uparrow+\gamma_\downarrow},
\qquad
p_\downarrow^{(\ast)}
=
\frac{\gamma_\downarrow}{\gamma_\uparrow+\gamma_\downarrow} .
\end{equation}
Since the coherences decay to zero, the stationary density matrix is diagonal,
\begin{equation}
\rho_\ast
=
\frac{\gamma_\uparrow}{\gamma_\uparrow+\gamma_\downarrow}
|\uparrow\rangle\langle\uparrow|
+
\frac{\gamma_\downarrow}{\gamma_\uparrow+\gamma_\downarrow}
|\downarrow\rangle\langle\downarrow| .
\end{equation}
This is the generic mixed stationary state of the qubit in the presence of both upward and downward jumps.

\paragraph{Thermal stationary state and detailed balance}

A particularly important case occurs when the environment is thermal. Then the stationary state should be the Gibbs state
\begin{equation}
\rho_\beta
=
\frac{e^{-\beta H}}{\Tr(e^{-\beta H})}
=
\frac{e^{-\beta\omega_0/2}|\uparrow\rangle\langle\uparrow|
+
e^{+\beta\omega_0/2}|\downarrow\rangle\langle\downarrow|}
{e^{-\beta\omega_0/2}+e^{+\beta\omega_0/2}} .
\end{equation}
This requires
\begin{equation}
\frac{p_\uparrow^{(\ast)}}{p_\downarrow^{(\ast)}}
=
e^{-\beta\omega_0} .
\end{equation}
Comparing with \eqref{qubitfixedpoint}, we obtain the detailed-balance condition
\begin{equation}\label{qubitdetailedbalance}
\frac{\gamma_\uparrow}{\gamma_\downarrow}
=
e^{-\beta\omega_0} .
\end{equation}
Thus a thermal bath does not permit arbitrary excitation and relaxation rates: they must be tuned relative to one another so that upward jumps are Boltzmann suppressed with respect to downward jumps. This is the qubit version of the more general KMS relation discussed earlier.

At zero temperature, $\beta\to\infty$, one has
\begin{equation}
\gamma_\uparrow \to 0 ,
\end{equation}
so the excited state decays irreversibly to the ground state. At infinite temperature, $\beta\to 0$, one finds
\begin{equation}
\gamma_\uparrow = \gamma_\downarrow ,
\end{equation}
and the stationary state becomes maximally mixed,
\begin{equation}
\rho_\ast = \frac12 \mathds{1}  .
\end{equation}


\paragraph{Bloch-vector form}

It is useful to translate the above discussion into Bloch-vector language. Writing
\begin{equation}
\rho = \frac12(\mathds{1}  + b_x\sigma_x+b_y\sigma_y+b_z\sigma_z) ,
\end{equation}
one finds
\begin{align}
\dot b_x
&=
-\left(\frac{\gamma_\downarrow+\gamma_\uparrow}{2}+2\gamma_\phi\right)b_x - \omega_0 b_y ,
\\
\dot b_y
&=
-\left(\frac{\gamma_\downarrow+\gamma_\uparrow}{2}+2\gamma_\phi\right)b_y + \omega_0 b_x ,
\\
\dot b_z
&=
-(\gamma_\downarrow+\gamma_\uparrow)b_z + (\gamma_\uparrow-\gamma_\downarrow) .
\end{align}
Thus the transverse components $b_x$ and $b_y$ decay exponentially while precessing around the $z$-axis, whereas the longitudinal component $b_z$ relaxes toward the stationary value
\begin{equation}
b_z^{(\ast)}
=
\frac{\gamma_\uparrow-\gamma_\downarrow}{\gamma_\uparrow+\gamma_\downarrow} .
\end{equation}
This makes the geometry very transparent: dissipation pushes the Bloch vector toward a preferred point on the $z$-axis, while decoherence shrinks its transverse projection. The stationary point lies on the south pole for zero-temperature relaxation, at the center for an infinite-temperature bath, and at an intermediate point for finite temperature.


\subsection{Example: the dissipative harmonic oscillator}

As a second example, let us consider a single quantum harmonic oscillator with Hamiltonian
\begin{equation}
H = \omega \left(a^\dagger a + \frac12\right) ,
\end{equation}
where the annihilation and creation operators satisfy
\begin{equation}
[a,a^\dagger]=1 .
\end{equation}
It is often convenient to introduce the number operator $  N := a^\dagger a  $, whose eigenstates $|n\rangle$ obey
\begin{equation}
N|n\rangle = n|n\rangle ,
\qquad
a|n\rangle = \sqrt{n}\,|n-1\rangle ,
\qquad
a^\dagger|n\rangle = \sqrt{n+1}\,|n+1\rangle .
\end{equation}
The energy eigenstates of the oscillator therefore coincide with the number eigenstates. This makes the oscillator particularly well suited for illustrating, in a simple but non-trivial setting, the notions of dissipation, excitation, dephasing, and thermalization.

The most important jump operators are the lowering operator $a$, the raising operator $a^\dagger$, and the number operator $N$. Respectively, these describe loss of quanta, gain of quanta, and pure dephasing in the number basis. We will discuss the corresponding processes in turn.


\paragraph{Pure damping}

Let us begin with the simplest dissipative choice,
\begin{equation}
L = \sqrt{\kappa}\,a ,
\end{equation}
with $\kappa\geq 0$. The Lindblad equation is
\begin{equation}\label{oscdamp}
\frac{d\rho}{dt}
=
-i[H,\rho]
+
\kappa
\left(
a\rho a^\dagger
-\frac12\{a^\dagger a,\rho\}
\right) .
\end{equation}
This is the oscillator analogue of qubit relaxation. It describes irreversible loss of excitations to the environment.

A particularly efficient way to understand the dynamics is through the Heisenberg-picture master equation. Using the adjoint Liouvillian, one finds
\begin{align}
\frac{da}{dt}
&=
-i[a,H]
+
\kappa
\left(
a^\dagger a a
-\frac12\{a^\dagger a,a\}
\right)
=
-\left(i\omega+\frac{\kappa}{2}\right)a ,
\\
\frac{da^\dagger}{dt}
&=
\left(i\omega-\frac{\kappa}{2}\right)a^\dagger .
\end{align}
Therefore
\begin{equation}
\langle a(t)\rangle
=
e^{-i\omega t}e^{-\kappa t/2}\langle a(0)\rangle .
\end{equation}
Thus the oscillator amplitude rotates with frequency $\omega$ and decays with rate $\kappa/2$.

The occupation number obeys
\begin{align}
\frac{dN}{dt}
&=
\kappa
\left(
a^\dagger N a - \frac12\{a^\dagger a,N\}
\right)
\nonumber\\
&=
-\kappa N ,
\end{align}
so that
\begin{equation}
\langle N(t)\rangle = e^{-\kappa t}\langle N(0)\rangle .
\end{equation}
Thus the mean number of quanta decays to zero exponentially in time. The unique stationary state is therefore the vacuum,
\begin{equation}
\rho_\ast = |0\rangle\langle 0| .
\end{equation}
This is a genuinely dissipative process: it changes the populations in the energy basis and drives the system toward the lowest-energy state.

The master equation also induces a closed rate equation for the diagonal matrix elements
\begin{equation}
p_n(t):=\langle n|\rho(t)|n\rangle .
\end{equation}
A short computation gives
\begin{equation}\label{oscPauliDamp}
\dot p_n
=
\kappa\bigl[(n+1)p_{n+1}-np_n\bigr] .
\end{equation}
This is the Pauli master equation for damping. Population flows from higher occupation numbers to lower ones, and eventually accumulates in the vacuum state.

The effect on the off-diagonal elements is also simple. Writing
\begin{equation}
\rho_{mn}(t):=\langle m|\rho(t)|n\rangle ,
\end{equation}
the Hamiltonian part gives the phase factor $e^{-i\omega(m-n)t}$, while the dissipator damps the matrix elements. In particular, coherences between widely separated number states decay faster than low-lying ones. Thus damping produces both dissipation and decoherence.


\paragraph{Thermal bath}

The natural extension of the previous example is to include both lowering and raising jumps,
\begin{equation}
L_- := \sqrt{\kappa(N_{\rm th}+1)}\,a ,
\qquad
L_+ := \sqrt{\kappa N_{\rm th}}\,a^\dagger ,
\end{equation}
The real parameter $N_{\rm th}\geq 0$, which characterizes the tuning between these two operators, will later be identified with the thermal occupation number of the bath at frequency $\omega$. The Lindblad equation becomes
\begin{align}\label{oscthermal}
\frac{d\rho}{dt}
&=
-i[H,\rho]
+
\kappa(N_{\rm th}+1)
\left(
a\rho a^\dagger
-\frac12\{a^\dagger a,\rho\}
\right)
\nonumber\\
&\hspace{1cm}
+
\kappa N_{\rm th}
\left(
a^\dagger\rho a
-\frac12\{aa^\dagger,\rho\}
\right) .
\end{align}
This is the oscillator analogue of the qubit with both relaxation and excitation. Using the Heisenberg-picture equation, one finds again
\begin{equation}
\frac{da}{dt}
=
-\left(i\omega+\frac{\kappa}{2}\right)a ,
\end{equation}
so the mean oscillator amplitude still decays with the same rate $\kappa/2$. The number operator, however, now obeys
\begin{align}
\frac{dN}{dt}
&=
\kappa(N_{\rm th}+1)
\left(
a^\dagger N a - \frac12\{a^\dagger a,N\}
\right)
+
\kappa N_{\rm th}
\left(
a N a^\dagger - \frac12\{aa^\dagger,N\}
\right)
\nonumber\\
&=
\kappa(N_{\rm th}+1)
\left(
a^\dagger N a - N^2
\right)
+
\kappa N_{\rm th}
\left(
a N a^\dagger - N(N+1)
\right)
\nonumber\\
&=
\kappa(N_{\rm th}+1)
\left(
a^\dagger a(N-1) - N^2
\right)
+
\kappa N_{\rm th}
\left(
aa^\dagger N + aa^\dagger - N(N+1)
\right)
\nonumber\\
&=
\kappa(N_{\rm th}+1)
\left(
N(N-1)-N^2
\right)
+
\kappa N_{\rm th}
\left(
(N+1)N + (N+1) - N(N+1)
\right)
\nonumber\\
&=
-\kappa(N_{\rm th}+1)N+\kappa N_{\rm th}(N+1)
=
-\kappa N+\kappa N_{\rm th}
\nonumber\\
&=
-\kappa (N-N_{\rm th}) .
\end{align}
and therefore
\begin{equation}
\langle N(t)\rangle
=
N_{\rm th}
+
e^{-\kappa t}\bigl(\langle N(0)\rangle-N_{\rm th}\bigr) .
\end{equation}
Thus the system no longer relaxes to the vacuum, but instead to a stationary state with mean occupation number $N_{\rm th}$.

The diagonal matrix elements satisfy the rate equation
\begin{align}\label{oscPauliThermal}
\dot p_n
&=
\kappa(N_{\rm th}+1)\bigl[(n+1)p_{n+1}-np_n\bigr]
\nonumber\\
&\hspace{0.5cm}
+\kappa N_{\rm th}\bigl[np_{n-1}-(n+1)p_n\bigr] .
\end{align}
The first line describes downward transitions, while the second describes upward transitions. This is a direct infinite-dimensional analogue of the two-level Pauli equation studied for the qubit.

The stationary solution is geometric,
\begin{equation}
p_n^{(\ast)}
=
\frac{1}{N_{\rm th}+1}
\left(
\frac{N_{\rm th}}{N_{\rm th}+1}
\right)^n ,
\end{equation}
which is precisely the Gibbs distribution
\begin{equation}
\rho_\beta
=
\frac{e^{-\beta H}}{\Tr(e^{-\beta H})}
\end{equation}
provided one identifies
\begin{equation}\label{NthBose}
N_{\rm th}=\frac{1}{e^{\beta\omega}-1} .
\end{equation}
Equivalently,
\begin{equation}
\frac{\kappa N_{\rm th}}{\kappa(N_{\rm th}+1)}
=
e^{-\beta\omega} .
\end{equation}
This is the oscillator version of detailed balance. Upward and downward jumps are not independent if the bath is thermal.

In the zero-temperature limit one has
\begin{equation}
N_{\rm th}\to 0 ,
\end{equation}
and the thermal master equation reduces to pure damping, which we studied previously. In the high-temperature limit, $N_{\rm th}\gg 1$, upward and downward rates become nearly equal and the stationary state is highly mixed.


\paragraph{Pure dephasing}

A third instructive process is obtained by taking
\begin{equation}
L_\phi := \sqrt{\gamma_\phi}\,N ,
\end{equation}
where $\gamma_\phi\geq 0$. The corresponding Lindblad equation is
\begin{equation}\label{oscdephasing}
\frac{d\rho}{dt}
=
-i[H,\rho]
+
\gamma_\phi
\left(
N\rho N
-\frac12\{N^2,\rho\}
\right) .
\end{equation}
Since $N$ is diagonal in the energy basis, this process does not change the populations. Indeed,
\begin{equation}
\dot p_n = 0 .
\end{equation}
However, the off-diagonal elements decay:
\begin{equation}
\frac{d}{dt}\rho_{mn}
=
-i\omega(m-n)\rho_{mn}
-\frac{\gamma_\phi}{2}(m-n)^2\rho_{mn} .
\end{equation}
Hence
\begin{equation}
\rho_{mn}(t)
=
e^{-i\omega(m-n)t}
e^{-\frac{\gamma_\phi}{2}(m-n)^2 t}
\rho_{mn}(0) .
\end{equation}
This is the clean oscillator analogue of qubit dephasing. There is no dissipation, since the energy populations remain fixed, but there is decoherence, because the off-diagonal elements in the number basis are exponentially suppressed. The pointer basis is therefore the number basis.


\paragraph{Gaussian Lindblad equation} 

It is useful to briefly mention the most general \textit{Gaussian} Lindblad equation for a single oscillator. Here Gaussian means that Gaussian states are mapped into Gaussian states, or equivalently that the evolution of the state is completely determined by the first and second moments of the canonical variables \(x\) and \(p\). This happens when the Hamiltonian is at most quadratic in \(x\) and \(p\), and the jump operators are at most linear in them. In that case, after suitable redefinitions of the quadratic Hamiltonian, the most general one-mode Gaussian Lindblad equation may be written as\footnote{Here we are using the fact that the jump operators are Hermitian to express the dissipator in the double commutator form \eqref{doublecommutator}.}
\begin{align}\label{generalLindbladian}
\frac{d\rho}{dt}
&=
-i[H,\rho]
-\frac{i\kappa}{2}[x,\{p,\rho\}]
-D_{pp}[x,[x,\rho]]
-D_{xx}[p,[p,\rho]]
\nonumber\\
&\hspace{1cm}
+D_{xp}\bigl([x,[p,\rho]]+[p,[x,\rho]]\bigr) ,
\end{align}
where \(H\) is quadratic in \(x\) and \(p\), \(\kappa\) is a friction coefficient, and the three real coefficients \(D_{xx}\), \(D_{xp}=D_{px}\), and \(D_{pp}\) form the symmetric diffusion matrix. The terminology Gaussian reflects the fact that this equation closes on the mean values and covariance matrix of \(x\) and \(p\), without generating higher connected moments if they are absent initially. Complete positivity imposes a constraint on the diffusion matrix, namely
\begin{equation}
D_{pp}D_{xx}-D_{xp}^{2}\geq \frac{\kappa^{2}}{16} ,
\end{equation}
in units with \(\hbar=1\). The simple damping and thermal examples discussed above are special cases of this general Gaussian structure. For example, pure damping corresponds to
\begin{equation}
D_{xx}=\frac{\kappa}{8\omega},
\qquad
D_{pp}=\frac{\kappa\omega}{8},
\qquad
D_{xp}=0 ,
\end{equation}
which saturates the complete-positivity bound. For the thermal bath one finds
\begin{equation}
D_{xx}=\frac{\kappa}{8\omega}(2N_{\rm th}+1),
\qquad
D_{pp}=\frac{\kappa\omega}{8}(2N_{\rm th}+1),
\qquad
D_{xp}=0 ,
\end{equation}
satisfying but not saturating the bound for $  N_{\rm th} \neq 0 $. Dephasing instead is not Gaussian.


\paragraph{Entropy and stationary states}

These three cases illustrate very clearly the distinction between dissipation and decoherence. Damping by $a$ changes the populations and drives the system toward the pure vacuum, so the entropy need not be monotonic: it may first increase as the state becomes mixed and later decrease as the vacuum is approached. The thermal bath with both $a$ and $a^\dagger$ drives the system toward a mixed Gibbs state, and the late-time entropy approaches the thermal entropy. Pure dephasing leaves all populations unchanged but reduces coherence, so it typically increases the entropy unless the initial state was already diagonal in the number basis.

It is also worth emphasizing the relation to the discussion of populations and coherences developed above. In the oscillator example, the number basis is the natural energy basis, and the jump operators $a$ and $a^\dagger$ induce downward and upward transitions between adjacent levels. The resulting Pauli equation has exactly the expected birth--death form. By contrast, the jump operator $N$ carries zero transition frequency and therefore generates pure dephasing without any redistribution of populations.


\section{The Schwinger-Keldysh formalism}\label{sec:4}

So far we have described open-system dynamics in the \textit{operator language} of density matrices and master equations. We now introduce a complementary and extremely powerful formulation based on \textit{path integrals}: the Schwinger--Keldysh, or closed-time-path, formalism \cite{Schwinger:1960qe,Keldysh:1964ud}. This framework is designed to compute real-time expectation values and reduced density matrices, and is therefore ideally suited both to closed systems evolving unitarily and to open systems obtained after integrating out an environment. In particular, it will provide the natural bridge between the quantum-mechanical discussion of the previous sections and the effective field theory perspective that we will later apply to cosmology.

The main idea is to double the fields and evolve them along a forward and a backward time contour, so that probabilities, expectation values, and reduced dynamics can all be encoded in a single functional integral. In this section we first review the construction for closed systems and explain the meaning of the closed-time path and the Keldysh basis. We then show how integrating out an environment leads to the Feynman--Vernon influence functional and discuss the general constraints that unitarity, Hermiticity, and (complete) positivity impose on the resulting effective action. Finally, we illustrate the formalism in simple examples, including the damped harmonic oscillator, Gaussian environments, the semiclassical limit, and a dissipative scalar field.


\subsection{Closed systems, unitary evolution, and the doubled contour}

We now begin our discussion of the Schwinger--Keldysh formalism. The first important point is that this formalism is not intrinsically tied to open systems. It already arises naturally for a closed quantum system, i.e. one that undergoes Hamiltonian evolution. Open-system effects enter only later, when one separates the full Hilbert space into system and environment and integrates out the environmental degrees of freedom.

The natural starting point is therefore the Hamiltonian evolution of a density matrix. Let the Hilbert space of the full closed system be $\mathcal H$, and let the Hamiltonian be $H$. The density matrix evolves as
\begin{equation}
\rho(t) = U(t,t_0)\,\rho(t_0)\,U^\dagger(t,t_0) ,
\qquad
U(t,t_0):=e^{-iH(t-t_0)} ,
\end{equation}
or, equivalently,
\begin{equation}
i\frac{d}{dt}U(t,t_0)=H\,U(t,t_0) ,
\qquad
-i\frac{d}{dt}U^\dagger(t,t_0)=U^\dagger(t,t_0)\,H .
\end{equation}
This already exhibits the basic doubling that lies at the heart of Schwinger--Keldysh (SK): the density matrix has a ket and a bra and therefore evolves both forward in time with $U$, for the ket, and backward in time with $U^\dagger$, for the bra. This should be contrasted with other objects we are familiar with, such as the wavefunction and time-ordered correlation functions, usually computed in the context of scattering amplitudes, both of which only display forward-in-time evolution. In jargon, we say that the density matrix is an \textit{in-in} object, while the wavefunction or time-ordered correlators are \textit{in-out} objects. 

For pedagogical reasons, we will introduce the SK path integral in three steps:
\begin{itemize}
\item First, we focus on computing the expectation value of a single simple operator.
\item Second, we introduce the formalism of external currents $ J$.
\item Third, we discuss the insertion of operators at different times. 
\end{itemize}


\paragraph{The SK path integral for a simple operator} The expectation value of an Heisenberg-picture operator\footnote{As we will see shortly, in this section there will be many different objects denoted by the letter ``$ x$''. For clarity, we sometimes use a hat when $ \hat x$ is a quantum operator.} $ \hat x(t)^{n}$ in the Schr\"odinger picture reads
\begin{equation}\label{Ooperator}
\langle \hat x(t)^{n} \rangle
=
\Tr\!\left(U(t,t_0)\,\rho(t_0)\,U^\dagger(t,t_0)\,\hat x^{n}\right) ,
\end{equation}
where
\begin{equation}
U(t,t_0)
:=
T\exp\!\left[
-i\int_{t_0}^{t} dt\,H \right] .
\end{equation}
and $T$ denotes time ordering. The Schwinger--Keldysh path integral is simply a path-integral representation of this same real-time evolution.

For simplicity, let us consider a single degree of freedom $  x $ with Lagrangian $  L(x,\dot x) $. The key idea is to use the resolution of the identity,
\begin{align}
\mathds{1}=\int_{\mathbb{R}} dx \ket{x}\bra{x}\,,
\end{align} 
three times inside \eqref{Ooperator}. This gives
\begin{align}\label{3resolutions}
\langle \hat x(t)^{n}\rangle&=\int_{\mathbb{R}} dx_{+,0} \int_{\mathbb{R}} dx_{-,0} \int_{\mathbb{R}} dx_{f} \\
& \qquad \times \Tr\left[ U(t,t_0)\ket{x_{+,0}}\bra{x_{+,0}}\rho(t_0)\ket{x_{-,0}}\bra{x_{-,0}} U^\dagger(t,t_0)\ket{x_{f}}\bra{x_{f}} \hat x(t)^{n}
\right] \nonumber 
\end{align}
We note that $ \hat x \ket{x_{f}}=x_{f}\ket{x_{f}}$ with $ x_{f}$ a number. Hence, we can use the cyclicity of the trace to move $ \ket{x_{f}} $ from the left-hand to the right-hand side\footnote{Alternatively, we perform the trace as a fourth integral $  \int dx \ket{x}\bra{x}$ and use $ \bra{x_{f}}\ket{x}=\delta(x_{f}-x)$ to perform this integral, giving the same result.}. One object that appears is
the matrix element of the time-evolution operator $ U$. This is represented by the usual path integral,
\begin{equation}\label{pathintegral}
\langle x_f|U(t,t_0)|x_{+,0}\rangle
=
\int_{x_{+}(t_0)=x_{+,0}}^{x_{+}(t)=x_f}\!\!\!\!\!\!\!\!\!\!\!\!\!\! \mathcal D x_{+}\;
e^{ iS[x_{+}] },
\end{equation}
where $ S$ is the standard action, which we choose to write in Lagrangian language as
\begin{equation}
S[x]:=\int_{t_0}^{t}dt\,L(x,\dot x) .
\end{equation}
The other, related matrix element is
\begin{align}\label{backpathintegral}
\langle x_{-,0}|U^{\dagger}(t,t_0)|x_f\rangle
&=\langle x_{f}|U(t,t_0)|x_{-,0}\rangle^{\ast} = \int_{x_{-}(t_0)=x_{-,0}}^{x_{-}(t)=x_f}\!\!\!\!\!\!\!\!\!\!\!\!\!\! \mathcal D x_{-}\;
e^{ -iS[x_{-}] } ,
\end{align}
Note that $ x_{+}(t)  $ and $ x_{-}(t)  $ are just integration variables for the path integrals. We could call them whatever we want. Here we have chosen to call them different names so we can easily distinguish which one refers to the first path integral describing forward time evolution of the ket and which one describes backward time evolution of the bra. The fact that two path integral variables $ x_{+}$ and $ x_{-}$ are needed to describe a single quantum operator $ \hat x$ is sometimes referred to as the ``doubling of fields'' in the SK formalism. Here we understand that this doubling is just a book-keeping trick to avoid confusion between the two path-integration variables.

Inserting these representations into \eqref{3resolutions}, and writing the initial density matrix in the position basis as
\begin{equation}
\rho(x_{+,0},x_{-,0};t_0):=\langle x_{+,0}|\rho(t_0)|x_{-,0}\rangle ,
\end{equation}
we obtain
\begin{align}
\ex{\hat x(t)^{n}}
&=
\int dx_{+,0}\,dx_{-,0}\,dx_f\;
\rho(x_{+,0},x_{-,0};t_0)
\nonumber\\
&\qquad\times
\int_{x_+(t_0)=x_{+,0}}^{x_+(t)=x_f}\!\!\!\!\!\!\!\!\!\!\!\!\!\! \mathcal D x_+\;
e^{ iS[x_+] } \, \int_{x_-(t_0)=x_{-,0}}^{x_-(t)=x_f}\!\!\!\!\!\!\!\!\!\!\!\!\!\! \mathcal D x_-\;
e^{ -iS[x_-] } \, x_{\pm}(t)^{n}.
\end{align}
This is the Schwinger--Keldysh path integral in the $\pm$ basis. Note that since $ x_{+}(t)=x_{-}(t)=x_{f}$, it doesn't matter if we write $ x_{+}(t)^{n}$ or $ x_{-}(t)^{n}$ above. The difference between these two choices is crucial for unequal time correlators as we discuss in detail in the next subsection. 

We see explicitly that the path integral contains two copies of the dynamical variable, $x_+$ and $x_-$, corresponding to the forward and backward branches. Their final values are identified because of the trace over the final Hilbert space. Averaging over $ x_{f}$ effectively stitches the forward and backward contours together (see Appendix A of \cite{Polchinski:1998rq} for a nice discussion of this). 
\\

Two aspects of the notation are worth stressing. First, there are three very different objects, all involving the letter ``$x $'': 
\begin{itemize}
\item $ \hat x$ is a quantum operator,
\item  $x_{\pm}(t)$ are arbitrary functions of time (not quantum operators!) over which we path-integrate,
\item $ x_{+,0}$, $ x_{-,0}$ and $ x_{f}$ are real numbers over which we integrate.
\end{itemize} 
Second, it's important to appreciate the difference between the standard 1d integral  $\int dx_{\pm,0} $ over the value of the functions $ x_{\pm}(t)$ at time $ t_{0}$ and the two path integrals $ \mathcal{D}x_{\pm}$ over all possible histories $ x_{\pm}(t)$ of the functions $ x_{\pm}(t)$.\\

To keep the notation light, it is often convenient not to display explicitly the integrations over the initial and final boundary data. We will therefore use the shorthand \(\int_{\rm I.C.}\) to denote the path integral with Initial Conditions (I.C.) weighted by the initial density matrix $ \rho(x_{+,0},x_{-,0};t_0) $, it being understood that the two branches are sewn together continuously at the final time $ t$, i.e. $ x_{+}(t)=x_{-}(t)$ as required by the trace. In other words, \(\int_{\rm I.C.}\) stands for the explicit integrations over \(x_{+,0}\), \(x_{-,0}\), and over the common final value \(x_f\), together with the corresponding boundary conditions on the path integral histories. Because of this, the values of $ x_{+}$ and $ x_{-}$ at $ t$ are equal but arbitrary. 

With this convention, the Schwinger--Keldysh representation of the expectation value may be written more compactly as
\begin{align}\label{Zhere}
\ex{\hat x(t)^{n}}
&=
\int_{\rm I.C.}\mathcal D x_+\,
\int_{\rm I.C.}\mathcal D x_-\;
\exp\!\Biggl[
iS[x_+] - iS[x_-] \Biggr] x_{\pm}(t)^{n}.
\end{align}
It is worth emphasizing again the relation to the operator formalism. The path integral has not introduced any new dynamics and nothing intrinsically dissipative has happened yet: this is still just Hamiltonian evolution of a closed system, written in a way adapted to expectation values rather than transition amplitudes. It is merely a different representation of the same operator expression \eqref{Ooperator}. The forward branch encodes the action of $U$, the backward branch encodes the action of $U^\dagger$, and the trace over the final state produces the closed contour. In this sense, the Schwinger--Keldysh path integral is the path-integral representation of the Hamiltonian evolution of the density matrix.


\paragraph{External sources} Instead of defining a different path integral for each expectation value that we want to compute, we can define a single \textit{generating functional}\footnote{This is sometimes also called a \textit{partition function} because of the close analogy to statistical physics for a thermal state with some conserved charges. This nomenclature is perhaps more common in Euclidean signature, while the term generating functional is more common in Lorentzian signature. However, in practice, they are often used interchangeably.} $ Z$ and compute from it all possible expectation values. As we do in the usual path integral, this is achieved by coupling our variables to external sources. Since the SK path integral has two branches, we need sources both on the forward and backward branch. 

For the generating function we define
\begin{equation}\label{Zpmop}
Z[J_+,J_-]
:=
\Tr\!\left(
U_{J_+}(t_f,t_0)\,\rho(t_0)\,U_{J_-}^\dagger(t_f,t_0)
\right) ,
\end{equation}
where, in the operator formalism, the time evolution in the presence of sources is
\begin{equation}
U_{J_\pm}(t_f,t_0)
:=
T\exp\!\left[
-i\int_{t_0}^{t_f} dt\,\bigl(H-J_\pm(t)x\bigr)
\right] .
\end{equation}
Here $ t_{f}$ is just some arbitrary time. We'll argue around \eqref{tf} that physical predictions are independent of $t_{f} $ as long as it is chosen to be later than any insertion of operators. 
The source $J_+(t)$ appears on the forward branch, while $J_-(t)$ on the backward branch. Again, let's use the example of a single quantum mechanical particle in 1 dimension. We will move to quantum field theory later on. For a single coordinate $x$, the time evolution is represented by a path integral, this time with a source 
\begin{align} 
\langle x_f|U_J(t_f,t_0)|x_{+,0}\rangle
&=
\int_{x_{+}(t_0)=x_{+,0}}^{x_{+}(t_f)=x_f}\!\!\!\!\!\!\!\!\!\!\!\!\!\! \mathcal D x_{+}\;
\exp\!\left[
iS[x_{+}] + i\int dt\,J(t)x_{+}(t)
\right] , \\
\langle x_{-,0}|U_J^{\dagger}(t_f,t_0)|x_f\rangle
&=\langle x_{f}|U_J(t_f,t_0)|x_{-,0}\rangle^{\ast} \\
&=
\int_{x_{-}(t_0)=x_{-,0}}^{x_{-}(t_f)=x_f}\!\!\!\!\!\!\!\!\!\!\!\!\!\! \mathcal D x_{-}\;
\exp\!\left[
-iS[x_{-}] - i\int dt\,J(t)x_{-}(t)
\right] ,
\end{align}
We can think of $J_{\pm} $ as just another coupling constant in the theory\footnote{In fact, one can introduce sources for all possible operators. For example, introducing a source for $ x_{\pm}^{2}$ leads, eventually, to the so-called 2PI effective action, while, in field theory, coupling to a spacetime metric $ g_{\mu\nu}$ is tantamount to introducing a source for the energy-momentum tensor.}, except that we allow it to be time-dependent (and spacetime dependent in field theory). Using the condensed notation introduced previously we can write the generating functional as 
\begin{align}
Z[J_+,J_-]
&=
\int_{\rm I.C.}\mathcal D x_+\,
\int_{\rm I.C.}\mathcal D x_-\;
\exp\!\Biggl[
iS[x_+] - iS[x_-]
\nonumber\\
&\hspace{1.5cm}
+ i\int dt\,J_+(t)x_+(t)
- i\int dt\,J_-(t)x_-(t)
\Biggr] .
\end{align}
The point is that functional differentiation of $Z[J_+,J_-]$ with respect to $ J_{\pm}$ generates various correlation functions. To write this down explicitly and avoid confusion, we need a small bit of useful notation. We will use ordinary angle brackets $\langle\cdots\rangle$ for quantum expectation values of operators in Hilbert space, as we have done throughout so far. Moreover, we introduce double angle brackets $\langle\!\langle\cdots\rangle\!\rangle$ for averages computed by the Schwinger--Keldysh path integral. Thus, for example,
\begin{equation}
\langle \O\rangle := \Tr \bigl(\rho(t)\,\O\bigr)
\end{equation}
is an operator expectation value, whereas
\begin{equation}
\langle\!\langle F[x_+,x_-]\rangle\!\rangle
:=
\int \mathcal D x_+ \mathcal D x_- \,
F[x_+,x_-]\,
e^{\,iS[x_+]-iS[x_-]}
\end{equation}
is a path-integral average over the doubled variables. As we discuss in detail in the next subsection, the two averages are related but they should not be confused: $x_\pm$ are classical commuting variables, while $\O\sim \hat x(t)^{n}$ denotes quantum operators. 

Varying $ Z$ with respect to its sources around $ J_{\pm}=0$ gives 
\begin{align}
\left[ \prod_{i} \frac{\delta}{J_{+}(t_{i})}\prod_{j} \frac{\delta}{J_{-}(t_{j})}Z[J_{+},J_{-}] \right]_{J_{\pm}=0}=\expi{\left[ \prod_{i} (ix_{+}(t_{i})) \right] \left[ \prod_{j} (-ix_{-}(t_{j})) \right]}\,.
\end{align}
We will see in the next section what this path integral average means in the operator language. 

Before concluding it is worth noticing that setting $J_+=J_-$ gives
\begin{equation}
Z[J,J]=\Tr \rho(t_0) = 1 ,
\end{equation}
provided the density matrix is normalized. This simple identity will later become the basic normalization condition of the Schwinger-Keldysh action. Note that this doubling of the sources and the subsequent doubling of the fields is simply bookkeeping: it allows us to  compute path integral averages involving both $  x_{+}(t)$ and $ x_{-}(t)$.


\subsection{Path ordering, time ordering and the Keldysh basis}

Having introduced the Schwinger--Keldysh generating functional in the $\pm$ basis, we now discuss three related issues. First, we show that the SK path integral is independent of the choice of turning point of the contour under some assumptions. Second,  we translate path integral averages into expectation values of quantum operators. Third, we introduce the Keldysh or $r/a$ basis, in which response and fluctuation are more transparently organized. 

\paragraph{Turn-around point} Before, we imposed time evolution until the time $t $ at which we inserted the operator $ \hat x(t)^{n}$, while later we extended the path integral to an arbitrary time $t_{f} $ in defining the generating functional $ Z$. Here we observe that we are indeed allowed to extend the contour to any time $t_f $ as long as this is later than the time $t $ of operator insertion. This simply follows from the fact that unitary forward time evolution without an operator insertion is exactly cancelled by unitary backwards evolution,
\begin{align}\label{tf}
 U(t_{f},t)U^{\dagger}(t_{f},t)=\mathds{1}\,.
\end{align}
In the case when many operators are inserted at different times, the turn-around point $ t_{f} $ must be later than the latest time of insertion. These observartions allow us to cleanly separate the time of insertion of operators from the turning point of the contour. 

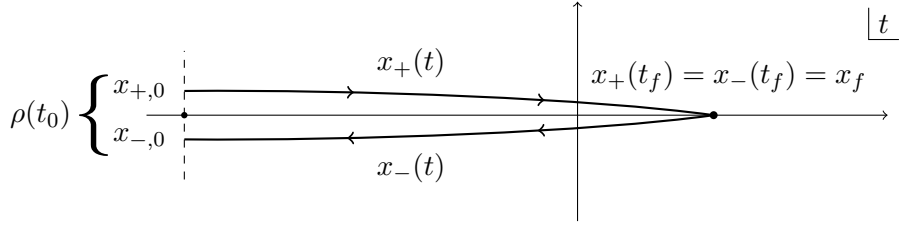
\begin{figure}
\centering
\begin{tikzpicture}[
    x=1cm,
    y=1cm,
    contour/.style={
        thick,
        postaction={decorate}
    },
    forward/.style={
        contour,
        decoration={
            markings,
            mark=at position 0.32 with {\arrow{>}},
            mark=at position 0.68 with {\arrow{>}}
        }
    },
    backward/.style={
        contour,
        decoration={
            markings,
            mark=at position 0.32 with {\arrow{<}},
            mark=at position 0.68 with {\arrow{<}}
        }
    }
]

\draw[->] (-0.5,0) -- (9.3,0);
\draw[->] (5.2,-1.4) -- (5.2,1.5);

\draw (8.05+1,1.42) -- (8.05+1,1.00) -- (8.45+1,1.00);
\node at (9.25,1.20) {$t$};

\draw[dashed] (0,-0.85) -- (0,0.85);

\fill (0,0) circle (1.2pt);

\node at (-1.9,0) {$\rho(t_0)$};
\node[scale=2.8] at (-1.2,0) {$\{$};

\node[left=3pt] at (0,0.32) {$x_{+,0}$};
\node[left=3pt] at (0,-0.32) {$x_{-,0}$};

\draw[forward]
    (0,0.32)
    .. controls (2.3,0.34) and (5.4,0.22) ..
    (7,0);

\draw[backward]
    (0,-0.32)
    .. controls (2.3,-0.34) and (5.4,-0.22) ..
    (7,0);

\node[above=2pt] at (3.0,0.29) {$x_+(t)$};
\node[below=2pt] at (3.0,-0.29) {$x_-(t)$};

%

\fill (7,0) circle (1.5pt);

\node[above=5pt, anchor=south] at (7.2,0)
    {$x_+(t_{f})=x_-(t_{f})=x_f$};
    
\end{tikzpicture}
\caption{The SK contour on the complex $ t$ plane. The parenthesis after $ \rho(t_{0})$ indicates that the boundary conditions $ x_{\pm,0}$ of the path integral should be averaged over the initial density matrix. The forward and backward contours are slightly separated in the imaginary direction for clarity. \label{fig:SK}}
\end{figure}

Henceforth, in the condensed notation of \eqref{Zhere}, we will simply assume that the upper boundary condition of the path integral is imposed at a sufficiently late time $ t_{f}$, the choice of which does not affect the result of the calculation. The resulting contour is depicted in the Figure \ref{fig:SK}. 


\paragraph{The closed-time contour} The goal in life of the path integral is to compute path-ordered correlation functions. If the path runs straight in time, as it is the case for the path integral we use to compute in-out correlators used in the calculation of amplitudes, then path and time ordering coincide. If the path turns around and comes back in time, as for the SK path integral, then \textit{path ordering} and \textit{time ordering} are distinct. On the forward branch, path ordering coincides with the usual time ordering; on the backward branch, path ordering coincides with anti-time ordering. When one insertion lies on each branch, there is no ordinary time ordering between them: the contour itself decides the ordering. 

To make this distinction clear it sometimes useful to think of the two branches in the SK path integral as just two parts of a single path known as the \textit{closed-time contour}. This is possible because forward and backward branches are stitched together by the regularity condition $ x_{+}(t_{f})=x_{-}(t_{f})$ at the turning point $ t_{f}$. In other words, the contour is closed in time by the trace. Indeed, one could simply rewrite the two-branch path integral as a single path integral
\begin{align}
\int_{\rm I.C.}\mathcal D x_+\,
\int_{\rm I.C.}\mathcal D x_- \to \int_{\rm I.C.}^{\rm I.C.}\mathcal{D}X
\end{align}
where $ X(\lambda)$ is a new integration variable encompassing both $ x_{+}$ and $ x_{-}$ and $  \lambda$ is a path variable that is related to time. More concretely, if we conveniently shift time such that the turning point is at $ t_{f}=0\geq t_{0}$, and all operator insertions are prior to that, then we have $ t_{0}\leq \lambda \leq |t_{0}|$ and the identification
\begin{equation}
\begin{cases}
X(\lambda)=x_{+}(\lambda) \text{ for } t_{0}\leq \lambda \leq 0  \\
X(\lambda)=x_{-}(-\lambda) \text{ for } 0 \leq \lambda \leq |t_{0}| \,.
\end{cases}
\end{equation}
Now path ordering on the $ X(\lambda)$ variable is the usual one with insertions at an earlier $ \lambda_{1}$ coming to the right of an insertion at a later $ \lambda_{2}> \lambda_{1}$. This simple path ordering in $ \lambda$ leads to a rich set of different time orderings.\\

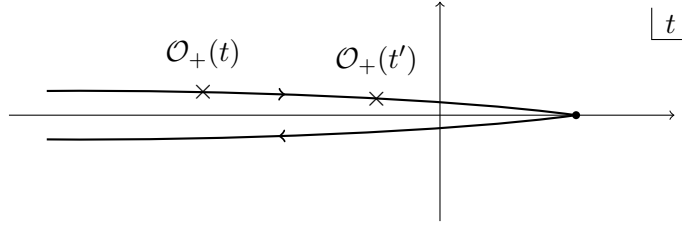
\begin{figure}
\centering

\begin{tikzpicture}[
    x=1cm,
    y=1cm,
    contour/.style={
        thick,
        postaction={decorate}
    },
    forward/.style={
        contour,
        decoration={
            markings,
            mark=at position 0.45 with {\arrow{>}}
        }
    },
    backward/.style={
        contour,
        decoration={
            markings,
            mark=at position 0.45 with {\arrow{<}}
        }
    }
]

\draw[->] (-0.5,0) -- (8.3,0);
\draw[->] (5.2,-1.4) -- (5.2,1.5);

\draw (8.0,1.42) -- (8.0,1.00) -- (8.4,1.00);
\node at (8.25,1.20) {$t$};


\draw[forward]
    (0,0.32)
    .. controls (2.3,0.34) and (5.4,0.22) ..
    coordinate[pos=0.28] (opA)
    coordinate[pos=0.58] (opB)
    (7,0);

\draw[backward]
    (0,-0.32)
    .. controls (2.3,-0.34) and (5.4,-0.22) ..
    (7,0);

\fill (7,0) circle (1.5pt);

\node at (opA) {$\times$};
\node[above=5pt] at (opA) {$\mathcal{O}_{+}(t)$};

\node at (opB) {$\times$};
\node[above=5pt] at (opB) {$\mathcal{O}_{+}(t')$};

\end{tikzpicture}
\caption{The SK contour featuring the insertion of two operators on the plus branch corresponding to a time-ordered 2-point function. All labels of Figure \ref{fig:SK} are left implicit for clarity. \label{fig:SKinsertion}}
\end{figure}

Let's see this in an explicit example. For two-point functions one finds
\begin{align}\label{++}
\langle\!\langle x_+(t)\,x_+(t')\rangle\!\rangle
&=
\langle T\,x(t)x(t')\rangle ,
\\
\langle\!\langle x_-(t)\,x_-(t')\rangle\!\rangle
&=
\langle \bar T\,x(t)x(t')\rangle ,
\\
\langle\!\langle x_+(t)\,x_-(t')\rangle\!\rangle
&=
\langle x(t')x(t)\rangle ,
\\
\langle\!\langle x_-(t)\,x_+(t')\rangle\!\rangle
&=
\langle x(t)x(t')\rangle .
\end{align}
Thus fields on the $+$ branch are time ordered among themselves, fields on the $-$ branch are anti-time-ordered among themselves, and fields on different branches give non-time-ordered correlators, often called Wightman correlators. As an example, the graphical representation of \eqref{++} is shown in Figure \ref{fig:SKinsertion}. 

More generally, we can consider the average over any number of operators on the plus and minus branch. This gives the product of a time-ordered block and an anti-time ordered block
\begin{align}
\expi{\prod_{i=1}^{n}\O_{+}(t_{i})\prod_{j=1}^{m}\O_{-}(t_{j})}=\ex{\bar T \left[ \prod_{j=1}^{m}\O(t_{j}) \right] T \left[ \prod_{i=1}^{n}\O(t_{i}) \right]}\,.
\end{align}
Even more generally, one may be interested in calculating correlators of operators in an arbitrary order \cite{Aleiner:2016eni,Haehl:2017qfl}. These are sometimes referred to as Out-of-Time-Order Correlators (OTOC) and have attracted recent attention as a diagnostic of quantum chaos. These more general orderings can be computed from a generalization of the SK path integral with more and more branches that alternate backward and forward evolution, but each stitched together in a continuous way to the previous and following branch. 

\paragraph{The retarded and advanced basis} We now introduce the \textit{Keldysh basis},
\begin{equation}
x_r := \frac{x_+ + x_-}{2},
\qquad
x_a := x_+ - x_- ,
\end{equation}
with inverse relations
\begin{equation}
x_+ = x_r + \frac{x_a}{2},
\qquad
x_- = x_r - \frac{x_a}{2} .
\end{equation}
The variable $x_r$ is the mean of the two contour copies and is often called the \textit{retarded} component. The variable $x_a$ is their difference and is often called the \textit{advanced} component. We will soon discuss how and when they are related to the classical and quantum contributions, but we will avoid using that terminology in these notes. In this basis, the non-equilibrium constraints discussed earlier take a particularly simple form, and the distinction between fluctuation and response becomes much more transparent. 

The two-point functions in the $r/a$ basis are obtained by simple linear combinations of the $\pm$ correlators. First,
\begin{align}
\langle\!\langle x_r(t)x_r(t^{\prime})\rangle\!\rangle
&=
\frac{1}{4}
\langle\!\langle
\bigl(x_+(t)+x_-(t)\bigr)
\bigl(x_+(t^{\prime})+x_-(t^{\prime})\bigr)
\rangle\!\rangle
\nonumber\\
&=
\frac{1}{4}
\Bigl[
\langle\!\langle x_+(t)x_+(t^{\prime})\rangle\!\rangle
+
\langle\!\langle x_+(t)x_-(t^{\prime})\rangle\!\rangle
\nonumber\\
&\hspace{1.5cm}
+
\langle\!\langle x_-(t)x_+(t^{\prime})\rangle\!\rangle
+
\langle\!\langle x_-(t)x_-(t^{\prime})\rangle\!\rangle
\Bigr]
\nonumber\\
&=
\frac{1}{4}
\Bigl[
\langle  T\,\hat x(t)\hat x(t^{\prime})\rangle
+
\langle \hat x(t^{\prime})\hat x(t)\rangle
\nonumber\\
&\hspace{1.5cm}
+
\langle \hat x(t)\hat x(t^{\prime})\rangle
+
\langle \widetilde{ T}\,\hat x(t)\hat x(t^{\prime})\rangle
\Bigr]
\nonumber\\
&=
\frac{1}{4}
\Bigl[
\theta(t-t^{\prime})
\langle \hat x(t)\hat x(t^{\prime})\rangle
+
\theta(t^{\prime}-t)
\langle \hat x(t^{\prime})\hat x(t)\rangle
\nonumber\\
&\hspace{1.5cm}
+
\langle \hat x(t^{\prime})\hat x(t)\rangle
+
\langle \hat x(t)\hat x(t^{\prime})\rangle
\nonumber\\
&\hspace{1.5cm}
+
\theta(t^{\prime}-t)
\langle \hat x(t)\hat x(t^{\prime})\rangle
+
\theta(t-t^{\prime})
\langle \hat x(t^{\prime})\hat x(t)\rangle
\Bigr]
\nonumber\\
&=
\frac{1}{2}
\Bigl[
\langle \hat x(t)\hat x(t^{\prime})\rangle
+
\langle \hat x(t^{\prime})\hat x(t)\rangle
\Bigr]=
\frac{1}{2}
\langle
\{\hat x(t),\hat x(t^{\prime})\}
\rangle .
\end{align}
namely the anti-commutator of the fields. This is the symmetrized, or \textit{Keldysh correlator}. Next, we can calculate the other two-point function in a similar fashion. Leaving the intermediate steps as an exercise for the reader, one finds 
\begin{align}
\langle\!\langle x_r(t)x_a(t')\rangle\!\rangle
&=
\frac12
\langle\!\langle
\bigl(x_+(t)+x_-(t)\bigr)
\bigl(x_+(t')-x_-(t')\bigr)
\rangle\!\rangle
\nonumber\\
&=
\theta(t-t')\,\langle [\hat x(t),\hat x(t')]\rangle .
\end{align}
This is the retarded correlator, up to the conventional factor of $-i$ often included in field theory. Similarly,
\begin{align}
\langle\!\langle x_a(t)x_r(t')\rangle\!\rangle
&=
\frac12
\langle\!\langle
\bigl(x_+(t)-x_-(t)\bigr)
\bigl(x_+(t')+x_-(t')\bigr)
\rangle\!\rangle
\nonumber\\
&=
\theta(t'-t)\,\langle [\hat x(t'),\hat x(t)]\rangle ,
\end{align}
which is the advanced correlator. Notice that the Heaviside theta places the time appearing in the advanced variable always before the time appearing in the retarded variable, hence justifying their names.  Finally,
\begin{align}
    \expi{x_{a}(t)x_{a}(t')}&= \expi{\left[ x_{+}(t)-x_{-}(t) \right] \left[ x_{+}(t')-x_{-}(t') \right]}\\
    &=\ex{T \hat x(t)\hat x(t')}+\ex{\widetilde T \hat x(t)\hat x(t')}-\ex{  \hat x(t)\hat x(t')}-\ex{  \hat x(t')\hat x(t)}=0\,.
\end{align}
Thus the four basic two-point functions in the $r/a$ basis are
\begin{align}
G_K(t,t') := \langle\!\langle x_r(t)x_r(t')\rangle\!\rangle
&= \frac12\langle \{\hat x(t),\hat x(t')\}\rangle ,
\\
G_R(t,t') := \langle\!\langle x_r(t)x_a(t')\rangle\!\rangle
&= \theta(t-t')\langle [\hat x(t),\hat x(t')]\rangle ,
\\
G_A(t,t') := \langle\!\langle x_a(t)x_r(t')\rangle\!\rangle
&= \theta(t'-t)\langle [\hat x(t'),\hat x(t)]\rangle ,
\\
\langle\!\langle x_a(t)x_a(t')\rangle\!\rangle &=0 .
\end{align}
These are the Keldysh, retarded, advanced, and vanishing $aa$ propagators, respectively. This is one of the main reasons the $r/a$ basis is so useful: response functions are immediately identified with commutators multiplied by Heaviside functions, while fluctuations are captured by the symmetrized correlator.  

To compute path-integral expectation values of the product of a retarded and advanced fields, it is convenient to replace the sources \(J_\pm\) coupled to the two branches by their Keldysh combinations
\begin{equation}\label{Jra}
J_r := \frac{J_+ + J_-}{2},
\qquad
J_a := J_+ - J_- ,
\end{equation}
or equivalently
\begin{equation}
J_+ = J_r + \frac{J_a}{2},
\qquad
J_- = J_r - \frac{J_a}{2} .
\end{equation}
Here \(J_r\) is called the retarded, or average, source, while \(J_a\) is called the advanced, or difference, source. In the Keldysh basis the source term becomes
\begin{equation}
J_+x_+ - J_-x_-
=
J_a x_r + J_r x_a ,
\end{equation}
so that the Schwinger--Keldysh generating functional may be written as
\begin{equation}
Z[J_r,J_a]
=
\int \mathcal D x_r\,\mathcal D x_a\;
\exp\!\left(
iS_{\rm SK}[x_r,x_a]
+
i\int dt\,\bigl(J_a x_r + J_r x_a\bigr)
\right) .
\end{equation}
A somewhat confusing consequence of the reasonable convention in \eqref{Jra} is that insertions of \(x_r\) are generated by differentiation with respect to \(J_a\), while insertions of \(x_a\) are generated by differentiation with respect to \(J_r\).

The vanishing of the $aa$ two-point function is not an accident. More generally, correlation functions built only from advanced fields vanish:
\begin{equation}\label{aeq0}
\langle\!\langle x_a(t_1)\cdots x_a(t_n)\rangle\!\rangle = 0
\qquad
\text{for all } n\geq 1 .
\end{equation}
This follows directly from the normalization condition of the Schwinger--Keldysh generating functional\footnote{An alternative derivation uses the relation $x_a=x_+ -x_-$. Expanding a product of \(n\) advanced fields gives a signed sum over all \(2^n\) ways of assigning each insertion to the forward or backward branch. The result \eqref{aeq0} is then equivalent to the identity that this alternating sum of contour-ordered correlators vanishes
\begin{equation}
\expi {\prod_{m=1}^n x_a(t_m) }
=
\sum_{I\subseteq \{1,\dots,n\}}
(-1)^{\,n-|I|}
\left\langle
\bar T \prod_{j\notin I}\hat x(t_j)\;
T \prod_{i\in I}\hat x(t_i)
\right\rangle
=0 .
\end{equation}}. In the \(r/a\) basis one has 
\[Z[J_r,J_a=0]=1 \] 
for arbitrary \(J_r\). Since insertions of \(x_a\) are generated by differentiation with respect to \(J_r\), \eqref{aeq0} follows immediately.

At higher order, the $r/a$ basis organizes correlation functions into nested commutators. The most important general family is obtained by taking one $r$ insertion and any number of $a$ insertions. These are the higher retarded correlators. For ordered times
\begin{equation}
t_1>t_2>\cdots>t_n ,
\end{equation}
one finds
\begin{equation}\label{higherretarded}
\langle\!\langle x_r(t_1)x_a(t_2)\cdots x_a(t_n)\rangle\!\rangle
=
\left\langle
[\cdots[[\hat x(t_1),\hat x(t_2)],\hat x(t_3)]\cdots,\hat x(t_n)]
\right\rangle ,
\end{equation}
while for arbitrary time arguments this is completed by the appropriate product of Heaviside functions enforcing retarded support,
\begin{align}
\langle\!\langle x_r(t_1)x_a(t_2)\cdots x_a(t_n)\rangle\!\rangle
&=
\theta(t_1-t_2)\theta(t_2-t_3)\cdots\theta(t_{n-1}-t_n)
\nonumber\\
&\qquad\times
\left\langle
[\cdots[[\hat x(t_1),\hat x(t_2)],\hat x(t_3)]\cdots,\hat x(t_n)]
\right\rangle
\;+\; \text{permutations} .
\end{align}
Thus each additional $a$ insertion adds one further level of nesting in the commutator. This is the precise sense in which the advanced field is a response field: differentiating with respect to $  J_{r} $ generates response functions.

To summarize, the $\pm$ basis makes the contour structure manifest: it directly encodes time ordering on the forward branch, anti-time ordering on the backward branch, and Wightman functions across branches. The $r/a$ basis, on the other hand, makes response and fluctuation manifest: the $rr$ correlator gives the symmetrized two-point function, the $ra$ and $ar$ correlators give the retarded and advanced propagators, and correlators built only from $a$ fields vanish identically. 


\subsection{Integrating out the environment and the Feynman--Vernon influence functional}

We now move from a closed system to an open one, i.e. from Hamiltonian to generically non-Hamiltonian evolution. The conceptual starting point is still a fully Hamiltonian theory, but now the total Hilbert space is split into a subsystem of interest $ S$ and an environment $ E$,
\begin{equation}
\mathcal H = \mathcal H_S \otimes \mathcal H_E .
\end{equation}
We denote by \(x\) the degrees of freedom of the system and by \(q\) those of the environment. Without loss of generality, the total action is correspondingly decomposed as
\begin{equation}
S_{\rm tot}[x,q]
=
S_S[x] + S_E[q] + S_{\rm int}[x,q] .
\end{equation}
At this stage the full theory is still closed and evolves unitarily according to some Hamiltonian.

The Schwinger--Keldysh generating functional $  Z[J_+,J_-] $ for the full system is therefore
\begin{align}
&Z=\int dx_{+,0}\,dx_{-,0}\,dq_{+,0}\,dq_{-,0}\,dx_f\,dq_f\;
\rho_{SE}(x_{+,0},q_{+,0};x_{-,0},q_{-,0};t_0)
\nonumber\\
& \int_{x_+(t_0)=x_{+,0},\,q_+(t_0)=q_{+,0}}^{x_+(t_f)=x_f,\,q_+(t_f)=q_f}
\mathcal D x_+\,\mathcal D q_+ 
\exp\!\Bigl[
iS_S[x_+] + iS_E[q_+] + iS_{\rm int}[x_+,q_+]
+ i\int dt J_+x_+
\Bigr]
\nonumber\\
&  \int_{x_-(t_0)=x_{-,0},\,q_-(t_0)=q_{-,0}}^{x_-(t_f)=x_f,\,q_-(t_f)=q_f}
\mathcal D x_-\,\mathcal D q_- 
\exp\!\Bigl[
-iS_S[x_-] - iS_E[q_-] - iS_{\rm int}[x_-,q_-]
- i\int dt J_-x_-
\Bigr] .\nonumber
\end{align}
As we did previously, to avoid excessively large formulas, we will henceforth omit the upper boundary value of the path integral, leaving it implicit that it should always be continuous between the plus and minus branch. We will also use the shorthand notation $ \int_{\rho}$ to indicate the fact that one should average over all possible lower boundary conditions of the path integral with weight given by the density matrix $ \rho$. So we write more compactly
\begin{align}
Z[J_+,J_-]
&=
\int_{\rho_{SE}(t_0)} \mathcal D x_+\,\mathcal D x_-\,\mathcal D q_+\,\mathcal D q_-\;
\exp\Bigl(
iS_S[x_+] - iS_S[x_-]
+iS_E[q_+] - iS_E[q_-]
\nonumber\\
&\hspace{1.5cm}
+iS_{\rm int}[x_+,q_+] - iS_{\rm int}[x_-,q_-]
+i\int dt\,J_+ x_+ - i\int dt\,J_- x_-
\Bigr) .
\end{align}

Here \(\rho_{SE}(t_0)\) is the initial density matrix of the full system. In the simplest and most common setup one assumes that the initial state factorizes,
\begin{equation}\label{factorisedrho}
\rho_{SE}(t_0)=\rho_S(t_0)\otimes \rho_E(t_0) ,
\end{equation}
although the formal manipulations below can be generalized.

The crucial step is now to integrate out the environment variables \(q_\pm\). This defines the Feynman--Vernon \textit{influence functional}\footnote{Sometimes one uses the word Feynman--Vernon influence \textit{functional} to denote $ e^{iS_{\mathrm{IF}}}$, where $S_{\mathrm{IF}}$ is called the influence \textit{action}. I do not make this distinction in these notes.} \cite{FEYNMAN1963118}. Explicitly, we define
\begin{align}
e^{\,iS_{\rm IF}[x_+,x_-]}
&:=
\int_{\rho_{E}(t_{0})} \mathcal D q_+\,\mathcal D q_-\;
\exp\Bigl(
iS_E[q_+] - iS_E[q_-]
+iS_{\rm int}[x_+,q_+]
-iS_{\rm int}[x_-,q_-]
\Bigr)\,.
\end{align}
This object depends only on the system's histories \(x_+\) and \(x_-\), but it contains the full effect of the environment on the subsystem. As we will see in the next subsection, the influence functional can be written as an overlap amplitude between two environment states evolved in the presence of two different histories of the system. 
Once this definition is made, the generating functional becomes
\begin{align}
Z[J_+,J_-]
&=
\int \mathcal D x_+\,\mathcal D x_-\;
\exp\Bigl(
iS_S[x_+] - iS_S[x_-]
+iS_{\rm IF}[x_+,x_-]
+i\int dt\,J_+x_+
-i\int dt\,J_-x_-
\Bigr)
\nonumber\\
&\hspace{2.3cm}
\times
\rho_S(x_{+,0},x_{-,0};t_0) .
\end{align}
It is therefore natural to define the open-system effective action
\begin{equation}
S_{\rm eff}[x_+,x_-]
:=
S_S[x_+] - S_S[x_-] + S_{\rm IF}[x_+,x_-] .
\end{equation}

This is the central result. For a closed system the Schwinger--Keldysh action is simply the difference \(S[x_+]-S[x_-]\). For an open system, integrating out the environment generates the additional term \(S_{\rm IF}[x_+,x_-]\), which in general couples the two branches of the contour. These branch-mixing terms have no counterpart in a purely Hamiltonian closed-system evolution, and they are precisely what encode the effects of the environment on the reduced dynamics.

The influence functional has a clear physical meaning. It summarizes, in a single functional of \(x_+\) and \(x_-\), all effects mediated by the environment. Depending on the problem, these include shifts of the system parameters, dissipation, stochastic fluctuations, and memory effects. In general, \(S_{\rm IF}\) is nonlocal in time: the environment remembers part of the past history of the system, and the reduced dynamics is therefore non-Markovian. In this sense, the Feynman--Vernon formalism provides an exact real-time description of open quantum dynamics before any Markovian approximation is made. In Section \ref{sec:4p7} we will explicitly compute the influence functional in a simple setup. 

It is often convenient to pass immediately to the Keldysh basis,
\begin{equation}
x_r := \frac{x_+ + x_-}{2},
\qquad
x_a := x_+ - x_- .
\end{equation}
Then the influence functional becomes \(S_{\rm IF}[x_r,x_a]\), and its structure can be interpreted more directly, as we will see shortly.


\subsection{Constraints on the Schwinger--Keldysh effective action}

Before studying explicit examples, it is important to understand the general constraints that any Schwinger--Keldysh effective action must satisfy in order to describe the evolution of a density matrix. These constraints follow directly from the defining properties of a density matrix,
\begin{equation}
\Tr \rho = 1,
\qquad
\rho^\dagger = \rho,
\qquad
\rho \ge 0 .
\end{equation}
They are the path-integral counterparts of trace preservation, Hermiticity preservation, and positivity. We will postpone the stronger requirement of complete positivity to the next subsection. In this subsection we derive the corresponding conditions on the effective action, namely
 \begin{equation}
S_{\rm eff}[x_+,x_+]=0,
\qquad
S_{\rm eff}[x_+,x_-]= -\,S_{\rm eff}^*[x_-,x_+],
\qquad
\Im S_{\rm eff}[x_+,x_-]\geq 0 .
\end{equation}
These constraints are trivially satisfied for Hamiltonian evolution where \(S_{\rm eff}=S[x_+]-S[x_-]\) with \(S\) a real functional of its variables by unitarity. The non-trivial fact is that these conditions are satisfied by the influence functional $  S_{\rm IF} $, on which we focus henceforth.

The logic of the proof is the following. One starts from a unitary evolution of the full system, made of the subsystem of interest \(x\) and an environment \(q\), and then rewrites the influence functional as an overlap of two environment states evolved with \(x_+\) and \(x_-\) treated as external sources. In this form, the standard Schwinger--Keldysh constraints follow almost immediately. This derivation is a straightforward adaptation of Appendices A of Ref. \cite{Glorioso:2016gsa,Salcedo:2024smn}.

Assuming an initially factorized density matrix as in \eqref{factorisedrho}, the influence functional is 
\begin{equation}
e^{\,iS_{\rm IF}[x_+,x_-]}
=
\int dq\,dq_i\,dq_i'\;
\int_{q_i}^{q}\mathcal D q_+\;
\int_{q_i'}^{q}\mathcal D q_-\;
e^{\,iS_{\rm tot}[x_+,q_+] - iS_{\rm tot}[x_-,q_-]}\,
\rho_q^{(0)}(q_i,q_i') .
\end{equation}
The crucial observation is now to regard \(x_+\) and \(x_-\) as external sources for the environment. Denoting by \(U_q(t,t_0;\{x_+\})\) the environment evolution operator in the presence of the prescribed background history \(x_+\), and similarly for \(x_-\), one may rewrite the two environment path integrals as matrix elements of sourced evolution operators,
\begin{equation}
\int_{q_i}^{q}\mathcal D q_+\; e^{\,iS_{\rm tot}[x_+,q_+]}
=
\langle q|\,U_q(t,t_0;\{x_+\})\,|q_i\rangle ,
\end{equation}
\begin{equation}
\int_{q_i'}^{q}\mathcal D q_-\; e^{-\,iS_{\rm tot}[x_-,q_-]}
=
\Bigl[\langle q|\,U_q(t,t_0;\{x_-\})\,|q_i'\rangle\Bigr]^\dagger .
\end{equation}
To keep the discussion as simple as possible, in the following we will assume that the environment is initially in a pure state 
\[   \rho_q^{(0)}(q_i,q_i') =  \ket{\Omega_q}\bra{\Omega_q}. \]
Substituting the above expressions into the influence functional and inserting the identity in the environment Hilbert space then gives
\begin{equation}
e^{\,iS_{\rm IF}[x_+,x_-]}
=
\langle \Omega_q|\,
U_q^\dagger(t,t_0;\{x_-\})\,
U_q(t,t_0;\{x_+\})\,
|\Omega_q\rangle .
\end{equation}
Equivalently, defining the sourced-evolved environment states
\begin{equation}
|\Omega_q^{\{x_+\}}(t)\rangle
:=
U_q(t,t_0;\{x_+\})\,|\Omega_q\rangle ,
\qquad
\langle \Omega_q^{\{x_-\}}(t)|
:=
\langle\Omega_q|\,U_q^\dagger(t,t_0;\{x_-\}) ,
\end{equation}
one finds the compact expression
\begin{equation}\label{eq:influence-overlap}
e^{\,iS_{\rm IF}[x_+,x_-]}
=
\langle \Omega_q^{\{x_-\}}(t)\,|\,\Omega_q^{\{x_+\}}(t)\rangle .
\end{equation}
Thus the influence functional is an overlap amplitude between two environment states evolved with two different histories of the system. This is the key result from which the constraints follow. 

The first consequence is positivity of the imaginary part. Since the states in \eqref{eq:influence-overlap} are normalized, the Cauchy--Schwarz inequality implies
\begin{equation}
\bigl|e^{\,iS_{\rm IF}[x_+,x_-]}\bigr|
=
\bigl|\langle \Omega_q^{\{x_-\}}(t)\,|\,\Omega_q^{\{x_+\}}(t)\rangle\bigr|
\le 1 .
\end{equation}
Hence
\begin{equation}
e^{-\Im S_{\rm IF}[x_+,x_-]} \le 1  \then 
\Im S_{\rm IR}[x_+,x_-] \ge 0 .
\end{equation}
This is the path-integral form of the positivity constraint. 

The second consequence is Hermiticity. Taking the complex conjugate of \eqref{eq:influence-overlap}, one obtains
\begin{equation}
e^{-\,iS_{\rm IF}^*[x_+,x_-]}
=
\langle \Omega_q^{\{x_+\}}(t)\,|\,\Omega_q^{\{x_-\}}(t)\rangle
=
e^{\,iS_{\rm IF}[x_-,x_+]} ,
\end{equation}
and therefore
\begin{equation}
S_{\rm IF}[x_+,x_-]
=
-\,S_{\rm IF}^*[x_-,x_+] .
\end{equation}
This is the Schwinger--Keldysh reality condition implied by Hermiticity of the reduced density matrix.  

The third consequence is the normalization or trace-preservation condition. If the two histories coincide, \(x_+=x_-\), then the two sourced evolutions of the environment are identical, and \eqref{eq:influence-overlap} reduces to the norm of a single state,
\begin{equation}
e^{\,iS_{\rm IF}[x_+,x_+]}
=
\langle \Omega_q^{\{x_+\}}(t)\,|\,\Omega_q^{\{x_+\}}(t)\rangle
=
1 .
\end{equation}
Hence
\begin{equation}
S_{\rm IF}[x_+,x_+] = 0 .
\end{equation}
This is the precise derivation of the statement that the Schwinger--Keldysh effective action vanishes when the two branches are identified. 

Passing to the Keldysh basis, and combining $  S_{\rm IF} $ with the Hamiltonian part of the evolution to form the full SK action $  S_{\rm eff} $, these constraints become
\begin{align}\label{constraintsra}
S_{\rm eff}[x_r,x_a=0] &= 0 ,\\
S_{\rm eff}[x_r,x_a] &= -\,S_{\rm eff}^*[x_r,-x_a] ,\\
\Im S_{\rm eff}[x_r,x_a] &\ge 0 .
\end{align}
They hold non-perturbatively and are inherited directly from the underlying Hamiltonian evolution of the full theory. They are the basic unitarity constraints on any Schwinger--Keldysh effective action, and they are sometimes referred to as non-equilibrium constraints. A longer but perhaps more accurate name would be ``constraints from the unitarity of the full system'', highlighting the fact that if these constraints are violated, then the open system dynamics cannot arise from a unitary full system by integrating out an environment, under the assumption \eqref{factorisedrho}.


\subsection{Gaussian environment linearly coupled to the system}\label{sec:4p7}

We now discuss an explicit model in which the influence functional can be computed exactly. The purpose of this example is to show in a concrete setting how integrating out the environment generates a non-trivial Schwinger--Keldysh effective action for the system, and how the characteristic branch-mixing structures arise. We consider a system degree of freedom \(x\) linearly coupled to an environment described by a Gaussian field \(q\).

Let the total action be
\begin{equation}
S_{\rm tot}[x,q]
=
S_S[x]+S_E[q]+S_{\rm int}[x,q] ,
\end{equation}
with interaction
\begin{equation}
S_{\rm int}[x,q]
=
\int dt\, g\,x(t)\,q(t) .
\end{equation}
For definiteness, we may think of \(q\) either as a single harmonic oscillator or, more generally, as a collection of harmonic oscillators. The crucial assumption is that the environment action \(S_E[q]\) is quadratic in \(q\), so that the path integral over \(q\) is Gaussian and can be performed exactly.

In the Schwinger--Keldysh formalism, the interaction term becomes
\begin{equation}
S_{\rm int}[x_+,q_+]-S_{\rm int}[x_-,q_-]
=
g\int dt\,\bigl(x_+(t)q_+(t)-x_-(t)q_-(t)\bigr) .
\end{equation}
The influence functional is therefore
\begin{align}
e^{\,iS_{\rm IF}[x_+,x_-]}
&=
\int_{\rho_E(t_0)} \mathcal D q_+\,\mathcal D q_-\;
\exp\Biggl\{
iS_E[q_+]-iS_E[q_-]
\nonumber\\
&\hspace{2cm}
+i g\int dt\,\bigl(x_+(t)q_+(t)-x_-(t)q_-(t)\bigr)
\Biggr\} .
\end{align}
Since the environment is Gaussian, this is a Gaussian functional integral with linear sources. It can therefore be evaluated exactly by completing the square. The result is quadratic in the system histories \(x_+\) and \(x_-\):
\begin{align}\label{IFpmGaussian}
S_{\rm IF}[x_+,x_-]
=
i\frac{g^2}{2}
\int dt\,dt'\,
\Big[
x_+(t)\,G_{++}(t,t')\,x_+(t')
-x_+(t)\,G_{+-}(t,t')\,x_-(t')
\nonumber\\
\hspace{3.7cm}
-x_-(t)\,G_{-+}(t,t')\,x_+(t')
+x_-(t)\,G_{--}(t,t')\,x_-(t')
\Big] ,
\end{align}
where \(G_{ab}(t,t')\), with \(a,b\in\{+,-\}\), are the Schwinger--Keldysh two-point functions of the environment operator \(q\)
\begin{align}
G_{++}(t,t')
&:=
\langle\!\langle q_+(t)q_+(t')\rangle\!\rangle
=
\langle T\,\hat q(t)\hat q(t')\rangle ,
\\
G_{+-}(t,t')
&:=
\langle\!\langle q_+(t)q_-(t')\rangle\!\rangle
=
\langle \hat q(t')\hat q(t)\rangle ,
\\
G_{-+}(t,t')
&:=
\langle\!\langle q_-(t)q_+(t')\rangle\!\rangle
=
\langle \hat q(t)\hat q(t')\rangle ,
\\
G_{--}(t,t')
&:=
\langle\!\langle q_-(t)q_-(t')\rangle\!\rangle
=
\langle \bar T\,\hat q(t)\hat q(t')\rangle .
\end{align}
In other words, the exact influence functional is determined entirely by the environment two-point correlators. This is the simplest explicit illustration of the general fact that a Gaussian environment linearly coupled to the system produces a quadratic influence functional.\\

It is now useful to pass to the Keldysh basis. After some algebra, the influence functional takes the standard form
\begin{align}\label{IFraGaussian}
S_{\rm IF}[x_r,x_a]
=
- g^2\int dt\,dt'\,
x_a(t)\,D_R(t,t')\,x_r(t')
+\frac{i g^2}{2}\int dt\,dt'\,
x_a(t)\,N(t,t')\,x_a(t') .
\end{align}
Before expressing \(D_R\) and \(N\) in terms of correlators of \(q\), we can already note a few properties. First, any antisymmetric part of \(N(t,t')\) does not contribute to \eqref{IFraGaussian}. Second, we may decompose
\begin{equation}
D_R(t,t')
=
D_R^{(S)}(t,t')+D_R^{(A)}(t,t'),
\end{equation}
where
\begin{equation}
D_R^{(S)}(t,t')=D_R^{(S)}(t',t),
\qquad
D_R^{(A)}(t,t')=-D_R^{(A)}(t',t).
\end{equation}
Using \(x_a=x_+-x_-\) and \(x_r=(x_++x_-)/2\), the contribution of the symmetric part becomes
\begin{align}
\int dt\,dt'\,
x_a(t)D_R^{(S)}(t,t')x_r(t')
&=
\frac{1}{2}\int dt\,dt'\,
\left[
x_+(t)D_R^{(S)}(t,t')x_+(t')
-
x_-(t)D_R^{(S)}(t,t')x_-(t')
\right]. \nonumber
\end{align}
Here the mixed $ +  $/$ -  $ terms cancel because \(D_R^{(S)}\) is symmetric. This contribution is therefore of the form
\begin{equation}
S_{\rm cons}[x_+]-S_{\rm cons}[x_-],
\end{equation}
and can be absorbed into a redefinition of the conservative system action.

By contrast, the diagonal terms involving \(D_R^{(A)}\) vanish, while the mixed terms add:
\begin{align}
- g^2\int dt\,dt'\,
x_a(t)D_R^{(A)}(t,t')x_r(t')
&=
- g^2\int dt\,dt'\,
x_+(t)D_R^{(A)}(t,t')x_-(t').
\end{align}
This term couples the two branches of the Schwinger--Keldysh contour and cannot be written as the difference of two single-branch actions. It therefore represents the genuinely dissipative part of the influence functional.

Equation \eqref{IFpmGaussian} to \eqref{IFraGaussian} one finds
\begin{equation}
D_R(t,t')
:=
-i\,\theta(t-t')\,\langle [q(t),q(t')]\rangle\,,
\end{equation}
which is the retarded kernel of the environment. It is real $ D_R(t,t')^* = D_R(t,t')   $. Also
\begin{equation}
N(t,t')
:=
\frac12\langle \{q(t),q(t')\}\rangle
\end{equation}
which is its Keldysh correlator. It is real and symmetric, 
\[ N(t,t')=N(t',t)=N^{*}(t,t') \,. \] 
Up to our conventions and the overall factor of \(g^2\), these are the two basic kernels that appear in the effective action.

Equation \eqref{IFraGaussian} is worth pausing over. The exact Gaussian influence functional contains two distinct structures. First, there is a term mixing \(x_a\) and \(x_r\), weighted by the retarded kernel \(D_R\). Second, there is a term quadratic in \(x_a\), weighted by the symmetric kernel \(N\). Their interpretation in terms of drift and noise will be explained soon.

A key feature of \eqref{IFraGaussian} is its \textit{nonlocality} in time. Even though the microscopic theory may be local, integrating out the environment generates memory kernels \(D_R(t,t')\) and \(N(t,t')\), so that the reduced dynamics at time \(t\) depends on the past history of the system. In general, therefore, the exact open-system effective action is non-Markovian. The local-in-time Schwinger--Keldysh actions discussed in the previous subsection arise only after an additional approximation, for example when the environment correlation time is much shorter than the characteristic time scale of the system.

In a suitable local limit, the kernels may be approximated by local distributions. If moreover the environment state is stationary, so that these kernels depend only on the time difference \(t-t'\), this immediately constrains their local derivative expansion. The Keldysh correlator admits only even powers of derivatives\footnote{In principle, negative powers of the derivative may also appear. },
\begin{equation}
N(t,t')
\sim
n_0\,\delta(t-t')
+n_2\,\partial_t^2\delta(t-t')
+\cdots ,
\end{equation}
The local derivative expansion of the retarded kernel takes the form
\begin{equation}
D_R(t,t')
\sim
c_0\,\delta(t-t')
+c_1\,\partial_t\delta(t-t')
+c_2\,\partial_t^2\delta(t-t')
+c_3\,\partial_t^3\delta(t-t')
+\cdots .
\end{equation}
Under the exchange \(t\leftrightarrow t'\), the terms containing an even number of
derivatives are symmetric, whereas those containing an odd number of derivatives
are antisymmetric. As shown above, the symmetric terms describe conservative local corrections,
such as shifts of the potential and of the kinetic terms, and may therefore be
absorbed into a redefinition of the conservative part of the effective action.
After doing so, the genuinely dissipative part of the retarded kernel admits an
expansion involving only odd derivatives,
\begin{equation}
D_R^{(A)}(t,t')
\sim
c_1\,\partial_t\delta(t-t')
+c_3\,\partial_t^3\delta(t-t')
+\cdots .
\end{equation}
This is why, in a local approximation, the dissipative \(x_a x_r\) sector naturally begins with a single time derivative and leads to a dissipative term such as \(x_a\dot x_r\), while the \(x_a x_a\) sector naturally begins with a zeroth-order term and leads to a local fluctuation term proportional to \(i\,x_a^2\). 

Finally, let us stress that in general the kernels \(D_R\) and \(N\) are independent. Only if the environment is in thermal equilibrium are they related by a fluctuation--dissipation relation, or equivalently by KMS conditions. Since our eventual interest includes cosmological applications far from equilibrium, we do not impose such relations here.

 
\section{The Schwinger Keldysh dynamical map}\label{sec:SKmap}


\subsection{Cutting open the contour and the SK dynamical map}

So far we have mostly used the Schwinger--Keldysh path integral on the closed-time contour, where the two branches are sewn together at the final time and one computes traces of the density matrix times operators, i.e. expectation values. To discuss complete positivity (CP), CP divisibility and the relation to Lindblad evolution, it is useful to take one step back. First, we will introduce the Schwinger--Keldysh path integral with a cut-open contour as a dynamical map that evolves the reduced density matrix. Second, we will prove that if the SK action is local in time, then this map obeys the composition law. Finally, we will extract from the SK action the Liouvillian generator $  \L_{t} $, which is generically explicitly time dependent, and show that the SK path integral obeys CP divisibility if and only if this generator is of the GKSL form.\\

\begin{figure}
\centering
\begin{tikzpicture}[
    x=1cm,
    y=1cm,
    contour/.style={
        thick,
        postaction={decorate}
    },
    forward/.style={
        contour,
        decoration={
            markings,
            mark=at position 0.50 with {\arrow{>}}
        }
    },
    backward/.style={
        contour,
        decoration={
            markings,
            mark=at position 0.50 with {\arrow{<}}
        }
    }
]

\draw[dashed] (0,-1) -- (0,1.15);
\draw[dashed] (6.4,-1) -- (6.4,1.15);

\node[above=4pt] at (0,1.15) {$t_0$};
\node[above=4pt] at (6.4,1.15) {$t$};

\fill (0,0.55) circle (1.5pt);
\fill (0,-0.55) circle (1.5pt);
\fill (6.4,0.55) circle (1.5pt);
\fill (6.4,-0.55) circle (1.5pt);

\draw[forward]  (0,0.55) -- (6.4,0.55);
\draw[backward] (0,-0.55) -- (6.4,-0.55);

\node at (3.2,0)
    {$\displaystyle
    \int_{x_{+,0}}^{x_{+,t}}\mathcal{D}x_+
    \int_{x_{-,0}}^{x_{-,t}}\mathcal{D}x_-\,
    e^{iS_{\mathrm{eff}}[x_+,x_-]}
    $};

\node[left=5pt] at (0,0.55) {$x_{+,0}$};
\node[left=5pt] at (0,-0.55) {$x_{-,0}$};

\node[right=5pt] at (6.4,0.55) {$x_{+,t}$};
\node[right=5pt] at (6.4,-0.55) {$x_{-,t}$};

\node at (0,-1.60)
    {$\rho(t_0)^{x_{+,0}}_{x_{-,0}}$};

\draw[->,thick]
    (1.65,-1.60) -- (4.75,-1.60)
    node[midway,above=0pt]
    {$\displaystyle \Phi_{t,t_0}$};

\node at (6.4,-1.60)
    {$\rho(t)^{x_{+,t}}_{x_{-,t}}$};

\end{tikzpicture}
\caption{Cutting open the closed-time path turns the SK path integral into a dynamical map that evolves the density matrix, $\Phi_{t,t_0}:\rho(t_0)\mapsto\rho(t)$. Note that, for a generic open system, the time evolution on the two branches is not independent but may be mixed by the influence functional. In this precise sense, the two branches of the SK path integral evolve the two indices of the density matrix.\label{fig:SKdynamicalmap}}
\end{figure}
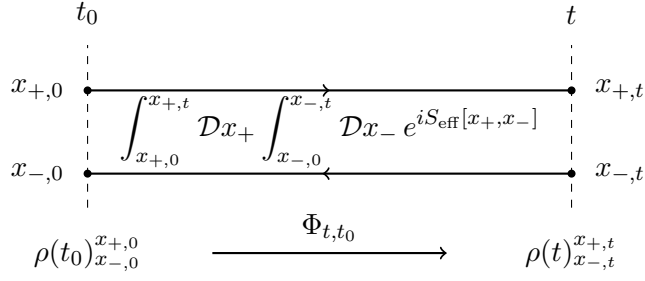

\paragraph{The SK dynamical map} 

The initial and final reduced density matrices at times \(t_0\) and \(t_{f}\) can be related as in \eqref{Krausoperatorrepresentation}, where we used the Kraus representation. Using the notation of this section this relation becomes
\begin{align}\label{rho-kernel}
\rho_S(x_f^+,x_f^-;t_f)
&=
\int dx_i^+\,dx_i^- \;
\mathcal K(t_f,t_0;x_f^+,x_f^-;x_i^+,x_i^-)\,
\rho_S(x_i^+,x_i^-;t_0) ,
\end{align}
for some kernel \(\mathcal K\). Since the Kraus operators are matrix elements of the time evolution operator, see \eqref{def:kraus}, they can be represented by a path integral. Then, the kernel above can be written as a path integral on a cut-open contour
\begin{align}\label{Kkernel}
\mathcal K(t_f,t_0;x_f^+,x_f^-;x_i^+,x_i^-)
&:=
\int_{x_i^+}^{x_f^+}\mathcal D x_+
\int_{x_i^-}^{x_f^-}\mathcal D x_-\;
e^{\,iS_{\rm eff}[x_+,x_-]} .
\end{align}
Here the notation indicates that the \(+\) and \(-\) histories begin at \(x_i^\pm\) and end at \(x_f^\pm\), with no sewing imposed at the final time. In this sense, one has ``cut open'' the closed-time contour. We still refer to this as a ``Schwinger-Keldysh path integral'', except that it takes place over a different, in this case disjoint, contour. 

Equation \eqref{rho-kernel} makes it clear that the Schwinger--Keldysh path integral does not merely compute expectation values, it may also be used to define a dynamical map on density matrices. Indeed, writing
\begin{equation}
\rho_S(t_f)=:\Phi_{t_f,t_0}\bigl[\rho_S(t_0)\bigr] ,
\end{equation}
the kernel \(\mathcal K\) is simply the field-basis representation of the map \(\Phi_{t_f,t_0}\), which we will call the \textit{Schwinger-Keldysh (SK) dynamical map}. This is depicted in Figure \ref{fig:SKdynamicalmap}.


\paragraph{The composition law} We recall from previous sections that in the presence of explicit time-dependence, the semi-group property does not apply. Instead now the dynamical map has two time arguments and one should impose the composition law in \eqref{compositionlaw}, which we prove here follows directly if the effective action is local in time. Suppose that
\begin{equation}\label{localSKaction}
S_{\rm eff}[x_+,x_-]
=
\int_{t_0}^{t_f} dt\,
L
\bigl(
x_+(t),x_-(t),\dot x_+(t),\dot x_-(t),\ldots;t
\bigr) \,,
\end{equation}
for some function $ L$. Then, for any intermediate time \(t_1\) with \(t_f>t_1>t_0\), the action splits additively,
\begin{equation}
S_{\rm eff}^{[t_0,t_f]}[x_+,x_-]
=
S_{\rm eff}^{[t_0,t_1]}[x_+,x_-]
+
S_{\rm eff}^{[t_1,t_f]}[x_+,x_-] .
\end{equation}
In the path integral, one may therefore insert a resolution of the identity at time \(t_1\), integrate over the intermediate boundary data \(\tilde x^\pm:=x_\pm(t_1)\), and factorize the kernel into two consecutive pieces. One obtains
\begin{align}\label{compositionkernel}
\mathcal K(t_f,t_0;x_f^+,x_f^-;x_i^+,x_i^-)
&=
\int d\tilde x^+\,d\tilde x^- \;
\mathcal K(t_f,t_1;x_f^+,x_f^-;\tilde x^+,\tilde x^-)
\nonumber\\
&\hspace{1.8cm}\times
\mathcal K(t_1,t_0;\tilde x^+,\tilde x^-;x_i^+,x_i^-) .
\end{align}
Hence, at the operator level, one finds
\begin{equation}
\Phi_{t_f,t_0}=\Phi_{t_f,t_1}\circ \Phi_{t_1,t_0} .
\end{equation}
Thus locality in time of the Schwinger--Keldysh effective action is a sufficient condition for the composition property of the dynamical map.

Notice that this is weaker than the semigroup property. Only in the further special case in which the local effective action is time-translation invariant does one recover the semigroup property discussed in the previous section.

\paragraph{CP divisibility} Once locality in time has been assumed, it is natural to extract from the short-time kernel a time-local generator, as we did in the proof of the GKSL theorem. Let \(\delta t\) be infinitesimal. Then the kernel over the interval \([t,t+\delta t]\) takes the form
\begin{align}
\mathcal K(t+\delta t,t;x_f^+,x_f^-;x_i^+,x_i^-)
&=
\delta(x_f^+-x_i^+)\,\delta(x_f^--x_i^-)
\nonumber\\
&\qquad
+\,
\delta t\,
\mathcal L_t(x_f^+,x_f^-;x_i^+,x_i^-)
+
\mathcal O(\delta t^2) ,
\end{align}
where \(\mathcal L_t\) is the kernel representation of a time-local generator, which we may call the time-dependent Liouvillian. Equivalently, assuming differentiability of the map, one defines
\begin{equation}\label{timelocalgeneratorSK}
\partial_{t_f}\Phi_{t_f,t_0}
=
\mathcal L_{t_f}\circ \Phi_{t_f,t_0} .
\end{equation}
As we have seen, the solution is the time-ordered exponential of the superoperator $ \L$:
\begin{equation}
\Phi_{t_f,t_0}
=
T\exp\!\left(
\int_{t_0}^{t_f} ds\,\mathcal L_s
\right) .
\end{equation}
To establish CP divisibility, one must then check whether this generator is of GKSL form for every time \(t\). Under the standard assumptions of differentiability and invertibility of the map, CP divisibility is equivalent to the statement that
\begin{equation}
\mathcal L_t[\rho]
=
-i[H(t),\rho]
+
\sum_{\alpha,\beta} c_{\alpha\beta}(t)
\left(
L_\alpha(t)\rho L_\beta^\dagger(t)
-\frac12\{L_\beta^\dagger(t)L_\alpha(t),\rho\}
\right) ,
\end{equation}
with the Kossakowski matrix \(c_{\alpha\beta}(t)\) positive semidefinite for every \(t\), for some jump operators $  L_{\alpha} $. We will give an explicit example of this procedure for a noisy and dissipative harmonic oscillator in a later section. 


\subsection{Example: the damped harmonic oscillator}\label{ssec:dampedharmonic}

To make the connection between the Schwinger--Keldysh action and the time-local generator more explicit, let us consider a single quantum-mechanical degree of freedom \(x\). Since the GKSL theorem is naturally formulated in operator phase space, it is convenient to begin directly with a first-order action for the canonical variables \(x\) and \(p\). We set the mass equal to one throughout, since it plays no essential role in the discussion. We restrict to a local-in-time Gaussian theory with at most one time derivative acting on each variable. A convenient representative is\footnote{One may integrate out the momentum variables and obtain an equivalent second-order action for \(x_r\) and \(x_a\). In particular, integration over \(p_r\) imposes
\(
p_a
=
\dot x_a-\kappa x_a.
\)
The phase-space noise terms then generate terms proportional to \(x_a^2\), \(\dot x_a^2\), and \(x_a\dot x_a\). For constant coefficients, the last of these is a total derivative and can be traded for a boundary term. Consequently, in the second-order description the information is distributed between bulk coefficients and boundary or contact terms. The complete-positivity constraints are therefore most transparent in the first-order formulation, which we use throughout the remainder of this subsection.}
\begin{align}\label{GaussianSKfirstorder}
S_{\rm SK}[x_r,x_a,p_r,p_a]
=
\int dt \Bigg[
&p_a \dot x_r + p_r \dot x_a
-p_r p_a
-\Omega^2(t)\,x_r x_a
-\kappa(t)\,x_a p_r
\nonumber\\
&\qquad
+\frac{i}{2}N_x(t)\,x_a^2
+\frac{i}{2}N_p(t)\,p_a^2
+iN_{xp}(t)\,x_a p_a
\Bigg].
\end{align}
Here \(\Omega(t)\) is the instantaneous oscillator frequency, \(\kappa(t)\) is a dissipative coefficient, and \(N_x(t)\), \(N_p(t)\), and \(N_{xp}(t)\) are the three independent entries of the symmetric phase-space noise matrix. The terms proportional to \(N_x\) and \(N_p\) describe noise in the two canonical directions, while \(N_{xp}\) describes their cross-correlation.

Over an infinitesimal interval \([t,t+\delta t]\), the kernel takes the form
\begin{align}\label{dictionary}
\mathcal K(t+\delta t,t)
=
\mathds{1}
+\delta t\,\mathcal L_t
+\mathcal O(\delta t^2),
\end{align}
where \(\mathcal L_t\) is the time-local generator. We compare this with the expansion of the Schwinger--Keldysh kernel,
\begin{equation}
e^{\,iS_{\rm eff}^{[t,t+\delta t]}}
=
1+iS_{\rm eff}^{[t,t+\delta t]}
+\mathcal O(\delta t^2).
\end{equation}
The term linear in \(S_{\rm eff}^{[t,t+\delta t]}\) determines the Liouvillian \(\mathcal L_t\). The resulting expressions may be translated into operator language using the Schwinger--Keldysh dictionary
\begin{align}
x_a=x_+-x_-
&\longleftrightarrow
[x,\rho],
\\
p_a=p_+-p_-
&\longleftrightarrow
[p,\rho],
\\
x_r=\frac{x_++x_-}{2}
&\longleftrightarrow
\frac{1}{2}\{x,\rho\},
\\
p_r=\frac{p_++p_-}{2}
&\longleftrightarrow
\frac{1}{2}\{p,\rho\}.
\end{align}
Thus one factor of an advanced variable translates into a commutator, while one factor of a retarded variable translates into an anticommutator. When dealing with products of advanced and retarded variables, one needs to specify an operator ordering to translate path integral variables into an operator statement. For us, this is necessary only for the term $ x_{a}p_{r}$ discussed below, in which case we adopt Weyl ordering and use the above dictionary for the symmetrised product.

In particular, the imaginary terms quadratic in the advanced fields give
\begin{align}
\frac{i}{2}N_x(t)\,x_a^2
&\longrightarrow
-\frac{N_x(t)}{2}[x,[x,\rho]],
\\
\frac{i}{2}N_p(t)\,p_a^2
&\longrightarrow
-\frac{N_p(t)}{2}[p,[p,\rho]],
\\
iN_{xp}(t)\,x_a p_a
&\longrightarrow
-\frac{N_{xp}(t)}{2}
[x,[p,\rho]]
=-\frac{N_{xp}(t)}{2}
[p,[x,\rho]]
\end{align}
where we included the correct factor of $i $ in going from the action to the Liouvillian and we massaged the last expression using Jacobi identities and $ [x,p]=i$. Note that no operator ordering specification is needed in any of these three expressions. 
The mixed term is represented by the trace-preserving symbol\footnote{This
relation may be taken as the definition of the operator ordering for the symbol associated with
$x_a p_r$. Equivalently, it follows by imposing Weyl ordering on $x_a$
and $p_r$ \textit{and} including a contact term required by trace preservation.}
\begin{equation}
-\kappa(t)\,x_a p_r
\;\longrightarrow\;
-\frac{i\kappa(t)}{2}[x,\{p,\rho\}] .
\end{equation}
It is convenient to decompose this term as
\begin{equation}
-\frac{i\kappa(t)}{2}[x,\{p,\rho\}]
=
-i\left[\frac{\kappa(t)}{4}\{x,p\},\rho\right]
-\frac{i\kappa(t)}{2}\bigl(x\rho p-p\rho x\bigr)
+\frac{\kappa(t)}{2}\rho .
\end{equation}
The master equation can therefore be written as
\begin{align}
\label{GaussianLindbladian}
\frac{d\rho}{dt}
&=
-i[H(t),\rho]
-\frac{i\kappa(t)}{2}\bigl(x\rho p-p\rho x\bigr)
+\frac{\kappa(t)}{2}\rho
\nonumber\\
&\quad
-\frac{N_x(t)}{2}[x,[x,\rho]]
-\frac{N_p(t)}{2}[p,[p,\rho]]
-\frac{N_{xp}(t)}{2}
\Bigl(
[x,[p,\rho]]
+
[p,[x,\rho]]
\Bigr),
\end{align}
where
\begin{equation}
H(t)
=
\frac{p^2}{2}
+
\frac{1}{2}\Omega^2(t)x^2
+
\frac{\kappa(t)}{4}\{x,p\}.
\end{equation}

We may now compare \eqref{GaussianLindbladian} with the general Gaussian
GKSL generator obtained in \eqref{generalLindbladian}, allowing the
Hamiltonian and all coefficients to depend on time. In the Hermitian
operator basis
\begin{equation}
F_1=p,
\qquad
F_2=x,
\end{equation}
the dissipative part of \eqref{GaussianLindbladian} is generated by the
Kossakowski matrix
\begin{equation}
\label{GaussianKossakowski}
C(t)
=
\begin{pmatrix}
N_p(t)
&
N_{xp}(t)+\dfrac{i}{2}\kappa(t)
\\[6pt]
N_{xp}(t)-\dfrac{i}{2}\kappa(t)
&
N_x(t)
\end{pmatrix}.
\end{equation}
The real off-diagonal part produces the mixed diffusion term
proportional to $N_{xp}$, while the imaginary antisymmetric part
produces the non-Hamiltonian part of the friction term,
\begin{equation}
-\frac{i\kappa(t)}{2}\bigl(x\rho p-p\rho x\bigr)
+\frac{\kappa(t)}{2}\rho .
\end{equation}
Together with the contribution
$\kappa(t)\{x,p\}/4$ included in $H(t)$, it reconstructs
$-i\kappa(t)[x,\{p,\rho\}]/2$.

Complete positivity requires
\begin{equation}
C(t)\geq0
\end{equation}
at every time. For a \(2\times2\) Hermitian matrix, this is equivalent to
\begin{equation}
\label{GaussianCPbound}
N_x(t)\geq0,
\qquad
N_p(t)\geq0,
\qquad
N_x(t)N_p(t)-N_{xp}(t)^2
\geq
\frac{\kappa(t)^2}{4}.
\end{equation}
These inequalities are the Gaussian open-system manifestation of
complete positivity and, when they hold instantaneously for a
time-local evolution, of CP divisibility. They show explicitly that a
non-zero friction coefficient requires a minimum amount of accompanying
noise: friction cannot arise in a completely positive Markovian
evolution without sufficiently large fluctuations.

It is useful to contrast this result with the weaker constraint obtained directly from Schwinger--Keldysh unitarity. The condition
\begin{equation}
\Im S_{\rm eff}\geq0
\end{equation}
requires the real symmetric noise matrix
\begin{equation}
N(t)
=
\begin{pmatrix}
N_p(t) & N_{xp}(t)
\\
N_{xp}(t) & N_x(t)
\end{pmatrix}
\end{equation}
to be positive semidefinite, and hence implies only
\begin{equation}
N_x(t)\geq0,
\qquad
N_p(t)\geq0,
\qquad
N_x(t)N_p(t)-N_{xp}(t)^2\geq0.
\end{equation}
Complete positivity of the time-local generator is stronger because it constrains the noise relative to the dissipative coefficient \(\kappa(t)\), leading to the lower bound in \eqref{GaussianCPbound}. In the second-order formulation the same constraint is present, but it is distributed among bulk terms, endpoint contributions, and contact terms generated by integrating out the momentum. This is why its phase-space origin is most clearly visible in the first-order action.


\subsection{Jump operators and positivity-improved influence functionals}

In realistic applications, the environmental degrees of freedom are usually too complicated to be integrated out exactly as we did in Section \ref{sec:4p7}. One must therefore adopt the Effective Field Theory (EFT) philosophy and model the SK functional using the degrees of freedom of the system, the symmetries of the problem, and an appropriate power counting. In this subsection we explain how the jump-operator structure familiar from the Lindblad equation provides a systematic parametrization of local-in-time influence functionals, which can be used as an alternative to a direct expansion of the full SK functional. Besides making complete positivity constraints transparent, this organization can preserve exact positivity under truncation and provides a physically admissible starting point for late-time resummation.


\paragraph{From the GKSL equation to the influence functional.}
Consider a local Markovian master equation of the form
\begin{equation}
\dot{\rho}
=
-i[H,\rho]
+
C_{\alpha\beta}
\left(
L_{\beta}\rho L_{\alpha}^{\dagger}
-
\frac{1}{2}
\left\{
L_{\alpha}^{\dagger}L_{\beta},
\rho
\right\}
\right),
\end{equation}
where repeated indices are summed and the Kossakowski matrix satisfies
\begin{equation}\label{Cprop}
C_{\alpha\beta}
=
C_{\beta\alpha}^{*},
\qquad
C\geq 0.
\end{equation}
For simplicity, suppose that the jump operators can be rewritten in terms of local-in-time functions $L_{\alpha}(x)$ of the path integral variables, as it was the case in the toy model in Section \ref{ssec:dampedharmonic}. Denoting
\begin{equation}
L_{\alpha,+}
:=
L_{\alpha}(x_{+}),
\qquad
L_{\alpha,-}
:=
L_{\alpha}(x_{-}),
\end{equation}
the dissipative part of the infinitesimal evolution and using the dictionary $ iS_{\rm IF} \sim \int dt \L  $, discussed around \eqref{dictionary}, gives the Schwinger--Keldysh influence functional
\begin{equation}
S_{\mathrm{IF}}
=
-i
\int dt\,
C_{\alpha\beta}
\left[
L_{\alpha,-}^{*}L_{\beta,+}
-
\frac{1}{2}
L_{\alpha,+}^{*}L_{\beta,+}
-
\frac{1}{2}
L_{\alpha,-}^{*}L_{\beta,-}
\right].
\end{equation}
This expression automatically vanishes when the two branches coincide, \( S_{\mathrm{IF}}[x,x]
=
0 \), as required by the normalization of the Schwinger--Keldysh generating functional.

To make the separation of $ S_{\rm IF}$ into real and imaginary parts more explicit, it is useful to define
\begin{equation}
\Delta L_{\alpha}
:=
L_{\alpha,+}
-
L_{\alpha,-}.
\end{equation}
Consider expanding the following quadratic form:
\begin{align}
\Delta L_{\alpha}^{*}
C_{\alpha\beta}
\Delta L_{\beta}
={}&
L_{\alpha,+}^{*}
C_{\alpha\beta}
L_{\beta,+}
+
L_{\alpha,-}^{*}
C_{\alpha\beta}
L_{\beta,-}-
L_{\alpha,+}^{*}
C_{\alpha\beta}
L_{\beta,-}
-
L_{\alpha,-}^{*}
C_{\alpha\beta}
L_{\beta,+}.
\end{align}
Since the Kossakowski matrix is Hermitian,
\(
C_{\alpha\beta}
=
C_{\beta\alpha}^{*},
\) one has
\begin{equation}
L_{\alpha,+}^{*}
C_{\alpha\beta}
L_{\beta,-}
=
\left(
L_{\alpha,-}^{*}
C_{\alpha\beta}
L_{\beta,+}
\right)^{*}.
\end{equation}
It follows that
\begin{align}
&
L_{\alpha,-}^{*}
C_{\alpha\beta}
L_{\beta,+}
-
\frac{1}{2}
L_{\alpha,+}^{*}
C_{\alpha\beta}
L_{\beta,+}
-
\frac{1}{2}
L_{\alpha,-}^{*}
C_{\alpha\beta}
L_{\beta,-}
\nonumber\\
&\hspace{1cm}
=
-\frac{1}{2}
\Delta L_{\alpha}^{*}
C_{\alpha\beta}
\Delta L_{\beta}
+
\frac{1}{2}
\left[
L_{\alpha,-}^{*}
C_{\alpha\beta}
L_{\beta,+}
-
\left(
L_{\alpha,-}^{*}
C_{\alpha\beta}
L_{\beta,+}
\right)^{*}
\right]
\nonumber\\
&\hspace{1cm}
=
-\frac{1}{2}
\Delta L_{\alpha}^{*}
C_{\alpha\beta}
\Delta L_{\beta}
+
i\,
\operatorname{Im}
\left(
L_{\alpha,-}^{*}
C_{\alpha\beta}
L_{\beta,+}
\right).
\end{align}
Substituting this identity into the influence functional yields
\begin{equation}
S_{\mathrm{IF}}
=
\int dt\,
\operatorname{Im}
\left(
L_{\alpha,-}^{*}
C_{\alpha\beta}
L_{\beta,+}
\right)
+
\frac{i}{2}
\int dt\,
\Delta L_{\alpha}^{*}
C_{\alpha\beta}
\Delta L_{\beta}.
\end{equation}
The first term is manifestly real, while the second is purely imaginary by virtue of \eqref{Cprop}. Moreover, since $C\geq0$,
\begin{equation}
\operatorname{Im}S_{\mathrm{IF}}
=
\frac{1}{2}
\int dt\,
\Delta L_{\alpha}^{*}
C_{\alpha\beta}
\Delta L_{\beta}
\geq0.
\end{equation}
We discovered that the last constraint in \eqref{constraintsra}, imposed by the unitarity of the full system, follows directly from the positivity of the Kossakowski matrix, $C\geq0$. 

Notice that for a single Hermitian jump operator, or more generally for
Hermitian jump operators with a real symmetric Kossakowski matrix, the real
part of the influence functional vanishes. For general non-Hermitian jump
operators, and also for Hermitian jumps before absorbing the imaginary
antisymmetric part of the Kossakowski matrix into the Hamiltonian, one may
instead have
\begin{equation}
\operatorname{Re}S_{\mathrm{IF}}
=
\int dt\,
\operatorname{Im}
\left(
L_{\alpha,-}^{*}
C_{\alpha\beta}
L_{\beta,+}
\right).
\end{equation}
Note that the argument of the imaginary part is real when $ x_{+}=x_{-}$, as it should be.


\paragraph{Expansion of jump operators versus expansion of the influence action.}
Let $d$ denote the short-distance scale suppressing nonlinear operators, and suppose that the jump operators admit an expansion in powers of $x/d$. As a simple example, consider
\begin{equation}
L_{1}(x)
=
x+\frac{x^{2}}{d},
\qquad
L_{2}(x)
=
\frac{x^{2}}{d},
\end{equation}
and assume, for simplicity, a diagonal Kossakowski matrix with positive rates $\gamma_{1}$ and $\gamma_{2}$. The corresponding imaginary part of the influence action is
\begin{equation}
\operatorname{Im}S_{\mathrm{IF}}
=
\frac{1}{2}
\int dt\,
\left[
\gamma_{1}\left|\Delta L_{1}\right|^{2}
+
\gamma_{2}\left|\Delta L_{2}\right|^{2}
\right].
\end{equation}
Suppose that we wish to construct the theory only through order $d^{-1}$. There are two natural prescriptions. The first is to truncate the expansion of the jump operators, discarding $L_{2}$ but retaining the full quadratic contribution of $L_{1}$:
\begin{equation}
\operatorname{Im}S_{\mathrm{IF}}^{\mathrm{imp}}
=
\frac{\gamma_{1}}{2}
\int dt\,
\left[
\Delta x
+
\frac{1}{d}\Delta(x^{2})
\right]^{2}.
\end{equation}
Expanding this expression gives
\begin{equation}
\operatorname{Im}S_{\mathrm{IF}}^{\mathrm{imp}}
=
\frac{\gamma_{1}}{2}
\int dt\,
\left[
(\Delta x)^{2}
+
\frac{2}{d}
\Delta x\,\Delta(x^{2})
+
\frac{1}{d^{2}}
\left(\Delta(x^{2})\right)^{2}
\right].
\end{equation}
Although the final term is formally of order $d^{-2}$, it is retained because it completes the positive square.

The second prescription is to expand the influence action itself and truncate it directly at order $d^{-1}$. One then obtains
\begin{equation}
\operatorname{Im}S_{\mathrm{IF}}^{(1/d)}
=
\frac{\gamma_{1}}{2}
\int dt\,
\left[
(\Delta x)^{2}
+
\frac{2}{d}
\Delta x\,\Delta(x^{2})
\right].
\end{equation}
The two prescriptions agree through order $d^{-1}$ and differ only at order $d^{-2}$. Neither expression is the complete answer at order $ d^{-2}$: additional jump operators, such as $L_{2}$, as well as other symmetry-allowed terms, may also contribute.

The first prescription nevertheless has a structural advantage. By retaining the complete square associated with the truncated jump operator, it guarantees
\(
\operatorname{Im}S_{\mathrm{IF}}^{\mathrm{imp}}
\geq0
\)
for arbitrary field configurations. Equivalently, the corresponding truncated Liouvillian remains of GKSL form and generates a completely positive and CP-divisible evolution. By contrast, the influence action truncated directly at order $d^{-1}$ is not sign definite for arbitrary values of $x_{+}$ and $x_{-}$; its positivity is meaningful only perturbatively, within the regime in which $x/d$ is small.

One may refer to the first prescription as a \emph{positivity-improved} truncation. It does not determine the full answer at order $d^{-2}$, but it provides a particular higher-order completion that preserves complete positivity exactly. This distinction is immaterial in a strictly fixed-order calculation, but can become useful when the truncated generator is evolved non-perturbatively in time, as in Section \ref{ssec:resummation}, since the resulting resummation then remains completely positive. This situation is structurally similar to what happens when one uses the renormalization group to improve some fixed-order loop calculations.


\subsection{Semiclassical limit, HS transform, and Langevin dynamics}

Let us begin with a general local Schwinger--Keldysh action $ S_{\rm eff}[x_r,x_a]   $ written in the \(r/a\) basis. We note that 
\begin{align}
\langle \! \langle x_{r} \rangle \! \rangle &= \frac{1}{2} (\langle \! \langle x_{+} \rangle \! \rangle+\langle \! \langle x_{-} \rangle \! \rangle)=\ex{x}\,, & \langle \! \langle x_{a} \rangle \! \rangle  &= \langle \! \langle x_{+} \rangle \! \rangle-\langle \! \langle x_{-} \rangle \! \rangle=0 \,.
\end{align}
This suggests that $ x_{r} $ captures the average, classical part of the operator $ x $ while $ x_{a}$ captures the fluctuation in the path integral away from the classical saddle, which can be quantum or stochastic in origin. 

To make this more concrete, we further note that, because of the normalization condition \(S_{\rm eff}[x_r,0]=0\), the action contains at least one power of \(x_a\). In a regime in which \(x_a\) is small, it is natural to expand the action as
\begin{align}\label{aeexpansion}
S_{\rm eff}[x_r,x_a]
&=
\int dt\, x_a(t)\,\mathcal E[x_r](t)
+\frac{i}{2}\int dt\,dt'\,x_a(t)\,N(t,t')\,x_a(t') +\mathcal O(x_a^3) .
\end{align}
Here \(\mathcal E[x_r](t)\) is some functional of the retarded field, and \(N(t,t')\) is a real symmetric kernel. The term linear in \(x_a\) is the part that survives if one varies with respect to the advanced field and then sets \(x_a=0\). The quadratic term is the leading imaginary contribution allowed by the Schwinger--Keldysh positivity conditions. Higher powers of \(x_a\) encode genuinely quantum or non-Gaussian corrections to the effective dynamics.

The semiclassical stochastic limit consists in truncating \eqref{aeexpansion} at quadratic order in \(x_a\),
\begin{equation}\label{quadratictruncate}
S_{\rm eff}[x_r,x_a]
\approx
\int dt\, x_a(t)\,\mathcal E[x_r](t)
+\frac{i}{2}\int dt\,dt'\,x_a(t)\,N(t,t')\,x_a(t') .
\end{equation}
At this stage, the Hubbard--Stratonovich transformation becomes useful. Since the kernel \(N(t,t')\) is positive semidefinite, one may use the Gaussian identity
\begin{align}\label{HSidentity}
\exp\!\left[
-\frac12\int dt\,dt'\,x_a(t)\,N(t,t')\,x_a(t')
\right]
&\propto
\int \mathcal D\xi\;
\exp\Biggl[
-\frac12\int dt\,dt'\,\xi(t)\,N^{-1}(t,t')\,\xi(t')
\nonumber\\
&\hspace{2.4cm}
+i\int dt\,\xi(t)\,x_a(t)
\Biggr] .
\end{align}
Here \(\xi(t)\) is an auxiliary field. The transformation \eqref{HSidentity} is exact, provided the action has been truncated to quadratic order in \(x_a\). After inserting it into the path integral, the effective action becomes linear in the advanced field,
\begin{equation}
S_{\rm eff}[x_r,x_a,\xi]
=
\int dt\,x_a(t)\bigl(\mathcal E[x_r](t)-\xi(t)\bigr) ,
\end{equation}
up to the Gaussian weight for \(\xi\). The path integral over \(x_a\) can then be performed exactly and produces a functional Dirac delta, which imposes
\begin{equation}\label{LangevinGeneral}
\mathcal E[x_r](t)=\xi(t) .
\end{equation}
This is the Langevin-type stochastic equation associated with the truncated Schwinger--Keldysh action.

The auxiliary field \(\xi\) is Gaussian distributed with zero mean and two-point function
\begin{equation}
\langle \xi(t)\xi(t')\rangle_\xi = N(t,t') ,
\end{equation}
where the average is taken with respect to the Gaussian measure appearing in \eqref{HSidentity}. Thus the kernel \(N(t,t')\) becomes the noise correlator of the stochastic process. In this way, the term quadratic in the advanced field is reinterpreted as Gaussian noise, while the term linear in \(x_a\) becomes the deterministic drift or equation of motion.

This logic may be illustrated with the Gaussian example discussed above. Suppose the local effective action takes the form
\begin{equation}
S_{\rm eff}[x_r,x_a]
=
\int dt\,
\left[
x_a\bigl(\ddot x_r+\kappa \dot x_r+\Omega^2 x_r\bigr)
+\frac{i}{2}N\,x_a^2
\right] ,
\end{equation}
where, for simplicity, we have written a local white-noise kernel \(N(t,t')=N\,\delta(t-t')\). The Hubbard--Stratonovich transformation then introduces a Gaussian field \(\xi(t)\) with correlator
\begin{equation}
\langle \xi(t)\xi(t')\rangle_\xi = N\,\delta(t-t') ,
\end{equation}
and the path integral over \(x_a\) yields
\begin{equation}
\ddot x_r+\kappa \dot x_r+\Omega^2 x_r=\xi(t) .
\end{equation}
This is the ordinary \textit{Langevin equation} for a damped harmonic oscillator driven by Gaussian white noise. The Schwinger--Keldysh description therefore reproduces the familiar stochastic description once the effective action is truncated to quadratic order in the advanced field.

It is important to keep track of what is exact and what is approximate in this construction. The Hubbard--Stratonovich transformation is an exact identity for a quadratic form. What is approximate is the truncation of the Schwinger--Keldysh action to at most quadratic order in \(x_a\). If higher powers of \(x_a\) are present, then the simple Gaussian noise representation is no longer exact. Such terms correspond to non-Gaussian fluctuations or genuinely quantum corrections beyond ordinary Langevin dynamics.


\paragraph{From the Langevin equation to the Fokker--Planck equation} For each realization of the noise $ \xi(t)$, the Langevin equation provides a realization of the time evolution of $ x(t)$. For a distribution of realizations of the noise, this gives a probability distribution $ P(x,t)$ for the value of $ x$ at a given time $ t$. Here we derive the differential equation satisfied by $ P(x,t)$. For simplicity, restrict ourselves to a first-order Langevin equation with additive Gaussian white noise, to be specified below. This restriction allows us to avoid the distinction between the It\^o and Stratonovich prescriptions, which becomes relevant for multiplicative noise. Langevin equations containing higher time derivatives may be brought into first-order form by enlarging the space of dynamical variables, for example by treating the position and velocity as independent variables.

Consider the stochastic equation
\begin{equation}
\dot{x}(t)
=
A\bigl(x(t)\bigr)
+
\xi(t),
\label{firstorderlangevin}
\end{equation}
where \(A(x)\) is the deterministic drift and \(\xi(t)\) is Gaussian white noise satisfying
\begin{equation}
\langle \xi(t) \rangle_{\xi}
=
0,
\qquad
\langle \xi(t)\xi(t') \rangle_{\xi}
=
N\,\delta(t-t').
\label{additivewhitenoise}
\end{equation}
The Langevin equation describes individual stochastic trajectories. Our goal is instead to determine the probability density \(P(x,t)\) for the stochastic variable to take the value \(x\) at time \(t\).

Over a short interval \(\Delta t\), equation \eqref{firstorderlangevin} gives
\begin{equation}
\Delta x
=
A(x)\Delta t
+
\Delta W,
\label{shorttimeincrement}
\end{equation}
where
\begin{equation}
\Delta W
\equiv
\int_{t}^{t+\Delta t}dt'\,\xi(t')
\end{equation}
is a Gaussian random variable with
\begin{equation}
\langle \Delta W\rangle_{\xi}
=
0,
\qquad
\langle (\Delta W)^{2}\rangle_{\xi}
=
N\Delta t.
\end{equation}
It follows that
\begin{align}
\langle \Delta x\rangle_{\xi}
&=
A(x)\Delta t,
\\
\langle (\Delta x)^{2}\rangle_{\xi}
&=
N\Delta t
+
\mathcal O\bigl((\Delta t)^{2}\bigr).
\end{align}
Higher moments of \(\Delta x\) are of higher order than \(\Delta t\) and do not contribute in the continuum limit.

Let \(f(x)\) be an arbitrary smooth test function. During the short interval \(\Delta t\),
\begin{equation}
f(x+\Delta x)
=
f(x)
+
f'(x)\Delta x
+
\frac{1}{2}f''(x)(\Delta x)^{2}
+
\cdots .
\end{equation}
Averaging over the noise and keeping terms of order \(\Delta t\), we obtain
\begin{equation}
\left\langle
f(x+\Delta x)-f(x)
\right\rangle_{\xi}
=
\left[
A(x)f'(x)
+
\frac{N}{2}f''(x)
\right]
\Delta t
+
\mathcal O\bigl((\Delta t)^{2}\bigr).
\end{equation}
Taking the continuum limit gives
\begin{equation}
\frac{d}{dt}
\left\langle
f(x)
\right\rangle
=
\left\langle
A(x)f'(x)
+
\frac{N}{2}f''(x)
\right\rangle.
\label{testfunctionevolution}
\end{equation}

The expectation value of \(f\) may also be written in terms of the probability density,
\begin{equation}
\left\langle
f(x)
\right\rangle
=
\int dx\,P(x,t)f(x).
\end{equation}
Therefore,
\begin{equation}
\frac{d}{dt}
\left\langle
f(x)
\right\rangle
=
\int dx\,
f(x)
\frac{\partial P(x,t)}{\partial t}.
\label{expectationfromP}
\end{equation}
On the other hand, using \eqref{testfunctionevolution},
\begin{equation}
\frac{d}{dt}
\left\langle
f(x)
\right\rangle
=
\int dx\,P(x,t)
\left[
A(x)f'(x)
+
\frac{N}{2}f''(x)
\right].
\end{equation}
Integrating by parts and assuming that the boundary terms vanish,
\begin{equation}
\frac{d}{dt}
\left\langle
f(x)
\right\rangle
=
\int dx\,f(x)
\left[
-
\frac{\partial}{\partial x}
\left(
A(x)P(x,t)
\right)
+
\frac{N}{2}
\frac{\partial^{2}P(x,t)}{\partial x^{2}}
\right].
\end{equation}
Since \(f(x)\) is arbitrary, comparison with \eqref{expectationfromP} yields the \textit{Fokker--Planck equation}
\begin{equation}
\frac{\partial P(x,t)}{\partial t}
=
-
\frac{\partial}{\partial x}
\left[
A(x)P(x,t)
\right]
+
\frac{N}{2}
\frac{\partial^{2}P(x,t)}{\partial x^{2}}.
\label{fokkerplanckgeneral}
\end{equation}

The first term transports probability according to the deterministic drift \(A(x)\), while the second term describes diffusion generated by the stochastic noise. We will use this result to describe light scalar fields in de Sitter in Section \ref{sec:stochastic}.

It is often useful to write \eqref{fokkerplanckgeneral} as a continuity equation,
\begin{equation}
\frac{\partial P}{\partial t}
=
-
\frac{\partial J}{\partial x},
\end{equation}
with probability current
\begin{equation}
J(x,t)
=
A(x)P(x,t)
-
\frac{N}{2}
\frac{\partial P(x,t)}{\partial x}.
\end{equation}

For completeness, consider a system of first-order Langevin equations
\begin{equation}
\dot{X}^{I}
=
A^{I}(X)
+
\xi^{I}(t),
\end{equation}
with additive Gaussian noise
\begin{equation}
\left\langle
\xi^{I}(t)\xi^{J}(t')
\right\rangle_{\xi}
=
N^{IJ}\delta(t-t').
\end{equation}
The associated probability density \(P(X,t)\) obeys
\begin{equation}
\frac{\partial P}{\partial t}
=
-
\frac{\partial}{\partial X^{I}}
\left[
A^{I}(X)P
\right]
+
\frac{1}{2}
N^{IJ}
\frac{\partial^{2}P}
{\partial X^{I}\partial X^{J}}.
\end{equation}
A higher-order Langevin equation may be treated by introducing additional variables until the dynamics is written as such a first-order system.


\paragraph{The Martin--Siggia--Rose construction}

It is useful to briefly review the Martin--Siggia--Rose construction \cite{PhysRevA.8.423} for a generic stochastic ordinary differential equation, in order to better understand its relation to the semiclassical limit of the SK path integral. Consider a set of real variables \(\phi^I(t)\), with \(I=1,\dots,N_r\), obeying
\begin{equation}
E^i[\phi](t)=\xi^i(t) ,
\end{equation}
where \(E^i[\phi]\), for \(i=1,\dots,N_a\), is some, possibly non-linear, differential operator and \(\xi^i(t)\) is a stochastic noise. For definiteness, let the noise be Gaussian with vanishing mean and covariance
\begin{equation}
\langle \xi^i(t)\,\xi^j(t')\rangle
=
N^{ij}(t,t') .
\end{equation}
One is then interested in predicting statistical properties of the fields \(\phi^I\), which must be averaged over the realizations of the noise. There is no quantum mechanics in this setting. This is a classical stochastic problem.

The central idea of Martin, Siggia, and Rose is to rewrite this stochastic average as a functional integral. One first inserts into the noise average a functional delta imposing the stochastic equation\footnote{For simplicity, in this derivation we now assume $  N_{r}=N_{a} $, so that the Jacobian is just a determinant. However, the MSR formalism can be generalized to $  N_{a}\neq N_{r} $.},
\begin{equation}
1=
\int \mathcal D\phi\;
\delta\!\bigl(E[\phi]-\xi\bigr)\,
\det\!\left(\frac{\delta E}{\delta \phi}\right) ,
\end{equation}
where the determinant is the Jacobian associated with the change of variables from \(\phi\) to \(E[\phi]\). The delta functional may then be exponentiated by introducing an auxiliary field \(\hat\phi_i(t)\), often called the response field,
\begin{equation}
\delta\!\bigl(E[\phi]-\xi\bigr)
\propto
\int \mathcal D\hat\phi\;
\exp\!\left[
i\int dt\,\hat\phi_i(t)\bigl(E^i[\phi](t)-\xi^i(t)\bigr)
\right] .
\end{equation}
At this stage the path integral is still exact.

If the noise is Gaussian, one may perform the average over \(\xi\) explicitly. This produces a term quadratic in the response field and yields the Martin--Siggia--Rose action
\begin{align}
S_{\rm MSR}[\phi,\hat\phi]
&=
\int dt\,\hat\phi_i(t)\,E^i[\phi](t)
+\frac{i}{2}\int dt\,dt'\,
\hat\phi_i(t)\,N^{ij}(t,t')\,\hat\phi_j(t')
\nonumber\\
&\qquad
+S_{\rm Jac}[\phi,\hat\phi] .
\end{align}
Here \(S_{\rm Jac}\) denotes the contribution of the Jacobian, which in many simple cases may be absorbed into a normalization or represented by ghosts if needed. The resulting functional integral generates the stochastic correlation functions of the original Langevin system. The term linear in \(\hat\phi_i\) imposes the deterministic part of the equations of motion, while the quadratic term encodes the statistics of the noise. In this sense, the response fields are Lagrange multipliers dressed by the noise average.

The construction is more general than the simplest Langevin equation. It applies to non-linear differential equations, to multiplicative noise with appropriate care about discretization, and to non-local kernels \(N^{ij}(t,t')\). The key point is always the same: the stochastic dynamics is represented by a path integral over the original variables together with one auxiliary response field for each equation, and the noise average generates terms quadratic and, for non-Gaussian noise, higher order in the response fields.

It is worth emphasizing the relation to the semiclassical limit of the Schwinger--Keldysh formalism. In that context, one expands the effective action in powers of the advanced fields. Truncating the action to terms linear and quadratic in those fields yields precisely the same structural form as the Martin--Siggia--Rose action: the linear term imposes an effective equation of motion, while the quadratic term can be interpreted as Gaussian noise after a Hubbard--Stratonovich transformation. In this sense, the Martin--Siggia--Rose formalism may be viewed as the stochastic semiclassical counterpart of the Schwinger--Keldysh path integral.


\subsection{A dissipative Gaussian scalar field and its propagators}

From this point onward, it is convenient to pass from a single variable $ x(t)$ to a field $\phi(t,\mathbf x) $. If the background is homogeneous and isotropic, it is natural to Fourier transform in space,
\begin{equation}
\phi(t,\mathbf x)
=
\int \frac{d^d k}{(2\pi)^d}\,
e^{i\mathbf k\cdot \mathbf x}\,\phi_{\mathbf k}(t) .
\end{equation}
Then different momenta decouple in a quadratic theory, and each Fourier mode \(\phi_{\mathbf k}(t)\) behaves as the analogue of the single variable discussed in the previous subsections. In this sense, a free or Gaussian field theory may be viewed as a collection of decoupled oscillators labelled by \(\mathbf k\). Of course, quantum field theory comes with additional complications on top of quantum mechanics, and we may mention some of them, such as loop divergences and renormalization, when we get to it. 

Accordingly, the Schwinger--Keldysh doubled variables become \(\phi_\pm(t,\mathbf x)\), or equivalently
\begin{equation}
\phi_r(t,\mathbf x):=\frac{\phi_+(t,\mathbf x)+\phi_-(t,\mathbf x)}{2},
\qquad
\phi_a(t,\mathbf x):=\phi_+(t,\mathbf x)-\phi_-(t,\mathbf x) .
\end{equation}
We now consider the simplest dissipative Gaussian model in Minkowski spacetime. In the \(r/a\) basis, let the quadratic Schwinger--Keldysh action be
\begin{align}\label{DissipativeFieldAction}
S_{\rm SK}
=
\int dt\,\frac{d^d k}{(2\pi)^d}\,
\Bigg[
&\phi_a(-\mathbf k,t)
\Bigl(
\partial_t^2+\gamma\,\partial_t+\omega_{\mathbf k}^2
\Bigr)\phi_r(\mathbf k,t)
\nonumber\\
&\qquad
+\frac{i}{2}\,\mathcal N_{\mathbf k}\,
\phi_a(-\mathbf k,t)\phi_a(\mathbf k,t)
\Bigg] ,
\end{align}
where
\begin{equation}
\omega_{\mathbf k}^2:=\mathbf k^2+m^2 .
\end{equation}
Here \(\gamma\) is a damping coefficient, while \(\mathcal N_{\mathbf k}\geq 0\) is the Keldysh or noise kernel. Because of the constraints inherited from a unitary full system, in particular 
\[ -S_{\rm SK}[\phi_{r},-\phi_{a}]^{*}=S_{\rm SK}[\phi_{r},\phi_{a}]\,, \]
both $ \gamma $ and $ \mathcal{N}$ must be real. 
This action is local in time, quadratic, and translation invariant in space and time. It therefore provides a particularly simple laboratory in which all propagators can be computed explicitly.

In frequency space,
\begin{equation}
\phi(\omega,\mathbf k)
=
\int dt\,e^{i\omega t}\phi(t,\mathbf k) ,
\end{equation}
the action becomes
\begin{align}
S_{\rm SK}
=
\int \frac{d\omega}{2\pi}\frac{d^d k}{(2\pi)^d}
\Bigg[
\phi_a(-\omega,-\mathbf k)\,
\mathcal D_R(\omega,\mathbf k)\,
\phi_r(\omega,\mathbf k)
+\frac{i}{2}\mathcal N_{\mathbf k}\,
\phi_a(-\omega,-\mathbf k)\phi_a(\omega,\mathbf k)
\Bigg] ,
\end{align}
with retarded inverse kernel
\begin{equation}\label{DRkernel}
\mathcal D_R(\omega,\mathbf k)
=
-\omega^2-i\gamma\omega+\omega_{\mathbf k}^2 .
\end{equation}
The advanced kernel is its conjugate,
\begin{equation}
\mathcal D_A(\omega,\mathbf k)
=
-\omega^2+i\gamma\omega+\omega_{\mathbf k}^2
=
\mathcal D_R(\omega,\mathbf k)^* .
\end{equation}

Since the theory is Gaussian, the propagators are obtained by inverting the quadratic kernel. This is worked out in detail in Appendix \ref{appA}. In the \(r/a\) basis one finds\footnote{Notice a difference with the free-field propagators in closed quantum field theory. There, the inverse quadratic operator typically develops zeros on the mass shell, so its inversion requires an additional prescription, such as the familiar \(i\epsilon\). Here, by contrast, the retarded inverse kernel already contains a finite imaginary part proportional to \(\gamma  \), so the denominator does not vanish for real kinematics, and the poles are automatically displaced away from the real axis into the lower half-plane. The inversion is therefore unambiguous.}
\begin{align}
G_R(\omega,\mathbf k)
&=
\langle\!\langle \phi_r(\omega,\mathbf k)\phi_a(-\omega,-\mathbf k)\rangle\!\rangle
=
\frac{1}{\mathcal D_R(\omega,\mathbf k)}
=
\frac{1}{-\omega^2-i\gamma\omega+\omega_{\mathbf k}^2} ,
\label{GRfreq}
\\
G_A(\omega,\mathbf k)
&=
\langle\!\langle \phi_a(\omega,\mathbf k)\phi_r(-\omega,-\mathbf k)\rangle\!\rangle
=
\frac{1}{\mathcal D_A(\omega,\mathbf k)}
=
\frac{1}{-\omega^2+i\gamma\omega+\omega_{\mathbf k}^2} ,
\label{GAfreq}
\\
G_K(\omega,\mathbf k)
&=
\langle\!\langle \phi_r(\omega,\mathbf k)\phi_r(-\omega,-\mathbf k)\rangle\!\rangle
=
-\,i\,\mathcal N_{\mathbf k}\,
G_R(\omega,\mathbf k)\,G_A(\omega,\mathbf k)
\nonumber\\
&=
-\,i\,\mathcal N_{\mathbf k}\,
\frac{1}{\bigl(-\omega^2-i\gamma\omega+\omega_{\mathbf k}^2\bigr)
\bigl(-\omega^2+i\gamma\omega+\omega_{\mathbf k}^2\bigr)} ,
\label{GKfreq}
\\
G_{aa}(\omega,\mathbf k)
&=
\langle\!\langle \phi_a(\omega,\mathbf k)\phi_a(-\omega,-\mathbf k)\rangle\!\rangle
=
0 .
\end{align}
Thus the retarded and advanced propagators are determined entirely by the deterministic part of the action, which is linear in $ \phi_{a}$, while the Keldysh propagator depends also on the noise kernel \(\mathcal N_{\mathbf k}\). 

It is useful to rewrite these propagators in the \(\pm\) basis:
\begin{align}
G_{++}
&:=
\langle\!\langle \phi_+\phi_+\rangle\!\rangle
=
G_K+\frac12 G_R+\frac12 G_A ,
\\
G_{--}
&:=
\langle\!\langle \phi_-\phi_-\rangle\!\rangle
=
G_K-\frac12 G_R-\frac12 G_A ,
\\
G_{+-}
&:=
\langle\!\langle \phi_+\phi_-\rangle\!\rangle
=
G_K-\frac12 G_R+\frac12 G_A ,
\\
G_{-+}
&:=
\langle\!\langle \phi_-\phi_+\rangle\!\rangle
=
G_K+\frac12 G_R-\frac12 G_A .
\end{align}
In other words, once \(G_R\), \(G_A\), and \(G_K\) are known, all contour-ordered propagators in the \(\pm\) basis are determined.

Let us now discuss the time-domain retarded propagator. By definition, it is the retarded Green's function of
\begin{equation}
\Bigl(\partial_t^2+\gamma\partial_t+\omega_{\mathbf k}^2\Bigr)G_R(t,\mathbf k)
=
\delta(t) ,
\qquad
G_R(t,\mathbf k)=0 \quad \text{for } t<0 .
\end{equation}
By Fourier transforming back to the time domain%
\footnote{For a finite physical damping coefficient $\gamma>0$, the poles of $G_R$ are displaced into the lower half of the complex $\omega$-plane and, except for the massless zero mode, they do not lie on the real-frequency integration contour. No additional $i\epsilon$ prescription is therefore required for the Fourier transform. If $\gamma=0$, the poles return to the real axis and the Fourier transform must instead be defined with the usual retarded prescription,
\begin{equation}
\mathcal D_R(\omega,\mathbf k)
=
-\omega^2+\mathbf k^2+m^2-i\epsilon\omega,
\qquad
\epsilon\to0^+.
\end{equation}
Thus the retarded $i\epsilon$ prescription may be viewed as introducing an infinitesimal positive damping coefficient.}%
(see Appendix \ref{appA} for details), one finds, in the underdamped regime $ \gamma \leq 2\omega_{k}$,
\begin{equation}\label{GRtime}
G_R(t,\mathbf k)
=
\theta(t)\,
e^{-\gamma t/2}
\frac{\sin(\Omega_{\mathbf k}t)}{\Omega_{\mathbf k}},
\qquad
\Omega_{\mathbf k}
:=
\sqrt{\mathbf k^2+m^2-\frac{\gamma^2}{4}}.
\end{equation}
and in the overdamped regime $ \gamma \geq 2\omega_{k}$, 
\begin{equation}
G_R(t,\mathbf k)
=
\theta(t)\,
\frac{
e^{-(\gamma/2-\Lambda_{\mathbf k})t}
-
e^{-(\gamma/2+\Lambda_{\mathbf k})t}
}{
2\Lambda_{\mathbf k}
}, \qquad  \Lambda_{\mathbf k}
:=
\sqrt{\frac{\gamma^2}{4}-\mathbf k^2-m^2}.
\end{equation}
Thus every mode with $\mathbf k^2+m^2>0$ retains only exponentially suppressed memory of the past. In the underdamped regime the envelope decays on the timescale $2/\gamma$, whereas in the overdamped regime the late-time decay is controlled by the mode-dependent rate
\begin{equation}
\Gamma_{\rm slow}
=
\frac{\gamma}{2}-\Lambda_{\mathbf k}
\simeq
\frac{\mathbf k^2+m^2}{\gamma},
\end{equation}
where the last expression applies for $\mathbf k^2+m^2\ll\gamma^2$. The corresponding memory time therefore becomes arbitrarily long as $\mathbf k^2+m^2\to0$. For the exactly massless homogeneous mode, $ m=k=0$, one finds
\begin{equation}
G_R(t,\mathbf 0)
=
\theta(t)\,
\frac{1-e^{-\gamma t}}{\gamma},
\end{equation}
which approaches a constant rather than decaying at late times.


\begin{figure}
\centering
\includegraphics[width=0.9\textwidth]{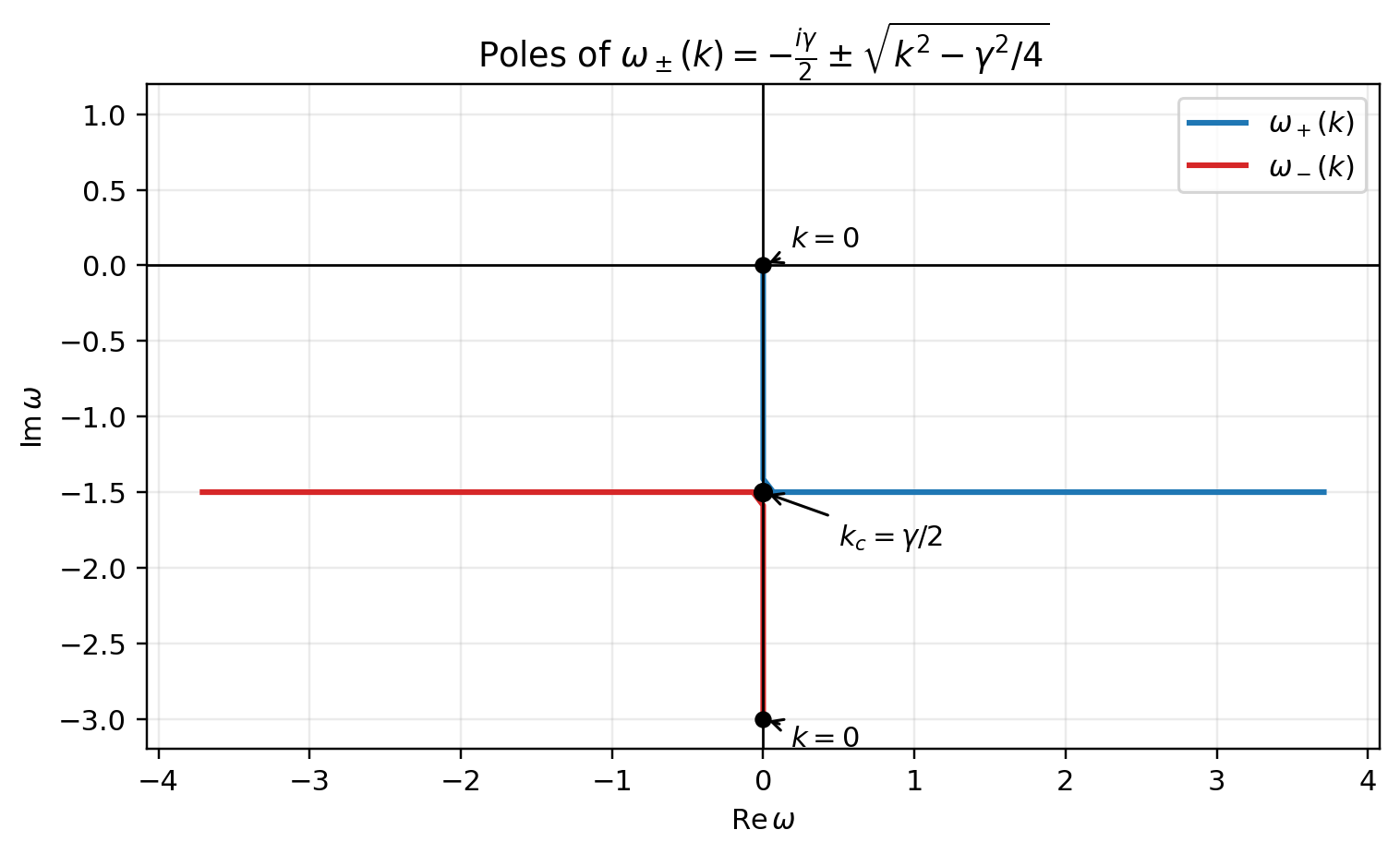}
\caption{Poles of the retarded propagator in the complex frequency plane for the massless dissipative mode. For \(k<\gamma/2\), both poles lie on the negative imaginary axis. At \(k=\gamma/2\) they merge, while for \(k>\gamma/2\) they move horizontally with constant imaginary part \(-\gamma/2\). \label{dispersion}}
\end{figure}

\paragraph{The dispersion relation} The poles of the retarded propagator, which define the relevant dispersion relation, are given by the zeros of \(\mathcal D_R(\omega,\mathbf k)\),
\begin{equation}
-\omega^2-i\gamma\omega+\omega_{\mathbf k}^2=0 .
\end{equation}
They are therefore
\begin{equation}\label{poles}
\omega_\pm(\mathbf k)
=
-\frac{i\gamma}{2}
\pm
\sqrt{\omega_{\mathbf k}^2-\frac{\gamma^2}{4}} .
\end{equation}
These poles always lie in the lower half of the complex \(\omega\)-plane, as required for a retarded propagator. This is shown in Figure \ref{dispersion} in the massless case. Their precise form distinguishes different regimes.

In the underdamped regime,
\begin{equation}
\gamma<2\omega_{\mathbf k},
\end{equation}
the square root in \eqref{poles} is real, and the poles have non-zero real parts. The corresponding modes oscillate with frequency \(\Omega_{\mathbf k}\) while decaying at rate \(\gamma/2\).

At critical damping,
\begin{equation}
\gamma=2\omega_{\mathbf k},
\end{equation}
the two poles coincide,
\begin{equation}
\omega_+(\mathbf k)=\omega_-(\mathbf k)=-i\gamma/2 .
\end{equation}

In the overdamped regime,
\begin{equation}
\gamma>2\omega_{\mathbf k},
\end{equation}
the square root becomes purely imaginary, and both poles are purely imaginary:
\begin{equation}
\omega_\pm(\mathbf k)
=
-\frac{i\gamma}{2}
\pm
i\sqrt{\frac{\gamma^2}{4}-\omega_{\mathbf k}^2} .
\end{equation}
In this case the two modes are no longer oscillatory. They are purely relaxational. In the low-momentum limit and for $  m=0 $ we find a low-frequency ``diffusive'' mode,
\begin{equation}
\omega(\mathbf k)\sim -\,iD\,\mathbf k^2+\cdots .
\end{equation}

It is also useful to comment on the real and imaginary parts of the retarded propagator. Writing
\begin{equation}
G_R(\omega,\mathbf k)
=
\frac{\omega_{\mathbf k}^2-\omega^2+i\gamma\omega}
{\bigl(\omega_{\mathbf k}^2-\omega^2\bigr)^2+\gamma^2\omega^2} ,
\end{equation}
one finds
\begin{align}
\Re G_R(\omega,\mathbf k)
&=
\frac{\omega_{\mathbf k}^2-\omega^2}
{\bigl(\omega_{\mathbf k}^2-\omega^2\bigr)^2+\gamma^2\omega^2} ,
\\
\Im G_R(\omega,\mathbf k)
&=
\frac{\gamma\omega}
{\bigl(\omega_{\mathbf k}^2-\omega^2\bigr)^2+\gamma^2\omega^2} .
\end{align}
The real part is dispersive and changes sign near the resonance, while the imaginary part is odd in \(\omega\) and measures dissipative broadening. In particular, the finite width of the pole is controlled by \(\gamma\).

Finally, the Keldysh propagator takes the explicit form
\begin{equation}
G_K(\omega,\mathbf k)
=
-\,i\,\mathcal N_{\mathbf k}\,
\frac{1}
{\bigl(\omega_{\mathbf k}^2-\omega^2\bigr)^2+\gamma^2\omega^2} .
\end{equation}
It is even in frequency and its amplitude is directly proportional to the noise kernel. 

At this stage \(\mathcal N_{\mathbf k}\) is arbitrary. Only if one further assumes thermal equilibrium does it become related to the retarded and advanced propagators by a fluctuation--dissipation or KMS relation, a point that we will discuss later.

To summarize, the dissipative Gaussian scalar field provides a simple but instructive example in which all Schwinger--Keldysh propagators can be computed explicitly. The retarded and advanced propagators are controlled by the inverse kernels \(\mathcal D_R\) and \(\mathcal D_A\), the Keldysh propagator encodes the noise spectrum, and the contour-ordered propagators in the \(\pm\) basis are reconstructed from these. The location of the retarded poles reveals whether the excitations are underdamped, critically damped, or overdamped, while a genuinely diffusive mode appears only in an additional low-frequency and low-momentum limit.


\section{Cosmology and inflation}\label{sec:5}

We now turn to the cosmological setting that will be the main arena for the rest of these notes. Our ultimate goal is to understand how the tools of open quantum systems, master equations, and Schwinger--Keldysh effective actions can be applied to the primordial universe, and in particular to inflation. Before doing so, however, it is useful to review the corresponding closed-system description. Inflation provides a remarkable playground for quantum field theory: the background spacetime is time dependent, the relevant observables are late-time correlators rather than scattering amplitudes, and the fluctuations generated during the early universe are believed to be the seeds of the structures observed in the sky today.

The aim of this section is therefore to introduce the basic cosmological ingredients and the formalism needed for later developments. We begin with a brief review of homogeneous cosmological backgrounds and the motivations for inflation, then discuss quasi-de Sitter expansion and simple single-field realizations. We next explain how cosmological observables are computed as equal-time correlators, introduce the corresponding perturbative Feynman rules, and briefly review primordial non-Gaussianity through the bispectrum. This will set the stage for the effective field theory of inflation in the next section, where we will first study the unitary case and then generalize it to an open-system framework.

 
\subsection{Background cosmology}

We begin with a brief review of the homogeneous cosmological backgrounds relevant for inflation. In general relativity, the spacetime metric is dynamical and obeys the Einstein equations
\begin{equation}
R_{\mu\nu}-\frac12 g_{\mu\nu}R=\frac{1}{M_{\rm Pl}^2}T_{\mu\nu},
\end{equation}
which follow from the action
\begin{equation}
S=\int d^4x\,\sqrt{-g}\,\left[\frac{M_{\rm Pl}^2}{2}R+\mathcal L_M\right].
\end{equation}
The energy-momentum tensor $ T_{\mu\nu}$ is defined as the response of the matter action to a variation of the spacetime metric,
\begin{equation}
T^{\mu\nu}
:=
\frac{2}{\sqrt{-g}}\,
\frac{\delta S_M}{\delta g_{\mu\nu}} \, .
\end{equation}
Exact solutions are in general difficult to obtain, but the high degree of symmetry observed on sufficiently large scales suggests focusing on homogeneous and isotropic spacetimes.

The simplest such metric is the Friedmann--Lema\^itre--Robertson--Walker geometry,
\begin{equation}
ds^2=-dt^2+a(t)^2\,\frac{\delta_{ij}\,dx^i dx^j}{(1+Kx^2/4)^2},
\end{equation}
where \(a(t)\) is the scale factor and \(K\) parameterizes the spatial curvature. We use the convention that the scale factor is set to unity today 
\begin{equation}
a(t_{0})=a_0=1 .
\end{equation}
It is often convenient to introduce conformal time \(\tau\), defined by \(dt=a\,d\tau\), so that
\begin{equation}
ds^2=a(\tau)^2\left[-d\tau^2+\frac{\delta_{ij}\,dx^i dx^j}{(1+Kx^2/4)^2}\right].
\end{equation}
The rate of expansion is measured by the Hubble parameter
\begin{equation}
H:=\frac{\dot a}{a}.
\end{equation}

Homogeneity and isotropy require the background energy-momentum tensor to take the perfect-fluid form
\begin{equation}
T^\mu{}_\nu={\rm diag}(-\rho,p,p,p),
\end{equation}
with energy density \(\rho(t)\) and pressure \(p(t)\). Covariant conservation of \(T_{\mu\nu}\) then implies the continuity equation
\begin{equation}
\dot\rho+3H(\rho+p)=0.
\end{equation}
For a component with equation of state
\begin{equation}
p=w\rho ,
\end{equation}
the continuity equation
\begin{equation}
\dot\rho+3H(\rho+p)=0
\end{equation}
integrates to
\begin{equation}
\rho(a)=\rho_0\,a^{-3(1+w)} ,
\end{equation}
where \(\rho_0\) is the value of the energy density today. This immediately gives the familiar scalings
\begin{equation}
\rho_{\rm rad}(a)=\rho_{{\rm rad},0}\,a^{-4},
\qquad
\rho_{\rm mat}(a)=\rho_{{\rm mat},0}\,a^{-3},
\qquad
\rho_\Lambda(a)=\rho_{\Lambda,0} .
\end{equation}

The \(00\) component of Einstein's equations gives the \textit{Friedmann equation}
\begin{equation}
3M_{\rm Pl}^2\left(H^2+\frac{K}{a^2}\right)=\sum_\alpha \rho_\alpha,
\end{equation}
while a second useful combination yields the acceleration equation
\begin{equation}
M_{\rm Pl}^2\,\frac{\ddot a}{a}=-\frac16(\rho+3p),
\end{equation}
and, equivalently the \textit{second Friedmann equation}
\begin{equation}
-\,M_{\rm Pl}^2 \left(  \dot H -\frac{K}{a^{2}}\right)=\frac12(\rho+p).
\end{equation}
These equations summarize the homogeneous dynamics of the universe.
In the standard cosmological history, different components dominate at different times. Radiation domination extends from the end of reheating until matter-radiation equality, which occurs at redshift
\begin{equation}
z_{\rm eq}\simeq 3.4\times 10^{3},
\end{equation}
when the universe was about $t_{\rm eq}\sim 5\times 10^{4}\ {\rm yr} $ old. During this era one has \(w=1/3\), and therefore
\begin{equation}
a(t)=\left(\frac{t}{t_0}\right)^{1/2}.
\end{equation}
Matter domination then lasts approximately from \(z_{\rm eq}\) down to the epoch at which matter and dark energy become comparable,
\begin{equation}
z_{\Lambda{\rm eq}}\sim 0.3,
\end{equation}
corresponding to a cosmic time of order $t_{\Lambda{\rm eq}}\sim 1\times 10^{10}\ {\rm yr} $. During matter domination one has \(w\simeq 0\), and so
\begin{equation}
a(t)=\left(\frac{t}{t_0}\right)^{2/3}.
\end{equation}
Finally, in the late-time dark-energy dominated era, which takes over after \(z\sim 0.3\), the equation of state is approximately \(w=-1\), so the scale factor grows approximately exponentially,
\begin{equation}
a(t)=\exp\!\bigl[H(t-t_0)\bigr].
\end{equation}
The transition to accelerated expansion actually begins somewhat earlier, at redshift \(z\sim 0.7\). Here \(t_0\simeq 1.4\times 10^{10}\,{\rm yr}\) denotes the present age of the universe.

\paragraph{Motivations for inflation} Despite the success of the hot Big Bang description after nucleosynthesis, it leaves several striking questions unanswered. The first is the \textit{horizon problem}: in a purely decelerating universe, regions of the Cosmic Microwave Background that appear statistically similar were never in causal contact. The second is the \textit{curvature problem}, namely the fact that current observations are consistent with a universe that is extremely close to spatially flat
\begin{equation}
\Omega_{K,0}=0.000\pm 0.005 .
\end{equation}
On the other hand, in a decelerating expanding universe the curvature contribution grows relative to radiation and matter
\begin{equation}
\frac{d}{dt}|\Omega_K|
=
-2\,|\Omega_K|\frac{\ddot a}{\dot a}
=
|\Omega_K|\,H\,(1+3w),
\end{equation}
Therefore, for ordinary radiation- or matter-dominated evolution, \(\Omega_K\) increases with time, so the fact that it is so small today implies that it must have been even smaller in the early universe. The curvature problem is precisely the question of why the initial curvature had to be so extraordinarily close to zero.  

The third is the \textit{phase coherence problem}: cosmological perturbations are observed to oscillate coherently even on scales that would be super-horizon in the standard hot Big Bang evolution.

These three puzzles all point toward the existence of a primordial phase of accelerated expansion, before the hot big bang, which we call inflation, during which $ \dot a>0$ and $ \ddot a >0$.

\begin{figure}
\centering
\includegraphics[width=0.7\textwidth]{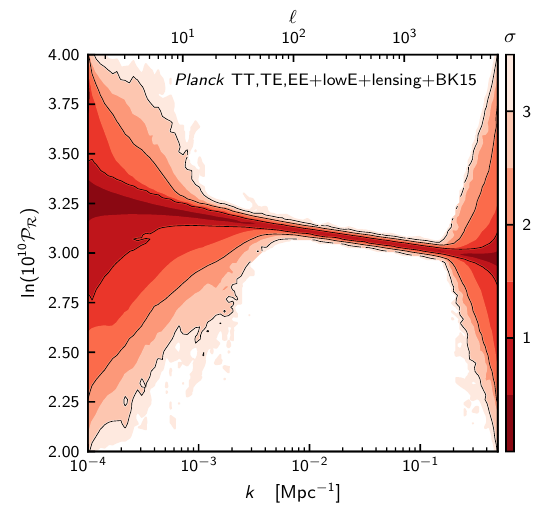}
\caption{Observations from Planck 2018 \cite{Planck:2018jri}. The plot shows the constraints on the primordial power spectrum, which is scale invariant up to a $  4\% $ ``red tilt''.\label{fig:scale}}
\end{figure}

The last motivation for inflation is the scale invariance problem. Observations indicate that the primordial perturbations have approximately the same amplitude on all cosmological scales that we can probe. In the language of equal-time correlators, scale invariance means that for every \(\lambda >0\),
\begin{equation}
\langle \phi(x_1)\phi(x_2)\cdots \phi(x_n)\rangle
=
\langle \phi(\lambda x_1)\phi(\lambda x_2)\cdots \phi(\lambda x_n)\rangle .
\end{equation}
In particular, the observed two-point function is very close to scale invariant. This is shown, for example, in Figure \ref{fig:scale}. Note that this is extremely surprising. Our general intuition, built upon the calculation of, for example, correlators in Minkowski, would tell us that the two-point function should decay as two points are separated. Scale invariance says that the correlation is independent of distance, an extremely remarkable fact that deserves a clear explanation. 

A very attractive possibility is that this property follows from a symmetry of the primordial background rather than from an unexplained coincidence. A simple and elegant possibility is that the early universe was approximately described by de Sitter spacetime,
\begin{equation}
ds^2
=
-\frac{d\tau^2-dx^i dx^i}{H^2\tau^2}
=
-dt^2+e^{2Ht}dx^i dx^i ,
\end{equation}
which is maximally symmetric. In particular, de Sitter in flat slicing admits the dilation isometry
\begin{equation}
\tau \to \lambda \tau,
\qquad
x^i \to \lambda x^i .
\end{equation}
If the other background quantities respect this symmetry at least approximately, then it provides a natural explanation for the approximate scale invariance of primordial correlators. This is In the following, we will assume an extended period during which the spacetime is well approximated by de Sitter. 


\subsection{A prolonged phase of quasi-de Sitter expansion and an example}

The problems reviewed above suggest that the early universe underwent a prolonged phase of accelerated expansion, together with an approximately scale-invariant background. The spacetime with these properties is de Sitter,
\begin{equation}
ds^2
=
-\frac{d\tau^2-dx^i dx^i}{H^2\tau^2}
=
-dt^2+e^{2Ht}dx^i dx^i ,
\end{equation}
whose scale factor is
\begin{equation}
a(\tau)=-\frac{1}{H\tau}=e^{Ht} .
\end{equation}
However, exact de Sitter is eternal, whereas inflation must end. We are therefore led to consider an approximately de Sitter phase, in which the Hubble parameter varies slowly with time. 

It is useful to rewrite the condition of accelerated expansion as
\begin{equation}
\frac{\ddot a}{a}=\dot H+H^2=H^2(1-\epsilon)>0 ,
\end{equation}
where we introduced the \textit{first Hubble slow-roll parameter}
\begin{equation}
\epsilon := -\frac{\dot H}{H^2} .
\end{equation}
Thus accelerated expansion requires \(\epsilon<1\), while a quasi-de Sitter stage corresponds to
\begin{equation}
0<\epsilon\ll 1 .
\end{equation}
This condition captures the idea that \(H\) varies only slowly while the universe expands rapidly.  

A second important requirement is that inflation last long enough. To quantify this, one introduces the \textit{number of e-foldings} \(N\),
\begin{equation}
dN := d\log a = H\,dt ,
\end{equation}
so that
\begin{equation}
N-N_0=\log\!\left(\frac{a}{a_0}\right) .
\end{equation}
Solving the horizon and curvature problems requires a sufficiently long interval of accelerated expansion,
\begin{equation}
\Delta N_{\rm infl}:=N_e-N_i
>
50+\log\!\left(\frac{T_{\rm reh}}{10^{10}\,{\rm GeV}}\right) ,
\end{equation}
where $ T_{\rm reh}$ is temperature of the universe at the end of reheating, a.k.a. the beginning of the hot big bang. For phenomenologically relevant reheating temperatures this is often summarized as
\begin{equation}
\Delta N_{\rm infl}\sim 25\text{--}60 .
\end{equation}
In rough estimates one often simply uses \(\Delta N_{\rm infl}\sim 50\).  

Since observations indicate approximate scale invariance over several e-foldings, it is natural to require that the departure from de Sitter remain small during the whole inflationary era. This is conveniently expressed in terms of the higher Hubble slow-roll parameters,
\begin{equation}
\epsilon = -\partial_N\log H ,
\qquad
\eta = \frac{\dot\epsilon}{H\epsilon}=\partial_N\log\epsilon ,
\qquad
\xi_{n\ge 3}:=\partial_N\log\xi_{n-1} ,
\end{equation}
with \(\xi_2\equiv \eta\). Expanding \(\epsilon(N)\) around some reference time \(N_\ast\), one finds schematically
\begin{align}
\epsilon(N)-\epsilon(N_\ast)
&=
\left.\frac{\partial\epsilon}{\partial N}\right|_{N_\ast}(N-N_\ast)
+\left.\frac{\partial^2\epsilon}{\partial N^2}\right|_{N_\ast}\frac{(N-N_\ast)^2}{2}
+\cdots
\nonumber\\
&=
\epsilon\left[
\eta(N-N_\ast)
+\eta(\eta+\xi_3)\frac{(N-N_\ast)^2}{2}
+\cdots
\right] .
\end{align}
This motivates the standard slow-roll conditions
\begin{equation}
\epsilon,\eta,\xi_n \ll 1 .
\end{equation}
These conditions ensure not only accelerated expansion, but also that the background remains close to de Sitter over many e-foldings. It should be noted that this description of inflation is somewhat different from what is encountered in introductory courses. In particular, note that we have not yet invoked a scalar field or a potential. Moreover, the slow-roll parameters have nothing to do with those ingredients and have been introduced here exclusively as a property of the spacetime. In this sense, this description is so far agnostic about what drove inflation. \\

\paragraph{A class of inflationary models} A broad class of explicit realizations is provided by single-field \(P(X,\phi)\) theories,
\begin{equation}
S=\int d^4x\,\sqrt{-g}\left[\frac12 M_{\rm Pl}^2R+P(X,\phi)\right] ,
\end{equation}
where $ X$ denotes the familiar canonical kinetic term 
\begin{equation}
X:=-\frac12 g^{\mu\nu}\partial_\mu\phi\,\partial_\nu\phi
=
\frac12\left(\dot\phi^2-\frac{\partial_i\phi\,\delta^{ij}\partial_j\phi}{a^2}\right) .
\end{equation}
A canonical scalar corresponds to the special case
\begin{equation}
P(X,\phi)=X-V(\phi) .
\end{equation}
We will not discuss here more general non-minimal couplings such as \(R\phi^2\) or curvature couplings with additional derivatives. 

The energy-momentum tensor derived from this action is
\begin{equation}
T_{\mu\nu}=P_{,X}\partial_\mu\phi\,\partial_\nu\phi+g_{\mu\nu}P(X,\phi) .
\end{equation}
This takes the perfect-fluid form
\begin{equation}
T_{\mu\nu}=(\rho+p)u_\mu u_\nu + p\,g_{\mu\nu}
\end{equation}
under the identifications
\begin{equation}
\rho = 2X P_{,X}-P ,
\qquad
p = P ,
\qquad
u_\mu = -\frac{\partial_\mu\phi}{\sqrt{2X}} .
\end{equation}
Thus \(P(X,\phi)\) theories provide a natural field-theoretic realization of the homogeneous cosmological fluid.  

Focusing on a homogeneous background \(\bar\phi(t)\), the scalar equation of motion becomes
\begin{equation}
\ddot{\bar\phi}\,\bigl(P_{,X}+2X P_{,XX}\bigr)
+
3H\dot{\bar\phi}\,P_{,X}
+
\bigl(2X P_{,X\phi}-P_{,\phi}\bigr)
=0 ,
\end{equation}
while the Friedmann and acceleration equations read
\begin{equation}
3M_{\rm Pl}^2H^2 = 2X P_{,X}-P ,
\qquad
- M_{\rm Pl}^2\dot H = X P_{,X} .
\end{equation}
The detailed background evolution depends on the specific choice of the function \(P\), but for our purposes we will not need to solve these equations explicitly.

It is enough to note that the first slow-roll parameter can be written as
\begin{equation}
\epsilon
=
\frac{3X P_{,X}}{2X P_{,X}-P} .
\end{equation}
This shows that many choices of \(P(X,\phi)\) can support a prolonged phase of slow-roll inflation, namely a regime with \(\epsilon,\eta\ll 1\). In this sense, \(P(X,\phi)\) theories provide a simple and flexible framework in which the quasi-de Sitter inflationary backgrounds discussed above can be realized dynamically.  


\subsection{Cosmology and in-in correlators}

Here we explain that the main objects of interest in cosmology are so-called cosmological correlators, namely expectation values of the product of local operators at the same time, usually taken to be the end of inflation. Luckily for us, this is exactly the object that is computed by the Schwinger-Keldysh formalism. In this section, we describe this formalism for a closed system in a pure state. This is the standard assumption in the vast majority of the literature. In the next section, we will extend this to open systems and density matrices, where we will see that the diagrammatic Feynman rules of perturbation theories are essentially the same.\\


In cosmology, we are interested in expectation values of the form
\begin{equation}
\langle \Omega | \, O(\tau)\, | \Omega \rangle ,
\end{equation}
where \(O(\tau)\) is a product of local Heisenberg-picture operators evaluated at the same conformal time \(\tau\), and \(|\Omega\rangle\) denotes the interacting vacuum. In inflationary applications one is often interested in the late-time limit
\begin{equation}
\tau \to 0^- ,
\end{equation}
which in de Sitter corresponds to future infinity, or equivalently to the future conformal boundary. This is the natural cosmological analogue of sending asymptotic times to \(\pm\infty\) in scattering theory.

To compute such correlators, it is convenient to work in the interaction picture. For a closed unitary system, let the Hamiltonian be split as
\begin{equation}
\hat H(\tau)=\hat H_0(\tau)+\hat H_{\rm int}(\tau) .
\end{equation}
The free part \(\hat H_0\) defines the interaction-picture fields, while \(\hat H_{\rm int}\) is treated perturbatively. If \(|0\rangle\) denotes the free vacuum, then the interacting vacuum \(|\Omega\rangle\) may be obtained by adiabatically switching on the interactions in the infinite past. The resulting expectation value may be written as
\begin{align}
\langle O(\tau)\rangle
&:=
\langle \Omega |\, O(\tau)\, | \Omega \rangle \\ \nonumber
&=
\frac{
\langle 0|\, \bar T \exp\!\left(i\int_{-\infty}^{\tau} d\tau' \,\hat H_{\rm int}(\tau')\right)
\, O(\tau)\,
T \exp\!\left(-i\int_{-\infty}^{\tau} d\tau' \,\hat H_{\rm int}(\tau')\right)
|0\rangle
}{
\langle 0|\, \bar T \exp\!\left(i\int_{-\infty}^{\tau} d\tau' \,\hat H_{\rm int}(\tau')\right)
T \exp\!\left(-i\int_{-\infty}^{\tau} d\tau' \,\hat H_{\rm int}(\tau')\right)
|0\rangle
} .
\end{align}
Here \(T\) and \(\bar T\) denote time ordering and anti-time ordering, respectively.

The denominator is in fact equal to one once the correct projection onto the interacting vacuum is implemented. To make this projection precise, one slightly deforms the integration contour into the complex plane. The final result is
\begin{equation}\label{inin-factorized}
\langle O(\tau)\rangle
=
\langle 0|
\left[
\bar T e^{\left(i\int_{-\infty(1+i\epsilon)}^{\tau} d\tau' \,\hat H_{\rm int}(\tau')\right)
}\right]
O(\tau)
\left[
T e^{ \left(-i\int_{-\infty(1-i\epsilon)}^{\tau} d\tau' \,\hat H_{\rm int}(\tau')\right)
}\right]
|0\rangle .
\end{equation}
The small imaginary tilt of the contour ensures that the free vacuum in the far past is projected onto the interacting vacuum. This expression can be rewritten in terms of a density matrix simply as 
\begin{align}
\ex{O(\tau)}=\Tr(\rho(\tau) O(\tau))\,,
\end{align}
where 
\begin{align}
\rho(\tau)&=U(\tau,-\infty)\ket{0}\bra{0}U^{\dagger}(\tau,-\infty)\,,\\
U(\tau,-\infty)&=T \exp\!\left(-i\int_{-\infty(1-i\epsilon)}^{\tau} d\tau' \,\hat H_{\rm int}(\tau')\right)\,.
\end{align}
In the previous section, we have translated this operator expression into the SK path integral formalism. Here below we show an equivalent way to study it while remaining in the operator formalism, which is what is often done in the cosmology literature and referred to as \textit{in-in formalism}. We do this to further clarify the connection between these two formalisms. The end point is the derivation of diagrammatic Feynman rules for perturbative calculations. Those, of course, are the same irrespective of whether they are derived in the operator or path integral formalism. \\

Expanding the exponentials perturbatively gives
\begin{align} \nonumber
\langle O \rangle
&=
\Bigg\langle
\bar T
\left[
1+i\int H_{\rm int}
-\frac12\left(\int H_{\rm int}\right)^2+\cdots
\right]
O \,
T
\left[
1-i\int H_{\rm int}
-\frac12\left(\int H_{\rm int}\right)^2+\cdots
\right]
\Bigg\rangle ,
\end{align}
where the shorthand \(\int H_{\rm int}\) denotes the appropriate time integral of \(\hat H_{\rm int}\). Writing out the first few terms explicitly, one finds schematically
\begin{align}\label{ptexp}
\langle O(\tau)\rangle
&=
\langle O(\tau)\rangle_0
+i\left\langle \int H_{\rm int}\, O(\tau)-O(\tau)\int H_{\rm int}\right\rangle_0
\\ \nonumber
&
-\frac12\left\langle
\left(\int H_{\rm int}\right)^2 O(\tau)
+O(\tau)\left(\int H_{\rm int}\right)^2
\right\rangle_0
+\left\langle
\left(\int H_{\rm int}\right) O(\tau)\left(\int H_{\rm int}\right)
\right\rangle_0
+\cdots ,
\end{align}
where \(\langle \cdots \rangle_0\) denotes expectation values in the free vacuum. This is sometimes called the \textit{factorized form} of the perturbative expansion. 


\paragraph{The commutator form} There is an equivalent representation that is often more convenient. It may be written as
\begin{align}\label{inin-commutator}
\langle O(\tau)\rangle
&=
\sum_{N=0}^{\infty}
i^N
\int_{-\infty}^{\tau} d\tau_N
\int_{-\infty}^{\tau_N} d\tau_{N-1}
\cdots
\int_{-\infty}^{\tau_2} d\tau_1\\ 
& \quad \langle 0|\,
[\hat H_{\rm int}(\tau_1),
[\hat H_{\rm int}(\tau_2),
\ldots
[\hat H_{\rm int}(\tau_N),O(\tau)]
\ldots
]]
\,|0\rangle . \nonumber
\end{align}
This is the \textit{commutator form} of the in-in perturbative expansion. The first two orders make the equivalence particularly transparent. At zeroth order,
\begin{equation}\label{commutatorform}
\langle O(\tau)\rangle^{(0)}=\langle 0|\,O(\tau)\,|0\rangle .
\end{equation}
At first order,
\begin{align}
\langle O(\tau)\rangle^{(1)}
&=
\left\langle 0\left|
\left[i\int d\tau' \,\hat H_{\rm int}(\tau')\right] O(\tau)
+
O(\tau)
\left[-i\int d\tau' \,\hat H_{\rm int}(\tau')\right]
\right|0\right\rangle
\nonumber\\
&=
i\int_{-\infty}^{\tau} d\tau' \,
\langle 0|\, [\hat H_{\rm int}(\tau'),O(\tau)] \,|0\rangle .
\end{align}
Thus the two branches of the factorized formula combine into a single commutator. At higher orders the same mechanism produces the nested commutators appearing in \eqref{inin-commutator}. The equivalence to \eqref{ptexp} can be proven by induction \cite{Weinberg:2005vy}. 

Using the formalism of open systems, there is a nice way to think about the above expression. First, recall that in the interaction picture the density matrix obeys
\begin{align}
\rho'=-i[H_{\rm int},\rho]=:\L \rho\,,
\end{align}
where we introduce the Liouvillian superoperator $ \L$. The formal solution is
\[ \rho(\tau)=T \exp\left(  \int_{\tau_{0}}^{\tau}\L\right)\rho(\tau_{0})\,.\]
Inserting into the trace we can write
\begin{align}
\ex{O(\tau)}&=\Tr\left(  \rho(\tau)O(\tau)\right)=\Tr\left(  T \exp\left(  \int_{\tau_{0}}^{\tau}\L\right)\rho(\tau_{0}) O(\tau)\right)\\
&=\Tr\left(  \rho(\tau_{0}) \bar T \exp\left(  \int_{\tau_{0}}^{\tau}\L^{\dagger}\right) O(\tau)\right)\,,
\end{align}
where the adjoint superoperator $ \L^{\dagger}$ now act on $ O(\tau)$ as $ i[H_{\rm int}, \dots]$. Using the pure state initial condition $ \rho(\tau_{0})=\ket{0}\bra{0}$ as $ \tau_{0}\to-\infty$ and using the cyclicity of the trace this reproduces \eqref{inin-commutator}.

The factorized form \eqref{inin-factorized} and the commutator form \eqref{inin-commutator} are completely equivalent, but each is useful in different circumstances. The factorized form is often closer to diagrammatic perturbation theory and to the Schwinger--Keldysh contour description, while the commutator form makes causality and cancellations between the two branches more manifest. In practice, one chooses whichever form is most convenient for the correlator under consideration.


\subsection{Feynman rules for in-in correlators: the $  \pm $ basis}\label{sec5p4}

As in the case of scattering amplitudes, perturbative calculations of cosmological correlators are most efficiently organized in terms of diagrams. The resulting diagrammatics are sometimes referred to as
Feynman--Witten rules, reflecting their resemblance to Witten diagrams in AdS. The rest of this section follows closely the discussion in \cite{Goodhew:2023bcu,PajerFieldTheoryCosmologyNotes}. 

Because of translation invariance, any momentum-space correlator contains one or more delta
functions imposing momentum conservation. Correlators proportional to more than one delta
function are called disconnected. By subtracting all possible products of lower-point functions,
one isolates the connected part, which carries a single overall delta function. For connected
correlators we write
\begin{equation}
\left\langle \prod_{a=1}^n \phi(\mathbf k_a)\right\rangle_c
=
(2\pi)^3\delta^{(3)}\!\left(\sum_{a=1}^n \mathbf k_a\right)\,
B_n(\mathbf k_1,\dots,\mathbf k_n) .
\end{equation}
The main object of interest is then the reduced correlator \(B_n\).

The perturbative rules for \(B_n\) are derived simply by performing all possible Wick contractions of the interaction picture operators appearing in the expansion we discussed in the previous section. Since in this closed system setting the time evolution is Hamiltonian, the final Feynman rule will have a natural interpretation in terms of the plus and minus basis of the path integral, as we see shortly. In the next section, we will develop the equivalent Feynman rules in the language of retarded and advanced fields, starting from the path integral and generalizing to open systems. 

\begin{enumerate}
\item
For an \(n\)-point correlator, draw all diagrams with \(n\) external lines ending on the late-time boundary \(\tau\to 0\), together with some number \(V\) of bulk interaction vertices and \(I\) internal lines connecting pairs of vertices. Time runs from the infinite past \(\tau\to-\infty\) to the future conformal boundary \(\tau\to 0\).

\item
Each interaction vertex can lie on either branch of the in-in contour. We therefore label each vertex either as a \(+\) vertex, associated with the time-ordered evolution operator, or as a \(-\) vertex, associated with the anti-time-ordered evolution operator. One must sum over all \(2^V\) assignments of \(+\)/\(-\) labels to the \(V\) vertices.

\item
Assign to each external line a momentum \(\mathbf k_a\), with \(a=1,\dots,n\), all taken to flow from the bulk vertex toward the future boundary. Assign to each internal line an integration momentum \(\mathbf p_m\), with \(m=1,\dots,I\), and integrate over all such momenta.

\item
For each bulk vertex, include the corresponding interaction Hamiltonian density, evaluated at the time \(\tau_i\) of that vertex, together with an integral over \(\tau_i\). A \(+\) vertex contributes a factor
\begin{equation}
-i\int_{-\infty(1-i\epsilon)}^0 d\tau_i\, H_{\rm int}(\tau_i),
\end{equation}
whereas a \(-\) vertex contributes
\begin{equation}
+i\int_{-\infty(1+i\epsilon)}^0 d\tau_i\, H_{\rm int}(\tau_i).
\end{equation}
The opposite signs and opposite \(i\epsilon\) prescriptions reflect the origin of these terms from $ U$ and $ U^{\dagger}$ in the time evolution of the bra and the ket. In the language of the Schwinger-Keldysh path integral, these are the two branches of the closed time contour.

\item
For every external line connecting a bulk vertex at time \(\tau_i\) to the boundary operator at time \(\tau=0\), insert the appropriate bulk-to-boundary propagator $  G_{\pm} $. Concretely, if the field operator is expanded in free modes as
\begin{equation}
\phi_{\mathbf k}(\tau)=u_k(\tau)a_{\mathbf k}+u_k^*(\tau)a^\dagger_{-\mathbf k},
\end{equation}
then an external line contributes either \(u_k(\tau_i)\) or \(u_k^*(\tau_i)\), depending on whether it is attached to a \(+\) or \(-\) vertex
\begin{equation}
G_+(k;\tau_i,0)=u_k(\tau_i)\,u_k^*(0),
\qquad
G_-(k;\tau_i,0)=u_k^*(\tau_i)\,u_k(0) .
\end{equation}

\item
For every internal line joining vertices at times \(\tau_i\) and \(\tau_j\), insert the free two-point contraction between the corresponding interaction-picture fields. The precise propagator depends on whether the two endpoints lie on the same branch or on different branches of the contour:
\begin{align}
G_{++}(\tau_i,\tau_j;\mathbf p) &= \langle T\,\phi_{\mathbf p}(\tau_i)\phi_{-\mathbf p}(\tau_j)\rangle,\\
G_{--}(\tau_i,\tau_j;\mathbf p) &= \langle \bar T\,\phi_{\mathbf p}(\tau_i)\phi_{-\mathbf p}(\tau_j)\rangle,\\
G_{+-}(\tau_i,\tau_j;\mathbf p) &= \langle \phi_{-\mathbf p}(\tau_j)\phi_{\mathbf p}(\tau_i)\rangle,\\
G_{-+}(\tau_i,\tau_j;\mathbf p) &= \langle \phi_{\mathbf p}(\tau_i)\phi_{-\mathbf p}(\tau_j)\rangle .
\end{align}
We will later see how one may work in the \(r/a\) basis, where the same information is encoded in the retarded, advanced and Keldysh propagators.

\item
At each vertex impose momentum conservation. After all vertex delta functions have been used, a single overall factor
\begin{equation}
(2\pi)^3\delta\!\left(\sum_{a=1}^n \mathbf k_a\right)
\end{equation}
remains, as required for a connected correlator.

\item
Divide by the appropriate symmetry factor of the diagram. For details see \cite{Goodhew:2023bcu,PajerFieldTheoryCosmologyNotes}.

\item
Finally, sum over all distinct diagrams and over all assignments of \(+\)/\(-\) labels.
\end{enumerate}

These rules are the cosmological analogue of ordinary Feynman rules. Their main novelty is the doubled contour structure, which forces one to distinguish \(+\) and \(-\) vertices and to keep track of the corresponding propagators and signs. In practice, the factorized form of the in-in expansion leads naturally to the \(+\)/\(-\) basis, while the \(r/a\) basis is often more efficient for organizing causal properties and studying the semiclassical limit. 

A number of examples of the application of these rules can be found in \cite{PajerFieldTheoryCosmologyNotes}. Here we will not repeat these and move on to the introduction of the main class of theories we will study. 


\subsection{Primordial non-Gaussianity and the bispectrum}

A Gaussian random field is entirely characterized by its two-point function. Equivalently, all connected correlators beyond the two-point function vanish, while the full higher even-point functions are determined by Wick's theorem in terms of the two-point function. Non-Gaussianity is therefore encoded in connected correlators beyond the two-point function, the leading example being the three-point function, or bispectrum. Here we briefly discuss and summarize the main aspects of the study of non-Gaussianity produced during inflation, often known as primordial non-Gaussianity. \\

For any connected \(n\)-point function of a statistically homogeneous field \(\zeta\), translation invariance implies the momentum-space form
\begin{equation}
\langle \zeta_{\mathbf k_1}\zeta_{\mathbf k_2}\cdots \zeta_{\mathbf k_n}\rangle
=
(2\pi)^3\delta^{(3)}\!\left(\mathbf k_1+\mathbf k_2+\cdots+\mathbf k_n\right)\,
B_n(\mathbf k_1,\mathbf k_2,\dots,\mathbf k_n) .
\end{equation}
The overall delta function enforces momentum conservation, while the reduced correlator \(B_n\) contains the physical information. Since the \(n\) momenta carry \(3n\) real components, translation invariance removes three of them. Rotational invariance removes three more. Therefore, before imposing any additional symmetry, the reduced \(n\)-point function depends on $ 3n-6  $ independent variables, for $ n\geq 3$.

If the field is approximately scale invariant, then the correlators satisfy
\begin{equation}
\langle \zeta(\lambda \mathbf x_1)\zeta(\lambda \mathbf x_2)\cdots \zeta(\lambda \mathbf x_n)\rangle
=
\langle \zeta(\mathbf x_1)\zeta(\mathbf x_2)\cdots \zeta(\mathbf x_n)\rangle ,
\end{equation}
or equivalently in momentum space
\begin{equation}
B_n(\lambda \mathbf k_1,\lambda \mathbf k_2,\dots,\lambda \mathbf k_n)
=
\lambda^{-3(n-1)}B_n(\mathbf k_1,\mathbf k_2,\dots,\mathbf k_n) .
\end{equation}
Thus scale invariance removes one further degree of freedom, leaving $ 3n-7  $ independent variables. 

\paragraph{The bispectrum} In particular, for the bispectrum \(B_3\), the most general rotationally invariant configuration is specified by three numbers, which may be taken to be the lengths of the three sides of a triangle,
\begin{equation}
k_1:=|\mathbf k_1|,
\qquad
k_2:=|\mathbf k_2|,
\qquad
k_3:=|\mathbf k_3|,
\qquad
\mathbf k_1+\mathbf k_2+\mathbf k_3=0 .
\end{equation}
Under the additional assumption of scale invariance, only two independent shape variables remain, which may be chosen for example as the ratios \(k_2/k_1\) and \(k_3/k_1\) \cite{Babich:2004gb}.

Because the field \(\zeta\) is identical in each insertion, the bispectrum is also symmetric under permutations of \(k_1\), \(k_2\), and \(k_3\). It is therefore natural to think of the bispectrum as a function on the space of momentum triangles modulo permutations. One often separates the overall amplitude from the dependence on the shape of the triangle by writing
\begin{equation}
B_3(k_1,k_2,k_3)=A\,\frac{S(k_1,k_2,k_3)}{(k_{1}k_{2}k_{3})^{2}} ,
\end{equation}
where \(A\) is an overall amplitude and \(S\) is a shape function. A normalized shape function may then be defined by fixing its value in a preferred configuration, for example by requiring that
\begin{equation}
S(k,k,k)=1
\end{equation}
in the equilateral configuration. In this way, the amplitude and the dependence on triangle shape are clearly disentangled. A variety of normalization conditions have been chosen in the literature, so one should check the convention being used before interpreting any discussion about the value of $ A  $. \\

Different physical mechanisms generate bispectra that peak in different regions of triangle space. A useful way to organize the discussion is therefore to focus on a few special limits.

The first is the \textit{squeezed limit},
\begin{equation}
k_1 \ll k_2 \simeq k_3 .
\end{equation}
This probes the coupling between a long-wavelength mode and two short-wavelength modes. A bispectrum that is large in this limit indicates strong long-short mode coupling. In simple single-clock attractor models this limit is highly constrained \cite{Maldacena:2002vr,Creminelli:2004yq,Flauger:2013hra,Creminelli:2011sq,Hinterbichler:2012nm,Hinterbichler:2013dpa,Creminelli:2012ed,Pajer:2013ana,Dai:2015rda,Creminelli:2013cga,Cabass:2016cgp,Pajer:2017hmb,Finelli:2017fml}, whereas a sizeable signal in the squeezed configuration can point to the presence of additional light degrees of freedom during inflation, or more generally to a departure from the simplest single-field adiabatic picture.

The second is the \textit{equilateral limit},
\begin{equation}
k_1=k_2=k_3 .
\end{equation}
This configuration is naturally enhanced by derivative self-interactions of the inflaton sector. As we will see later, the effective field theory of inflation in the decoupling limit gives rise precisely to this type of non-Gaussianity.

The third is the flattened, or \textit{folded limit},
\begin{equation}
k_1+k_2=k_3 ,
\end{equation}
up to permutations. In this case the area of the momentum triangle tends to zero. Signals peaking in such configurations can arise from non-standard physical effects, for example from non-trivial initial states \cite{Holman:2007na,Agarwal:2012mq} or, as we will later discuss, from open-system effects in an effective theory of inflation \cite{LopezNacir:2011kk,Salcedo:2024smn,Salcedo:2026sdn,Green:2020whw,Creminelli:2023aly}.

We will see that the standard effective field theory of inflation predicts a bispectrum of predominantly equilateral type, while later we will study how this picture is modified in open effective theories of inflation, where contributions to the folded limit generically arise.


\section{The Open Effective Field Theory of Inflation}\label{sec:6}

We are now ready to combine the two main threads developed so far: inflation and the general framework of open quantum systems. In the unitary Effective Field Theory (EFT) approach to inflation, one focuses on the Goldstone boson of broken time translations and studies its dynamics as a closed system, possibly coupled to gravity. In practice, however, the observable sector may interact with additional degrees of freedom that are heavy, short-lived, or otherwise inaccessible. It is therefore natural to ask how the inflationary dynamics is modified when the Goldstone mode is treated as an open system interacting with an unknown environment. The purpose of this section is to construct the corresponding open effective field theory and to study its leading observational consequences.

Our goal in the first two subsections is to derive the EFT of inflation for the field $ \pi$ alone, \eqref{goal}, where gravity is a fixed, non-dynamical background. To reach this goal, one can work first in flat spacetime and construct the theory of the Goldstone boson $ \pi$ of spontaneously broken time translations. Then one can minimally couple this to a fixed FLRW background spacetime. This approach, which we outline in Section \ref{sec:6p1}, is very intuitive but misses some constraints on the Wilsonian coefficients from the background Einstein equations. A longer but more precise derivation starts from a general theory of the metric alone, where time diffs are spontaneously broken. Here, $ \pi$ emerges as a St\"uckelberg field that makes the theory 4d-diff invariant. The final action \eqref{goal} is then obtained by invoking the decoupling of $ \pi$ from metric fluctuations.

Next, we generalize this construction to an open-system setting by doubling the fields, imposing the non-equilibrium consistency conditions, discussing the fate of global symmetries on the Schwinger-Keldysh contour and organizing the resulting theory in the retarded/advanced basis. Finally, we study the quadratic theory, its propagators and power spectrum, and then turn to the cubic interactions and the bispectrum. This will allow us to identify characteristic signatures of open inflationary dynamics, in particular the enhancement of the signal near folded configurations.


\subsection{The Goldstone boson of broken time translations in flat space}\label{sec:6p1}

Before introducing the effective field theory of inflation in the presence of gravity, it is useful to consider a simpler non-gravitational system in flat spacetime\footnote{This subsection introduces the construction of the EFT of inflation \cite{Creminelli:2006xe,Cheung:2007st} paralleling the discussion in \cite{Finelli:2018upr}.}. The purpose of this warm-up is to isolate the symmetry logic behind the appearance of the Goldstone mode associated with the spontaneous breaking of time translations.

Consider a Poincar\'e invariant theory of a single scalar field \(U(x)\), with the most general local Lagrangian
\begin{equation}
\mathcal L=\mathcal L\bigl(U,\partial_\mu U,\partial_\mu\partial_\nu U,\dots\bigr) .
\end{equation}
Suppose that this theory admits a homogeneous but time-dependent background solution \(\bar U(t)\). Perturbations around this background are then conveniently parameterized as
\begin{equation}
U(x)=\bar U\bigl(t+\pi(x)\bigr).
\end{equation}
The field \(\pi(x)\) is the Goldstone boson associated with the spontaneous breaking of time translations. It \textit{transforms non-linearly}  under the broken spacetime symmetries.

Under a Poincar\'e transformation \(x^\mu\to \Lambda^\mu{}_\nu x^\nu+\alpha^\mu\), the scalar field \(U(x)\) transforms linearly, but the Goldstone field \(\pi\) transforms non-linearly. Writing the result explicitly,
\begin{equation}\label{Poincareonpi}
\pi(x)\to \pi(\Lambda x+\alpha)+\alpha^0+\Lambda^0{}_\mu x^\mu-t .
\end{equation}
This expression makes manifest that the background has broken time translations and boosts, while spatial translations and rotations remain linearly realized. This non-linear transformation should be contrasted with the more familiar transformation of a covariant scalar field
\[ \phi(x) \to \phi(\Lambda x +\alpha) \,, \]
where the transformed expression is \textit{linear} in the field $  \phi $. This structure is the hallmark of spontaneous symmetry breaking. 

The effective action for \(\pi\) is obtained by writing the most general local action compatible with these non-linearly realized symmetries, organized in a derivative expansion.  Assuming that \(\bar U(t)\) is monotonic, one may always perform a field redefinition so that
\begin{equation}
\bar U(t)=t .
\end{equation}
Then, at lowest order in derivatives, the first covariant building block is
\begin{equation}
\partial_\mu U\,\partial^\mu U 
=
-1-2\dot\pi+\partial_\mu\pi\,\partial^\mu\pi .
\end{equation}
It is often convenient to remove the background by adding 1 to the above building block so that this quantity starts linear in perturbations.
The second covariant building block is $ U=t+\pi$, so the Wilson coefficients may be arbitrary functions of  $  (t+\pi(t,{\bf x})) $. For any Poincar\'e transformation acting on $  \pi $ as in \eqref{Poincareonpi} one finds
\begin{align}
t+\pi(t,{\bf x})\to t'+\pi(t',{\bf x}')\,.
\end{align} 
In building the effective action all Wilsonian coefficients can be arbitrary functions of $  (t+\pi(t) ) $. Therefore, up to terms with at most one derivative per field, the effective action takes the form
\begin{equation}
S
=
\int d^4x\,
\sum_{n=0}^{\infty}
\frac{d_n(t+\pi)}{n!}
\Bigl(-2\dot\pi+\partial_\mu\pi\,\partial^\mu\pi\Bigr)^n .
\end{equation}
More generally, one may also include higher-derivative operators. Following the notation of the paper, one useful example is the quantity
\begin{equation}
K
=
\frac{\Box U}{\sqrt{-\partial_\mu U\partial^\mu U}}
+
\frac{\partial^\mu U\,\partial^\nu U\,\partial_\mu\partial_\nu U}
{\bigl(-\partial_\mu U\partial^\mu U\bigr)^{3/2}} ,
\end{equation}
which, through quadratic order in \(\pi\), becomes
\begin{equation}
K
=
-\partial_i\partial_i\pi
+\dot\pi\,\partial_i\partial_i\pi
+2\,\partial_i\pi\,\partial_i\dot\pi
+\mathcal O(\pi^3) .
\end{equation}
Including such terms, the effective action may be written as
\begin{equation}
S_{\rm h.d.}=
\int d^4x
\sum_{n=0}^{\infty}
\frac{\hat d_n(t+\pi)}{n!}
\Bigl(-2\dot\pi+\partial_\mu\pi\,\partial^\mu\pi\Bigr)^n K
+\cdots
\end{equation}
Henceforth we will drop these higher-derivative terms.

It is useful to rewrite the action so that two facts become manifest: first, tadpole cancellation implies that the action begins at quadratic order in perturbations; second, when the coefficients are time independent, the Goldstone mode decouples at zero momentum, equivalently \(\pi={\rm const}\) is always a solution. In this form one finds
\begin{align}
S=
\int d^4x\,
\Bigg[
&d_1(t+\pi)\,\partial_\mu\pi\,\partial^\mu\pi
+
\sum_{n=2}^{\infty}
\frac{d_n(t+\pi)}{n!}
\Bigl(-2\dot\pi+\partial_\mu\pi\,\partial^\mu\pi\Bigr)^n\Bigg]
\end{align}
This action describes the universal low-energy sector of any flat-space system that spontaneously breaks time translations and boosts, independently of the microscopic details.

The lesson for cosmology is immediate. In any cosmology, the homogeneous background also breaks time translations, and the corresponding Goldstone boson will again be introduced by shifting the clock variable. The main difference is that in the gravitational system time translations are part of diffeomorphism invariance, so the Goldstone mode is most naturally introduced after fixing unitary gauge and then restoring time diffeomorphisms through a St\"uckelberg field. The flat-space construction above provides the simplest setting in which this logic can be seen without the additional complications of gravity.


\subsection{The EFT of inflation in the decoupling limit}

We now introduce the class of theories for which we will perform explicit calculations: the effective field theory of inflation. In this subsection we restrict throughout to the unitary, closed-system theory, and we keep only the operators with the lowest number of derivatives per field. In particular, we neglect operators involving the extrinsic curvature and higher-derivative geometric quantities.

The conceptual starting point is that an inflationary background selects a preferred slicing of spacetime, namely the slices of constant background clock field. As a result, time diffeomorphisms are spontaneously broken by the background, while spatial diffeomorphisms remain unbroken. The effective field theory is therefore the most general action built from geometric quantities that respect spatial diffeomorphisms only.

One may choose to use time diffeomorphisms to set the fluctuation of the clock field to zero, so that all scalar perturbations are encoded in the metric. This is called \textit{unitary gauge}. In this gauge, the theory is simply described by a theory of the spacetime metric that is invariant under time-dependent spatial diffeomorphisms, but not necessarily under time diffeomorphisms. Restricting to the lowest-derivative sector, the action takes the form
\begin{equation}\label{EFTI-unitary}
S
=
\int d^4x\,\sqrt{-g}\,
\left[
\frac{M_{\rm Pl}^2}{2}R
+d_0(t)
+d_1(t)\,\delta g^{00}
+\sum_{n=2}^{\infty}\frac{d_n(t)}{n!}\bigl(\delta g^{00}\bigr)^n
\right] ,
\end{equation}
where we work in cosmological time $  t $, so that $  \bar g_{00}=-1 $, and we removed the background contribution by defining 
\begin{equation}
\delta g^{00}:=g^{00}-\bar g^{00}=g^{00}+1 .
\end{equation}
The first two terms after the Einstein-Hilbert piece are tadpoles. Their role is simply to ensure that the chosen FLRW background solves the equations of motion. They are therefore fixed by the background evolution and are not independent interaction coefficients.

For a flat FLRW background,
\begin{equation}
ds^2=-dt^2+a(t)^2\,d\mathbf x^2 ,
\end{equation}
the tadpole coefficients are fixed by the Friedmann equations. It is convenient to rewrite them as
\begin{equation}
d_0(t)=-M_{\rm Pl}^2\bigl(3H^2+2\dot H\bigr),
\qquad
d_1(t)=M_{\rm Pl}^2\dot H ,
\end{equation}
so that the action becomes
\begin{equation}\label{EFTI-unitary-standard}
S
=
\int d^4x\,\sqrt{-g}\,
\left[
\frac{M_{\rm Pl}^2}{2}R
-M_{\rm Pl}^2(3H^2+\dot H)
+M_{\rm Pl}^2\dot H\,g^{00}
+\sum_{n=2}^{\infty}\frac{d_n(t)}{n!}\bigl(\delta g^{00}\bigr)^n
\right] .
\end{equation}
At lowest order in derivatives, all departures from the minimal slow-roll theory are encoded in the coefficients \(d_n(t)\).

While one could work directly with this action, it is very convenient to rewrite it in a way that is invariant under the full four-dimensional diffs, Which makes contact with the way we usually study general relativity. One can achieve this through the St\"uckelberg trick. As we will see shortly, this achieves much more, though. The St\"uckelberg trick makes explicit the presence of an additional scalar degree of freedom and the high-energy regime in which it decouples from the gravitational fluctuations. 

In practice, one performs a broken time diffeomorphism with a gauge parameter $  \pi(x) $ that is a new field of the theory,
\begin{equation}
t\to t+\pi(x) .
\end{equation}
We interpret \(\pi(x)\) as the Goldstone boson of the broken time translations. Under this transformation, the metric component \(g^{00}\) transforms non-linearly
\begin{equation}\label{Deltadeltag00}
\delta g^{00}
\;\to\;
\delta g^{00}
+2g^{0\mu}\partial_\mu \pi
+g^{\mu\nu}\partial_\mu \pi\,\partial_\nu \pi\,.
\end{equation}

\paragraph{The decoupling limit} A particularly useful simplification is the \emph{decoupling limit}. In the full theory, the Goldstone mode \(\pi\) mixes with metric fluctuations because time diffeomorphisms are broken by the inflationary background. However, at sufficiently high energies this mixing becomes negligible, and one may study the dynamics of \(\pi\) on the unperturbed FLRW background alone. In practice, this means that one evaluates the metric on its homogeneous background and restores the Goldstone through the replacement \(t\to t+\pi(x)\), without keeping the fluctuations of the lapse, shift, or spatial metric. The resulting action captures the leading scalar dynamics while greatly simplifying the calculation.

The decoupling limit is expected to be reliable at energies of order \(H\) or larger. Physically, this is because the mixing of \(\pi\) with gravity is controlled by the same scales that govern the slow time dependence of the background, whereas the self-interactions of \(\pi\) that we retain remain unsuppressed in this regime. Deviations from the decoupling limit are therefore expected to be slow-roll suppressed, and hence to produce corrections that are parametrically smaller than the leading effects we keep here, under our assumptions. In the decoupling limit, one may simply evaluate the transformation \eqref{Deltadeltag00} on the unperturbed FLRW metric. One then finds
\begin{equation}\label{dg00-Stuck}
\delta g^{00}
\;\to\;
-2\dot\pi-\dot\pi^2+\frac{(\partial_i\pi)^2}{a^2} .
\end{equation}
This single replacement is enough to determine the Goldstone action to the order relevant for us. Substituting \eqref{dg00-Stuck} into \eqref{EFTI-unitary-standard} and expanding to cubic order in \(\pi\), one obtains
\begin{align}\label{goal}
S_\pi
=
\int d^4x\,a^3
\Bigg[
&-M_{\rm Pl}^2\dot H
\left(
\dot\pi^2-\frac{(\partial_i\pi)^2}{a^2}
\right)
+2d_2\,\dot\pi^2
\nonumber\\
&\qquad
+\left(2d_2-\frac{4}{3}d_3\right)\dot\pi^3
-2d_2\,\dot\pi\,\frac{(\partial_i\pi)^2}{a^2}
+\cdots
\Bigg] .
\end{align}
The ellipsis denotes higher-order terms and terms suppressed in the decoupling-limit expansion.

It is customary to parametrize the quadratic action in terms of the speed of sound \(c_s\). Combining the first two quadratic terms, one may write
\begin{equation}
S_\pi^{(2)}
=
\int d^4x\,a^3
\left[
-M_{\rm Pl}^2\dot H\,c_s^{-2}\,\dot\pi^2
+M_{\rm Pl}^2\dot H\,\frac{(\partial_i\pi)^2}{a^2}
\right] ,
\end{equation}
with
\begin{equation}\label{cs-def}
\frac{1}{c_s^2}
=
1-\frac{2d_2}{M_{\rm Pl}^2\dot H} .
\end{equation}
Since during inflation one has \(\dot H<0\), the coefficient of the kinetic term is positive for stable theories.

Using \eqref{cs-def}, the quadratic action becomes
\begin{equation}
S_\pi^{(2)}
=
\int d^4x\,a^3
\left[
-M_{\rm Pl}^2\dot H
\left(
\frac{1}{c_s^2}\dot\pi^2-\frac{(\partial_i\pi)^2}{a^2}
\right)
\right] .
\end{equation}
This shows that a non-trivial coefficient \(d_2\) modifies the propagation speed of the Goldstone mode away from the relativistic value \(c_s=1\).

At cubic order, the two leading interactions are
\begin{equation}\label{EFTI-cubic}
S_\pi^{(3)}
=
\int d^4x\,a^3
\left[
\left(2d_2-\frac{4}{3}d_3\right)\dot\pi^3
-2d_2\,\dot\pi\,\frac{(\partial_i\pi)^2}{a^2}
\right] .
\end{equation}
The operator \(\dot\pi(\partial_i\pi)^2\) is not an independent addition to the theory once we restrict to the lowest-derivative EFT built from powers of \(\delta g^{00}\). Rather, it is fixed by the same non-linearly realized symmetry that dictated the replacement \eqref{dg00-Stuck}. In particular, it arises from the same operator \((\delta g^{00})^2\) that modifies the quadratic kinetic term and hence the speed of sound. This is why its coefficient is determined by \(c_s\), whereas the coefficient of \(\dot\pi^3\) also depends on the next EFT parameter \(d_3\).

Using \eqref{cs-def}, the mixed cubic interaction may be written as
\begin{equation}
-2d_2\,\dot\pi\,\frac{(\partial_i\pi)^2}{a^2}
=
-M_{\rm Pl}^2\dot H
\left(1-\frac{1}{c_s^2}\right)
\dot\pi\,\frac{(\partial_i\pi)^2}{a^2} .
\end{equation}
Thus lowering the speed of sound automatically enhances this interaction.

Finally, the Goldstone mode \(\pi\) is directly related to the curvature perturbation \(\zeta\) at linear order,
\begin{equation}
\zeta = -H\pi + \cdots ,
\end{equation}
so the action above provides the starting point for computing inflationary correlators of the observable scalar perturbations. In the next subsection, we will use the in-in formalism developed above to evaluate the correlators generated by the cubic interactions in \eqref{EFTI-cubic}.


\subsection{Primordial non-Gaussianity from the leading EFT operators}

We now compute the tree-level bispectrum generated by the two leading cubic interactions in the decoupling-limit EFT of inflation,
\begin{equation}
S_\pi^{(3)}
=
\int d^4x\,a^3
\left[
C_{\dot\pi^3}\,\dot\pi^3
+
C_{\dot\pi(\partial\pi)^2}\,
\dot\pi\,\frac{(\partial_i\pi)^2}{a^2}
\right] ,
\end{equation}
where
\begin{equation}
C_{\dot\pi(\partial\pi)^2}
=
-\,M_{\rm Pl}^2\dot H\left(1-\frac{1}{c_s^2}\right) ,
\end{equation}
while \(C_{\dot\pi^3}\) depends on the next EFT coefficient as discussed in the previous subsection.

For these purely unitary tree-level computations, the \(+\)/\(-\) basis is the most economical one, because the answer is simply the sum of a \(+\) contact diagram and a \(-\) contact diagram. The corresponding calculation in the \(r/a\) basis is possible but less compact, since the interaction is split into several terms. By contrast, the \(r/a\) basis will become much more useful later on in the open-system setting.

At this order only bulk-to-boundary propagators are needed. Using the free massless de Sitter mode function for the Goldstone field,
\begin{equation}
u_k(\tau)
=
\frac{i}{2M_{\rm Pl}\sqrt{\epsilon c_s\,k^3}}
\left(1+ikc_s\tau\right)e^{-ikc_s\tau} ,
\end{equation}
the bulk-to-boundary propagator is
\begin{equation}
G_{+}(k;\tau)
:=
u_k(\tau)\,u_k^*(0) =G_{-}(k;\tau)^{\ast}.
\end{equation}
We also need its time and space derivatives,
\begin{equation}
\partial_\tau u_k(\tau)
=
-\frac{i}{2M_{\rm Pl}\sqrt{\epsilon c_s\,k^3}}\,
c_s^2k^2\tau\,e^{-ikc_s\tau},
\qquad
\partial_i \to ik_i
\quad
\text{in Fourier space}.
\end{equation}
Finally, at linear order
\begin{equation}
\zeta=-H\pi ,
\end{equation}
so once the three-point function of \(\pi\) has been computed, the corresponding bispectrum of \(\zeta\) follows immediately.

Since the \(+\) and \(-\) contact diagrams are complex conjugates of each other, the total result is twice the real part of the \(+\) contribution. Equivalently, one may write it as
\begin{equation}
\langle \zeta_{\mathbf k_1}\zeta_{\mathbf k_2}\zeta_{\mathbf k_3}\rangle'
=
2\,\mathrm{Re}\left[
-i
\int_{-\infty(1-i\epsilon)}^0 d\tau\,
\langle
\zeta_{\mathbf k_1}(0)\zeta_{\mathbf k_2}(0)\zeta_{\mathbf k_3}(0)\,
H_{\rm int}(\tau)
\rangle
\right] ,
\end{equation}
where the prime indicates that the overall momentum-conserving delta function has been stripped off. The small \(i\epsilon\) tilt guarantees convergence at \(\tau\to-\infty\).

\paragraph{The \(\dot\pi^3\) interaction.}
For
\begin{equation}
H_{\rm int}^{\dot\pi^3}(\tau)
=
-\int d^3x\,a^3\,C_{\dot\pi^3}\,\dot\pi^3 ,
\end{equation}
the factors of the scale factor cancel against the three time derivatives, so the \(+\) diagram is proportional to
\begin{equation}
-i\,C_{\dot\pi^3}
\int_{-\infty(1-i\epsilon)}^0 d\tau\,
\partial_\tau G_{+}(\tau,k_1)\,
\partial_\tau G_{+}(\tau,k_2)\,
\partial_\tau G_{+}(\tau,k_3) .
\end{equation}
Using the explicit mode functions, the time integral is elementary and one finds
\begin{equation}
B_\zeta^{\dot\pi^3}(k_1,k_2,k_3)
=
\mathcal A_{\dot\pi^3}\,
\frac{1}{(k_1k_2k_3)^3}\,
S_{\dot\pi^3}(k_1,k_2,k_3) ,
\end{equation}
where \(\mathcal A_{\dot\pi^3}\) is an overall amplitude depending on \(C_{\dot\pi^3}\), \(H\), \(\epsilon\), \(c_s\), and \(M_{\rm Pl}\), and
\begin{equation}
S_{\dot\pi^3}(k_1,k_2,k_3)
=
\frac{k_1^2k_2^2k_3^2}{k_{T}^3},
\qquad
k_{T}:=k_1+k_2+k_3 .
\end{equation}
The symmetric combination $  k_{T} $ of the lengths of the three momenta is known as the \textit{total energy}. Thus the shape generated by \(\dot\pi^3\) is supported when the three momenta are comparable.

\paragraph{The \(\dot\pi(\partial_i\pi)^2\) interaction.}
For
\begin{equation}
H_{\rm int}^{\dot\pi(\partial\pi)^2}(\tau)
=
-\int d^3x\,a^3\,
C_{\dot\pi(\partial\pi)^2}\,
\dot\pi\,\frac{(\partial_i\pi)^2}{a^2},
\end{equation}
the scale factors again cancel, and the \(+\) contact diagram is proportional to
\begin{equation}
-i\,C_{\dot\pi(\partial\pi)^2}
\int_{-\infty(1-i\epsilon)}^0 d\tau\,
\left[
\partial_\tau G_{+}(\tau,k_1)\,
\bigl(-\mathbf k_2\!\cdot\!\mathbf k_3\bigr)\,
G_{+}(\tau,k_2)G_{+}(\tau,k_3)
+
2\ \text{perms.}
\right] .
\end{equation}
Performing the time integral and using
\begin{equation}
\mathbf k_2\!\cdot\!\mathbf k_3
=
\frac12\left(k_1^2-k_2^2-k_3^2\right)
\end{equation}
together with permutations, one obtains
\begin{equation}
B_\zeta^{\dot\pi(\partial\pi)^2}(k_1,k_2,k_3)
=
\mathcal A_{\dot\pi(\partial\pi)^2}\,
\frac{1}{(k_1k_2k_3)^3}\,
S_{\dot\pi(\partial\pi)^2}(k_1,k_2,k_3) ,
\end{equation}
with
\begin{equation}
S_{\dot\pi(\partial\pi)^2}(k_1,k_2,k_3)
=
-\frac{1}{k_{T}}\sum_{i<j}k_i^2k_j^2
+\frac{1}{2k_{T}^2}\sum_{i\neq j}k_i^2k_j^3
+\frac18\sum_i k_i^3 .
\end{equation}
Again, the overall amplitude \(\mathcal A_{\dot\pi(\partial\pi)^2}\) depends on \(C_{\dot\pi(\partial\pi)^2}\), \(H\), \(\epsilon\), \(c_s\), and \(M_{\rm Pl}\), but the shape dependence is fixed. Both shapes of non-Gaussianity, as well as all other tree-level shapes to any order in derivatives, can be also obtained using the boostless cosmological bootstrap \cite{Pajer:2020wxk,Jazayeri:2021fvk}. 

\paragraph{Squeezed limit.}
To see that both operators are predominantly equilateral, let us study the squeezed limit
\begin{equation}
k_1=q,
\qquad
k_2=k_3=k,
\qquad
x:=\frac{q}{k}\ll 1 .
\end{equation}
Dropping the overall scale \(k^3\), the first shape becomes
\begin{equation}
S_{\dot\pi^3}(x,1,1)
=
\frac{x^2}{(2+x)^3}
=
\frac{x^2}{8}
-\frac{3x^3}{16}
+\mathcal O(x^4) .
\end{equation}
Thus both the constant and linear terms vanish exactly.

For the second operator one finds
\begin{equation}
S_{\dot\pi(\partial\pi)^2}(x,1,1)
=
-\frac{1+2x^2}{2+x}
+\frac{1+x^2+x^3}{(2+x)^2}
+\frac{2+x^3}{8}
=
-\frac{11}{16}x^2
+\frac{9}{16}x^3
+\mathcal O(x^4) .
\end{equation}
Again, the constant and linear terms vanish exactly. In this sense, neither operator produces a local-type signal in the squeezed configuration. Both shapes are instead mostly supported on configurations in which the three momenta are comparable, namely on equilateral triangles. We conclude that, at lowest order in derivatives and in the decoupling limit, the unitary EFT of inflation generates bispectra of predominantly equilateral type. 


\subsection{Constructing the open effective field theory of inflation}

We now move from the unitary effective field theory of inflation to its open-system generalization. The physical motivation is that the Goldstone boson of broken time translations, which we denote by \(\pi\), may interact with additional degrees of freedom that are not directly observed. This idea has been discussed many times in the literature. Earlier on open system dynamics emerged from warm inflation \cite{Berera:1995ie,Berera:1995wh,Berera:1999ws,Gupta:2002kn,Berera:2008ar,Bastero-Gil:2017yzb}. An EFT description of open dynamics in a different formalism was developed in \cite{LopezNacir:2011kk}. Here we follow the discussion of \cite{Salcedo:2024smn}, which builds upon \cite{Hongo:2018ant,Hongo:2019qhi,Akyuz:2023lsm,Lau:2024mqm}.\\

After tracing over an unknown environment, the appropriate object is no longer an ordinary action for a single field, but a Schwinger--Keldysh effective functional depending on two copies of the Goldstone field,
\begin{equation}
S_{\rm eff}[\uppi_+,\uppi_-]
=
S_\uppi[\uppi_+]-S_\uppi[\uppi_-]+S_{\rm IF}[\uppi_+,\uppi_-] ,
\end{equation}
where \(S_\uppi\) denotes the unitary part of the dynamics and \(S_{\rm IF}\) is the influence functional induced by the environment. Here we are using the different font $  \uppi $ for the Greek letter pi because we will reserve the other font $  \pi $ for the normalized version of the field, which we will introduce shortly. 

It is convenient to pass to the Keldysh basis,
\begin{equation}
\uppi_r:=\frac{\uppi_++\uppi_-}{2},
\qquad
\uppi_a:=\uppi_+-\uppi_-,
\qquad
\uppi_\pm=\uppi_r\pm \frac12 \uppi_a .
\end{equation}
As usual, the open effective functional must satisfy three general non-equilibrium constraints in \eqref{constraintsra}.
These conditions encode, respectively, normalization, Hermiticity, and positivity of the reduced density matrix. In practice, they imply that the effective action starts at linear order in \(\uppi_a\), that odd powers of \(\uppi_a\) come with real coefficients, and that even powers of \(\uppi_a\) come with purely imaginary coefficients.

The symmetry structure is also modified relative to the unitary theory. Before tracing out the environment, the doubled closed system admits independent time translations on the \(+\) and \(-\) branches. Once the environment is integrated out, only the diagonal subgroup survives. As a result, in the Keldysh basis the diagonal time translation acts as
\begin{equation}
\uppi_r(t,\mathbf x)\to \uppi_r(t+\epsilon_r,\mathbf x)+\epsilon_r ,
\qquad
\uppi_a(t,\mathbf x)\to \uppi_a(t+\epsilon_r,\mathbf x) .
\end{equation}
Similarly, under the diagonal Lorentz boosts one has
\begin{equation}
\uppi_r(x)\to \uppi_r(\Lambda_r x)+\Lambda_r{}^0{}_\mu x^\mu-t ,
\qquad
\uppi_a(x)\to \uppi_a(\Lambda_r x) .
\end{equation}
The key point is that \(\uppi_r\) transforms non-linearly, as the Goldstone boson of broken time translations, whereas \(\uppi_a\) transforms linearly, like an ordinary matter field. This is the basic symmetry principle underlying the open EFT.

To construct the effective action, we assume locality in space and time and work directly in the decoupling limit. We further restrict to operators with at most one derivative per field. The invariant building blocks are then \(\uppi_a\), its derivative \(\partial_\mu\uppi_a\), and the combination\footnote{The notation here is chosen to match that of \cite{Salcedo:2024smn}.}
\begin{equation}
P_\mu:=\partial_\mu\bigl(t+\uppi_r\bigr)=\delta_\mu^0+\partial_\mu\uppi_r .
\end{equation}
From \(P_\mu\) one can form the usual invariant
\begin{equation}
P_\mu P^\mu+1
=
-2\dot\uppi_r+(\partial_\mu\uppi_r)^2 ,
\end{equation}
which is the direct open-system analogue of the unitary EFT building block. The most general local effective Lagrangian may then be organized as an expansion in powers of \(\uppi_a\),
\begin{equation}
\mathcal L_{\rm eff}
=
\sum_{n=1}^{\infty}\mathcal L_n,
\qquad
\mathcal L_n=\mathcal O(\uppi_a^n) .
\end{equation}

At first order in \(\uppi_a\), the most general local structure with at most one derivative per field is
    \begin{align}
        \mathcal{L}_1 = & \sum_{n=0}^\infty \left( P_\mu P^\mu  +1\right)^{n} \left[ \upgamma_n \uppi_{a} -\upalpha_n P^\mu \partial_\mu \uppi_{a}  \right].
    \end{align}
The coefficients \(\upalpha_n\) and \(\upgamma_n\) are real. The terms proportional to \(\upalpha_n\) contain the unitary kinetic structure and its higher-order symmetry completions, while the terms proportional to \(\upgamma_n\) encode dissipative effects. In particular, \(\alpha_0\) is the analogue of the usual kinetic coefficient, \(\upalpha_1\) controls the sound speed, and \(\upgamma_1\) gives the leading dissipative term. Expanding to quadratic and cubic order, one finds the contributions 
    \begin{align}
        \mathcal{L}_1^{(2)} &= \left( \upalpha_0 -2 \upalpha_1\right) \dot{\uppi}_{r} \dot{\uppi}_{a} - \upalpha_0 \partial_i \uppi_{r}  \partial^i \uppi_{a} - 2 \upgamma_1  \dot{\uppi}_{r} \uppi_a \\
                \mathcal{L}_1^{(3)} &=  \left(4 \upalpha_2 -3\upalpha_1 \right) \dot{\uppi}^2_{r} \dot{\uppi}_{a} + \upalpha_1 \left(\partial_i \uppi_{r} \right)^2 \dot{\uppi}_a + 2 \upalpha_1  \dot{\uppi}_{r}	\partial_i \uppi_{r}  \partial^i \uppi_{a} \Big.\nonumber \\
        &\qquad + \left(4 \upgamma_2 -\upgamma_1 \right)  \dot{\uppi}^2_{r} \uppi_{a}	+ \upgamma_1 	\left(\partial_i \uppi_{r} \right)^2 \uppi_a 
    \end{align}

At second order in \(\uppi_a\), the most general local structure takes the schematic form
\begin{align}\label{eq:diffquad}
    \mathcal{L}_2 &= i \Big[\upbeta_1 \uppi_{a}^2 + \upbeta_2 \left( \partial_\mu \uppi_{a}\right)^2 +\upbeta_3 \left(-\dot{\uppi}_{a} + \partial^\mu \uppi_{r} \partial_\mu \uppi_{a}\right)\uppi_{a} +\upbeta_4 \left(-\dot{\uppi}_{a} + \partial^\mu \uppi_{r} \partial_\mu \uppi_{a}\right)^2 + \cdots \Big]
\end{align}
where the coefficients \(\upbeta_i\) are real and the overall factor of \(i\) follows from the non-equilibrium constraints. These are the leading noise terms of the EFT. The coefficient \(\upbeta_1\) corresponds to the usual local noise, while \(\upbeta_2\) and \(\upbeta_4\) generate derivative corrections to the noise kernel. Expanding again to quadratic and cubic order, one finds the contributions 
\begin{align}
\mathcal{L}_2^{(2)} &= i \left[\upbeta_1 \uppi_a^2 - \left(\upbeta_2 - \upbeta_4\right)\dot{\uppi}_a^2 + \upbeta_2 \left(\partial_i \uppi_{a} \right)^2\right]\,,\label{eq:piq2free}\\
\mathcal{L}_2^{(3)} &= i \Big[-(\upbeta_3 -2 \upbeta_7)\dot{\uppi}_{r} \dot{\uppi}_{a} \uppi_a + \upbeta_3 \partial_i \uppi_{r}  \partial^i \uppi_{a} \uppi_a +2 (\upbeta_4+\upbeta_6 - \upbeta_8) \dot{\uppi}_{r} \dot{\uppi}_a^2 \nonumber \\
& \qquad \quad - 2 \upbeta_4   \partial_i \uppi_{r}  \partial^i \uppi_{a} \dot{\uppi}_{a} - 2\upbeta_5 \dot{\uppi}_{r} \uppi_a^2  - 2 \upbeta_6 \dot{\uppi}_{r}(\partial_i\uppi_a)^2\Big].
\end{align}
Finally, we can write the term that is cubic in the advanced field, 
    \begin{align}
        \mathcal{L}_3 &= \updelta_1 \uppi_a^3 + \updelta_2 (\partial_\mu \uppi_a)^2 \uppi_a + \updelta_3 \left(-\dot{\uppi}_{a} + \partial^\mu \uppi_{r} \partial_\mu \uppi_{a}\right) \uppi_a^2 + \updelta_4 \left(-\dot{\uppi}_{a} + \partial^\mu \uppi_{r} \partial_\mu \uppi_{a}\right) (\partial_\nu \uppi_a)^2 \Big.  \nonumber\\
        &\qquad \qquad + \updelta_5 \left(-\dot{\uppi}_{a} + \partial^\mu \uppi_{r} \partial_\mu \uppi_{a}\right)^2 \uppi_a  + \updelta_6 \left(-\dot{\uppi}_{a} + \partial^\mu \uppi_{r} \partial_\mu \uppi_{a}\right)^3 + \cdots,
    \end{align}
This contributes only to cubic order, giving 
     \begin{align}
        \mathcal{L}_3^{(3)} &=\updelta_1 \uppi_a^3 + (\updelta_5- \updelta_2)  \dot{\uppi}_{a}^2 \uppi_a  + \updelta_2  (\partial_i \uppi_{a})^2 \uppi_a - \updelta_4  (\partial_i \uppi_{a})^2 \dot{\uppi}_a + (\updelta_4-\updelta_6)\dot{\uppi}_{a}^3 ,
    \end{align}


\subsection{The quadratic action, propagators and the power spectrum} 

Combining all the quadratic terms we have found so far, we obtain the action
\begin{align}
&\quad S_{\mathrm{eff}}^{(2)} = \int \dd^4 x \Big\{\left( \upalpha_0 -2 \upalpha_1\right) a^2 \uppi'_{r} \uppi'_{a} - \upalpha_0 a^2 \partial_i \uppi_{r}  \partial^i \uppi_{a}   \\
-& 2 a^3 \upgamma_1 \uppi'_{r} \uppi_a + i \left[\upbeta_1 a^4 \uppi_a^2 - \left(\upbeta_2 - \upbeta_4\right) a^2\uppi_a^{\prime2} + \upbeta_2 a^2 \left(\partial_i \uppi_{a} \right)^2\right] \Big\}, \nonumber
\end{align}
Because this is the most general quadratic open theory that is local in time, it must also contain the unitary EFT of inflation, which we discussed in previous sections. Indeed, that unitary theory is obtained by setting to zero the second line and re-defining
\begin{align}
c_{s}^{2} &= \frac{\upalpha_0}{\upalpha_{0}-2\upalpha_1} \,.
\end{align}
It is convenient to work with canonically normalized fields. To that end, we define the energy scale
\begin{align}
f_\pi^4 \equiv \upalpha_0 -2 \upalpha_1\,.
\end{align}
We use this scale to define a canonically normalized Goldstone boson
\begin{equation}\label{eq:can}
\pi \equiv f_\pi^2 \uppi.
\end{equation}
Then, we also rescale the coefficients by
\begin{align}
c_s^{2} \equiv \frac{\upalpha_{0}}{f_\pi^4}\;,\quad\; \gamma \equiv
\frac{2\upgamma_{1}}{f_\pi^4}\;,\quad\;\beta_{i} \equiv \frac{\upbeta_{i}}{f_{\pi}^{4}}\;\quad\mathrm{for}\quad i = 1, \ldots, 8.
\end{align}
to obtain the quadratic action in its final form
\begin{align}\label{eq:canonorm} 
&\quad S_{\mathrm{eff}}^{(2)} = \int \dd^4 x \Big\{ a^2 \pi'_{r} \pi'_{a} - c_{s}^{2} a^2 \partial_i \pir  \partial^i \pia  \\
-&  a^3 \gamma \pi'_{r} \pia + i \left[\beta_{1} a^4 \pia^2 - \left(\beta_2 - \beta_4\right) a^2\pia^{\prime 2} + \beta_2 a^2 \left(\partial_i \pia \right)^2\right] \Big\}. \nonumber
\end{align}


\paragraph{Propagators} After integrating the quadratic action by parts, one may write it in bilinear form as
\begin{equation}
S_{\rm eff}^{(2)}
=
-\frac12
\int d^4x\,
\begin{pmatrix}
\pi_r & \pi_a
\end{pmatrix}
\begin{pmatrix}
0 & \widehat D_A \\
\widehat D_R & -2i\,\widehat D_K
\end{pmatrix}
\begin{pmatrix}
\pi_r \\
\pi_a
\end{pmatrix} .
\end{equation}
The corresponding differential operators are
\begin{equation}
\widehat D_A
=
a^2\partial_\eta^2
+
a^3\left(\frac{2H}{a}-\gamma\right)\partial_\eta
-
a^2 c_s^2 \partial_i^2
-
3a^3H\gamma ,
\end{equation}
\begin{equation}
\widehat D_R
=
a^2\partial_\eta^2
+
a^3\left(\frac{2H}{a}+\gamma\right)\partial_\eta
-
a^2 c_s^2 \partial_i^2 ,
\end{equation}
\begin{equation}
\widehat D_K
=
a^4\beta_1
+
a^2(\beta_2-\beta_4)\bigl(\partial_\eta^2+2H\partial_\eta\bigr)
-
a^2\beta_2\,\partial_i^2 .
\end{equation}
In the following, and as in the paper, it is convenient to set \(c_s=1\) for simplicity.

The retarded and advanced propagators are defined by
\begin{equation}
\widehat D_R(x)\,G_R(x,y)=-i\delta(x-y),
\qquad
\widehat D_A(x)\,G_A(x,y)=-i\delta(x-y),
\end{equation}
with
\begin{equation}
G_R(x,y)=G_A(y,x) .
\end{equation}
In Fourier space, the retarded Green function obeys
\begin{equation}
\left(
\partial_{\eta_1}^2
-
\frac{2+\gamma/H}{\eta_1}\partial_{\eta_1}
+k^2
\right)
G_R(k;\eta_1,\eta_2)
=
H^2\eta_1^2\,\delta(\eta_1-\eta_2) .
\end{equation}
Defining
\begin{equation}
\nu_\gamma:=\frac32+\frac{\gamma}{2H},
\qquad
z_i:=-k\eta_i ,
\end{equation}
the solution may be written as
\begin{equation}
G_R(k;\eta_1,\eta_2)
=
\frac{\pi^2 H^2}{k^3}
\left(\frac{z_1}{z_2}\right)^{\nu_\gamma}
z_2^3
\Bigl[
Y_{\nu_\gamma}(z_1)J_{\nu_\gamma}(z_2)
-
J_{\nu_\gamma}(z_1)Y_{\nu_\gamma}(z_2)
\Bigr]
\theta(\eta_1-\eta_2) ,
\end{equation}
Where $ Y  $ and $ J  $ are Bessel functions. The Keldysh propagator is not a Green function of a homogeneous equation, but is instead determined by the noise kernel,
\begin{equation}
G_K(x,y)
=
2i\int d^4z\,\sqrt{-g(z)}\,
G_R(x,z)\,\widehat D_K(z)\,G_R(y,z)\,.
\end{equation}
Restricting to the first noise term proportional to \(\beta_1\), one finds in Fourier space
\begin{equation}
G_{K,1}(k;\eta_1,\eta_2)
=
\frac{i\pi^2\beta_1}{4k^3}
(z_1 z_2)^{\nu_\gamma}
\Bigl[
Y_{\nu_\gamma}(z_1)Y_{\nu_\gamma}(z_2)A^{(1)}_{\nu_\gamma}(z_2)
+
J_{\nu_\gamma}(z_1)J_{\nu_\gamma}(z_2)C^{(1)}_{\nu_\gamma}(z_2)
\Bigr.
\end{equation}
\begin{equation}
\Bigl.
-
\bigl(
J_{\nu_\gamma}(z_1)Y_{\nu_\gamma}(z_2)
+
J_{\nu_\gamma}(z_2)Y_{\nu_\gamma}(z_1)
\bigr)
B^{(1)}_{\nu_\gamma}(z_2)
\Bigr]
+
(1\leftrightarrow 2) ,
\end{equation}
where
\begin{equation}
A^{(1)}_{\nu_\gamma}(z)
:=
\int_z^\infty dz'\,z'^{\,2-2\nu_\gamma}J_{\nu_\gamma}(z')^2 ,
\end{equation}
\begin{equation}
B^{(1)}_{\nu_\gamma}(z)
:=
\int_z^\infty dz'\,z'^{\,2-2\nu_\gamma}J_{\nu_\gamma}(z')Y_{\nu_\gamma}(z') ,
\end{equation}
\begin{equation}
C^{(1)}_{\nu_\gamma}(z)
:=
\int_z^\infty dz'\,z'^{\,2-2\nu_\gamma}Y_{\nu_\gamma}(z')^2 .
\end{equation}


\paragraph{The power spectrum} The power spectrum is obtained from the coincident-time limit of the Keldysh propagator. Using
\begin{equation}
\zeta=-\frac{H}{f_\pi^2}\,\pi ,
\end{equation}
one defines
\begin{equation}
\Delta_\zeta^2(k)
:=
\frac{k^3}{2\pi^2}P_\zeta(k),
\qquad
\langle \zeta_{\mathbf k}\zeta_{\mathbf k'}\rangle
=
(2\pi)^3\delta(\mathbf k+\mathbf k')\,P_\zeta(k) .
\end{equation}
For the \(\beta_1\) contribution, the exact result is
\begin{equation}
P_\zeta(k)
=
\frac{\pi^2\beta_1}{8k^3}\,
\frac{H^2}{f_\pi^4}\,
z^{2\nu_\gamma}
\left[
Y_{\nu_\gamma}(z)^2A^{(1)}_{\nu_\gamma}(z)
+
J_{\nu_\gamma}(z)^2C^{(1)}_{\nu_\gamma}(z)
-
2J_{\nu_\gamma}(z)Y_{\nu_\gamma}(z)B^{(1)}_{\nu_\gamma}(z)
\right] .
\end{equation}
Explicit results for the contributions from other noise terms and the case of $  c_{s}\neq 1 $ can be found in \cite{Salcedo:2024smn}. In the super-Hubble regime \(z\ll 1\), the power spectrum freezes to
\begin{equation}
\Delta_\zeta^2(k)
=
\frac14\,
\frac{\beta_1}{H^2}\,
\frac{H^4}{f_\pi^4}\,
2^{2\nu_\gamma}
\frac{
\Gamma(\nu_\gamma-1)\Gamma(\nu_\gamma)^2
}{
\Gamma\!\left(\nu_\gamma-\frac12\right)
\Gamma\!\left(2\nu_\gamma-\frac12\right)
} .
\end{equation}
Expanding this result in the two opposite dissipation regimes gives
\begin{equation}
\Delta_\zeta^2(k)\propto
\begin{cases}
\displaystyle
\frac{\beta_1}{H^2}\,\frac{H^4}{f_\pi^4}
+\mathcal O\!\left(\frac{\gamma}{H}\right),
& \gamma\ll H, \\[1.2em]
\displaystyle
\frac{\beta_1}{H^2}\,\frac{H^4}{f_\pi^4}
\sqrt{\frac{H}{\gamma}}
\left[
1+\mathcal O\!\left(\frac{H}{\gamma}\right)
\right],
& \gamma\gg H .
\end{cases}
\end{equation}
If one further assumes thermal equilibrium of the environment, the fluctuation-dissipation relation imposes \(\beta_1=2\pi\gamma T\), so that in the strongly dissipative regime
\begin{equation}
\Delta_\zeta^2(k)
\propto
\frac{T}{H}\,
\frac{H^4}{f_\pi^4}\,
\sqrt{\frac{\gamma}{H}} .
\end{equation}
These expressions make the physical interpretation transparent: \(G_R\) captures the dissipative response of the Goldstone mode, while \(G_K\) captures the fluctuations injected by the environment. The late-time power spectrum is determined by the coincident-time Keldysh propagator and remains nearly scale invariant, but with an amplitude controlled by the interplay of noise and dissipation. Note that since the system has finite memory of the past, whatever initial conditions were imposed in the infinite past are quickly forgotten, and the power spectrum is driven to the above value by the noise fluctuations of the environment. This is a conceptually different paradigm for the generation of primordial perturbations and it stands in stark contrast to the more often discussed quantum fluctuations of the Bunch-Davies state. 


\subsection{Feynman rules for in-in correlators: the $r/a$ basis}

The same perturbative expansion discussed in the previous subsection can be reorganized in the Keldysh, or retarded/advanced, basis. Being just a change of basis, this is always completely equivalent to the \(+\)/\(-\) description, but it reshuffles the perturbative expansion into terms that make causality and the distinction between response and fluctuations more manifest. In particular, \(r\)-fields are associated with anti-commutators and statistical fluctuations, while \(a\)-fields are associated with commutators and retarded response. In a unitary closed system this is often a less economical basis for calculation unless one is particularly interested in the semiclassical limit or certain aspects of causality. In that case, the plus and minus basis is much more widely used for calculations. Conversely, the $  r/a $ basis is quite convenient for calculations in open theories, and therefore we briefly review it below. The rest of this section follows closely our previous discussion in Section \ref{sec5p4}.

The perturbative rules for \(B_n\) now follow by rewriting the interaction vertices and propagators in the \(r/a\) basis.
\begin{enumerate}
\item
For an \(n\)-point correlator, draw all diagrams with \(n\) external lines ending on the late-time boundary \(\tau\to 0\), together with some number \(V\) of bulk interaction vertices and \(I\) internal lines connecting pairs of vertices. Time runs from the infinite past \(\tau\to-\infty\) to the future conformal boundary \(\tau\to 0\).

\item
Rewrite every interaction vertex in the \(r/a\) basis. For a unitary closed system, each vertex contains an odd number of \(a\)-fields. For example, a cubic interaction gives vertices of the schematic form $ \phi_r^2\phi_a  $ and $  \phi_a^3 $, while a quartic interaction gives $ \phi_r^3\phi_a  $ and $ \phi_r\phi_a^3 $. Each such term is treated as a separate vertex, with the coefficient inherited from the expansion of the action. Notice that because of the unitarity constraints, every vertex has at least one power of the advanced fields.

\item
Assign to each external line a momentum \(\mathbf k_a\), with \(a=1,\dots,n\), all taken to flow from the bulk vertex toward the future boundary. Assign to each internal line an integration momentum \(\mathbf p_m\), with \(m=1,\dots,I\), and integrate over all such momenta.

\item
For each bulk vertex, include the corresponding interaction coefficient together with the integral over its time \(\tau_i\),
\begin{equation}
\int_{-\infty}^{0} d\tau_i \, .
\end{equation}
The \(i\epsilon\) prescription is inherited from the original \(+\)/\(-\) contour representation, but in the \(r/a\) basis it is usually left implicit once the retarded and Keldysh propagators are used.

\item
For every external line connecting a bulk vertex at time \(\tau_i\) to a boundary operator at \(\tau=0\), insert the appropriate bulk-to-boundary propagator in the \(r/a\) basis. Starting from
\begin{equation}
\phi_{\mathbf k}(\tau)=u_k(\tau)a_{\mathbf k}+u_k^*(\tau)a^\dagger_{-\mathbf k},
\end{equation}
one defines
\begin{align}
G_r(k;\tau_i,0)
&=
\frac12\Bigl(u_k(\tau_i)u_k^*(0)+u_k^*(\tau_i)u_k(0)\Bigr), \\
G_a(k;\tau_i,0)
&=
u_k(\tau_i)u_k^*(0)-u_k^*(\tau_i)u_k(0) .
\end{align}

\item
For every internal line joining vertices at times \(\tau_i\) and \(\tau_j\), insert the appropriate free two-point function in the \(r/a\) basis. These are
\begin{align}
G_K(\tau_i,\tau_j;\mathbf p)
&=
\langle\!\langle \phi_r(\tau_i,\mathbf p)\phi_r(\tau_j,-\mathbf p)\rangle\!\rangle =
\frac12
\langle \{\phi_{\mathbf p}(\tau_i),\phi_{-\mathbf p}(\tau_j)\}\rangle ,
\\
G_R(\tau_i,\tau_j;\mathbf p)
&=
\langle\!\langle\phi_r(\tau_i,\mathbf p)\phi_a(\tau_j,-\mathbf p)\rangle\!\rangle =
-i\,\theta(\tau_i-\tau_j)
\langle [\phi_{\mathbf p}(\tau_i),\phi_{-\mathbf p}(\tau_j)]\rangle .
\end{align}
Thus \(G_R\) encodes causal response, while \(G_K\) encodes statistical fluctuations.

\begin{figure}[t]
    \centering
\sidesubfloat[]{\includegraphics[width=0.4\textwidth]{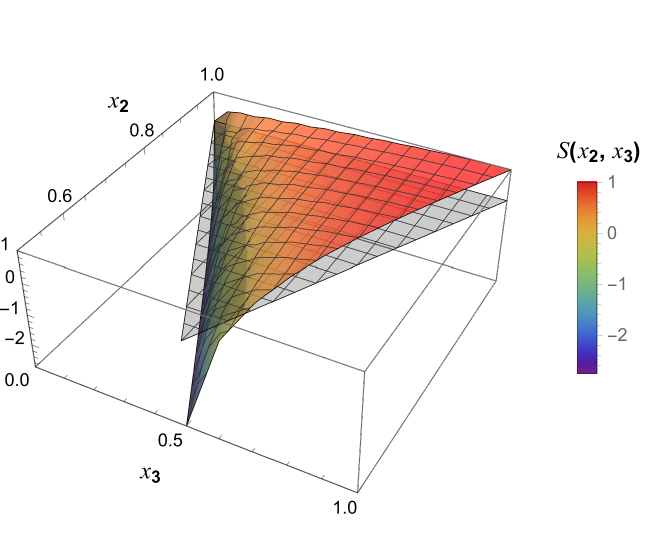}\label{fig:cbis}}
\hfil
\sidesubfloat[]{\includegraphics[width=0.4\textwidth]{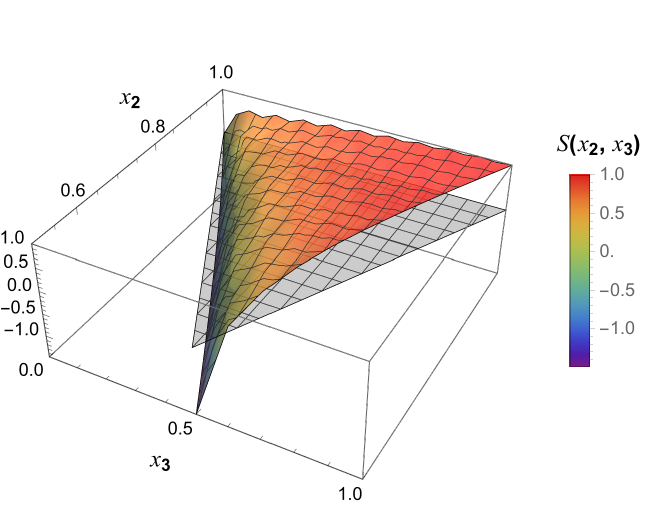}\label{fig:dbis}}

\medskip
\sidesubfloat[]{\includegraphics[width=0.4\textwidth]{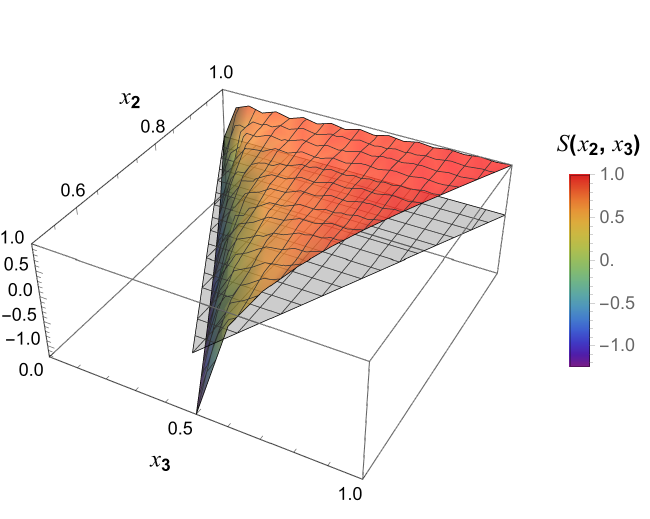}\label{fig:ebis}}
\hfil
\sidesubfloat[]{\includegraphics[width=0.4\textwidth]{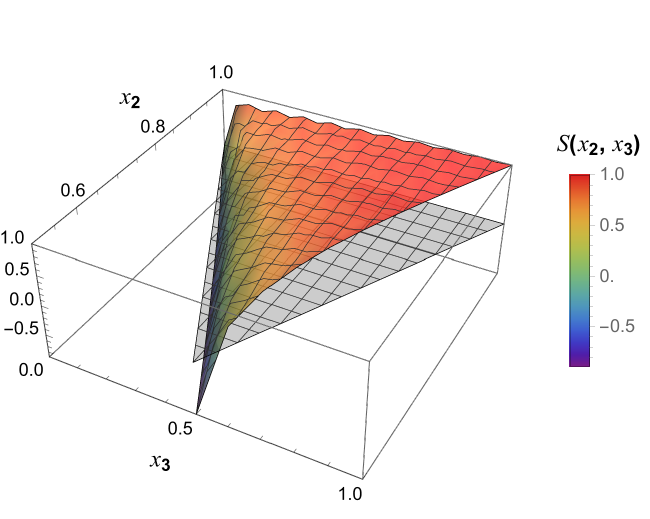}\label{fig:fbis}}
\caption{The figure shows numerical evaluations of the shape of the bispectrum for small dissipation, $\gamma= 0.001H$ (small oscillations are numerical artifacts). The axes are the ratio of external momenta, $  x_{2,3}:=k_{2,3}/k_{1} $. \textbf{a.} $\left(\partial_i \pir \right)^2 \pia$ operator; \textbf{b.} $\dot{\pi}^2_{r} \pia$ operator; \textbf{c.} $\dot{\pi}_{r} \pia^2$ operator; \textbf{d.} $\pia^3$ operator. Figure taken from \cite{Salcedo:2024smn}.}
    \label{fig:shapes_low}
    \end{figure}
    
\item
Because \(G_{aa}=0\), any diagram containing an internal \(a\)-\(a\) line vanishes. More generally, the \(r/a\) basis makes causality manifest: every non-vanishing connected diagram must contain enough retarded or advanced propagators to connect each \(a\)-field to the rest of the graph. In practice, this strongly reduces the number of non-zero diagrams compared with the \(+\)/\(-\) basis.

\item
At each vertex impose momentum conservation. After using all vertex delta functions, a single overall factor
\begin{equation}
(2\pi)^3\delta\!\left(\sum_{a=1}^n \mathbf k_a\right)
\end{equation}
remains, as required for a connected correlator.

\item
Divide by the appropriate symmetry factor of the diagram. As in the \(+\)/\(-\) basis, this removes the overcounting associated with identical vertices and identical internal contractions. For details see \cite{Goodhew:2023bcu,PajerFieldTheoryCosmologyNotes}.

\item
Finally, sum over all distinct non-vanishing \(r/a\) labelings of propagators obtained from the expansion of the interaction vertices.
\end{enumerate}
We will now use these Feynman rules to compute primordial non-Gaussianities in the open EFT of inflation. 


\subsection{Non-Gaussianity}

Combining all cubic terms we found in the construction of the open EFT of inflation and rescaling them to the canonical basis, we find the following cubic action
\begin{align}\label{eq:canonormcub}
S_{\mathrm{eff}}^{(3)} =   \frac{1}{f_\pi^2} \int & \dd^4 x \Big\{\Big[4 \alpha_2 -  \frac{3}{2} (c^2_s-1) \Big]  a \pir^{\prime2} \pi'_{a} +\frac{1}{2} (c^2_s-1) a \left[\left(\partial_i \pir \right)^2 \pi'_a + 2 \pi'_{r}	\partial_i \pir  \partial^i \pia \right] \\
& \qquad \quad + \left(4 \gamma_2 - \frac{\gamma}{2}\right)  a^2\pir^{\prime2} \pia	+ \frac{\gamma}{2} a^2	\left(\partial_i \pir \right)^2 \pia \nonumber \Big. \\
&+ i \Big[\left(2\beta_7-\beta_3\right) a^2 \pi'_{r} \pi'_{a} \pia + \beta_3 a^2 \partial_i \pir  \partial^i \pia \pia + 2(\beta_4+ \beta_6 - \beta_8) a \pi'_{r} \pia^{\prime2} \Big. \nonumber \\
& \qquad \quad - 2 \beta_4  a \partial_i \pir  \partial^i \pia \pi'_{a} - 2\beta_5 a^3 \pir^{\prime} \pia^2  - 2 \beta_6 a \pir^{\prime} (\partial_i\pia)^2 \Big]  \Big. \nonumber\\
&+\delta_1 a^4 \pia^3 + (\delta_5- \delta_2) a^2 \pia^{\prime2} \pia  + \delta_2 a^2 (\partial_i \pia)^2 \pia - \delta_4 a  (\partial_i \pia)^2 \pi'_a + (\delta_4-\delta_6) a \pia^{\prime3} \Big\},  \nonumber
\end{align}
\begin{figure}[t!]
    \centering
    \includegraphics[width=0.6\linewidth]{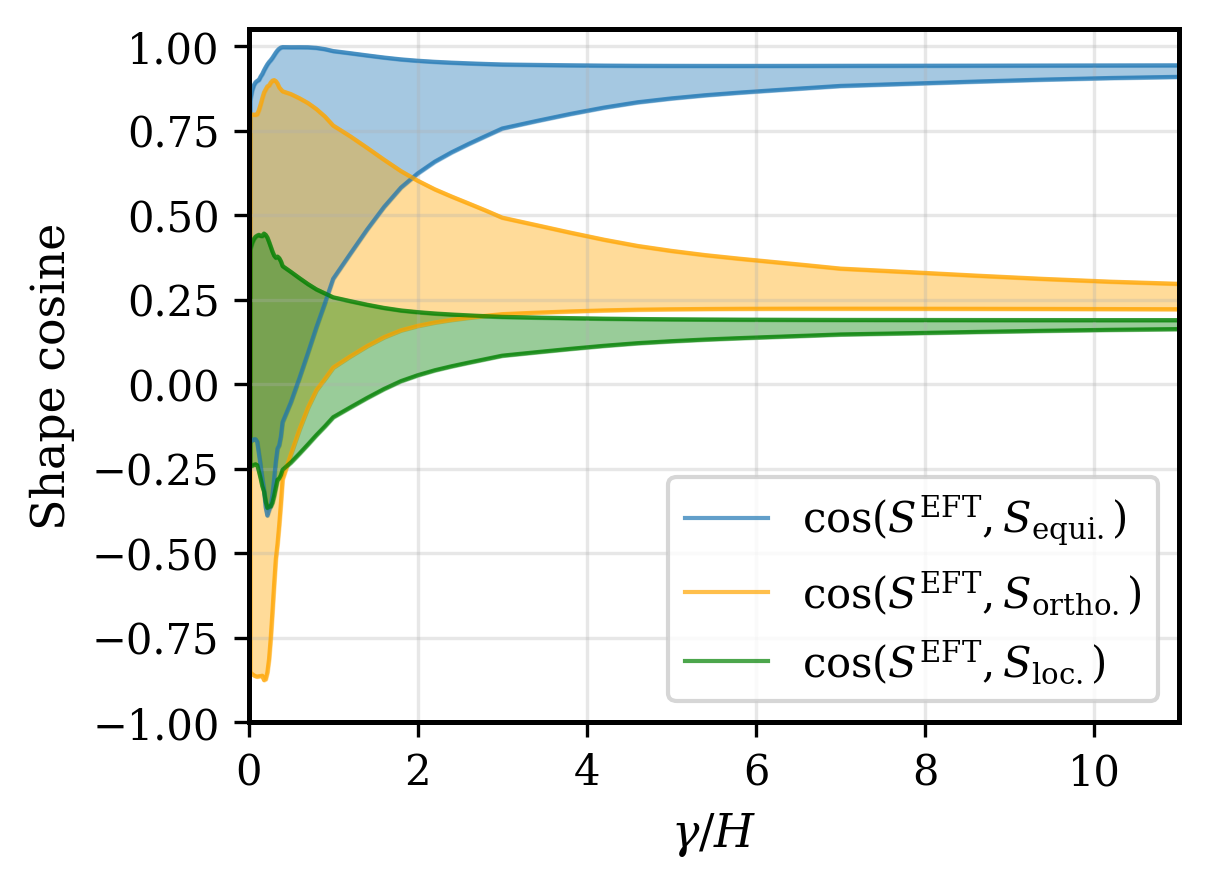}
    \caption{
    Shape correlation of the Open EFToI bispectrum with the standard \textit{equilateral}, \textit{orthogonal} and \textit{local} templates, as used in \textit{Planck}'s non-Gaussianity analysis. The shaded regions show the envelope obtained by varying $c_s$ between $0.01$ and $1$. At strong dissipation, the signal correlates with the equilateral template at the $90\%$ level.
    At weak dissipation, the signal peaks in a strongly folded configuration whose correlations with standard templates depend on $c_s$  and are not generically orthogonal. Figure taken from \cite{Salcedo:2026sdn}.}
    \label{fig:shapecorrel}
\end{figure}
This clearly contains a large number of new effective parameters to be constrained by data. However, a particularly remarkable fact is that the size of some of these interaction terms is fixed by coefficients appearing in the quadratic action, namely the speed of sound $  c_{s} $ and the dissipation $  \gamma $ \cite{LopezNacir:2011kk}. Focusing on these we have 
\begin{align}
    S^{(3)}_{\mathrm{eff}} &= \int \dd^4 x \sqrt{-g} \bigg\{\frac{c_{s}^{2}-1}{2\fpi^{2}}\Big[\left(\partial_{i}\pir\right)^{2} \dpia +2\dpir\partial_{i}\pir\partial^{i}\pia    - 3 \dpir^{2} \dpia\Big] +\frac{\gamma}{2\fpi^{2}}\left[\left(\partial_{i}\pir\right)^{2}-\dpir^{2}\right] \pia \bigg\}\,, \nonumber
\end{align}
The tree-level bispectrum can be computed using the Feynman rules in the previous section from this set of operators, or the enlarged set written above. Unfortunately, only the operators that are cubic in the advanced fields can be computed analytically, while for all others one needs to resort to numerics. Representative numerical results are shown below. In Figure \ref{fig:shapes_low}, we show some numerical results for a variety of cubic couplings. These shapes are particularly interesting because, in the weak-dissipation regime, they have a large signal for folded configurations. This makes them particularly different from the non-Gaussianity encountered from self and gravitational interactions of the inflaton in the  Bunch-Davies state with unitary evolution. This difference can be captured quantitatively using a cosine \cite{Babich:2004gb}. A high value, positive or negative, indicates very similar shapes, while a low value means shapes that would look very different in the data and can be discriminated effectively. The cosines with the most commonly studied shapes, which are those arising in unitary single-field slow-roll inflation with a Bunch-Davies state, are plotted in Figure \ref{fig:shapecorrel} \cite{Salcedo:2026sdn}.

 
\section{Stochastic inflation and late-time behaviour in de Sitter}\label{sec:stochastic}

This section introduces stochastic inflation as an effective description of the long-wavelength dynamics of light scalar fields in de Sitter space. The central idea is to separate the field into long- and short-wavelength modes using a time-dependent coarse-graining scale of order the Hubble radius. Because modes continuously cross this scale as the universe expands, the long-wavelength sector behaves as an open system: its evolution contains both a deterministic drift and a stochastic force generated by the short modes entering the coarse-grained sector. We first illustrate this mechanism for free massless and light massive fields, for which the exact mode functions make the secular growth and its eventual saturation transparent. We then derive the corresponding Langevin and Fokker--Planck equations and use them to study interacting theories, with particular emphasis on a quartic potential. This provides a simple setting in which perturbative secular growth, stochastic resummation, and the non-perturbative late-time equilibrium distribution can be compared explicitly. This review is elementary and captures exclusively the leading stochastic behavior. For a more in-depth discussion covering the literature, see for example \cite{Woodard:2025cez,Cruces:2022imf,Vennin:2020kng}. The open-system perspective adopted here and my own introduction to the subject owes a particular debt to C.~P.~Burgess, R.~Holman and collaborators, whose work connected inflationary
decoherence and stochastic inflation to master equations, open effective field theories and late-time resummation
\cite{Burgess:2006jn,Burgess:2014eoa,Burgess:2015ajz}.


\subsection{A time-dependent split between system and environment}
\label{sec:stochastic-secular-growth}

We consider a light spectator scalar field propagating on a fixed, exact de Sitter background. The scalar is assumed not to contribute appreciably to the stress-energy tensor and therefore does not affect the geometry. In spatially flat coordinates,
\begin{equation}
ds^{2}
=
-dt^{2}
+
a^{2}(t)\,d\mathbf{x}^{2},
\qquad
a(t)
=
e^{Ht},
\end{equation}
where the Hubble parameter \(H\) is constant. The action for the scalar is
\begin{equation}
S
=
-\int d^{4}x\,\sqrt{-g}
\left[
\frac{1}{2}
g^{\mu\nu}
\partial_{\mu}\phi
\partial_{\nu}\phi
+
V(\phi)
\right].
\end{equation}
We shall eventually be interested in the quartic potential
\begin{equation}
V(\phi)
=
\frac{\lambda}{4!}\phi^{4},
\end{equation}
but it is useful first to consider the free massless theory and then a free field with a small mass. The latter provides an exactly solvable example in which the origin and interpretation of secular terms can be seen particularly clearly.

To turn this Hamiltonian quantum field theory into an open system problem, we have to separate system and environment. Throughout these notes, we have considered only open system effects that emerge because of interactions between a system and its environment. However, there is a conceptually different way in which open system dynamics may emerge. The key idea is to consider a bipartition between system and environment that is time dependent. This is, in fact, the only element that is relevant to understanding the \textit{leading order} description of stochastic inflation as originally understood by Starobinsky and Yokoyama \cite{Starobinsky:1986fx,Starobinsky:1994bd}. Additional open system effects do arise from genuine interactions, and have been studied in an extensive body of literature. That is the challenging and hotly debated part of the problem and will not be discussed in these notes.

We separate the field into long- and short-wavelength components according to
\begin{align}
\phi(t,\bfx)
&=
\phi_{L}(t,\bfx)
+
\phi_{S}(t,\bfx), \\
\phi_{L}(t,\mathbf{x})
&=
\int\frac{d^{3}k}{(2\pi)^{3}}\,
W\left(\frac{k}{\epsilon aH}\right)
\phi_{\mathbf{k}}(t)
e^{i\mathbf{k}\cdot\mathbf{x}},
\end{align}
where \(W\) is a window function and \(\epsilon\lesssim 1\). The precise definition of the window will not be important for the leading infrared behavior. Now we define our system to consist of the long modes $ \phi_{L}$, and our environment to consist of the short modes $ \phi_{S}$. This separation is time dependent because of the argument of the window function $ W$, which roughly separates long modes $ k_{L}<\epsilon aH$ from short modes $ k_{S}>\epsilon aH$. Even if at any given time long and short modes did not interact, open effects still emerge from the fact that short modes drift into our system as the universe expands and the scale factor $ a$ grows. When neglecting system-environment interactions, this phenomenon is sufficiently simple that we will be able to describe it non-perturbatively and compare the result with the misleading results of perturbation theory. This whole section studies the effect of a time-dependent bipartition in a series of examples of increasing complexity. 


\paragraph{The massless free scalar} As the first example, we consider a massless free scalar in the Bunch--Davies state. The exact mode function is
\begin{equation}
\phi_{\mathbf{k}}(\eta)
=
\frac{H}{\sqrt{2k^{3}}}
\left(
1+ik\eta
\right)
e^{-ik\eta},
\qquad
\eta
=
-\frac{1}{aH}.
\end{equation}
Its absolute square is therefore
\begin{equation}
\left|
\phi_{\mathbf{k}}(\eta)
\right|^{2}
=
\frac{H^{2}}{2k^{3}}
\left(
1+k^{2}\eta^{2}
\right).
\end{equation}
At late times, when \(k\ll aH\), this approaches
\begin{equation}\label{Pk}
\left|
\phi_{\mathbf{k}}(t)
\right|^{2}
\simeq
\frac{H^{2}}{2k^{3}}.
\end{equation}
Our system comprises of the long-wavelength modes. To focus on the phenomenon that is relevant for us, namely the time dependence of the system-environment bipartition, we introduce an artificial deep infrared (IR) cutoff $ k_{\rm IR}$ and remove all modes longer than this from our system. The equal-point variance is then given by the integral
\begin{equation}
\left\langle
\phi_{L}^{2}(t,\mathbf{x})
\right\rangle
\simeq
\int_{k_{\mathrm{IR}}}^{\epsilon a(t)H}
\frac{d^{3}k}{(2\pi)^{3}}\,
\frac{H^{2}}{2k^{3}},
\end{equation}
and hence
\begin{equation}
\left\langle
\phi_{L}^{2}(t,\mathbf{x})
\right\rangle
\simeq
\frac{H^{2}}{4\pi^{2}}
\ln\left(
\frac{\epsilon a(t)H}{k_{\mathrm{IR}}}
\right).
\end{equation}
Equivalently, measuring the growth relative to some initial time \(t_{0}\),
\begin{equation}
\left\langle
\phi_{L}^{2}(t,\mathbf{x})
\right\rangle
-
\left\langle
\phi_{L}^{2}(t_{0},\mathbf{x})
\right\rangle
\simeq
\frac{H^{3}}{4\pi^{2}}
\left(
t-t_{0}
\right)
=
\frac{H^{2}}{4\pi^{2}}N,
\label{masslessvariancelinear}
\end{equation}
where
\begin{equation}
N
=
\ln\left(
\frac{a(t)}{a(t_{0})}
\right)
=
H(t-t_{0})
\end{equation}
is the number of elapsed e-folds. This result tells us that the two-point function grows without bound as time progresses, an instance of a secular effect.

The growth in \eqref{masslessvariancelinear} has a simple physical interpretation. During each Hubble time, a new shell of Fourier modes crosses the bipartition scale $ \epsilon a H$, leaves the environment and becomes part of our system. Each such shell contributes an approximately independent fluctuation of characteristic magnitude extracted from \eqref{Pk},
\begin{equation}
\delta\phi_{L}
\sim
\frac{H}{2\pi}.
\end{equation}
After \(N\) e-folds, the long field $ \phi_{L}$ in our system has therefore received approximately \(N\) independent kicks. As in an ordinary random walk, the mean displacement remains zero,
\(
\left\langle
\phi_{L}
\right\rangle
=
0,
\)
while the variance grows in proportion to the number of steps,
\begin{equation}
\left\langle
\phi_{L}^{2}
\right\rangle
\sim
N
\left(
\frac{H}{2\pi}
\right)^{2}.
\end{equation}
This reproduces \eqref{masslessvariancelinear}. The long-wavelength scalar thus undergoes diffusion in field space. In this free massless theory, the linear growth of the variance is an exact, non-perturbative result. It reflects the continual accumulation of independent modes in the long-wavelength sector. We stress again that, since this is a free theory on a homogeneous background, this effect emerges even in the absence of any instantaneous interaction between long (system) and short (environment) wavelength modes. 


\subsection{A light massive scalar}
\label{sec:light-massive-scalar}

We now consider a free scalar field with a small positive mass,
\begin{equation}
V(\phi)
=
\frac{1}{2}m^{2}\phi^{2},
\qquad
0<m^{2}\ll H^{2}.
\end{equation}
This theory is again free and hence provides a particularly simple example in which the late-time behavior can be determined exactly and can be contrasted with a perturbative treatment. It will later serve as a useful toy model for understanding how secular terms may arise when an otherwise regular result is expanded perturbatively.

We'll work in conformal time,
\(
\eta
=
-1/(aH)\) for \( -\infty<\eta<0\). Expanding the field as
\begin{equation}
\phi(\eta,\mathbf{x})
=
\int\frac{d^{3}k}{(2\pi)^{3}}
\left[
a_{\mathbf{k}}
\phi_{k}(\eta)
e^{i\mathbf{k}\cdot\mathbf{x}}
+
a_{\mathbf{k}}^{\dagger}
\phi_{k}^{*}(\eta)
e^{-i\mathbf{k}\cdot\mathbf{x}}
\right],
\end{equation}
the mode functions obey
\begin{equation}
\phi_{k}''
-
\frac{2}{\eta}\phi_{k}'
+
\left(
k^{2}
+
\frac{m^{2}}{H^{2}\eta^{2}}
\right)
\phi_{k}
=
0.
\end{equation}
The Bunch--Davies solution is
\begin{equation}
\phi_{k}(\eta)
=
\frac{\sqrt{\pi}}{2}
e^{i\frac{\pi}{2}\left(\nu+\frac{1}{2}\right)}
H(-\eta)^{3/2}
H_{\nu}^{(1)}(-k\eta),
\qquad
\nu
=
\sqrt{
\frac{9}{4}
-
\frac{m^{2}}{H^{2}}
}.
\label{massiveBDmode}
\end{equation}
The phase has been chosen so that at early times, when \(-k\eta\gg1\), the canonically normalized field approaches the positive-frequency Minkowski mode.

For a light scalar it is useful to introduce
\begin{equation}
\Delta
\equiv
\frac{3}{2}
-
\nu.
\end{equation}
When \(m^{2}\ll H^{2}\),
\begin{equation}
\Delta
=
\frac{m^{2}}{3H^{2}}
+
\mathcal{O}\left(
\frac{m^{4}}{H^{4}}
\right).
\label{lightmassDelta}
\end{equation}
Using the small-argument behavior of the Hankel function, the super-Hubble mode function becomes
\begin{equation}
\phi_{k}(\eta)
\simeq
-i\,
\frac{2^{\nu-1}\Gamma(\nu)}{\sqrt{\pi}}\,
\frac{H}{k^{3/2}}
\left(
\frac{k}{aH}
\right)^{\Delta},
\qquad
k\ll aH.
\end{equation}
Its absolute value is therefore
\begin{equation}
\left|
\phi_{k}(\eta)
\right|^{2}
\simeq
\frac{H^{2}}{2k^{3}}
\mathcal{C}(\nu)
\left(
\frac{k}{aH}
\right)^{2\Delta},
\label{massivesuperhorizonmode}
\end{equation}
where
\begin{equation}
\mathcal{C}(\nu)
=
\frac{2^{2\nu-1}\Gamma^{2}(\nu)}{\pi}\overset{m\to0}{\longrightarrow} 1\,.
\end{equation}
In the following we shall retain the leading infrared behavior and set \(\mathcal{C}(\nu)\simeq1\).

The important difference from the massless case is the factor
\begin{equation}
\left(
\frac{k}{aH}
\right)^{2\Delta}.
\end{equation}
At late times, the amplitude of a mode $ k$ is no longer constant but decays slowly as $ a^{-\Delta}$. The mass therefore causes long modes, which are inside our system, to contribute progressively less as time progresses. As we will see now, this provides a restoring force that leads to a finite late-time behavior. 

Let's consider again our bipartition between system and environment. Again field modes with
\(
k_{\rm IR}
<
k
<
\epsilon a(t)H,
\) define our system, with \(k_{\rm IR}\) a fixed infrared cutoff and \(\epsilon\lesssim1\). Using \eqref{massivesuperhorizonmode}, the variance for the modes in our system is
\begin{align}
\left\langle
\phi_{L}^{2}(t,\mathbf{x})
\right\rangle
&\simeq
\int_{k_{\rm IR}}^{\epsilon aH}
\frac{d^{3}k}{(2\pi)^{3}}\,
\frac{H^{2}}{2k^{3}}
\left(
\frac{k}{aH}
\right)^{2\Delta}=
\frac{H^{2}}{4\pi^{2}}
\int_{k_{\rm IR}}^{\epsilon aH}
\frac{dk}{k}
\left(
\frac{k}{aH}
\right)^{2\Delta}\\
&=
\frac{H^{2}}{8\pi^{2}\Delta}
\left[
\epsilon^{2\Delta}
-
\left(
\frac{k_{\rm IR}}{a(t)H}
\right)^{2\Delta}
\right]=\frac{H^{2}\epsilon^{2\Delta}}{8\pi^{2}\Delta}
\left[
1
-
e^{-2 N\Delta}
\right]
\label{massivevariancecutoff}
\end{align}
where in the last step we chose \( k_{\rm IR}
=
\epsilon a(t_{0})H \) for some sufficiently early time $ t_{0}$. At leading order in the light-mass limit, we may approximate \(\epsilon^{2\Delta}\simeq 1\) , yielding
\begin{equation}
\left\langle
\phi_{L}^{2}(t,\mathbf{x})
\right\rangle
\simeq
\frac{H^{2}}{8\pi^{2}\Delta}
\left[
1
-
e^{-2N\Delta }
\right].
\label{exactmassivevariance}
\end{equation}
Using \eqref{lightmassDelta}, this can equivalently be written as
\begin{equation}
\left\langle
\phi_{L}^{2}(t,\mathbf{x})
\right\rangle
\simeq
\frac{3H^{4}}{8\pi^{2}m^{2}}
\left[
1
-
\exp\left(
-\frac{2m^{2}}{3H}(t-t_{0})
\right)
\right].
\label{exactlightmassivevariance}
\end{equation}
This non-perturbative results, valid in the superHubble regime for a sharp window function, tell us that the late-time divergence we encountered in the massless free scalar, \eqref{masslessvariancelinear}, is absent in the presence of a small but finite mass. Intuitively, the restoring force of the quadratic potential balances the spreading due to new modes entering into the long-wavelength system from the short-wavelength environment at every time step.


\paragraph{Perturbative expansion and secular growth} We now compare the non-perturbative result above with what would have been obtained by treating the mass term perturbatively. The simplest way to do so is to expand our non-perturbative result in powers of \(\Delta\), or equivalently in powers of \(m^{2}/H^{2}\). To set the expectation, we notice that this is an expansion around $m = 0 $, which, as we just saw, has a genuine divergence at late times. This warns us that perturbation theory will likely feature these secular divergences as well. 

Expanding \eqref{exactmassivevariance} one finds
\begin{align}
\left\langle
\phi_{L}^{2}(t,\mathbf{x})
\right\rangle
&=
\frac{H^{2}}{4\pi^{2}}
\left[
N
-
 N^{2}\Delta
+
\frac{2}{3}N^{3}\Delta^{2}
+
\cdots
\right]
\label{perturbativemassivevariance}\\
&\simeq 
\frac{H^{2}}{4\pi^{2}}
\left[
N
-
\frac{m^{2}}{3H^{2}}N^{2}
+
\frac{2m^{4}}{27H^{4}}N^{3}
+
\cdots
\right],
\end{align}
where we used \(\Delta\simeq m^{2}/(3H^{2})\). At early times, when
\(
N\Delta 
\ll
1
\), this expansion agrees with the non-perturbative result. At later times, however, the secular powers of \(N\) compensate the smallness of the mass, and perturbation theory breaks down when
\begin{equation}
N\Delta 
\sim
1,
\qquad\text{or equivalently}\qquad
N
\sim
\frac{H^{2}}{m^{2}}.
\end{equation}
We conclude that a calculation performed perturbatively from the outset would therefore appear to predict an increasingly large correction to the massless result.

The non-perturbative solution shows that this apparent breakdown is not a physical instability but an artefact of perturbation theory. The full variance \eqref{exactlightmassivevariance} remains finite and approaches
\begin{equation}
\lim_{t\rightarrow\infty}
\left\langle
\phi_{L}^{2}(t,\mathbf{x})
\right\rangle
=
\frac{H^{2}}{8\pi^{2}\Delta}
\simeq
\frac{3H^{4}}{8\pi^{2}m^{2}}.
\end{equation}
The secular terms arise from expanding the factor
\(
e^{-2\Delta N}
\)
in powers of \(\Delta\). Keeping this exponential unexpanded resums the entire series of terms enhanced by powers of \(N\Delta \) and restores the correct regular late-time behavior. This provides a canonical example in which secular growth is an artefact of strict perturbation theory rather than a property of the theory.\\

So far, our discussion of light fields in de Sitter was connected to open systems because of the bipartition of the Hilbert space, but did not employ any of the techniques we have developed in these notes. This is because we have so far only studied three theories for which a non-perturbative solution is readily available.

We will now move on to interacting theories, or the full solution is unknown. To make progress there, we will develop a technique that sums the effect of the time-dependent partition between system and environment. As we mentioned above, we will effectively neglect all instantaneous interactions between system and environment, Since these do not contribute to leading order.


\subsection{A Langevin equation for the long-wavelength field}
\label{sec:stochastic-langevin}

We now turn to the massless interacting theory,
\begin{equation}
V(\phi)
=
\frac{\lambda}{4!}\phi^{4},
\qquad
\lambda>0.
\end{equation}
As in the massive example, strict perturbation theory produces terms that grow secularly with the number of elapsed e-folds. For instance, the leading infrared terms in the equal-point two-point function are \cite{Cespedes:2023aal}
\begin{align}
\left\langle
\phi_{L}^{2}
\right\rangle
&=
\frac{H^{2}}{4\pi^{2}}
\left[
N
-
\frac{\lambda}{36\pi^{2}}N^{3}
+
\frac{\lambda^{2}}{720\pi^{4}}N^{5}
-
\frac{53\lambda^{3}}{544320\pi^{6}}N^{7}
+
\cdots
\right].
\label{quarticsecularseries}
\end{align}
The first term is tree-level, the second is at one loop, and so on. More generally,
\begin{equation}
\left\langle
\phi_{L}^{2}
\right\rangle_{n\text{-loop}}
\sim
H^{2}\lambda^{n}N^{2n+1}.
\end{equation}
The effective expansion parameter is therefore \(\lambda N^{2}\), and perturbation theory ceases to be reliable when \(N\sim\lambda^{-1/2}\). We shall derive an evolution equation that captures these leading infrared effects without expanding the evolution perturbatively in time.


\paragraph{Coarse graining in phase space}

A systematic derivation keeps both the coarse-grained field and its velocity as phase-space variables \cite{Grain:2017dqa,Vennin:2020kng}; for a recent open-system formulation, see also \cite{Christie:2026qsi}. At leading order in the stochastic approximation, the short modes that cross the coarse-graining scale may be evaluated using the free Bunch--Davies mode functions. We define
\begin{align}
\phi_{L}(t,\mathbf{x})
&=
\int\frac{d^{3}k}{(2\pi)^{3}}\,
W_{t}(k)
\left[
a_{\mathbf{k}}u_{k}(t)e^{i\mathbf{k}\cdot\mathbf{x}}
+
a_{\mathbf{k}}^{\dagger}u_{k}^{*}(t)e^{-i\mathbf{k}\cdot\mathbf{x}}
\right],
\label{longfieldmovingwindow}
\\
v_{L}(t,\mathbf{x})
&=
\int\frac{d^{3}k}{(2\pi)^{3}}\,
W_{t}(k)
\left[
a_{\mathbf{k}}\dot u_{k}(t)e^{i\mathbf{k}\cdot\mathbf{x}}
+
a_{\mathbf{k}}^{\dagger}\dot u_{k}^{*}(t)e^{-i\mathbf{k}\cdot\mathbf{x}}
\right],
\end{align}
where
\begin{equation}
W_{t}(k)
=
W\left(\frac{k}{k_{c}(t)}\right),
\qquad
k_{c}(t)
=
\epsilon a(t)H.
\end{equation}
Because the window is time dependent, \(v_{L}\) is not equal to \(\dot\phi_{L}\). The microscopic Heisenberg field obeys
\begin{equation}
\ddot{\phi}(t,\mathbf{x})
+
3H\dot{\phi}(t,\mathbf{x})
-
\frac{1}{a^{2}(t)}
\nabla^{2}\phi(t,\mathbf{x})
+
V'\!\left(\phi(t,\mathbf{x})\right)
=
0.
\label{microscopic_field_equation}
\end{equation}
Differentiating the two definitions and using the microscopic field equation gives, at leading zeroth order in spatial gradients and in instantaneous long--short interactions,
\begin{align}
\dot\phi_{L}
&=
v_{L}
+
\hat\xi_{\phi},
\label{phase_space_langevin_phi}
\\
\dot v_{L}
&=
-3Hv_{L}
-
V'(\phi_{L})
+
\hat\xi_{v},
\label{phase_space_langevin_v}
\end{align}
where the two noise operators are
\begin{align}
\hat\xi_{\phi}(t,\mathbf{x})
&=
\int\frac{d^{3}k}{(2\pi)^{3}}\,
\dot W_{t}(k)
\left[
a_{\mathbf{k}}u_{k}(t)e^{i\mathbf{k}\cdot\mathbf{x}}
+
a_{\mathbf{k}}^{\dagger}u_{k}^{*}(t)e^{-i\mathbf{k}\cdot\mathbf{x}}
\right],
\label{noiseoperator}
\\
\hat\xi_{v}(t,\mathbf{x})
&=
\int\frac{d^{3}k}{(2\pi)^{3}}\,
\dot W_{t}(k)
\left[
a_{\mathbf{k}}\dot u_{k}(t)e^{i\mathbf{k}\cdot\mathbf{x}}
+
a_{\mathbf{k}}^{\dagger}\dot u_{k}^{*}(t)e^{-i\mathbf{k}\cdot\mathbf{x}}
\right].
\label{velocitynoiseoperator}
\end{align}
The noise is thus present throughout the phase-space evolution: it enters because modes cross the moving boundary between the short- and long-wavelength sectors.


\paragraph{Noise correlations}

It is convenient to write \(U_{\phi,k}=u_{k}\) and \(U_{v,k}=\dot u_{k}\). In the Bunch--Davies state the full phase-space noise kernel is
\begin{equation}
\begin{aligned}
\frac{1}{2}
\left\langle
\left\{
\hat\xi_{A}(t,\mathbf{x}),
\hat\xi_{B}(t',\mathbf{x}')
\right\}
\right\rangle
&=
\operatorname{Re}
\int\frac{d^{3}k}{(2\pi)^{3}}\,
\dot W_{t}(k)\dot W_{t'}(k)
\\
&\quad\times
U_{A,k}(t)U_{B,k}^{*}(t')
e^{i\mathbf{k}\cdot(\mathbf{x}-\mathbf{x}')},
\qquad
A,B\in\{\phi,v\}.
\end{aligned}
\label{genericnoisekernel}
\end{equation}
and \(\langle\hat\xi_{A}\rangle=0\).
For definiteness, consider the sharp window
\begin{equation}
W_{t}(k)
=
\Theta(k_{c}(t)-k),
\qquad
\dot W_{t}(k)
=
Hk_{c}(t)\delta(k-k_{c}(t)).
\end{equation}
The two delta functions in \eqref{genericnoisekernel} imply
\begin{equation}
\delta\left(k_{c}(t)-k_{c}(t')\right)
=
\frac{1}{Hk_{c}(t)}\delta(t-t'),
\end{equation}
so the noise is local in time. For the massless Bunch--Davies mode function, one obtains
\begin{align}
\frac{1}{2}
\left\langle
\left\{
\hat\xi_{\phi}(t,\mathbf{x}),
\hat\xi_{\phi}(t',\mathbf{x}')
\right\}
\right\rangle
&=
\frac{H^{3}}{4\pi^{2}}
\left(1+\epsilon^{2}\right)
\frac{\sin\left(k_{c}(t)r\right)}{k_{c}(t)r}
\delta(t-t'),
\qquad
r=|\mathbf{x}-\mathbf{x}'|.
\label{spatialnoisecorrelator}
\end{align}
Moreover, at the crossing scale \(k=\epsilon aH\),
\begin{equation}
\frac{\dot u_{k}}{Hu_{k}}
=
-\frac{\epsilon^{2}}{1-i\epsilon}.
\end{equation}
Consequently, the correlators involving \(\hat\xi_{v}\) are suppressed by powers of \(\epsilon^{2}\): if \(D_{AB}\) denotes the local phase-space noise matrix, then
\begin{equation}
\frac{D_{\phi v}}{H D_{\phi\phi}}
=
\mathcal O(\epsilon^{2}),
\qquad
\frac{D_{vv}}{H^{2}D_{\phi\phi}}
=
\mathcal O(\epsilon^{4}).
\label{velocitynoisesuppression}
\end{equation}
At coincident points and to leading order for \(\epsilon\ll1\), the field noise therefore satisfies
\begin{equation}
\left\langle
\xi_{\phi}(t,\mathbf{x})
\xi_{\phi}(t',\mathbf{x})
\right\rangle_{\xi}
=
\frac{H^{3}}{4\pi^{2}}
\delta(t-t'),
\label{classicalnoisecorrelator}
\end{equation}
where we have used the super-Hubble classicalization of the modes to replace the operator noise by a classical Gaussian stochastic variable.


\paragraph{Overdamped reduction}

Equations \eqref{phase_space_langevin_phi} and \eqref{phase_space_langevin_v} are the appropriate starting point for the long-wavelength dynamics. Combining them into a single second-order equation would instead give
\begin{equation}
\ddot\phi_{L}
+
3H\dot\phi_{L}
+
V'(\phi_{L})
=
\dot\xi_{\phi}
+
3H\xi_{\phi}
+
\xi_{v}.
\label{secondorderlangevin}
\end{equation}
For a sharp window, \(\xi_{\phi}\) is white noise, so it is preferable to perform the overdamped reduction directly in phase space rather than manipulate \(\dot\xi_{\phi}\).

The velocity relaxes on the Hubble timescale, while a light field satisfying \(|V''|\ll H^{2}\) evolves more slowly. Adiabatically eliminating \(v_{L}\) from \eqref{phase_space_langevin_v} gives
\begin{equation}
v_{L}
\simeq
-\frac{V'(\phi_{L})}{3H}
+
\frac{\xi_{v}}{3H}.
\end{equation}
Substituting this result into \eqref{phase_space_langevin_phi}, and using the suppression \eqref{velocitynoisesuppression}, yields the leading stochastic equation
\begin{equation}
\dot{\phi}_{L}(t,\mathbf{x})
=
-\frac{V'(\phi_{L})}{3H}
+
\xi(t,\mathbf{x}),
\qquad
\xi\equiv\xi_{\phi},
\label{stochasticinflationlangevin}
\end{equation}
with
\begin{equation}
\left\langle
\xi(t,\mathbf{x})
\right\rangle_{\xi}
=
0,
\qquad
\left\langle
\xi(t,\mathbf{x})
\xi(t',\mathbf{x})
\right\rangle_{\xi}
=
\frac{H^{3}}{4\pi^{2}}
\delta(t-t').
\label{stochasticinflationnoise}
\end{equation}
The drift describes the slow classical motion in the potential, while the noise describes the continual addition of modes across the time-dependent system--environment boundary. This is the leading Starobinsky Langevin equation \cite{Starobinsky:1986fx,Starobinsky:1994bd}.

\subsection{Fokker--Planck evolution in stochastic inflation}
\label{sec:stochastic-fokker-planck}

We have shown that, at leading order in the stochastic approximation, the long-wavelength field obeys the Langevin equation in \eqref{stochasticinflationlangevin} with Gaussian white noise. Henceforth we suppress the spatial dependence.

Using the general relation between a first-order Langevin equation with additive noise and the corresponding Fokker--Planck equation derived in \eqref{fokkerplanckgeneral}, the probability density \(P(\phi,t)\) obeys
\begin{equation}
\frac{\partial P(\phi,t)}{\partial t}
=
\frac{1}{3H}
\frac{\partial}{\partial\phi}
\left[
V'(\phi)P(\phi,t)
\right]
+
\frac{H^{3}}{8\pi^{2}}
\frac{\partial^{2}P(\phi,t)}{\partial\phi^{2}}.
\label{stochasticinflationFP}
\end{equation}
The first term describes the deterministic drift generated by the potential, while the second describes diffusion in field space due to modes continually entering the long-wavelength sector.

To gain intuition, we may write \eqref{stochasticinflationFP} as a continuity equation, where the probability current is
\begin{equation}
J(\phi,t)
=
-
\frac{V'(\phi)}{3H}
P(\phi,t)
-
\frac{H^{3}}{8\pi^{2}}
\frac{\partial P(\phi,t)}{\partial\phi}.
\label{FPcurrent}
\end{equation}
The first contribution transports probability along the classical slow-roll flow, while the second transports probability from regions of larger probability density to regions of smaller probability density.

Equation \eqref{stochasticinflationFP} is a first-order evolution equation in time,
\begin{equation}
\frac{\partial P}{\partial t}
=
\mathcal{L}_{\rm FP}P,
\end{equation}
with Fokker--Planck generator
\begin{equation}\label{FPeq}
\mathcal{L}_{\rm FP}
=
\frac{1}{3H}
\frac{\partial}{\partial\phi}
V'(\phi)
+
\frac{H^{3}}{8\pi^{2}}
\frac{\partial^{2}}{\partial\phi^{2}},
\end{equation}
where the first term is understood to act on the product \(V'(\phi)P\). Once this generator has been obtained, its evolution can be solved without expanding the solution perturbatively in time. This was the key insight of Starobinsky and allows us to understand the true nature of the secular effects encountered in perturbation theory. Note the similarity of \eqref{FPeq} to our discussion in Section \ref{ssec:resummation}, Which can be thought of as the stochastic equivalent of the quantum master equation.


\paragraph{Free fields}

As a first check, consider the free massless theory,
\(
V(\phi)
=
0.
\)
The Fokker--Planck equation reduces to the ordinary diffusion equation,
\begin{equation}
\frac{\partial P}{\partial t}
=
\frac{H^{3}}{8\pi^{2}}
\frac{\partial^{2}P}{\partial\phi^{2}}.
\label{masslessdiffusionequation}
\end{equation}
For the sharply localized initial condition
\begin{equation}
P(\phi,t_{0})
=
\delta(\phi-\phi_{0}),
\end{equation}
the solution is\footnote{The observant reader will recognise in \eqref{masslessdiffusionequation} the Wick rotation of the Schr\"odinger equation and in this solution the ground state of the quantum harmonic oscillator.}
\begin{equation}
P(\phi,t)
=
\frac{1}{
\sqrt{
4\pi D(t-t_{0})
}
}
\exp\left[
-
\frac{
(\phi-\phi_{0})^{2}
}{
4D(t-t_{0})
}
\right],
\qquad
D
=
\frac{H^{3}}{8\pi^{2}}.
\label{masslessGaussianP}
\end{equation}
The mean remains fixed,
\begin{equation}
\left\langle
\phi
\right\rangle
=
\phi_{0},
\end{equation}
while the variance grows as
\begin{equation}
\left\langle
(\phi-\phi_{0})^{2}
\right\rangle
=
2D(t-t_{0})
=
\frac{H^{3}}{4\pi^{2}}
(t-t_{0}).
\label{masslessFPvariance}
\end{equation}
For \(\phi_{0}=0\), this reproduces the secular growth obtained directly from the non-perturbative mode functions in \eqref{masslessvariancelinear}. The Fokker--Planck equation therefore reproduces the random-walk interpretation of the free massless field.

We next consider the free massive field with quadratic potential
\begin{equation}
V(\phi)
=
\frac{1}{2}m^{2}\phi^{2}.
\end{equation}
Equation \eqref{stochasticinflationFP} becomes
\begin{equation}
\frac{\partial P}{\partial t}
=
\frac{m^{2}}{3H}
\frac{\partial}{\partial\phi}
\left[
\phi P
\right]
+
\frac{H^{3}}{8\pi^{2}}
\frac{\partial^{2}P}{\partial\phi^{2}}.
\label{massiveOrnsteinUhlenbeck}
\end{equation}
This is the Fokker--Planck equation for the Ornstein--Uhlenbeck process.

Rather than solving for the full probability distribution, we may directly derive the evolution of the second moment. Multiplying \eqref{massiveOrnsteinUhlenbeck} by \(\phi^{2}\), integrating over \(\phi\), and assuming that the boundary terms vanish, one finds
\begin{equation}
\frac{d}{dt}
\left\langle
\phi^{2}
\right\rangle
=
-
\frac{2m^{2}}{3H}
\left\langle
\phi^{2}
\right\rangle
+
\frac{H^{3}}{4\pi^{2}}.
\label{massivesecondmomentequation}
\end{equation}
For the initial condition
\begin{equation}
\left\langle
\phi^{2}(t_{0})
\right\rangle
=
0,
\end{equation}
the solution is
\begin{equation}
\left\langle
\phi^{2}(t)
\right\rangle
=
\frac{3H^{4}}{8\pi^{2}m^{2}}
\left[
1
-
\exp\left(
-\frac{2m^{2}}{3H}
(t-t_{0})
\right)
\right].
\label{massiveFPvariance}
\end{equation}
This agrees with the result obtained directly from the exact massive mode functions in \eqref{exactlightmassivevariance}. The stochastic description therefore reproduces both the unbounded diffusion of the massless field and the finite late-time variance of the light massive field. The beauty of the Fokker-Planck equation is that it extends this ability of resuming the effects of a time-dependent system-environment split to an interacting theory. 


\paragraph{Stationary distribution}

We now return to a generic potential \(V(\phi)\). A stationary solution satisfies
\begin{equation}
\frac{\partial P_{\rm eq}}{\partial t}
=-\frac{\partial J_{\rm eq}}{\partial \phi} =0,
\end{equation}
and therefore has a probability current independent of \(\phi\). For a normalizable equilibrium distribution on the full real line, the current must vanish,
\(
J_{\rm eq}(\phi)
=
0.
\)
Using \eqref{FPcurrent}, this condition becomes
\begin{equation}
\frac{H^{3}}{8\pi^{2}}
\frac{dP_{\rm eq}}{d\phi}
=
-
\frac{V'(\phi)}{3H}
P_{\rm eq}(\phi).
\end{equation}
Solving this gives
\begin{equation}
P_{\rm eq}(\phi)
=
\frac{1}{Z}
\exp\left[
-
\frac{8\pi^{2}}{3H^{4}}
V(\phi)
\right],
\label{StarobinskyYokoyamaDistribution}
\end{equation}
where
\begin{equation}
Z
=
\int_{-\infty}^{+\infty}
d\phi\,
\exp\left[
-
\frac{8\pi^{2}}{3H^{4}}
V(\phi)
\right]
\end{equation}
is fixed by normalization.

A stationary probability distribution exists only if this integral converges. The free massless theory therefore has no normalizable equilibrium distribution, in agreement with its unbounded diffusion. By contrast, any potential that grows sufficiently rapidly as \(\lvert\phi\rvert\rightarrow\infty\) admits a normalizable stationary distribution. In such cases, the deterministic drift towards smaller values of the potential eventually balances the continual diffusion induced by modes entering the long-wavelength sector.


\subsection{The quartic potential}

Let's now focus on the  quartic potential
\begin{equation}
V(\phi)
=
\frac{\lambda}{4!}\phi^{4}.
\end{equation}
In this case, we can solve the Fokker-Planck equation in two complementary regimes: at early times in an expansion in $ \lambda N^{2}$, and at very late times, around the equilibrium distribution we found previously.


\paragraph{Perturbative solutions} The Fokker--Planck equation \eqref{stochasticinflationFP} implies a coupled hierarchy of equations for the moments of the probability distribution. Multiplying \eqref{stochasticinflationFP} by \(\phi^{n}\), integrating over \(\phi\), and assuming that boundary terms vanish, one finds
\begin{equation}\label{quarticmomenthierarchy}
\frac{d}{dN}
\left\langle
\phi^{n}
\right\rangle
=
-
\frac{n\lambda}{18H^{2}}
\left\langle
\phi^{n+2}
\right\rangle
+
\frac{n(n-1)H^{2}}{8\pi^{2}}
\left\langle
\phi^{n-2}
\right\rangle ,
\end{equation}
where we switched to the number of e-folds \(N=H(t-t_{0})\) as the time variable. This equation is not closed: the two-point function depends on the four-point function, whose evolution in turn depends on the six-point function, and so on. Nevertheless, the hierarchy may be solved recursively as an expansion in \(\lambda\).

For a distribution initially localized at \(\phi=0\), the free massless theory gives
\begin{equation}
\left\langle
\phi^{2n}
\right\rangle_{0}
=
(2n-1)!!
\left(
\frac{H^{2}N}{4\pi^{2}}
\right)^{n},
\qquad
\left\langle
\phi^{2n+1}
\right\rangle_{0}
=
0.
\end{equation}
Substituting these moments into \eqref{quarticmomenthierarchy} for $ n=2$ gives the first perturbative correction, and
the procedure may then be iterated order by order. For the two-point
function, the resulting expansion takes the form
\begin{equation}
\left\langle
\phi^{2}(N)
\right\rangle
=
H^{2}
\sum_{p=0}^{\infty}
c_{p}\lambda^{p}N^{2p+1},
\label{quarticperturbativeseries}
\end{equation}
where the coefficients \(c_{p}\) are determined recursively by the moment
hierarchy. The first terms are
\begin{equation}
\left\langle
\phi^{2}(N)
\right\rangle
=
\frac{H^{2}}{4\pi^{2}}
\left[
N
-
\frac{\lambda}{36\pi^{2}}N^{3}
+
\mathcal O\left(
\lambda^{2}N^{5}
\right)
\right].
\label{quarticfirstsecularterms}
\end{equation}

These terms reproduce the leading infrared secular contributions obtained in
perturbative quantum field theory. More generally, the stochastic hierarchy
generates at order \(\lambda^{p}\) the leading contribution proportional to
\begin{equation}
\lambda^{p}N^{2p+1}.
\end{equation}
As anticipated, the perturbative expansion is therefore controlled by the combination
\(
\lambda N^{2},
\)
rather than by \(\lambda\) alone. Even for arbitrarily small coupling,
strict perturbation theory eventually breaks down when
\(
N
\sim \lambda^{-1/2}
\).


\paragraph{Non-perturbative equilibrium solution}

The moment expansion describes the early-time solution of the Fokker--Planck
equation at fixed order in \(\lambda\), but breaks down at late times. From the non-perturbative derivation of the stationary distribution, we already suspect that these secular effects may be an artefact of perturbation theory.  To determine the late-time behavior therefore we instead look at the stationary distribution
\eqref{StarobinskyYokoyamaDistribution}. For the quartic potential, it is
\begin{equation}
P_{\rm eq}(\phi)
=
\frac{1}{Z}
\exp\left[
-
\frac{\pi^{2}\lambda}{9H^{4}}
\phi^{4}
\right],
\label{quarticequilibriumdistribution}
\end{equation}
which is normalizable for \(\lambda>0\), showing that the
quartic interaction arrests the unbounded diffusion of the free massless
field.

All equal-time moments of the long-wavelength field can now be evaluated
without expanding in \(\lambda\). Since the equilibrium distribution is even,
all odd moments vanish,
\begin{equation}
\left\langle
\phi^{2n+1}
\right\rangle_{\rm eq}
=
0.
\end{equation}
The even moments are
\begin{equation}
\left\langle
\phi^{2n}
\right\rangle_{\rm eq}
=
\left( \frac{\pi^{2}\lambda}{9H^{4}} \right)^{-n/2}
\frac{
\Gamma\left(
\frac{2n+1}{4}
\right)
}{
\Gamma\left(
\frac{1}{4}
\right)
}.
\label{quarticequilibriummoments}
\end{equation}
In particular,
\begin{equation}
\left\langle
\phi^{2}
\right\rangle_{\rm eq}
=
\frac{3H^{2}}{\pi\sqrt{\lambda}}
\frac{
\Gamma\left(
\frac{3}{4}
\right)
}{
\Gamma\left(
\frac{1}{4}
\right)
},
\label{quarticequilibriumtwopoint} \qquad 
\left\langle
\phi^{4}
\right\rangle_{\rm eq}
=
\frac{9H^{4}}{4\pi^{2}\lambda}.
\end{equation}

The equilibrium two-point function is finite, but it is nonanalytic in the
coupling. It therefore cannot be reproduced at any finite order in perturbation theory.
Instead, it represents the result of resumming the entire series of secular
terms generated by the moment hierarchy.

The relation between the perturbative and equilibrium descriptions can be
made manifest by dimensional analysis. The full time-dependent solution must
take the scaling form
\begin{equation}
\left\langle
\phi^{2}(N)
\right\rangle
=
\frac{H^{2}}{\sqrt{\lambda}}\,
F\left(
\sqrt{\lambda}\,N
\right),
\label{quarticscalingform}
\end{equation}
for some dimensionless function \(F\). At early times,
\(\sqrt{\lambda}\,N\ll1\), the perturbative hierarchy determines its Taylor
expansion,
\begin{equation}
F(z)
=
\frac{z}{4\pi^{2}}
-
\frac{z^{3}}{144\pi^{4}}
+
\cdots.
\end{equation}
At late times, the equilibrium distribution determines its asymptotic value,
\begin{equation}
F(z)
\xrightarrow{z\rightarrow\infty}
\frac{3}{\pi}
\frac{
\Gamma\left(
\frac{3}{4}
\right)
}{
\Gamma\left(
\frac{1}{4}
\right)
}.
\end{equation}
The full Fokker--Planck evolution provides the interpolation between these
two regimes. The time-dependent evolution close to the equilibrium distribution can be studied via the eigenvectors of the Fokker-Planck generator, but we omit this discussion here and refer the interested reader to the literature. 


\section{Conclusions and outlook}

These lectures developed two complementary descriptions of open quantum systems. In the operator formalism, density matrices, partial traces and completely positive dynamical maps led, under Markovian and time-homogeneous assumptions, to the Gorini--Kossakowski--Sudarshan--Lindblad master equation and its Hamiltonian and dissipative contributions.

We then introduced the Schwinger--Keldysh path integral. Its doubled fields encode time- and anti-time-ordered products, while the \(r/a\) basis separates response from fluctuations. Integrating out an environment yields the Feynman--Vernon influence functional, which describes dissipation, noise and decoherence. Although its quadratic semiclassical limit often admits a Langevin description, the full effective action may contain genuinely quantum interactions beyond Gaussian stochastic dynamics.

We applied these ideas to inflation by constructing an open effective field theory for the Goldstone boson of broken time translations. This framework systematically incorporates environmental dissipation and fluctuations and can be used to calculate the power spectrum and primordial non-Gaussianity, including interactions with no purely classical stochastic counterpart.

Finally, a time-dependent split between long and short modes in de Sitter led to stochastic evolution for light fields. For interacting fields, the resulting Fokker--Planck equation resums perturbative contributions that grow secularly with time and may approach a regular late-time distribution. Such secular growth need not signal a physical instability; it may instead indicate that an evolution equation has been expanded beyond the regime of validity of perturbation theory.

\paragraph{Operator and path-integral perspectives.}

One motivation for presenting the operator and path-integral formalisms together is that each makes manifest structures that are comparatively obscure in the other. The density-matrix language is particularly well adapted to questions concerning states, entropy, quantum channels and complete positivity. The Schwinger--Keldysh language, by contrast, is naturally suited to locality, symmetry, effective field theory and the computation of general multi-time correlation functions. In principle, the two descriptions encode the same physics, but they organize the relevant information in very different ways.

Several examples in these lectures illustrate what may be learned by bringing the two perspectives together. The elementary Schwinger--Keldysh conditions, such as the normalization of the generating functional and the positivity of the imaginary part of the effective action, already place important restrictions on an influence functional. Nevertheless, when a local-in-time influence functional admits a description in terms of a GKSL generator, complete positivity imposes stronger constraints. In the Gaussian examples studied above, positivity of the Kossakowski matrix relates the dissipative and noise coefficients and shows that a non-zero amount of friction requires a minimum amount of accompanying fluctuation. More generally, expanding the jump operators rather than directly expanding the influence functional offers a way of constructing positivity-improved effective theories in which complete positivity can remain exact even after truncation.

Conversely, the Schwinger--Keldysh formalism naturally generates a much broader class of observables than those most directly accessed by evolving a density matrix and evaluating equal-time expectation values. Functional differentiation with respect to the doubled sources produces correlators with different time orderings, while the $r/a$ basis efficiently organizes response and fluctuation functions. This is particularly valuable in cosmology, where the main observables are correlation functions of quantum fields rather than the density matrix itself. The operator formalism supplies powerful structural constraints on the admissible evolution, while the path integral provides the language in which these constraints can be combined with spacetime symmetries and effective field theory.

The applications considered here have been restricted almost entirely to the early universe and, more specifically, to inflation. They should therefore be viewed only as an initial illustration of the much wider role that open-system methods may play in cosmology. In many cosmological problems one observes only a restricted range of scales, species or collective variables, while the remaining degrees of freedom act as an environment. Coarse-graining then generically produces dissipative terms, fluctuations, memory effects and an evolution that need not be unitary when restricted to the retained variables.

An well studied example is the effective field theory of large-scale structure. Short-wavelength nonlinear modes are integrated out in order to obtain an effective description of the long-wavelength matter distribution. The result contains effective stresses and stochastic contributions and is therefore, in this sense, an open statistical theory. It is traditionally formulated directly in terms of equations of motion and their correlation functions, but an equivalent path-integral formulation should make it possible to organize its symmetries, response fields, stochastic terms and renormalization properties in a unified way.

There are many further possible applications throughout cosmology and astroparticle physics. The thermal history of the universe involves repeated examples of subsystems interacting with an evolving medium, including reheating, thermalization, particle production and freeze-out. Neutrino oscillations and transport, dark-matter kinetics, baryogenesis, gravitational-wave propagation through matter and the evolution of cosmological perturbations in the presence of additional sectors may all, in suitable regimes, benefit from an open-system description. Not every such problem requires a fully quantum treatment, but the operator and Schwinger--Keldysh methods provide a common framework in which quantum, statistical and semiclassical limits can be distinguished.

\paragraph{Open quantum field theories.}

At a more formal level, a central open problem is to develop a systematic understanding of open quantum field theories. Effective field theory teaches us to write the most general local action compatible with the degrees of freedom, symmetries and power counting of a problem. For an open theory this prescription must be supplemented by the Schwinger--Keldysh constraints associated with normalization, Hermiticity and positivity, and, whenever a Markovian description is appropriate, by the requirement of complete positivity. Additional constraints should follow from fundamental properties of the full system evolution, such as unitarity, locality and causality.

In particular, it would be useful to characterize which local Schwinger--Keldysh actions define admissible quantum evolutions. The condition
\(
\operatorname{Im}S_{\mathrm{eff}}\geq 0
\)
is necessary but is not, in general, sufficient to guarantee complete positivity of a corresponding dynamical map. The jump-operator construction discussed in these notes gives one possible starting point, but several questions remain. Complete positivity can become obscured by derivative expansions, field redefinitions, the integration of auxiliary variables or the truncation of an operator basis. It would be valuable to formulate criteria that are both sufficiently general for field theory and sufficiently practical to be imposed on an effective action.

Generic open systems are also non-Markovian. Integrating out an environment ordinarily produces kernels that are nonlocal in time and retain information about the past history of the system. A local master equation or local influence functional is justified only when the environmental correlation time is sufficiently short compared with the scales being probed. Understanding the systematic expansion around this limit, and developing an effective-field-theory power counting for memory effects, would extend the framework substantially. In cosmology this issue is particularly important because the background is time dependent and a clean separation of timescales may hold only during part of the evolution.

A related question concerns the dependence of the effective theory on the choice of system--environment split. Which variables are retained, and which are integrated out, is not unique. In stochastic inflation the separation between long and short modes evolves continuously as modes cross the coarse-graining scale. More generally, changing this split should induce a transformation of the effective dissipative, noise and interaction coefficients. Clarifying this dependence may lead to a notion of renormalization-group evolution acting not only on ordinary couplings but also on the choice of reduced description itself.

\paragraph{Gravity and gauge theories.}

A particularly important challenge is the systematic formulation of dynamical gravity as an open quantum theory. The Schwinger--Keldysh path integral appears to be a natural starting point because diffeomorphism invariance and spacetime locality can be implemented directly at the level of a doubled action. Nevertheless, the operator formalism suggests that additional constraints related to positivity and the structure of the reduced state must also be present.

Several conceptual difficulties arise already in defining the problem. In gravity, the division of the Hilbert space into a subsystem and an environment may be complicated by gauge constraints, and local regions do not generally admit a simple tensor-product decomposition. The Hamiltonian and momentum constraints must remain compatible with the reduced dynamics, while the distinction between system and environment may itself depend on the geometry. Horizons provide a natural source of inaccessible degrees of freedom, but they also introduce questions concerning observer dependence, entropy and information loss. A satisfactory open effective theory of gravity should reconcile these features with locality, causality, diffeomorphism invariance and, whenever appropriate, complete positivity.

Such a framework could have applications to semiclassical gravity, black-hole evaporation, cosmological horizons and the stochastic description of metric fluctuations. It could also help clarify when the semiclassical Einstein equations should be supplemented by noise and dissipation, and how those terms are constrained by the quantum state of matter. More ambitiously, it may provide an effective language for discussing situations in which only a restricted set of gravitational observables is operationally accessible.

\paragraph{Thermal systems, hydrodynamics and quantum information.}

Another major direction that we have not developed is the relation between open systems and thermal physics. When the environment is in equilibrium, the Kubo--Martin--Schwinger condition imposes nontrivial relations between response and fluctuation functions. At the level of the Schwinger--Keldysh effective action and for local equilibrium, these relations are encoded by a dynamical KMS symmetry. The resulting fluctuation--dissipation relations connect coefficients that would otherwise appear independent and provide an important additional organizing principle for open effective theories.

These ideas play a central role in dissipative hydrodynamics. Hydrodynamics is a universal effective description of systems close to local thermal equilibrium, including systems for which no weakly coupled microscopic description is available. Its Schwinger--Keldysh formulation naturally incorporates dissipation, fluctuations, conservation laws and the dynamical KMS symmetry. It therefore provides one of the most developed examples of an interacting open effective field theory. Since hydrodynamics also describes strongly coupled quantum field theories and admits a gravitational realization through holography, open-system methods have become increasingly relevant to the study of gauge--gravity duality.

There are equally important connections with quantum information. Decoherence, thermalization, scrambling and the growth of quantum chaos concern the way information is redistributed among degrees of freedom that may no longer be individually accessible. Quantum channels and reduced density matrices provide a natural language for these processes, while generalized Schwinger--Keldysh contours can be used to compute out-of-time-ordered correlators and other diagnostics of operator growth. It would be interesting to understand more systematically how the effective description of open quantum fields constrains entanglement generation, information transport and scrambling.

The quantum-to-classical transition deserves particular care in cosmological applications and has been repeatedly studied over the years. The emergence of a Langevin or Fokker--Planck equation does not by itself establish that the underlying fluctuations have become classical. Squeezing, decoherence, the suppression of interference and the emergence of a positive probabilistic description are related but distinct phenomena. A framework that keeps both the density matrix and the Schwinger--Keldysh effective action in view may help disentangle these notions and identify which genuinely quantum correlations are discarded in a semiclassical stochastic approximation.

\paragraph{Renormalization and universality.}

Finally, much remains to be understood about renormalization in open quantum field theory. Radiative corrections will generically generate every operator compatible with the symmetries and Schwinger--Keldysh constraints. It is therefore important to determine whether physically interesting subclasses of open theories are closed under renormalization. For example, one may ask whether Markovianity, complete positivity or a particular jump-operator structure can be preserved along a renormalization-group flow, or whether integrating out additional scales inevitably produces non-locality and more general multi-advanced-field interactions.

It would also be interesting to identify the endpoints of open-system renormalization-group flows. In closed quantum field theory, fixed points and universality classes provide a powerful classification of long-distance behaviour. Open systems may possess a richer set of possibilities, including thermal, driven, dissipative and genuinely non-equilibrium fixed points. One would like to know which quantities characterize these endpoints, whether analogues of monotonicity theorems exist, and how positivity, entropy production and information loss constrain the flow. Understanding these questions could eventually place open quantum field theory on a footing comparable to that of ordinary effective field theory.

The general lesson of these lectures is that open-system effects should not be regarded merely as small corrections caused by an imperfect isolation of an otherwise closed system. In many settings, including cosmology, the observables of interest intrinsically refer only to a subset of the available degrees of freedom. Their effective dynamics is then naturally described by a reduced state, an influence functional or both. The operator and path-integral formalisms emphasize different aspects of this description, and their combination offers a powerful route towards effective theories that incorporate dissipation, fluctuations and quantum correlations while remaining consistent with the fundamental principles of quantum mechanics. The examples studied here, which represent only a glimpse of what has been studies in literature, suggest that open quantum systems may provide an increasingly important language for quantum field theory, gravity and cosmology.


\section*{Acknowledgements} I am deeply grateful to Santiago Agui Salcedo, Thomas Colas and Lennard Dufner for the scientific journey on which we embarked together several years ago, and for our continuing collaboration in exploring how open-quantum-system techniques can illuminate problems in cosmology. I thank the organizers of the Scuola Tematica on ``Theoretical Physics of the Fundamental Interactions'' at the Scuola Galileiana di Studi Superiori in Padova, the 2026 IHES Summer School on ``Cosmological Correlators,'' and the La Ricotta Summer School on ``The Disordered Universe'' for the opportunity to present these lectures. I am especially thankful to the students for their patience, enthusiasm and incisive questions, which helped me sharpen both the presentation and my own understanding of the material. I also benefited greatly from the stimulating talks and discussions at \href{https://indico.global/event/16432/}{Contours 2026: Effective Field Theories Meet the Schwinger--Keldysh Formalism} in Cambridge, the MITP program \href{https://indico.mitp.uni-mainz.de/event/436/}{``Open Quantum Systems: Dissipation and Decoherence from Subatomic to Cosmic Scales,''} and the INT program \href{https://www.int.washington.edu/programs-and-workshops/25-3b}{``Open Quantum Systems: Dissipative Dynamics from Quarks to the Cosmos.''} Finally, I am grateful for illuminating conversations with Cliff Burgess, Simon Caron-Huot, Perseas Christodoulidis, Stefano Cusumano, Luca Delacr\'etaz, Felix Haehl, Amaury Jean, Austin Joyce, Greg Kaplanek, Enrica Lausdei, R.~Loganayagam, Scott Melville, Maria Mylova, Atsuhisa Ota, Riccardo Penco, Riccardo Rattazzi, Sarah Shandera, Andrew Tolley and Xi Tong.


\appendix


\section{Inverting the quadratic Schwinger--Keldysh kernel}\label{appA}

In this appendix we show in some detail how the propagators of a quadratic Schwinger--Keldysh theory are obtained by inverting the corresponding kernel. We work in momentum space, where translation invariance makes different Fourier modes independent. For simplicity we suppress the momentum label \((\omega,\mathbf k)\) whenever no confusion can arise.\\

Consider a generic Gaussian Schwinger--Keldysh action for a real field \(\phi\), written in the \(r/a\) basis as
\begin{align}\label{AppQuadAction}
S_{\rm SK}
=
\int \frac{d\omega}{2\pi}\frac{d^d k}{(2\pi)^d}\,
\Bigg[
\phi_a(-\omega,-\mathbf k)\,\mathcal D_R(\omega,\mathbf k)\,\phi_r(\omega,\mathbf k)
+\phi_r(-\omega,-\mathbf k)\,\mathcal D_A(\omega,\mathbf k)\,\phi_a(\omega,\mathbf k)
\nonumber\\
\qquad\qquad
+\frac{i}{2}\,\phi_a(-\omega,-\mathbf k)\,\mathcal N(\omega,\mathbf k)\,\phi_a(\omega,\mathbf k)
\Bigg] .
\end{align}
Here \(\mathcal D_R\) and \(\mathcal D_A\) are the inverse retarded and advanced kernels, while \(\mathcal N\) is the Keldysh or noise kernel. For a real field, one has
\begin{equation}
\mathcal D_A(\omega,\mathbf k)=\mathcal D_R(\omega,\mathbf k)^* ,
\qquad
\mathcal N(\omega,\mathbf k)^*=\mathcal N(\omega,\mathbf k) .
\end{equation}

It is convenient to combine the two fields into the column vector
\begin{equation}
\Phi(\omega,\mathbf k)
:=
\begin{pmatrix}
\phi_r(\omega,\mathbf k) \\[2mm]
\phi_a(\omega,\mathbf k)
\end{pmatrix} .
\end{equation}
Then the action may be written as
\begin{equation}\label{AppMatrixAction}
S_{\rm SK}
=
\frac12
\int \frac{d\omega}{2\pi}\frac{d^d k}{(2\pi)^d}\,
\Phi^\dagger(-\omega,-\mathbf k)\,
\mathbb K(\omega,\mathbf k)\,
\Phi(\omega,\mathbf k) ,
\end{equation}
with kernel matrix
\begin{equation}\label{AppKmatrix}
\mathbb K(\omega,\mathbf k)
=
\begin{pmatrix}
0 & \mathcal D_A(\omega,\mathbf k) \\[2mm]
\mathcal D_R(\omega,\mathbf k) & i\,\mathcal N(\omega,\mathbf k)
\end{pmatrix} .
\end{equation}
The vanishing of the \(rr\) entry is the usual Schwinger--Keldysh normalization constraint that $ S_{\rm SK}[\phi_{r},0]=0$.

\paragraph{Propagator matrix}

The propagators are obtained by inverting the quadratic kernel. Let
\begin{equation}
\mathbb G(\omega,\mathbf k)
:=
\begin{pmatrix}
G_K(\omega,\mathbf k) & G_R(\omega,\mathbf k) \\[2mm]
G_A(\omega,\mathbf k) & G_{aa}(\omega,\mathbf k)
\end{pmatrix} ,
\end{equation}
where
\begin{align}
G_K(\omega,\mathbf k)
&:=
\langle\!\langle \phi_r(\omega,\mathbf k)\phi_r(-\omega,-\mathbf k)\rangle\!\rangle ,
\\
G_R(\omega,\mathbf k)
&:=
\langle\!\langle \phi_r(\omega,\mathbf k)\phi_a(-\omega,-\mathbf k)\rangle\!\rangle ,
\\
G_A(\omega,\mathbf k)
&:=
\langle\!\langle \phi_a(\omega,\mathbf k)\phi_r(-\omega,-\mathbf k)\rangle\!\rangle ,
\\
G_{aa}(\omega,\mathbf k)
&:=
\langle\!\langle \phi_a(\omega,\mathbf k)\phi_a(-\omega,-\mathbf k)\rangle\!\rangle .
\end{align}
At the formal Gaussian level, \(\mathbb G\) is the inverse of \(\mathbb K\), up to the conventional factors of \(i\) already absorbed in our definitions. Thus
\begin{equation}
\mathbb K(\omega,\mathbf k)\,\mathbb G(\omega,\mathbf k)=1 .
\end{equation}
Writing this out explicitly gives
\begin{equation}
\begin{pmatrix}
0 & \mathcal D_A \\[1mm]
\mathcal D_R & i\mathcal N
\end{pmatrix}
\begin{pmatrix}
G_K & G_R \\[1mm]
G_A & G_{aa}
\end{pmatrix}
=
\begin{pmatrix}
1 & 0 \\[1mm]
0 & 1
\end{pmatrix} .
\end{equation}
Multiplying the matrices gives four equations:
\begin{align}
\mathcal D_A\,G_A &= 1 ,
\\
\mathcal D_A\,G_{aa} &= 0 ,
\\
\mathcal D_R\,G_K + i\mathcal N\,G_A &= 0 ,
\\
\mathcal D_R\,G_R + i\mathcal N\,G_{aa} &= 1 .
\end{align}
Assuming \(\mathcal D_R\) and \(\mathcal D_A\) are invertible, this immediately yields
\begin{align}
G_R(\omega,\mathbf k)
&=
\frac{1}{\mathcal D_R(\omega,\mathbf k)} ,
\\
G_A(\omega,\mathbf k)
&=
\frac{1}{\mathcal D_A(\omega,\mathbf k)} ,
\\
G_{aa}(\omega,\mathbf k)
&=0 ,
\\
G_K(\omega,\mathbf k)
&=
-\,i\,\mathcal N(\omega,\mathbf k)\,
\frac{1}{\mathcal D_R(\omega,\mathbf k)\mathcal D_A(\omega,\mathbf k)} .
\end{align}
This is the general Gaussian result quoted in the main text.


\subsection{Time-domain retarded propagator}\label{app:GK}

Let us now derive the retarded propagator in the time domain. Starting from
\begin{equation}
G_R(\omega,\mathbf k)
=
\frac{1}{-\omega^2-i\gamma\omega+\omega_{\mathbf k}^2},
\qquad
\omega_{\mathbf k}^2:=\mathbf k^2+m^2,
\end{equation}
we define
\begin{equation}
G_R(t,\mathbf k)
=
\int \frac{d\omega}{2\pi}\,
e^{-i\omega t}\,
G_R(\omega,\mathbf k).
\end{equation}
The poles are located at
\begin{equation}
\omega_\pm(\mathbf k)
=
-\frac{i\gamma}{2}
\pm
\sqrt{\omega_{\mathbf k}^2-\frac{\gamma^2}{4}},
\end{equation}
so that
\begin{equation}
G_R(\omega,\mathbf k)
=
-\frac{1}{(\omega-\omega_+)(\omega-\omega_-)}.
\end{equation}

For $\gamma>0$, both poles lie in the closed lower half of the complex
$\omega$-plane. Therefore, when $t<0$, the contour may be closed in the
upper half-plane and no poles are enclosed, giving
\begin{equation}
G_R(t,\mathbf k)=0,
\qquad
t<0.
\end{equation}
For $t>0$, the contour is instead closed in the lower half-plane and the
retarded propagator is obtained by summing the residues of the enclosed
poles. Its explicit form depends on whether the mode is underdamped or
overdamped. To make this explicit, we separate the different regimes.

\paragraph{Underdamped regime.}

Suppose first that
\begin{equation}
\omega_{\mathbf k}^2>\frac{\gamma^2}{4}.
\end{equation}
It is convenient to define
\begin{equation}
\Omega_{\mathbf k}
:=
\sqrt{
\omega_{\mathbf k}^2-\frac{\gamma^2}{4}
},
\end{equation}
so that
\begin{equation}
\omega_\pm
=
-\frac{i\gamma}{2}
\pm
\Omega_{\mathbf k}.
\end{equation}
The propagator may then be decomposed as
\begin{equation}
G_R(\omega,\mathbf k)
=
-\frac{1}{2\Omega_{\mathbf k}}
\left(
\frac{1}{\omega-\omega_+}
-
\frac{1}{\omega-\omega_-}
\right).
\end{equation}
For $t>0$, the residue theorem gives
\begin{align}
G_R(t,\mathbf k)
&=
-i\,\theta(t)
\left[
-\frac{1}{2\Omega_{\mathbf k}}e^{-i\omega_+t}
+
\frac{1}{2\Omega_{\mathbf k}}e^{-i\omega_-t}
\right]=
\theta(t)\,
e^{-\gamma t/2}
\frac{\sin(\Omega_{\mathbf k}t)}{\Omega_{\mathbf k}}.
\end{align}
Thus, in the underdamped regime, the response oscillates with frequency
$\Omega_{\mathbf k}$ while its amplitude decays as $e^{-\gamma t/2}$.

\paragraph{Overdamped regime.}

Suppose instead that
\begin{equation}
\omega_{\mathbf k}^2<\frac{\gamma^2}{4}.
\end{equation}
For clarity, we define
\begin{equation}
\Lambda_{\mathbf k}
:=
\sqrt{
\frac{\gamma^2}{4}-\omega_{\mathbf k}^2
} \in \mathbb{R},
\end{equation}
so that
\begin{equation}
\omega_\pm
=
-\frac{i\gamma}{2}
\pm
i\Lambda_{\mathbf k}
=
-i\left(
\frac{\gamma}{2}
\mp
\Lambda_{\mathbf k}
\right).
\end{equation}
The two poles are now purely imaginary and lie in the lower half-plane.
Repeating the same contour argument gives
\begin{align}
G_R(t,\mathbf k)
&=
\theta(t)\,
e^{-\gamma t/2}
\frac{\sinh(\Lambda_{\mathbf k}t)}{\Lambda_{\mathbf k}}
\nonumber\\
&=
\theta(t)\,
\frac{
e^{-(\gamma/2-\Lambda_{\mathbf k})t}
-
e^{-(\gamma/2+\Lambda_{\mathbf k})t}
}{
2\Lambda_{\mathbf k}
}.
\end{align}
The response is therefore a sum of two decaying exponentials, with decay
rates
\begin{equation}
\Gamma_\pm
=
\frac{\gamma}{2}
\pm
\Lambda_{\mathbf k}.
\end{equation}
The late-time behaviour is controlled by the slower rate
\begin{equation}
\Gamma_{\rm slow}
=
\frac{\gamma}{2}
-
\Lambda_{\mathbf k}.
\end{equation}
In the strongly overdamped regime,
$\omega_{\mathbf k}\ll\gamma$, this becomes
\begin{equation}
\Gamma_{\rm slow}
=
\frac{\gamma}{2}
-
\sqrt{
\frac{\gamma^2}{4}-\omega_{\mathbf k}^2
}
\simeq
\frac{\omega_{\mathbf k}^2}{\gamma}.
\end{equation}
For a massless field, this gives the diffusive scaling
\begin{equation}
\Gamma_{\rm slow}
\simeq
\frac{\mathbf k^2}{\gamma}.
\end{equation}

The critically damped case,
$\omega_{\mathbf k}^2=\gamma^2/4$, is obtained continuously by taking
$\Omega_{\mathbf k}\to0$ in the underdamped expression, or equivalently
$\Lambda_{\mathbf k}\to0$ in the overdamped expression. Using
$\sin z/z\to1$ or $\sinh z/z\to1$, one finds
\begin{equation}
G_R(t,\mathbf k)
=
\theta(t)\,t\,e^{-\gamma t/2}.
\end{equation}

These expressions make causality manifest through the factor $\theta(t)$.
For every mode with $\omega_{\mathbf k}^2>0$, the retarded response decays
at late times, although the relaxation timescale is mode dependent and can
become parametrically long in the overdamped regime. The exceptional
massless homogeneous mode has
\begin{equation}
m=0,
\qquad
\mathbf k=0,
\qquad
\omega_{\mathbf k}=0,
\end{equation}
for which
\begin{equation}
G_R(t,\mathbf 0)
=
\theta(t)\,
\frac{1-e^{-\gamma t}}{\gamma}.
\end{equation}
This approaches a constant rather than vanishing at late times.

\bibliographystyle{JHEP}
\bibliography{OQS_bibliography}

\end{document}